\documentclass[11pt]{amsart}
\usepackage[foot]{amsaddr} 
\usepackage{graphicx,color} % Required for inserting images
\usepackage{amssymb}
\usepackage{amsthm}
\usepackage{enumerate}
\usepackage{subcaption}
\usepackage{xcolor}
\IfFileExists{ulem.sty}{\usepackage{ulem}}{}
\usepackage{float}
\usepackage{placeins}
\usepackage{needspace}
\usepackage{cancel}
\usepackage{tablefootnote} 
\usepackage{array}
\usepackage{booktabs}
\usepackage{tikz}
\usetikzlibrary{arrows.meta,positioning,calc,fit,backgrounds}
\definecolor{edgeblue}{RGB}{34,86,170}
\definecolor{edgered}{RGB}{190,45,45}
\definecolor{connector}{RGB}{115,65,150}
\definecolor{panelbg}{RGB}{247,248,251}
\definecolor{green}{RGB}{0,120,55} 

\tikzset{
  vertex/.style={circle,draw=black,fill=white,inner sep=1.8pt,minimum size=7pt,line width=.8pt},
  center/.style={circle,draw=black,fill=yellow!16,inner sep=2.0pt,minimum size=8pt,line width=.9pt},
  auxv/.style={circle,draw=connector,fill=connector!10,inner sep=2.0pt,minimum size=8pt,line width=.9pt},
  origedge/.style={draw=black,line width=1.55pt},
  origrep/.style={origedge},
  auxedge/.style={draw=connector,line width=1.55pt},
  loopedge/.style={draw=connector,line width=1.3pt},
  rededge/.style={draw=edgered,line width=1.55pt},
  blueedge/.style={draw=edgeblue,line width=1.55pt},
  redrep/.style={draw=edgered,line width=1.55pt},
  bluerep/.style={draw=edgeblue,line width=1.55pt},
  connedge/.style={draw=connector,densely dotted,line width=1.55pt},
  oldedge/.style={draw=gray!65,dashed,line width=1.0pt},
  panel/.style={draw=gray!45,rounded corners=3pt,fill=panelbg,inner sep=8pt},
  lbl/.style={font=\small},
  tiny/.style={font=\scriptsize,fill=white,inner sep=1.2pt}
}
\usepackage{enumitem}
\usepackage{xcolor}
\usepackage{caption}
\usepackage{microtype}
\usepackage{hyperref}

\hypersetup{colorlinks=true,linkcolor=blue!55!black,urlcolor=blue!55!black}

\usepackage{hyperref}
\usepackage{tikz}
\usetikzlibrary{arrows,calc,patterns}
\usepackage{diagbox}
\usepackage{fullpage}
\IfFileExists{bbm.sty}{\usepackage{bbm}}{}

\usepackage{mathtools}
\usepackage{lipsum}
\usepackage{verbatim}
\IfFileExists{algorithm2e.sty}{%
\usepackage[ruled,vlined,linesnumbered]{algorithm2e}
\DontPrintSemicolon
\SetKwInOut{Input}{Input}
\SetKwInOut{Output}{Output}}{
\floatstyle{ruled}
\newfloat{algorithm}{htbp}{loa}
\floatname{algorithm}{Algorithm}}
\IfFileExists{algorithmic.sty}{\usepackage{algorithmic}}{}

\DeclareMathOperator{\R}{\mathbb{R}}
\DeclareMathOperator{\odd}{odd}
\newcommand{\rep}[1]{\operatorname{rep}(#1)}

\def\R{{\mathbb R}}

\def\int{\textup{int}}

\newcommand{\ALG}{\textsc{StarMatch}}
\newcommand{\SubtourLP}{\textsc{SubtourLP}}
\newcommand{\MSopt}{\operatorname{MS}_{\mathrm{OPT}}}

\newcommand{\LPone}{\textsc{StarDeletionLP}}
\newcommand{\LPtwo}{\textsc{StarReconnectionLP}}

\newcommand{\Pone}{\mathcal P^{\mathrm{SDL}}}
\newcommand{\PhiOne}{\Phi^{\mathrm{SDL}}}

\makeatother
\usepackage[capitalize,nameinlink]{cleveref}
\crefname{equation}{}{equations}
\crefname{chapter}{Appendix}{chapters}
\crefname{item}{}{items}
\crefname{enumi}{}{}
\newtheorem{theorem}{Theorem}[section]
\newtheorem{lemma}[theorem]{Lemma}

\newtheorem{corollary}[theorem]{Corollary}

\newtheorem{proposition}[theorem]{Proposition}

\newtheorem{definition}[theorem]{Definition}

\hypersetup{
 pdftitle={Approximating graphic TSP by b-matchings},
 pdfsubject={Star removable systems, bounded T-joins, circulation comparisons, and tour bounds},
 bookmarksnumbered=true
}
\tikzset{
 s10v/.style={circle,draw=black,fill=white,minimum size=5.5mm,inner sep=1pt,font=\small,line width=.7pt},
 s10c/.style={s10v,fill=yellow!14},
 s10z/.style={draw=connector,rounded corners=1.3pt,fill=connector!9,minimum height=6mm,minimum width=8mm,inner sep=2pt,font=\small,line width=.8pt},
 s10t/.style={draw=edgeblue,line width=1.2pt},
 s10b/.style={draw=edgered,line width=1.2pt},
 s10con/.style={draw=connector,densely dotted,line width=1.5pt},
 s10shore/.style={draw=gray!70,dashed,fill=gray!7,rounded corners=5pt,line width=.7pt},
 s10lab/.style={font=\small,fill=white,inner sep=2pt,align=center},
 s10note/.style={font=\small,align=center},
 s10box/.style={draw=gray!60,rounded corners=3pt,fill=panelbg,inner sep=6pt,font=\small,align=center},
 s10arr/.style={-{Latex[length=2mm]},draw=black!65,line width=.8pt}
}

\usepackage{ulem}
\definecolor{revisionpink}{RGB}{200,45,120}
\colorlet{figurepink}{pink} % Preserve the original figure color.
\colorlet{pink}{revisionpink} % Readable \textcolor{pink} revision marks.

\begin{document}
\title[Approximating graphic TSP by $b$-matchings]{Approximating graphic TSP by $b$-matchings}
\date{}
\author{Swati Gupta}
\address{MIT Sloan School of Management.}
\email{swatig@mit.edu}
\author{Mehrdad Sohrabi}
\address{Department of Combinatorics and Optimization, Faculty of Mathematics, University of Waterloo, Ontario, Canada. Research of this author was supported in part by a Discovery Grant from NSERC.}
\email{mehrdad.sohrabi@uwaterloo.ca}

\author{Levent Tun\c{c}el}
\address{Department of Combinatorics and Optimization, Faculty of Mathematics, University of Waterloo, Ontario, Canada. Research of this author was supported in part by a Discovery Grant from NSERC.}
\email{levent.tuncel@uwaterloo.ca}

\begin{abstract}
We give a polynomial-time algorithm for graphic TSP that returns a tour of length at most
$\frac{89}{64}=\frac{7}{5}-\frac{3}{320}$ times the value of the subtour relaxation.
This improves the long-standing ratio $\frac75$ of Seb\H{o} and Vygen and shows that the
integrality gap of the subtour LP on graphic instances is at most
$\frac{89}{64}\approx 1.3906$. We optimize a non-linear objective function over an odd-join polytope, which uses (i) a new removable system of edges, and (ii) reconnection of subtours if the graph is disconnected. This approach gives a $\frac{13}{10}$-approximation algorithm for $4$-regular
$4$-vertex-connected graphs, improving the $\frac43$ of Newman, and a
$\frac43$-approximation for $k$-vertex-connected graphs of maximum degree at most $k+1$,
for every $k\ge 3$. For the general graphic TSP, we construct a feasible point in the odd-join polytope, using an optimal vertex from the Subtour linear programming relaxation of TSP, which helps us bound the approximation factor by 89/64.

\end{abstract}

\maketitle

\section{Introduction}
\label{sec:approach-generalization}

For more than a decade, the best approximation ratio known for the graphic Traveling Salesperson Problem (graphic TSP) has been the $7/5$ of Seb\H{o} and Vygen~\cite{SeboVygen}. We improve it to $7/5-3/320$.
In graphic TSP, the input is a connected
unweighted graph $G=(V,E)$. The distance between two vertices is the number of edges in a shortest path between these two vertices in $G$. Then the goal is to find a shortest Hamiltonian circuit with respect to these distances. Equivalently, we want a connected spanning
Eulerian multigraph using copies of edges of $G$, with length equal to
its number of edge copies.
{Write $n=|V|$, $m=|E|$, and let $\text{\SubtourLP}(G)$ denote the optimal value of the graphic subtour relaxation:}
\begin{equation}
    \label{eqn:SEP}
 {\text{\SubtourLP}(G)}:=\min\bigl\{x(E):x\in\mathbb R_{+}^{E},\quad
 x(\delta_G(U))\ge2\text{ for every }
 \varnothing\ne U\subsetneq V\bigr\}.
\end{equation}
Here, $\delta_G(U)$ is the set of edges with exactly one endpoint in $U$, and $x(F):=\sum_{e\in F}x_e$. The relaxation~\eqref{eqn:SEP} is the graphic form of the subtour relaxation of Dantzig, Fulkerson, and Johnson~\cite{DFJ54}, whose optimal objective value equals the Held--Karp
bound~\cite{HK70}. By the parsimonious property of Goemans and
Bertsimas~\cite{GoemansBertsimas93}, its optimal value coincides with that of
the subtour relaxation for the metric closure of $G$; see
also~\cite{CornuejolsFonluptNaddef} and \cite[Theorem~2.31]{TraubVygen}.

Christofides~\cite{Chr76} and, independently, Serdyukov~\cite{Ser78} gave a
$3/2$-approximation for metric TSP in the 1970s.
Wolsey~\cite{Wol80} 
and, independently, Cunningham~\cite{Cun86} showed that
this tour has length at most $3/2$ times the value of the subtour relaxation;
see also Shmoys and Williamson~\cite{SW90}. For metric TSP, both the approximation ratio and this bound on the integrality gap
were improved only recently, to $3/2-\varepsilon$ for some
$\varepsilon>10^{-36}$, by Karlin, Klein, and Oveis
Gharan~\cite{KKO21,KKO22,KKO23}; Gurvits, Klein, and Leake~\cite{GKL24}
increased $\varepsilon$ to about $10^{-34}$. On the hardness side, metric TSP
is NP-hard to approximate within $123/122$~\cite{KLS15}, and graphic TSP is
NP-hard to approximate within $1153/1152$ even on cubic graphs~\cite{KS15}.

For general graphic TSP, Oveis Gharan, Saberi, and
Singh~\cite{GSS11} first improved this guarantee to
$3/2-\varepsilon$ for an absolute constant $\varepsilon>0$.
M\"omke and Svensson~\cite{MomkeSvensson} then obtained a $1.461$-approximation by allowing edges to be deleted as well as doubled.
Mucha~\cite{Mucha} improved their circulation analysis to achieve the ratio $13/9$,
and Seb\H{o} and Vygen~\cite{SeboVygen} obtained $7/5$ using ear
decompositions and matroid intersection methods. The $4/3$ conjecture for the
subtour relaxation, stated by Goemans~\cite[Conjecture~1]{Goemans95},
remains open even for graphic instances. The conjecture (see also~\cite{Williamson90}) has been verified computationally
for all instances with at most ten vertices by Benoit and
Boyd~\cite{BenoitBoyd2008}, and the underlying enumeration of extreme points of
the subtour polytope has been extended
in~\cite{BoydElliottMagwood2010,CookHougardyPetrich2026}. Schalekamp,
Williamson, and van Zuylen~\cite{SWZ14} conjectured that the integrality gap
is attained at half-integral solutions; for metric instances with a
half-integral optimal solution, ratios below $3/2$ are
known~\cite{KKO20half,GLLMNS24,KT25}, and $4/3$ or other improved bounds are
known for several structured classes~\cite{BC11,BS21,HN19,JKW25}.

Stronger guarantees are known for restricted graph classes. For cubic
$3$-edge-connected graphs, Gamarnik, Lewenstein, and
Sviridenko~\cite{GLS05} were the first to improve on $3/2$, and Agarwal, Garg,
and Gupta~\cite{AGG18} obtained $4/3$. Boyd, Sitters, van der Ster, and
Stougie~\cite{BoydCubic} proved $4/3$ for bridgeless cubic graphs and $7/5$
for bridgeless subcubic graphs, and the algorithm of M\"omke and
Svensson~\cite{MomkeSvensson} gives $4/3$ for subcubic graphs. For
bridgeless cubic graphs the tour length was subsequently brought below
$4n/3$~\cite{CLS15,vZ18} and to $1.3n$~\cite{DL18}, and the approximation
ratio for cubic graphs was lowered to $9/7$~\cite{DKM17} and to
$5/4$~\cite{WYY23}; see also~\cite{KR16,vZ18} for cubic bipartite graphs. For
graphs of maximum degree four, Newman~\cite{Newman} proved $25/18$, and $4/3$
when the value of the subtour relaxation equals $n$. For connected $d$-regular
graphs, the randomized algorithms of Vishnoi~\cite{Vis12} and of Feige, Ravi,
and Singh~\cite{FRS14} find tours of length $(1+o(1))n$ as $d\to\infty$. The
bound of~\cite{FRS14} is $(1+O(1/\sqrt{d}))n$, and it extends to graphs whose minimum and maximum degrees are close.

Our approach is straightforward, and strengthens the removable-pairing framework of
M\"omke and Svensson~\cite{MomkeSvensson} in two ways:
(i) we generalize removable pairs to \emph{removable stars},
each consisting of edges incident to a common vertex.
These stars are edge-disjoint, and we may delete edges
from them simultaneously while preserving connectivity,
provided at least one edge remains in each star;
(ii) we optimize parity correction and deletions together over a degree-bounded $T$-join polytope (rather than relying on circulations) using a \LPone. We further propose a \LPtwo\  that allows the join to select an entire star with a cost of two, which then allows us to restore connectivity by adding two copies of one of its edges. The latter allows us to push our approximation bounds beyond what was previously known.

Our key results are as follows: 

\begin{enumerate}
\item We give an 
$89/64=1.390625$ approximation for Graphic TSP, by proposing a $b$-matching based algorithm \ALG. This is the first improvement on 7/5-approximation in 14 years \cite{SeboVygen}. Our analysis uses \textsc{Star} removable systems of edges, and improves on Mucha's
$13/9\approx1.444$ guarantee for the circulation
method of M\"{o}mke-Svensson~\cite{Mucha}. 
\item We also obtain stronger bounds for several
classes of highly connected graphs: 
\begin{itemize}
\item For 4-regular 4-vertex-connected graphs, we give the first 1.3-approximation, improving upon the previous best bound of 4/3 by Newman \cite{Newman}. 
\item For $k$-vertex-connected graphs of maximum degree at most $k+1$, $k\ge 3$, we
give a deterministic $4/3$-approximation. The previous best deterministic
guarantees were $25/18$ for $k=3$~\cite{Newman} (with stronger bounds when the
graph is cubic~\cite{DKM17,WYY23}) and $7/5$ for $k\ge 4$~\cite{SeboVygen}.
For large $k$, the randomized algorithm of Feige, Ravi, and
Singh~\cite[Theorem~5]{FRS14} achieves the ratio $1+O(1/\sqrt{k})$, because
these graphs have minimum degree at least $k$. 
\end{itemize}
\end{enumerate}

\noindent
We include Table \ref{tab:intro-results} to summarize these results. We next give an overview of our techniques. 

\begin{table}[t]
\small
\setlength{\tabcolsep}{5pt}
\renewcommand{\arraystretch}{1.15}
\caption{Approximation factors for simple graphs.}\label{tab:intro-results}
\begin{tabular}{@{}
  >{\raggedright\arraybackslash}p{.38\textwidth}
  >{\raggedright\arraybackslash}p{.25\textwidth}
  >{\raggedright\arraybackslash}p{.24\textwidth}@{}}
\toprule
Graph class & Previous {Best} & {Our Results}\\
\midrule
Connected simple graphs
  & $7/5 = 1.4$ \cite{SeboVygen}
  & $89/64 = 1.390625$\\
\addlinespace
$4$-regular, $4$-vertex-connected
  & $4/3 \approx 1.3333$ \cite{Newman}
  & $13/10 = 1.3$\\
\addlinespace
$3$-vertex-connected, $\Delta\le4$
  & $25/18 \approx 1.3889$ \cite{Newman}$^{\dagger}$
  & $4/3 \approx 1.3333$\\
\addlinespace
$k$-vertex-connected, $k\ge4$,\newline $\Delta\le k+1$
  & $7/5 = 1.4$ \cite{SeboVygen}$^{\star}$
  & $4/3 \approx 1.3333$\\
\bottomrule
\addlinespace[3pt]
\multicolumn{3}{@{}l@{}}{\footnotesize $^{\dagger}$\,A variant of the argument in~\cite{Newman} also gives $4/3$. For the cubic
graphs in this class, $5/4$ is known~\cite{WYY23}.}\\
\addlinespace[1pt]
\multicolumn{3}{@{}l@{}}{\footnotesize
$^{\star}$For large $k$, a randomized $(1+O(1/\sqrt{k}))$-approximation follows
from~\cite[Theorem~5]{FRS14}.}
\end{tabular}
\end{table}

\section{An Overview of Our  Techniques}
We first briefly discuss the generalization to \textsc{Star} removable systems, and then discuss the two linear programs we use for parity correction and connectivity, {and the projection of the latter will give our final formulation.} 

\subsection{Star Removable Systems} 
\begin{figure}[!t]
\centering
\begin{tikzpicture}[x=1cm,y=.62cm,font=\footnotesize,
 ucopy/.style={fill=figurepink!60},
 vcopy/.style={fill=blue!22},
 starone/.style={draw=orange!85!black},
 startwo/.style={draw=teal!85!black},
 starthree/.style={draw=violet!85!black},
 otheredge/.style={draw=black!55,line width=1.55pt}]

\begin{scope}[xshift=-3.25cm]
\node at (0,2.9) {(a) $G=(V,E):$ before splitting};
\node[vertex,label=right:{$p$}] (p) at (0,2.2) {};
\node[center,ucopy,label=below left:{$u$}] (u) at (0,1.5) {};
\node[center,vcopy,label=below:{$v$}] (v) at (0,.4) {};
\node[vertex,label=right:{$w_1$}] (w1) at (-1,-.8) {};
\node[vertex,label=left:{$w_2$}] (w2) at (1,-.8) {};
\node[vertex,label=below:{$x_1$}] (x1) at (-1,-2.65) {};
\node[vertex,label=below:{$x_2$}] (x2) at (1,-2.65) {};

\draw[otheredge] (p)--(u);
\draw[otheredge] (w1)--(x1);
\draw[otheredge] (w2)--(x2);

\draw[blueedge,starone] (u)--(v);
\draw[rededge,starone] (u)
  .. controls (-2.7,1.15) and (-2.7,-1.85) .. (x1);
\draw[rededge,starone] (u)
  .. controls (2.7,1.15) and (2.7,-1.85) .. (x2);

\draw[blueedge,startwo] (v)--(w1);
\draw[rededge,startwo] (v)
  .. controls (-1.75,.05) and (-1.75,-1.95) .. (x1);

\draw[blueedge,starthree] (v)--(w2);
\draw[rededge,starthree] (v)
  .. controls (1.75,.05) and (1.75,-1.95) .. (x2);
\end{scope}

\begin{scope}[xshift=3.25cm]
\node at (0,2.9) {(b) $H = (W,A):$ after splitting};
\node[vertex,label=right:{$p$}] (p) at (0,2.2) {};

\node[center,ucopy,label=left:{$u$}] (u) at (0,1.65) {};
\node[auxv,ucopy,label=below right:{$z_{S_1}$}] (z1) at (0,.7) {};

\node[center,vcopy,label=below:{$v$}] (v) at (0,-.2) {};
\node[auxv,vcopy,label=above:{$z_{S_2}$}] (z2) at (-1,-.95) {};
\node[auxv,vcopy,label=above:{$z_{S_3}$}] (z3) at (1,-.95) {};

\node[vertex,label=left:{$w_1$}] (w1) at (-1,-1.7) {};
\node[vertex,label=right:{$w_2$}] (w2) at (1,-1.7) {};
\node[vertex,label=below:{$x_1$}] (x1) at (-1,-2.65) {};
\node[vertex,label=below:{$x_2$}] (x2) at (1,-2.65) {};

\draw[otheredge] (p)--(u);
\draw[otheredge] (w1)--(x1);
\draw[otheredge] (w2)--(x2);

\draw[connedge,starone] (u)
  --node[left,inner sep=2pt] {$c_{S_1}$}(z1);
\draw[blueedge,starone] (z1)--(v);
\draw[rededge,starone] (z1)
  .. controls (-2.9,.4) and (-2.9,-2.25) .. (x1);
\draw[rededge,starone] (z1)
  .. controls (2.9,.4) and (2.9,-2.25) .. (x2);
\draw[loopedge,starone] (z1)
  .. controls (.9,1.65) and (.9,1.05) .. (z1);
\node[anchor=west,inner sep=1pt] at (.8,1.1) {$\ell_{S_1}$};

\draw[connedge,startwo] (v)--(z2);
\draw[blueedge,startwo] (z2)--(w1);
\draw[rededge,startwo] (z2)
  .. controls (-.2,-1.2) and (-.2,-2.25) .. (x1);
\draw[loopedge,startwo] (z2)
  .. controls (-1.85,-.4) and (-1.85,-1.5) .. (z2);

\draw[connedge,starthree] (v)--(z3);
\draw[blueedge,starthree] (z3)--(w2);
\draw[rededge,starthree] (z3)
  .. controls (.2,-1.2) and (.2,-2.25) .. (x2);
\draw[loopedge,starthree] (z3)
  .. controls (1.85,-.4) and (1.85,-1.5) .. (z3);
\end{scope}
\end{tikzpicture}
\caption{The stars $S_1=\{uv,ux_1,ux_2\}$,
$S_2=\{vw_1,vx_1\}$, and $S_3=\{vw_2,vx_2\}$
are orange, teal, and purple, respectively.
Splitting $u$ gives the two pink vertices $u,z_{S_1}$;
splitting the two stars centered at $v$ gives the three blue
vertices $v,z_{S_2},z_{S_3}$.
Each star keeps its edge color after splitting, and its
connector and loop use the same color.
Straight solid edges follow the DFS tree; curved edges
between distinct vertices correspond to back edges.}
\label{proc:star-split}
\end{figure}
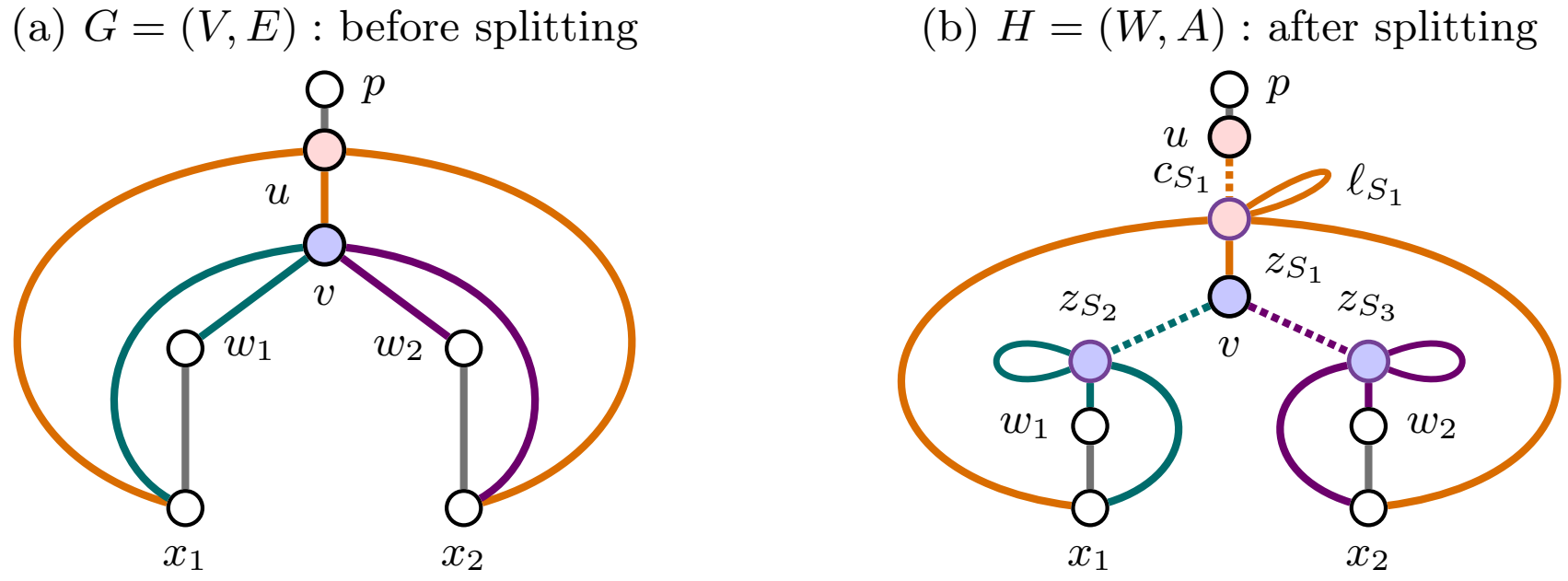

The removable-pairing method of M\"omke and Svensson specifies
edges that can be deleted while preserving connectivity, provided
at most one edge is deleted from each prescribed pair, that is defined using edges incident to the same vertex.
We allow larger collections of edges incident to a common vertex.
A \emph{\textsc{Star} removable system} is a family of edge-disjoint stars, so that deleting all but one edge from each star still keeps the graph connected. We show how to construct such a system using DFS trees. {See Figure \ref{proc:star-split}(a) for an example, where we depict 3 stars and one can check that removal of all but one edge from each star maintains connectivity.}

\subsubsection{A Parity-Correcting Formulation} 
A natural way to correct parity of each vertex, one can simply add an $\odd(G)$-join $J$, where $\odd(G)$ denotes the odd-degree
vertices of $G$, to the edge set $E$ of $G$. The resulting multigraph is connected and has even degrees. One can then delete both copies of each edge in a maximum
independent subset of $J$ in the cographic matroid, of size
$r^*(J)$, without disconnecting the graph or changing the
corrected parity of the vertices. Therefore, the optimum tour length is \begin{align}
\text{tsp}(G) &= \min_{\text{odd}(G)\text{-join } J} |E| + |J| - 2r^*(J).\label{eq:matroid-join}
\end{align}

This {\it matroid-join} formulation is a constrained non-monotone supermodular function minimization, and this is NP-hard to optimize. {Note that Christofides' algorithm \cite{Chr76} first approximates $r^*(J)$ by fixing a spanning tree $T^\prime$, and allowing the deletion of $J \cap (E\setminus T^\prime)$; here $|J \cap (E \setminus T^\prime)| \leq r^*(J)$. Approaches that use removable systems constrain the choice of $T$-joins (e.g., to take at most one edge from each removable pair), and then approximate $|J \cap R| \leq r^*(J)$, for example, the removable system $R_{MS}$ constructed using pairs of edges \cite{MomkeSvensson}. We do so by (i) giving a larger removable system $R^\prime \supseteq R_{MS}$, and bounding the degree of the T-join $J$ on every vertex (via the \LPone), in a way that deleting the removable edges $R'$ in the T-join allows for local connectivity at every vertex, while ensuring global connectivity, i.e., again $|J \cap R^\prime| \leq r^*(J)$. Since we impose these bounds only at individual vertices in an auxiliary graph (see Figure \ref{proc:star-split}(b)), the
degree-bounded join problem has an integral polytope formulation and any linear function can be optimized over these polytopes in polynomial
time~\cite[pp.564-568]{Schrijver2003}. We detail the construction in Appendix~\ref{subsec:framework-1}. We show the following:}

\begin{theorem}[\LPone]
Given a simple $k$-vertex-connected graph $G$, where $k\ge 3$
and with maximum degree at most $k+1$, the \LPone{} gives a tour of length at most $4n/3-2/3$.
\end{theorem}

As a comparison, we note that the algorithm of M\"omke and Svensson {(MS)}
\cite[Sections~3--4]{MomkeSvensson} uses a circulation to
choose a support graph and an MS removable pairing. On this
support, they construct a distribution of joins in which
each edge is selected with probability $1/3$. Mucha~\cite{Mucha} and Newman~\cite{Newman} use the subtour solution to improve the
circulation analysis; the subsequent join
distribution still has uniform marginals. Once the circulation
step removes edges, the subtour solution restricted to the
remaining graph need not satisfy the subtour cut constraints,
making it harder to use its individual edge values in the
parity correction. In our LPs, the Blossom cut inequalities give
a direct connection to the subtour cut constraints, so individual
subtour values can guide the join coordinates after the removable
system is fixed. We construct fractional points with improved
objective values, and integrality gives integral choices with
at most the same objective value.

\subsection{Directly Optimizing over T-joins} The \LPone{} does not allow the final tour to contain
a doubled edge belonging to a removable star, and dropping at least one edge in each star may be too restrictive over T-joins. {We next show how we can remove the restrictions on the T-join.} We introduce a slightly modified \LPtwo{}, which adds
self-loops to our auxiliary graph and assigns weight $-2$ to them (Figure \ref{proc:star-split}(b)). This modified auxiliary graph allows for the possibility that if the loop is not selected in the degree-bounded T-join, we will restore connectivity by adding two copies of one edge to connect the graph. {In the language of the matroid-join formulation \eqref{eq:matroid-join}, we are able to (i) optimize over {\it all} T-joins to find a candidate join $J$ in \LPtwo, (ii) choose $J'\subseteq J\cap R$ by omitting one edge from
each star fully contained in $J$. Then $E\setminus J'$ is
connected. After deleting $J\cap R$ and doubling $J\cap N$,
adding two copies of each edge in $(J\cap R)\setminus J'$
restores connectivity. This is the non-linear cost we are essentially able to bring into the formulation.  Therefore, we approximate $r^*(J)$ by $|J^\prime \cap R| \leq r^*(J)$. The parity arguments still work, as we indeed delete all edges in $J \cap R$, while charging 2 units to each tight star (where $J$ picks all incident edges of the star). This extra freedom of being able to optimize over all T-joins is crucial in our fractional constructions: we can use higher values on nonremovable edges when they help
satisfy the Blossom cut inequalities, while accounting for both
their added cost and the cost of reconnections. We construct feasible points in \LPtwo{} using subtour LP vertex $x^*$, and prove a bound on the approximation factor. This gives us the main result} of the paper.

\Needspace{4\baselineskip}
\begin{theorem}[General graphs]
\label{thm:intro-general}
{Let $G$ be a connected simple graph on $n$ vertices.}
We give a polynomial-time
algorithm \ALG\  based on the \LPtwo{} that returns a tour of length
\[
 \ell\le\frac{89}{64}{\text{\SubtourLP}(G)}
       =\left(\frac75-\frac3{320}\right){\text{\SubtourLP}(G)}.
\]
\end{theorem}
The proof of \Cref{thm:intro-general} is given in Appendix~\ref{sec:general-L}.
Higher connectivity allows further improvements.

\begin{samepage}
\begin{theorem}[Highly connected graphs]
\label{thm:intro-connected}
Tours with the following worst-case guarantees can be computed in polynomial time {using any DFS tree of the graph:} 
\begin{enumerate}[label=(\roman*),leftmargin=*]
\item If $G$ is simple, $4$-regular, and $4$-vertex-connected,
the \LPtwo{} gives a tour of length at most $13n/10-3/5$.
\item If $G$ is simple and $k$-vertex-connected, where $k\ge3$
and $\Delta(G)\le k+1$, the \LPone{} and the \LPtwo{}
both give a tour of length at most $4n/3-2/3$.
\end{enumerate}
\end{theorem}
\end{samepage}
The two parts of \Cref{thm:intro-connected} are proved in Appendices~\ref{sec:quartic-framework-2}
and~\ref{sec:k-connected-degree-bounded}, respectively. \\

We further note that every point feasible in the \LPone{} extends to feasible point in the \LPtwo{} with the same tour bound. For the \textsc{Star} removable system $\mathcal S$ constructed from $T$ for a given graph $G$, with subtour optimum ${\text{\SubtourLP}(G)}$, and optimum values $\text{\LPtwo}_{\mathrm{OPT}}$ and $\text{\LPone}_{\mathrm{OPT}}$ attained by our two LPs,  
we show the following theorem.  

\begin{theorem}[Comparison with the circulation method; see {Corollary~\ref{thm:framework-1-dominates-MS}} in
Appendix~\ref{sec:flow-to-framework-1}]
\label{thm:intro-circulation}
Let $G$ be a simple $2$-vertex-connected graph on $n$ vertices,
and let $T$ be a DFS tree rooted at a vertex $r$ with
$\deg_G(r)\ge3$. Let $\mathcal S$ be the \textsc{Star} removable system constructed
from $T$, as in Section~\ref{proc:systems}. Then
\[
 \begin{aligned}
 \operatorname{tsp}(G)
 &\le \text{\LPtwo}_{\mathrm{OPT}}(G,\mathcal S)\\
 &\le \min\left\{\text{\LPone}_{\mathrm{OPT}}(G,\mathcal S),
                  \min_f\MSopt(f)\right\}\\
 &\le \frac43n+{\frac23}\min_f C(f)-\frac23.
 \end{aligned}
\]
Here $f$ ranges over minimum-cost integral M\"omke--Svensson
circulations for $T$; $\MSopt(f)$ is the minimum
tour length obtained by optimizing the parity correction
for the support and fixed MS removable pairing produced from $f$
\cite[proof of Theorem~12.3]{TraubVygen}, and $C(f)$ is its
circulation cost \cite[Lemma~12.8]{TraubVygen}.
\end{theorem}
Thus, for the same DFS tree, both of our LPs give a tour bound at
least as strong as the optimized circulation bound, including
the guarantees obtained by Mucha~\cite{Mucha} and Newman~\cite{Newman}. In our algorithm \ALG, we will use the same DFS tree as \cite{MomkeSvensson}, and obtain a better guarantee using the \LPtwo. In fact, in Section~\ref{subsec:four-vertex-separation}, we give a graph
$G$ and a tree $T$ obtained from an optimal extreme subtour
solution by greedy DFS (as in \hyperref[proc:algorithm]{\ALG}
and the algorithms of M\"omke--Svensson and
Mucha~\cite{MomkeSvensson,Mucha}).
For the \textsc{Star} removable system $\mathcal S$ constructed from $T$, all the inequalities in Section~\ref{subsec:four-vertex-separation} are strict.
We obtain
\[
 \begin{aligned}
 {\text{\SubtourLP}(G)}=26
 &<\text{\LPtwo}_{\mathrm{OPT}}(G,\mathcal S)=27\\
 &<\text{\LPone}_{\mathrm{OPT}}(G,\mathcal S)=28\\
 &<\MSopt(f^*)=29 <\frac43n+\frac23C(f^*)-\frac23=\frac{92}{3},
 \end{aligned}
\]
where $f^*$ is an optimal integral solution of the circulation, shown in Section~\ref{subsec:four-vertex-separation}. Different choices of $f^*$ can lead to different bounds.

\subsection{Bounding the \LPtwo{} using the Subtour Solution}

To prove our approximation guarantee for general graphic TSP,
we provide a map from the subtour LP solution $x^*$ to a point feasible in the \LPtwo. Our construction allows removable edges
belonging to the DFS tree to be selected with higher probability than in the uniform join distribution of M\"omke and Svensson.  We use the subtour cut constraints to prove that this point satisfies the Blossom inequalities. We combine this point with a second feasible point obtained directly from the subtour solution. Using the sparsity of an optimal extreme subtour support and the structure of the DFS tree, we bound the resulting tour length and obtain our approximation guarantee.

\subsection{Outline}
Section~\ref{proc:systems} constructs \textsc{Star} removable
systems from DFS trees. Section~\ref{proc:repair-lp} presents
the \LPtwo{} and explains how its solutions give tours.
Section~\ref{proc:StarMatch} describes \ALG{}, and
Section~\ref{proc:value-n} develops its analysis when
$\SubtourLP(G)=n$. Section~\ref{proc:quartic} gives the
$13/10$ bound for $4$-regular, $4$-vertex-connected graphs.
The appendices provide detailed proofs, including the
analysis for arbitrary subtour value in
Appendix~\ref{sec:general-L}, the comparison with the
M\"omke--Svensson method, and the $4/3$ bound for
$k$-vertex-connected graphs of maximum degree at most $k+1$. 
\section{\textsc{Star} removable systems from a DFS tree}
\label{proc:systems}

Let $G=(V,E)$ be a connected simple graph, with $n=|V|$ and
$m=|E|$. A \emph{removable star} centered at $u$ is a nonempty
set $S$ of edges incident to $u$ such that $G-(S\setminus\{e\})$
is connected for every $e\in S$. A \emph{\textsc{Star} removable system}
$\mathcal S$ is a family of pairwise edge-disjoint removable
stars with the following property: for every tuple
$(f_S)_{S\in\mathcal S}\in\prod_{S\in\mathcal S}S$, obtained by
choosing one edge $f_S$ from each star $S$,
\[
 G-\bigcup_{S\in\mathcal S}(S\setminus\{f_S\})
 \quad\text{is connected}.
\]
Write $R:=\bigcup_{S\in\mathcal S}S$, $N:=E\setminus R$, and
$k_S:=|S|$. Thus, we may delete any subset of $R$ that leaves
at least one edge of each star; every edge of $N$ is retained. We call the edges in $R$ \emph{removable} and those in $N$
\emph{nonremovable}, with respect to $\mathcal S$.

We can obtain such a \textsc{Star} removable system from an arbitrary DFS tree as follows.
Fix a DFS tree $T$ rooted at $r$, and let $T_v$ denote the
subtree rooted at $v$. Every edge $wu$ outside $T$ joins a vertex {$w$}
to a proper ancestor $u$ and is called a \emph{back edge}. For a
tree edge $uv$, with parent $u$ and child $v$, consider the set of back edges from $T_v$ to $u$:
\[
 I(u,v) := \{{wu}\in E\setminus E(T):w\in V(T_v)\}.
\] 
If $I(u,v)\ne\varnothing$, form the star
$S(u,v):=\{uv\}\cup I(u,v)$ centered at $u$. This includes
stars centered at the root, and different children of a vertex
may give different stars.
Write $\mathcal S$ for the family of all these stars. See Figure \ref{fig:DFS-star} for an example.

\begin{lemma}[DFS \textsc{Star} removable system; Appendix~\ref{sec:DFS-removable-stars}]
\label{proc:dfs-system}
Given any DFS tree $T$ of $G=(V,E)$, the stars
$\mathcal S=\{S(u,v):uv\in E(T),\ I(u,v)\ne\varnothing\}$ form a \textsc{Star} removable system with $R = \bigcup_{S \in \mathcal{S}} S$. Moreover, keeping every edge
of $N = E\setminus R$ and any one edge of each star gives a spanning tree.
\end{lemma}

Every back edge belongs to exactly one star, and each star contains exactly one tree edge. Consequently,
\begin{equation}
\label{proc:dfs-counts}
 |N|+|\mathcal S|=n-1,
 \qquad
 m-\sum_{S\in\mathcal S}(k_S-1)=n-1.
\end{equation}

\section{The \LPtwo}
\label{proc:repair-lp}

For an $\odd(G)$-join $J$, doubling the edges of $J$ makes
every degree even. We would then like to delete both copies
of each selected removable edge. {If $J$ contains an entire
star, we charge the star with a cost of two units for reconnecting it back}. The
\LPtwo{} includes this cost of two and chooses the parity
correction to minimize the resulting tour bound.

For every star $S$ centered at $u$, introduce a vertex $z_S$
and replace each edge $uw\in S$ by the edge $z_Sw$.
Add the connector $c_S=uz_S$ and {a loop $\ell_S$ at $z_S$}. Keep every edge of $N$ with its endpoints unchanged, and let
$H=(W,A)$ be the resulting graph. { See Figure~\ref{proc:star-split}}. 

Define the injective map {$\operatorname{\bf rep}\colon E(G)\to E(H)$ that simply allows us to map edges in the original graph $G$ to the auxiliary graph $H$, i.e., every star edge $uw \in S$ is simply mapped to $z_S w$ in $H$, and nonremovable edges in $e\in N$ map to their counterpart. Formally, } 
\[
  \rep{e}:=
 \begin{cases}
  z_Sw,&e=uw\in S,\quad S\text{ is centered at }u,\\[3pt]
  e , &e\in N.
 \end{cases}
\]
{Let $S_H$, $R_H$,
and $N_H$ be the corresponding edges in the $H$ for $S$, $R$, and $N$.}

The feasible region of the \LPtwo{} is the polytope
${P_{\mathcal R}}(G,\mathcal S)$ defined by
\begin{equation}
\label{proc:repair-polytope}
\begin{aligned}
 y(S_H)+y_{c_S}+2y_{\ell_S}
 &\le k_S+1
 &&(S\in\mathcal S)
 &&{\text{(degree bounds)}},\\[4pt]
 y(C\setminus F)+|F|-y(F)
 &\ge1
 &&\left(
 \begin{gathered}
 U\subseteq W,\ C=\delta_H(U),\\[-2pt]
 F\subseteq C,\ |C\setminus F|\text{ odd}
 \end{gathered}
 \right)
 &&{\text{(Blossom cut inequalities)}},\\[4pt]
 0\le y_a&\le1
 &&(a\in E(H))
 &&{\text{(box constraints)}}.
\end{aligned}
\end{equation}
{When $G$ and $\mathcal S$ are fixed, we abbreviate this region
as $P_{\mathcal R}$}. The box constraints and Blossom cut inequalities define the
$\mathrm{odd}(H)$-join polytope, by the theorem of Edmonds and
Johnson~\cite{EJ73}; see~\cite[Chapter~29]{Schrijver2003}. {The additional degree bounds keep the polytope integral\footnote{The bound $k_S+1$ has the same parity as
$\deg_H(z_S)=k_S+3$, which is important for the integrality of the polytope.}, which permits polynomial-time
optimization via capacitated b-matching~\cite{Schrijver2003}. Note that loops} do not cross cuts and contribute twice to degrees.
For an integral point, selecting $\ell_S$ gives
\[
 y(S_H)+y_{c_S}\le k_S-1,
\]
so the join leaves an edge of $S$ unselected and no reconnection
is needed at this star. {A reconnection can therefore occur only
when $y_{\ell_S}=0$, and in this case, its cost is exactly 2 units} \footnote{{Note that for a feasible solution to $P_{\mathcal R}$, $y_{\ell_S}=0$ induces a cost of at most $2(1-y_{\ell_S})$.}}. Selecting a removable edge saves one copy, while selecting
a nonremovable edge adds one copy. Together with the reconnection
costs, this gives the following upper bound on the tour length 
\[
\Phi_H(y):= m-y(R_H)+y(N_H)+2\sum_{S\in\mathcal S}(1-y_{\ell_S}), 
\]
which can be directly obtained by optimizing the \LPtwo. {Note that in fact we can simplify \LPtwo{} by factoring out the loop variables $y_{\ell_S}$ and simply optimize the following (projected) system using $z \in \mathbb{R}^{E(H)\setminus \{\ell_S:S\in \mathcal S\}}$:
\begin{align}
\min_{y \in P_\mathcal R} \Phi_{H}(y) &= \min_{y \in P_\mathcal R} m-y(R_H)+y(N_H)+2\sum_{S\in\mathcal S}(1-y_{\ell_S}) \\
&= \min_{z\in \text{odd}(H^\circ)\text{-join polytope}} n-1+z(N_H)+
 \sum_{S\in\mathcal S}\max\{z_{c_S},\,k_S-1-z(S_H)\}\label{proc:join-cost}\\
 &=: \min_{z\in \odd (H^\circ)\text{-join polytope}} \Phi_{G, {\mathcal S}}(z),   
\end{align}}
where $H^\circ$ is the graph $H$ without the self-loops. {We will use $\mathcal P^\circ$ to denote the $\odd (H^\circ)$\text{-join polytope}.}

\begin{theorem}\label{proc:repair-bound}
{We can compute a tour of $G$
with length at most
$\min_{z\in\mathcal P^\circ}\Phi_{G,\mathcal S}(z)$ in polynomial time.
In particular,
\[
 \operatorname{tsp}(G)\le\Phi_{G,\mathcal S}(z)
 \qquad\text{for every }z\in\mathcal P^\circ.
\]}
\end{theorem}

{The proof is given in
Appendix~\ref{sec:framework-2}.}
{The only thing that we need to do is to bound {$\min_{z\in\mathcal P^\circ}\Phi_{G,\mathcal S}(z)$} with respect to the cost of the subtour LP.} 

\section{Graphic TSP using the \ALG{} Algorithm} \label{proc:StarMatch}
{We now discuss our main algorithm \ALG{}: Given a graph $G=(V,E)$, we compute an optimal extreme
point $x^*$ of the graphic subtour relaxation
\eqref{eqn:SEP}. This can be done in polynomial time by the ellipsoid method, because the
constraints of~(1) can be separated in polynomial time by minimum-cut
computations~\cite{GLS81}. It is known that the support $G^\prime$ of $x^*$ has at most $2n-3$ edges, due to laminarity of the tight cuts \cite{CornuejolsFonluptNaddef}. By the standard block reduction
\cite[Proposition~12.1]{TraubVygen}, it suffices to consider extreme points with support $2$-vertex-connected. Next, as in Newman \cite{Newman}, we also construct a subtour feasible point $x \in [0,1]^{E}$ that has the same cost as $x^*$ and the same support. This is our starting point, which we'll refer to as the box-constrained \SubtourLP{} solution.}

{Let $G$ be the support of $x$, and suppose $G$ is not a cycle (otherwise done). Let $r$ be any vertex of $G$ with degree at least 3. Construct a greedy depth-first search tree $T$, rooted at $r$, {i.e, visiting each new neighbor using largest $x_e$ edge possible. Construct the \textsc{Star}
removable system $\mathcal S$ and the corresponding auxiliary
graph $H$ as in Sections~\ref{proc:systems} and~\ref{proc:repair-lp}. Solve \LPtwo{} to obtain an optimal $\odd(H)$-join $J$. Let $J'=\operatorname{rep}^{-1}(J)$ be the projection of $J$ onto
$G$, obtained by contracting the connector edges. Let $Z$ be the Eulerian multigraph obtained from $G$ by deleting the edges in $J'\cap R$ and
adding a second copy of each edge in $J'\cap N$. Finally,
reconnect the components of $Z$ using doubled edges to obtain
a tour. See Algorithm \hyperref[proc:algorithm]{\ALG} for a complete description.} }

{To present the main ideas, we describe the construction
under the assumption $\text{\SubtourLP}(G)=n$.
Appendix~\ref{sec:general-L} proves the feasibility and cost
estimates for arbitrary subtour value, and
Proposition~\ref{prop:value-n-specialization} gives their
specialization to $\SubtourLP(G)=n$. The general analysis gives an
approximation ratio of at most $89/64$ for graphic TSP.}

\begin{algorithm}[!t]
\small\raggedright
\caption{\ALG}
\label{proc:algorithm}
\Input{A simple graph $G$ with $|V(G)|\ge3$, an optimal extreme point $x^*$ in \SubtourLP{} relaxation \eqref{eqn:SEP}, whose support is $2$-vertex-connected.}
\Output{A tour of $G$.}
\BlankLine
Find $x \in \{\SubtourLP(G) \mid x(E)=x^*(E), 0 \leq x_e \le 1 \text{ for } e\in E\}$ on the same support as $x^*$ and let $G =(V,E) $ be the \text{supp}$(x)$\;
\textbf{If} {$G$ is a cycle}, \textbf{return} $G$\;
Let root $r$ be any vertex of degree at least three in $V$\;
$T =$ Greedy DFS tree rooted at $r$, constructed by visiting each new neighbor using largest $x_e$ edge\;Let $\mathcal S$ = \textsc{Star} removable system from $T$, let $R:=\bigcup_{S\in\mathcal S}S$ and $N:=E\setminus R$ \tcp*[r]{see Section ~\ref{proc:systems}}
$J =$ \LPtwo{}$(G,\mathcal S)$, \tcp*[r]{$J$ is an odd$(H)$-join, $H$ is constructed from $G,\mathcal S$}
$Z = E \cup (J^\prime \cap N) \setminus (J^\prime \cap R)$, where $J^\prime = \operatorname{rep}^{-1}(J)$ \tcp*[r]{Note $Z$ is Eulerian}
Let $\mathcal T$ be the resulting tour obtained by reconnecting $Z$ using doubled edges\footnotemark\;
\Return{$\mathcal T$}\;
\end{algorithm}
\addtocounter{footnote}{-1}
\addtocounter{footnote}{1}
\footnotetext{One can improve the tour construction by first
adding a second copy of every edge of $J'$ to $G$, then
deleting both copies of every edge in a maximal independent
set of the cographic matroid of $G$ restricted to $J'$.
This preserves connectivity and even degrees and gives a
tour no longer than the one obtained above.}

\section{The case \texorpdfstring{${\text{\SubtourLP}(G)}=n$}{SubtourLP(G)=n}}
\label{proc:value-n}

{Let $x$ be the box-constrained \SubtourLP{} as defined in {Section \ref{proc:StarMatch}}}. Let $G=(V,E)$ 
$(|V|=n, |E|=m)$ be its support, and recall that $x \in [0,1]^{E}$ and $x(E)=x^*(E)=\text{\SubtourLP}(G)$. Although $x$ need not be extreme, the extremality of $x^*$
gives $m\le2n-3$ \cite{CornuejolsFonluptNaddef}. Let $T$ be the DFS tree rooted at $r$ constructed by \ALG{},
and let $\mathcal S$ be the \textsc{Star} removable system
constructed from $T$ as in Section~\ref{proc:systems}.
{As noted earlier, to bound the length of the tour returned by
\ALG{}, it suffices to construct a feasible point
$Y\in\mathcal P^\circ$ and bound $\Phi_{G,\mathcal S}(Y)$ in terms
of the cost of the subtour solution $x$. We construct $Y$ as a convex combination of two feasible
points, $Q^{(1)}$ and $Q^{(2)}$. Compared with $Q^{(1)}$,
the point $Q^{(2)}$ assigns larger values to the
representatives of back edges, which helps reduce the
second term in each star's maximum in
\eqref{proc:join-cost}. Its values on nonremovable edges
and connectors may, however, be larger. We choose the
convex combination to control both terms in each maximum. \\
We obtain $Q^{(1)}$ from a parametric construction $Q(d)$,
where $d\in\mathbb R_+^{\mathcal S}$. Each star
$S\in\mathcal S$ contains a unique DFS tree edge, and $d_S$
determines the increase in its weight. This flexibility is
one of the ingredients behind our improvement over the
M\"omke--Svensson bound: their matching construction uses
uniform $1/3$ marginals, {whereas our extension allows the weight on the removable tree
edge of each star $S$ to increase from $1/3$ to
$1/3+\frac23d_S$}. We choose a suitable
vector $\bar d$ and set $Q^{(1)}:=Q(\bar d)$.
The point $Q^{(2)}$ is obtained directly from $x$, by lifting the point with weight
$x_e/2$ on each tree edge and $1-x_e/2$ on each back edge to $\mathcal{P}^\circ$.
We then use the convex combination
\[
    Y:=\frac34 Q^{(1)}+\frac14 Q^{(2)}
    \in\mathcal P^\circ.
\]} 
{$Q^{(1)}$ tends to incur a higher loss on chains of certain ({\it positive stars}, defined in the next section) stars, that we can balance using $Q^{(2)}$.} We next discuss the construction of these points.

\subsection{The initial point and its excess}
\label{proc:value-n-points}

{Let $S_r$ denote the unique star centered at the root $r$.}
Define the initial point $Q^{(0)}\in\R_+^{E(H^\circ)}$ {by using box-constrained \SubtourLP{}  point $x$ in the following way}:
\[
 Q^{(0)}_a :=
 \begin{cases}
  1-\dfrac23x_b,
     &a=\rep{b},\quad b\notin E(T) \text{ {(non-tree edges)}},\\[5pt]
  1,&a=c_{S_r} \text{ {(connector of the root)}},\\[3pt]
  \dfrac13,&\text{otherwise}.
 \end{cases}
\]
{It is easy to check that the point $Q^{(0)}$ belongs to the odd-join polytope $\mathcal P^\circ$, by checking the Blossom inequalities for an arbitrary cut (see Proposition~\ref{prop:value-n-specialization}, applied with
$d=0$). } 

{For a star $S=S(u,v)$, let $a_S:=x(I(u,v))$ be the total
$x$-value of its back edges {from subtree under $v$ to $u$}. For each nonroot star $S$,
its contribution to {objective in} \eqref{proc:join-cost} is
\[
 \max\{Q^{(0)}_{c_S},\,|S|-1-Q^{(0)}(S_H)\}
 =\frac13+\frac23(a_S-1)_+,
\]
where $(z)_+:=\max\{z,0\}$.
Define the \emph{excess} $e_S:=(a_S-1)_+$ and call $S$
\emph{positive} when $e_S>0$. The root star contributes one,
so
\begin{equation}\label{proc:initial-cost}
 \Phi_{G,\mathcal S}(Q^{(0)})
 =\frac43n-\frac23+\frac23\sum_{S\ne S_r}e_S.
\end{equation}
The excess therefore measures the loss above $4n/3-2/3$, and note that this function, for the fixed DFS tree, depends only on $x$.
We reduce this loss by increasing the coordinates of the
removable tree edges, thereby decreasing the corresponding
terms $|S|-1-Y(S_H)$ in the objective.} {Note that subcubic graphs have excess 0 at every nonroot star, since at most one back edge enters any nonroot vertex. We can therefore certify a tour of length at most $4n/3-2/3$ for these graphs using $Q^{(0)}$.}

\subsection{Construction of $Q(d)$}
\label{proc:increases}

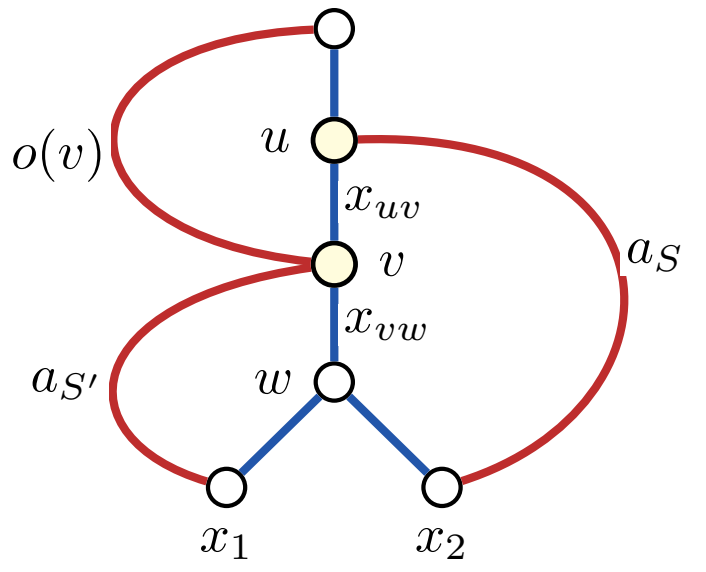
\begin{figure}[!t]
\centering
\begin{tikzpicture}[x=.95cm,y=.82cm,font=\small,
 nlab/.style={font=\small,fill=white,inner sep=1pt}]
\node[vertex] (h) at (0,2.25) {};
\node[center,label=left:{$u$}] (u) at (0,1.35) {};
\node[center,label=right:{$v$}] (v) at (0,.35) {};
\node[vertex,label=left:{$w$}] (w) at (0,-.6) {};
\node[vertex,label=below:{$x_1$}] (x1) at (-.75,-1.45) {};
\node[vertex,label=below:{$x_2$}] (x2) at (.75,-1.45) {};
\draw[blueedge] (h)--(u);
\draw[blueedge] (u)--node[nlab,right] {$x_{uv}$}(v);
\draw[blueedge] (v)--node[nlab,right] {$x_{vw}$}(w);
\draw[blueedge] (w)--(x1) (w)--(x2);
\draw[rededge] (v) .. controls (-2,.6) and (-2,2.15) ..
 node[nlab,pos=.45,left] {$o(v)$}(h);
\draw[rededge] (u) .. controls (2.5,1.45) and (2.5,-.8) ..
 node[nlab,pos=.45,right] {$a_S$}(x2);
\draw[rededge] (v) .. controls (-1.8,.05) and (-1.9,-1.05) ..
 node[nlab,pos=.55,left] {$a_{S'}$}(x1);
\node[anchor=north west,align=left,font=\footnotesize] at (3.5,2.3)
 {$S=S(u,v),\qquad S'=S(v,w)$\\[7pt]
  $a_S$: back-edge weight ending at $u$ from $T_v$\\[4pt]
  $a_{S'}$: back-edge weight ending at $v$ from $T_w$\\[4pt]
  $o(v)$: back-edge weight from $v$ to its ancestors};
\end{tikzpicture}
\caption{\footnotesize Quantities for consecutive stars $S=S(u,v)$ and
$S'=S(v,w)$ when $w$ is the only DFS child of $v$.
Blue edges belong to the greedy DFS tree $T$, and red curves
represent groups of back edges. Labels give edge weights or
total back edge weights under $x$ (e.g., $o(v)$ is the sum of the $x$-values of all back edges going between $v$ and its proper ancestors).} \label{proc:star-parameters}
\end{figure}

{Call a vector $d\in\R_+^{\mathcal S}$ \emph{admissible} if
$d_{S_r}=0$ and $d_S\le e_S$ for every nonroot star $S$.
For each vertex $v$, let $o(v)$ be the total $x$ value of
the back edges from $v$ to its proper ancestors in the box-constrained \SubtourLP{} point.
Figure~\ref{proc:star-parameters} illustrates the relevant
quantities for two consecutive stars. Our construction of $Q(d)$ will depend on the presence of consecutive stars.}

For $d\in\R_+^{\mathcal S}$, define
$Q(d)\in\R_+^{E(H^\circ)}$ as follows.
For each $e\in E(G)$, set
\begin{equation}\label{proc:increase-choice}
 Q(d)_{\rep{e}} :=
 \begin{cases}
  \dfrac13+\dfrac23d_S,
     &e=uv,\quad S=S(u,v)\ne S_r,\\[5pt]
  1-\dfrac23x_e,&e\notin E(T),\\[5pt]
  \dfrac13,&\text{otherwise}.
 \end{cases}
\end{equation}
For each star $S$, we further need to define the value on the connectors. We set these as: 
\begin{equation}\label{proc:following-connector}
 Q(d)_{c_{S}} :=
 \begin{cases}
  1,&S=S_r,\\[5pt]
  \dfrac13+\dfrac23(d_{S^{\mathrm{par}}}-o(v))_+,
    &\begin{array}{@{}l@{}}
       S=S(v,w),\quad v\text{ has exactly one DFS child }w,\\
       \text{and }S^{\mathrm{par}}=S(\mathrm{par}(v),v)\ne S_r\text{ exists},
      \end{array}\\[6pt]
  \dfrac13,&\text{otherwise}.
 \end{cases}
\end{equation}

By checking Blossom inequalities, one can verify that for every admissible $d \in \mathbb{R}_+^{\mathcal S}$, $Q(d)$ is in ${\mathcal P}^{\circ}$ (see Appendix~\ref{sec:general-L}, Proposition~\ref{prop:value-n-specialization}). Further, for an admissible vector $d$, the tour bound can be reduced to: 
\begin{equation}\label{proc:increased-cost}
 \Phi_{G,\mathcal S}(Q(d))
 =\frac43n-\frac23+
 \sum_{S\ne S_r}\max\left\{
     Q(d)_{c_S}-\frac13,\,
     \frac23(e_S-d_S)\right\}.
\end{equation}
{Note that for a nonroot star $S=S(u,v)$, increasing $d_S$ reduces
the term $e_S-d_S$, which helps the objective $\Phi_{G, \mathcal{S}}(Q(d))$. However, if there is a consecutive star $S^\prime$ centered at $v$ (see Figure~\ref{proc:star-parameters}) and if $v$ has exactly one child, then choosing $d_S>o(v)$ increases
the connector coordinate $Q(d)_{c_{S^\prime}}$. Therefore, increasing $d_S$ on stars that belong to such chains of consecutive stars may not have a monotone impact on the objective. This is why we construct a specific $d$ such that we can bound both the terms in the sum-of-max. On the other hand, if there is no consecutive chain of stars, then indeed increasing $d_S$ up to $e_S$ will reduce the objective.}

\subsection{Construction of $Q^{(2)}$}
\label{proc:mixing}
We construct a point
$Q^{(2)}\in\mathcal P^\circ$ with
\[
 Q^{(2)}_{\rep{e}} :=
 \begin{cases}
  \dfrac{x_e}{2},&e\in E(T),\\[5pt]
  1-\dfrac{x_e}{2},&e\notin E(T),
 \end{cases}
\]
and show that there exist connector values satisfying, for every nonroot star
$S=S(u,v)$,
\begin{equation}\label{proc:connector-bounds}
 Q^{(2)}_{c_S}\le
 \begin{cases}
  \dfrac{x_{uv}+a_S}{2},
     &\text{$u$ has odd degree in $T$},\\[6pt]
  \min\left\{\dfrac{x_{uv}+a_S}{2},
              1-\dfrac{x_{uv}+a_S}{2}\right\},
     &\text{$u$ has even degree in $T$}.
 \end{cases}
\end{equation}
The proof idea is to show that every integral $\odd(G)$-join
can be extended to an integral $\odd(H^\circ)$-join by choosing
$0/1$ values for the connectors. Taking convex combinations
then lifts every point in the $\odd(G)$-join polytope to a
point in $\mathcal P^\circ$, with each edge of $G$ and its
image receiving the same value
(see Lemma~\ref{s10n:join-lifting}).
The root connector has value one, and the bounds on the
remaining connectors follow from local Blossom inequalities;
see Appendix~\ref{s12pair:half-subtour}.
\subsection{Construction of the Final Certificate $Y$}
Now we take a convex combination of $Q^{(1)}= Q(d)$ for a specific $d$ and $Q^{(2)}$, to obtain a feasible point in $\mathcal{P}^\circ$, which will help us bound the approximation factor for graphic TSP. For each star $S=S(u,v)$, write $t_S:=|I(u,v)|$ for its
number of back edges. {Define $\bar d\in\R_+^{\mathcal S}$ as follows}\footnote{The Lean-verified analysis uses a different choice
of $\bar d$ to obtain stronger bounds.}:
\[
 \bar d_S:=
 \begin{cases}
  0,&S=S_r,\\[4pt]
  \left(e_S-\dfrac{x_{uv}}{2}-\dfrac{t_S}{32}\right)_+,
       &S\ne S_r,
 \end{cases}
\]
and set $Q^{(1)}:=Q(\bar d)$. We use the convex combination
\begin{equation}\label{proc:mixture}
 Y:=\frac34Q^{(1)}+\frac14Q^{(2)}.
\end{equation}
The vector $\bar d$ is admissible, so
Proposition~\ref{prop:value-n-specialization} gives
$Q^{(1)}\in\mathcal P^\circ$, and hence $Y\in\mathcal P^\circ$
by convexity. This choice of $\bar d$ yields the following
bound on each nonroot star. 

\begin{lemma}[{Proposition~\ref{prop:value-n-specialization}}]
\label{proc:local-star-bound}
Every nonroot star $S=S(u,v)$ satisfies
\[
 \max\{Y_{c_S},\,t_S-Y(S_H)\}
 \le\frac14+\frac{x_{uv}+a_S}{8}+\frac{t_S}{64}.
\]
\end{lemma}

{The proof of the above lemma is algebraic, and we include in Appendix~\ref{sec:general-L}. We next use the choice of box-constrained \SubtourLP{} point $x$ and the size of the support of optimal \SubtourLP{} $x^*$ to claim that $\sum_{S\in \mathcal S\setminus \{S_r\} } t_S \leq n-4$. 

By applying Lemma~\ref{proc:local-star-bound},
we obtain
\[
\begin{aligned}
 \Phi_{G,\mathcal S}(Y)
 &=n-1+Y(N_H)
   +\sum_{S\in\mathcal S}
      \max\{Y_{c_S},\,k_S-1-Y(S_H)\}\\[6pt] 
 &\le n+
   \underbrace{
     Y(N_H)+\sum_{S\ne S_r}
       \left(\frac14+\frac{x_{uv}+a_S}{8}\right)
   }_{=\,\frac38(n-2)}
   +\frac1{64}
    \underbrace{\sum_{S\ne S_r}t_S}_{\le\,n-4}\\[10pt]
 &\le n+\frac38(n-2)+\frac{n-4}{64}
  =\frac{89n-52}{64}.
\end{aligned}
\]}

{The details are given in Appendix~\ref{sec:general-L},
with the specialization to $L=n$ in
Proposition~\ref{prop:value-n-specialization}.}
Since $Y\in\mathcal P^\circ$, \ALG{} returns a tour
satisfying this bound. 

{For arbitrary subtour values ($\SubtourLP(G)>n$),
Appendix~\ref{sec:general-L} imposes the additional restriction
$d_S\le1-x_{uv}$ for $S=S(u,v)$ and chooses the connector
coordinates of $Q(d)$ using minimum cuts.
The construction of $Q^{(2)}$ remains unchanged.
A suitable convex combination of $Q(d)$ and $Q^{(2)}$
certifies a tour of length at most
\[
 \frac{89}{64}n
 +\frac98\bigl(\text{\SubtourLP}(G)-n\bigr)
 -\frac{13}{16}
 \le\frac{89}{64}\text{\SubtourLP}(G),
\]
obtained by \ALG{},
which proves
Theorem~\ref{thm:intro-general}.}

\section{\texorpdfstring{$4$}{4}-regular, \texorpdfstring{$4$}{4}-connected graphs}
\label{proc:quartic}

{Let $G=(V,E)$ be a simple, $4$-regular, $4$-vertex-connected
graph, and let $n:=|V|$. Fix any DFS tree $T$, rooted at $r$, and let
$\mathcal S$ be the \textsc{Star} removable system constructed
from $T$ as in Section~\ref{proc:systems}.
We use the same notation $H^\circ$ for the auxiliary graph
and $\Phi_{G,\mathcal S}$ for the objective.
Our goal is to construct a point in ${\mathcal P^\circ}$
whose value under $\Phi_{G,\mathcal S}$ is at most
$13n/10-3/5$. We use $S_r$ for the unique star at the root.} Each nonroot star has size two or three, and the root star
has size four. Define the initial vector $x^{\mathrm{init}}\in \R_+^{E(H^\circ)}$ by setting $x^{\mathrm{init}}_e:=1/4$ for every nonremovable edge $e\in N_H$ and
assigning the following values on the stars:
\[
\begin{array}{|c|ccc|}
\hline 
 &\text{edge $\rep{uv}$ of $S(u,v)$}&\text{edge $\rep{b}$, $b\in I(u,v)$}
 &\text{connector}\\ \hline
 |S|=2,\ S\ne S_r&1/4&1/2&1/4\\
 |S|=3,\ S\ne S_r&1/4&3/4&1/4\\
 S=S_r&1/4&3/4&1\\
 \hline
\end{array}
\]
These values need not satisfy every Blossom inequality. We correct
them by increasing selected nonloop edge values from $1/4$ to $1/2$,
and denote the vector after all corrections by $\bar x$.
Our goal is to bound the number of increases that raise the
objective expression in \eqref{proc:join-cost}. The initial vector satisfies
$x^{\mathrm{init}}(S_H)+x^{\mathrm{init}}_{c_S}\ge k_S-1$ for
every star. This inequality is preserved by every increase,
so the objective expression in \eqref{proc:join-cost} for the
final vector $\bar x$ reduces to
\[
 n-1+\bar x(N_H)+\sum_{S\in\mathcal S}\bar x_{c_S}.
\]
For
$x^{\mathrm{init}}$, the expression equals $5n/4-1/2$.
Increasing the coordinate of $\rep{uv}$ from $1/4$ to $1/2$
for a removable tree edge $uv$ leaves the objective unchanged.
The same increase on an edge of $N_H$ or a connector raises it
by $1/4$; we call this a \textit{paid increase}, as the objective value increases with this change.

{ We next show that $x^{\mathrm{init}}$ does
not satisfy the Blossom inequalities for the following two
types of cuts in $H^\circ$, and describe how to modify it
to obtain a point in $\mathcal P^\circ$. Note that, after the reductions in
Appendix~\ref{sec:quartic-framework-2}, cuts of size at least
four satisfy all the Blossom inequalities. Full proofs are
given in Appendix~\ref{sec:quartic-framework-2}.}

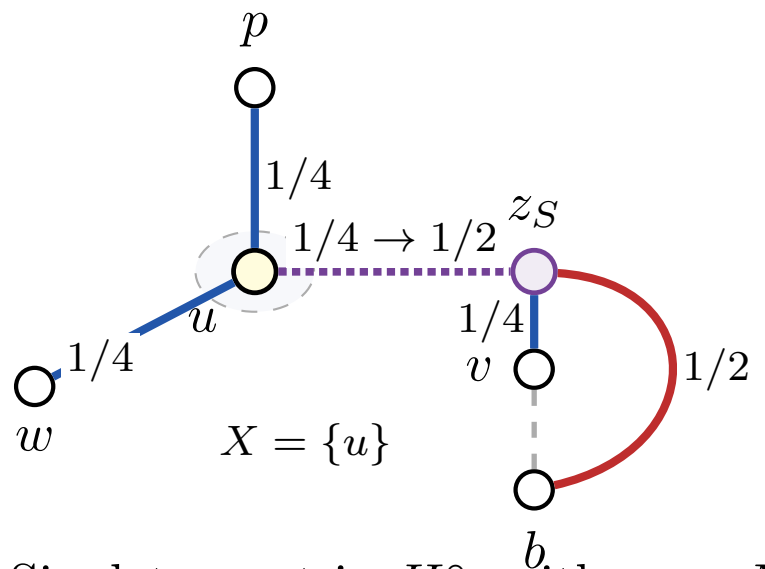
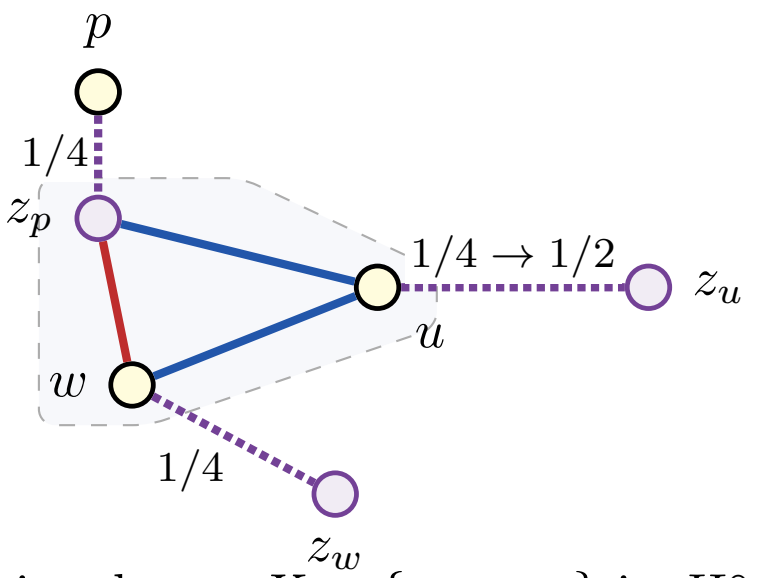
\begin{figure}[!t]
\renewcommand{\thesubfigure}{\alph{subfigure}}
\captionsetup[subfigure]{labelformat=parens,labelfont=normalfont}
\captionsetup{subrefformat=simple}
\centering

\begin{subfigure}[t]{.46\textwidth}
\centering
\begin{tikzpicture}[x=1.12cm,y=.76cm,font=\small,
 qlab/.style={font=\scriptsize,fill=white,inner sep=1pt}]
\useasboundingbox (-1.65,-2.2) rectangle (3,2.05);

\draw[draw=gray!65,dashed,fill=panelbg]
 (0,0) circle (.35);

\node[vertex,label=above:{$p$}] (p) at (0,1.6) {};
\node[center,label=below left:{$u$}] (u) at (0,0) {};
\node[vertex,label=below:{$w$}] (w) at (-1.3,-1) {};
\node[auxv,label=above:{$z_S$}] (z) at (1.65,0) {};
\node[vertex,label=left:{$v$}] (v) at (1.65,-.85) {};
\node[vertex,label=below:{$b$}] (b) at (1.65,-1.9) {};

\draw[blueedge] (p)--node[qlab,right] {$1/4$}(u);
\draw[blueedge] (u)--node[qlab,below left] {$1/4$}(w);
\draw[connedge] (u)--node[qlab,above] {$1/4\to1/2$}(z);

\draw[blueedge] (z)--node[qlab,left] {$1/4$}(v);
\draw[rededge] (z)
 .. controls (2.7,-.1) and (2.7,-1.55) ..
 node[qlab,pos=.5,right] {$1/2$}(b);
\draw[draw=gray!65,dashed,line width=1pt] (v)--(b);

\node[font=\scriptsize] at (.3,-1.5) {$X=\{u\}$};
\end{tikzpicture}
\caption{Singleton cut in $H^\circ$, with $pu\in N$.}
\label{proc:fig-quartic-singleton}
\end{subfigure}\hfill
\begin{subfigure}[t]{.51\textwidth}
\centering
\begin{tikzpicture}[x=1.12cm,y=.76cm,font=\small,
 qlab/.style={font=\scriptsize,fill=white,inner sep=1pt}]
\useasboundingbox (-1.65,-2.2) rectangle (3,2.05);

\draw[draw=gray!65,dashed,fill=panelbg,rounded corners=3pt]
 (-1,.9)--(.2,.9)--(1.35,.1)--(1.35,-.4)
 --(-.35,-1.25)--(-1,-1.25)--cycle;

\node[auxv,label=left:{$z_p$}] (z) at (-.65,.55) {};
\node[center,label=below right:{$u$}] (u) at (1,-.05) {};
\node[center,label=left:{$w$}] (w) at (-.45,-.9) {};
\node[center,label=above:{$p$}] (p) at (-.65,1.65) {};
\node[auxv,label=right:{$z_u$}] (zu) at (2.6,-.05) {};
\node[auxv,label=below:{$z_w$}] (zw) at (.75,-1.85) {};

\draw[blueedge] (z)--(u)--(w);
\draw[rededge] (w)--(z);
\draw[connedge] (p)--node[qlab,left] {$1/4$}(z);
\draw[connedge] (u)--node[qlab,above] {$1/4\to1/2$}(zu);
\draw[connedge] (w)--node[qlab,below left] {$1/4$}(zw);

\end{tikzpicture}
\caption{Triangle cut $X=\{z_p,u,w\}$ in $H^\circ$.}
\label{proc:fig-quartic-cuts}
\end{subfigure}

\caption{Two cuts whose Blossom inequalities are violated by
$x^{\mathrm{init}}$, both drawn in $H^\circ$.
Edge labels give the values of $x^{\mathrm{init}}$; on each
marked edge $e$, the label $1/4\to1/2$ shows the increase
from $x^{\mathrm{init}}_e$ to $\bar x_e$.
Each cut initially has three boundary edges of value $1/4$;
the marked increase raises their sum to one.
Shading identifies the vertices in $X$.
Blue and red edges correspond to tree and back edges of $G$,
respectively, via $\operatorname{rep}$; dotted violet edges are connectors.
The gray dashed segment in (a) represents an omitted path
in $H^\circ$.}
\label{proc:quartic-cut-configurations}
\end{figure}

\smallskip\noindent\textbf{A cut around one vertex.}
{In $G$, let $u$ be a nonroot vertex with
two DFS children $v,w$ and exactly one back edge $ub$ ending
at $u$, where $b\in V(T_v)\setminus\{v\}$.
Write $p=\operatorname{par}(u)$ and
$S:=\{uv,ub\}$ for the two-edge star centered at $u$.
In $H^\circ$, the edges of $S$ become $z_Sv$ and $z_Sb$,
and the connector is $c_S=uz_S$. The cut around the vertex $u$ (in $H^\circ$) is:
\[
 \delta_{H^\circ}(\{u\})
 =\{\rep{pu},\rep{uw},c_S\}.
\]
Each edge in this cut has value $1/4$, so
 $x^{\mathrm{init}}(\delta_{H^\circ}(\{u\}))
 =\frac34<1$,
violating its Blossom inequality.
If $pu$ is removable, increasing the value on $\rep{pu}$
to $1/2$ satisfies this inequality without a paid increase.
Otherwise, all three available increases are paid.
Figure~\ref{proc:quartic-cut-configurations}(\subref{proc:fig-quartic-singleton}) shows the choice
of increasing the connector $c_S$.}

\smallskip\noindent\textbf{A cut around a triangle.}
{Let $u$ and $w$ be as in the preceding case,
and suppose that $p,u,w$ form a triangle in $G$, where
$p=\operatorname{par}(u)\ne r$ and $u=\operatorname{par}(w)$.
Thus $pu$ and $uw$ are tree edges, while $pw$ is a back edge.
Suppose further that the star at $p$ is $\{pu,pw\}$ and that
$w$ centers a two-edge star.
In $H^\circ$, the images of $pu,uw,pw$ form a triangle
on $X:=\{z_p,u,w\}$, where $z_p$ is the auxiliary vertex
for the star at $p$. Its boundary consists of the connectors $c_p,c_u,c_w$
of these three stars, each initially of value $1/4$.
Recall that we increased $\rep{pu}$ when repairing the cut
around $u$. However, both endpoints of $\rep{pu}$ lie in $X$,
so that increase does not change the boundary sum of $X$. Thus the boundary sum therefore remains
\[
 x^{\mathrm{init}}(\delta_{H^\circ}(X))
 =x^{\mathrm{init}}_{c_p}
  +x^{\mathrm{init}}_{c_u}
  +x^{\mathrm{init}}_{c_w}
 =\frac34<1,
\]
violating the Blossom inequality for this cut.
Increasing any one of these connector coordinates from $1/4$
to $1/2$ makes the boundary sum one, this modification is a
\textit{paid increase}.}

\smallskip\noindent\textbf{Sharing the increases.}
{Two of these deficient cuts may share a boundary edge, allowing one
increase to satisfy both. For example, if two remaining singleton
cuts are centered at adjacent vertices, increasing their common
tree edge from $1/4$ to $1/2$ makes both boundary sums one.
After the increases on removable tree edges, form a graph with one node for each of these
singleton and triangle cuts, and an edge whenever two cuts share
such an increase.
This graph is a forest. A maximum matching pairs cuts that can be
handled together; each unmatched cut is handled by increasing one
connector.}

Appendix~\ref{sec:quartic-framework-2} shows that the increases
on removable tree edges, together with at most $(n-2)/5$ paid
increases chosen as above, give a feasible solution
$\bar x\in{\mathcal P^\circ}$.
Consequently,
\[
 \operatorname{tsp}(G)\le\text{\LPtwo}_{\mathrm{OPT}}(G,\mathcal S)
 \le\Phi_{G,\mathcal S}(\bar x)
 \le\frac54n-\frac12+\frac14\frac{n-2}{5}
 =\frac{13}{10}n-\frac35.
\]
This bound holds for an arbitrarily chosen DFS tree $T$.

\section{Discussion}

{We provided a new approximation algorithm for graphic TSP by
essentially constructing a point in a $T$-join polytope
using the subtour LP solution. We establish feasibility of the constructed point 
by relating the Blossom inequalities to the subtour cut
constraints. For fixed fractional values on the nonloop
edges, our degree-bounded-join formulation gives a decomposition
into integral joins which minimizes how often each star is
fully selected, and hence minimizes the expected reconnection
charge.

Our main result on graphic TSP also carries over to the path version of graphic TSP, in which one is given two distinct vertices $s$ and $t$ and seeks a shortest walk from $s$ to $t$ that visits every vertex. Traub, Vygen, and
Zenklusen~\cite{TVZ22} showed that a polynomial-time $\alpha$-approximation
algorithm for graphic TSP yields, for every fixed $\varepsilon>0$, a
polynomial-time $(\alpha+\varepsilon)$-approximation algorithm for the path
version. Together with Theorem~\ref{thm:intro-general}, this gives the ratio
$89/64+\varepsilon$, improving on the $7/5+\varepsilon$ that they derived from
the algorithm of Seb\H{o} and Vygen~\cite{SeboVygen}.

\noindent 
{\it Additional Evidence from Lean verification.} Though the main ideas for \ALG\  and for these approximations are human-generated and human-explained,
for the last column of Table~\ref{tab:discussion-results}, we formalized the proofs
in Lean~4~\cite{dMU21} (v4.19.0) over Mathlib~\cite{Mathlib20} (commit c44e0c8e), to find better constructions of the admissible $d$ vector, smarter choices of DFS trees, and other feasible points in $\mathcal{P}^\circ$. The Lean formalization verifies that our framework can improve the approximation bounds for Graphic TSP even further than those presented in the paper. These proofs show the potential of our techniques to bring down the approximation factor to $14524/10481\approx1.38575$ for graphic TSP, $9/7$ for 4-regular 4-vertex-connected graphs, $41944/31181\approx1.34518$ for subquartic graphs. We distill some of the techniques in Appendix~\ref{sec:further-improvements}. However, we emphasize that the readers do not need to review these proofs, and our main results are self-contained with careful exposition in the main body (and appendix) of the paper. We include Table \ref{tab:discussion-results} to summarize these results. 
\begin{table}[t]
\centering
\small
\setlength{\tabcolsep}{4pt}
\renewcommand{\arraystretch}{1.35}
\begin{tabular}{@{}p{.29\textwidth}p{.22\textwidth}p{.22\textwidth}p{.21\textwidth}@{}}
\hline
Graph class & Previously known & Presented here & Lean-verified\tablefootnote{For these bounds we only give some of the main ingredients of the proofs; complete proofs have been verified by Lean, see~\cite{github-location-lean}.}\\
& & with full proofs &\\
\hline
Connected simple graphs
 & $7/5=1.4$\newline\cite{SeboVygen}
 & $89/64$\newline$=1.390625$
 & $14524/10481$\newline$\approx1.38575$\\
$4$-regular, $4$-vertex-conn.
 & $4/3$ \quad \cite{Newman}
 & $13/10=1.3$
 & $9/7\approx1.28571$\\
Subquartic graphs
 & $25/18\approx1.3889$\newline\cite{Newman}
 & ---
 & $41944/31181$\newline$\approx1.34518$\\
$3$-vertex-connected, $\Delta\le4$
 & $25/18\approx1.3889$~\cite{Newman}
\newline{\scriptsize
An argument similar to that in~\cite{Newman} would also give $4/3$}
 & $4/3$
 & 4/3\\
$k$-vertex-connected, $k\ge4$,\newline $\Delta\le k+1$
 & $7/5=1.4$\newline\cite{SeboVygen}
 & $4/3$
 & $4/3$\\
\hline
\end{tabular}
\caption{Approximation factors for simple graphs.}
\label{tab:discussion-results}
\end{table}
}

\section*{Acknowledgements}
S.G. is grateful to Michel X. Goemans, Patrick Jaillet, Ahmad Abdi, and Naveen Garg for stimulating discussions on graphic TSP and T-joins.

\subsection*{AI methodology}
\label{sec:ai-methodology}
The authors developed the \LPone{}
and \LPtwo{} frameworks and their application to
$k$-vertex-connected graphs of maximum degree at most $k+1$.

For $4$-regular, $4$-vertex-connected graphs, the authors
developed the initial vector, the procedure for pairing
demands, and the use of shared corrections, where increasing
one edge satisfies two demands. The counting arguments and
computations leading to the final bound on paid increases
in Appendix~\ref{sec:quartic-framework-2} were developed by
the authors with assistance from OpenAI's GPT-5.6 Pro.

For general graphic TSP, the authors developed the
constructions of ${Q(d)}$ and ${Q^{(2)}}$, including the idea of
increasing the coordinates of edges in removable stars.
GPT-6 Astra assisted in refining the choice of the increase
parameters $d$ and deriving the bounds obtained from convex
{combinations of these points in
Appendix~\ref{sec:general-L}.} This assistance included developing
and checking the inequalities used to bound the contribution
of each star and combining these estimates to obtain the
final tour bounds.

AI tools were used in writing this manuscript,
including drafting and rewriting substantial portions of
the text. All the text has been reread and edited by the authors. AI tools also assisted with checking and simplifying
proofs, as well as developing Lean formalizations. The authors take full responsibility for the mathematical claims and 
proofs in this work.

\bibliographystyle{alpha}

\bibliography{GST2026a}

\clearpage
\appendix
\renewcommand{\theHsection}{appendix.\Alph{section}}
\renewcommand{\theHsubsection}{\theHsection.\arabic{subsection}}

\section{\textsc{Star} removable systems and parity polytopes}
\label{sec:preliminaries}

\subsection{Removable stars and \textsc{Star} removable systems}

\begin{definition}[Removable star]
Let $G=(V,E)$ be a connected simple graph.
A nonempty set $S\subseteq E$ is a \emph{removable star centered at $v$}
if every edge of $S$ is incident to $v$ and
\[
    G-(S\setminus\{e\})
    \quad\text{is connected for every }e\in S.
\]
\end{definition}

\begin{definition}[\textsc{Star} removable system]

Let $G=(V,E)$ be a connected simple graph.
A \emph{\textsc{Star} removable system} in $G$ is a family $\mathcal{S}$
of pairwise edge-disjoint removable stars such that, for every
choice of an edge $e_S\in S$ for each $S\in\mathcal{S}$,
\[
    G-\bigcup_{S\in\mathcal{S}}
    \bigl(S\setminus\{e_S\}\bigr)
    \quad\text{is connected}.
\]
Thus, simultaneously deleting all but any one edge of each star
preserves connectivity.
\end{definition}

For a \textsc{Star} removable system $\mathcal{S}$, let $R$ denote the set of removable edges:
\[
    R:=\bigcup_{S\in\mathcal{S}}S.
\]
The defining property also implies that $G-D$ is connected
for every $D\subseteq R$ satisfying
\[
    |D\cap S|\le |S|-1
    \qquad(S\in\mathcal{S}).
\]

\subsection{The \texorpdfstring{$b$}{b}-factor polytope and degree-bounded \texorpdfstring{$T$}{T}-joins}

We recall the definitions of $b$-matchings and their polytopes
from \cite{Schrijver2003}, using slightly different notation.

Let $G=(V,E)$ be an undirected multigraph. For $U\subseteq V$,
let $\delta_G(U)$ denote the set of edges with exactly one endpoint
in $U$, and write $\delta_G(v):=\delta_G(\{v\})$.
Let $L_G(v)$ denote the set of loops at $v$. The degree of $v$ is
\[
    \deg_G(v):=|\delta_G(v)|+2|L_G(v)|,
\]
so each loop counts twice.
For $x\in\mathbb{R}^{E}$ and $F\subseteq E$, write
\[
    x(F):=\sum_{e\in F}x_e,
    \qquad
    \deg_x(v):=x(\delta_G(v))+2x(L_G(v)).
\]
Fix vertex demands $b\in\mathbb{Z}_{+}^{V}$, and write
$b(U):=\sum_{v\in U}b_v$ for $U\subseteq V$.

\begin{definition}[Simple $b$-matching and $b$-factor]
A \emph{simple $b$-matching} in $G$ is a vector
$z\in\{0,1\}^{E}$ satisfying
\[
    \deg_z(v)\le b_v
    \quad\text{for every }v\in V.
\]
It is \emph{perfect}, also called a \emph{$b$-factor}, if
$\deg_z(v)=b_v$ for every $v\in V$.
The \emph{$b$-factor polytope} is
\[
    P_{\mathrm{PM}}(G,b)
    :=
    \operatorname{conv}
    \left\{
        z\in\{0,1\}^{E}:
        \deg_z(v)=b_v\text{ for every }v\in V
    \right\}.
\]
\end{definition}

\begin{theorem}[Edmonds--Johnson~\cite{EdmondsJohnson}]
\label{thm:b-factor}
The $b$-factor polytope $P_{\mathrm{PM}}(G,b)$ consists precisely
of the vectors $x\in\mathbb{R}^{E}$ satisfying
\[
    \begin{aligned}
        0\le x_e &\le 1
        &&\text{for every }e\in E,\\[3pt]
        \deg_x(v) &= b_v
        &&\text{for every }v\in V,\\[3pt]
        x\bigl(\delta_G(U)\setminus F\bigr)+|F|-x(F) &\ge 1
        &&\begin{aligned}[t]
            &\text{for every }U\subseteq V,\ F\subseteq\delta_G(U)\\
            &\text{with }b(U)+|F|\text{ odd}.
        \end{aligned}
    \end{aligned}
\]
The last family consists of the \emph{Blossom inequalities}.
In particular, every extreme point belongs to $\{0,1\}^{E}$.
\end{theorem}

\smallskip\noindent
\textbf{Degree-bounded $T$-joins.}
For a subset $F$ of a finite ground set, write $\mathbf{1}_F$ for
its incidence vector, with coordinate $(\mathbf{1}_F)_e=1$ exactly
when $e\in F$. We use $\mathbf{1}_{\{\mathcal E\}}$ for the scalar
indicator of a condition $\mathcal E$.
Let $T\subseteq V$ have even cardinality. For $J\subseteq E$, write
\[
    \deg_J(v):=\deg_{\mathbf{1}_{J}}(v),
    \qquad
    \odd(J):=\{v\in V:\deg_J(v)\text{ is odd}\}.
\]
An edge set $J\subseteq E$ is a \emph{$T$-join} if $\odd(J)=T$.

{For a graph $G=(V,E)$, an $\odd(G)$-join is an
edge set $J\subseteq E$ such that $\odd(J)=\odd(G)$.
The \emph{$\odd(G)$-join polytope} is the convex hull in
$\mathbb R^E$ of the incidence vectors of these joins.}

For each $v\in V$, fix an integer degree bound
$b_v\in\{0,\ldots,\deg_G(v)\}$. Define
\[
    P_{\mathrm{join}}(G,T,b)
    :=
    \operatorname{conv}
    \left\{
        \mathbf{1}_{J}:
        J\subseteq E,\;
        \odd(J)=T,\;
        \deg_J(v)\le b_v\text{ for every }v\in V
    \right\}.
\]
Put
\[
    t_v:=(\mathbf{1}_T)_v,
    \qquad
    \beta_v:=t_v+2\left\lfloor\frac{b_v-t_v}{2}\right\rfloor
    \qquad(v\in V).
\]
Write $\beta(U):=\sum_{v\in U}\beta_v$ for $U\subseteq V$.
Note that $\beta_v=b_v$ whenever
$b_v\equiv t_v\pmod 2$.

Integrality of degree-bounded $T$-join polyhedra whose bounds have the parity
prescribed by $T$ is known. It follows from the results of Edmonds and Johnson
on bidirected graphs, see~\cite[Chapter~36]{Schrijver2003}, and it was used
for bounded-degree subset TSP by Friggstad and Mousavi~\cite{FM22}.
Theorem~\ref{thm:bounded-T-join} gives the description in the form we need,
with the box
constraints $0\le x\le 1$ and with bounds of arbitrary parity. We include
a short proof from Theorem~\ref{thm:b-factor}.

\begin{theorem}[Degree-bounded $T$-join polytope]
\label{thm:bounded-T-join}
The polytope $P_{\mathrm{join}}(G,T,b)$ consists precisely of the
vectors $x\in\mathbb{R}^{E}$ satisfying
\begin{equation}
\label{eq:bounded-T-join-polytope}
    \begin{aligned}
        0\le x_e &\le 1
        &&\text{for every }e\in E,\\[3pt]
        \deg_x(v) &\le \beta_v
        &&\text{for every }v\in V,\\[3pt]
        x\bigl(\delta_G(U)\setminus F\bigr)+|F|-x(F) &\ge 1
        &&\begin{aligned}[t]
            &\text{for every }U\subseteq V,\ F\subseteq\delta_G(U)\\
            &\text{with }|T\cap U|+|F|\text{ odd}.
        \end{aligned}
    \end{aligned}
\end{equation}
In particular, this system is integral, and every extreme point
is the incidence vector of a $T$-join satisfying the degree bounds.
\end{theorem}

\begin{proof}
If $\beta_v<0$, then no $T$-join
can satisfy the degree bound at $v$. Thus $P_{\mathrm{join}}(G,T,b)=\varnothing$, and the displayed
system is also infeasible because it requires
$0\le\deg_x(v)\le\beta_v<0$.

Form $G^+$ by adding a set $\mathcal L(v)$ of distinct
\emph{slack loops} at each $v\in V$, where
\[
    |\mathcal L(v)|=r_v:=\left\lfloor\frac{\beta_v}{2}\right\rfloor,
\]
and consider $P_{\mathrm{PM}}(G^+,\beta)$.
Write $x$ for the coordinates of the edges of $G$ and $y$ for
the coordinates of the added loops. The degree equations are
\begin{equation}
\label{eq:T-join-lift-degrees}
    \deg_x(v)+2y(\mathcal L(v))=\beta_v
    \qquad(v\in V).
\end{equation}
An integral factor projects to a $T$-join with degrees at most $b$.
Conversely, every such $T$-join extends to an integral factor by
selecting exactly $(\beta_v-\deg_J(v))/2$ added loops at each vertex.
Consequently, projecting $P_{\mathrm{PM}}(G^+,\beta)$ onto the
coordinates indexed by $E(G)$ gives $P_{\mathrm{join}}(G,T,b)$.

The singleton Blossom inequalities give $\deg_x(v)\ge1$ for
$v\in T$, so the degree bounds imply
$0\le\beta_v-\deg_x(v)\le2r_v$ at every vertex; hence the
added loops can supply the required degree. The added loops do not cross any cut, and
$\beta(U)\equiv|T\cap U|\pmod2$. Together with the degree
equations, this shows that a vector $x$ satisfies
\eqref{eq:bounded-T-join-polytope} if and only if it can be
extended to a point of $P_{\mathrm{PM}}(G^+,\beta)$.
Thus \eqref{eq:bounded-T-join-polytope} describes exactly
$P_{\mathrm{join}}(G,T,b)$.
\end{proof}

\textbf{Polynomial-time optimization.}
Linear optimization over $P_{\mathrm{join}}(G,T,b)$ reduces in
polynomial time to a weighted perfect $b$-matching problem via the
slack-loop construction above, and can therefore be solved in polynomial time~\cite{EdmondsJohnson}. For
instance, Tutte's reduction~\cite{Tut54} turns it into a weighted perfect
matching problem, which is solved by the algorithm of
Edmonds~\cite{Edm65} (see~\cite{Gab83,Ans87} for direct algorithms).

\textbf{Upper bounds on cuts.}
The following proposition shows that imposing upper bounds on
cuts instead of degrees makes the integral optimization problem
NP-hard, even for pairwise edge-disjoint cuts of size four.

\begin{proposition}\label{prop:cut-bounded-joins-hardness}
Finding a maximum-cardinality $T$-join subject to
\[
 |J\cap C|\le3\qquad(C\in\mathcal C)
\]
is NP-hard, even when $T=\varnothing$, the graph is simple
and connected, and $\mathcal C$ consists of pairwise
edge-disjoint cuts of size four.
\end{proposition}

\begin{proof}
We reduce from maximum stable set. Let $G$ have $n$ vertices.
For each $u\in V(G)$, construct a cycle $C_u$ of length
$2n+1$. Identify one vertex of all these cycles as a common
vertex $r$, and keep all other vertices distinct. Let $H$
be the resulting graph.

Each cycle has $n$ pairwise nonadjacent vertices different
from $r$. Assign a distinct such vertex $p_{u,v}$ to each
edge $uv$ incident to $u$. For every $uv\in E(G)$, impose
\[
 |J\cap\delta_H(\{p_{u,v},p_{v,u}\})|\le3.
\]
These cuts have size four and are pairwise edge-disjoint.

Every vertex of $H$ other than $r$ has degree two, so a
$\varnothing$-join contains either all or none of each cycle.
Conversely, every union of these cycles is a $\varnothing$-join.
The constraint for $uv$ excludes precisely the choice of both
$C_u$ and $C_v$. Thus feasible joins correspond to stable sets
$I\subseteq V(G)$, with join cardinality $(2n+1)|I|$.
The construction has polynomial size, proving the claim.
\end{proof}

\section{The \LPone: keeping an edge of every removable star}
\label{subsec:framework-1}

Let $G=(V,E)$ be a connected simple graph with $m:=|E|$ edges
and a \textsc{Star} removable system $\mathcal S$. For each $S\in\mathcal S$,
let $v_S$ denote its center and put $k_S:=|S|$. Let
\[
    R:=\bigcup_{S\in\mathcal S}S,
    \qquad
    N:=E\setminus R.
\]
We call the edges in $R$ \emph{removable} and those in $N$
\emph{nonremovable} respectively.
A \emph{tour} of $G$ is a connected spanning Eulerian multigraph
whose edges are copies of edges of $G$. Its length is its number
of edge copies, and $\operatorname{tsp}(G)$ denotes the minimum
length of such a tour.

We choose an $\odd(G)$-join $J$ so that deleting $J\cap R$
and doubling $J\cap N$ produces a tour. To preserve connectivity,
we require $J$ to leave at least one edge of every star unselected.
We express these restrictions as degree bounds in an auxiliary
graph, allowing us to use the integral degree-bounded join
polytope defined above.

\subsection{Constructing the auxiliary graph}
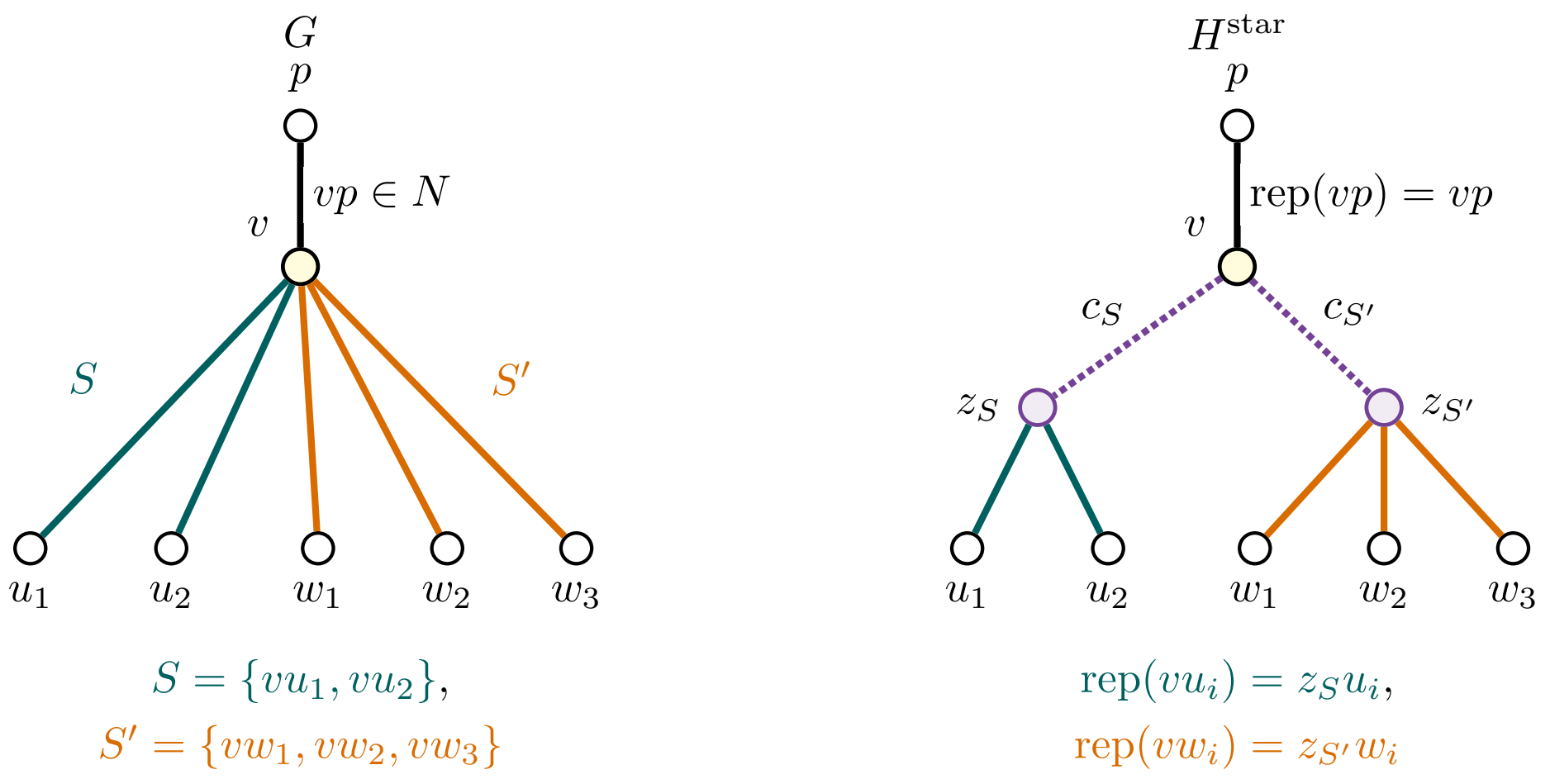
\begin{figure}[!t]
\centering
\begin{tikzpicture}[
 x=.94cm,y=.94cm,font=\small,
 starS/.style={origedge,draw=teal!75!black},
 starSp/.style={origedge,draw=orange!85!black}
]
\begin{scope}[xshift=-3.75cm]
\node at (0,3.25) {$G$};
\node[vertex,label=above:{$p$}] (p) at (0,2.45) {};
\node[center,label=above left:{$v$}] (v) at (0,1.25) {};
\node[vertex,label=below:{$u_1$}] (u1) at (-2.3,-1.15) {};
\node[vertex,label=below:{$u_2$}] (u2) at (-1.1,-1.15) {};
\node[vertex,label=below:{$w_1$}] (w1) at (.15,-1.15) {};
\node[vertex,label=below:{$w_2$}] (w2) at (1.25,-1.15) {};
\node[vertex,label=below:{$w_3$}] (w3) at (2.35,-1.15) {};

\draw[origedge] (p)--node[right,fill=white,inner sep=2pt]
  {$vp\in N$}(v);
\draw[starS] (v)--(u1);
\draw[starS] (v)--(u2);
\draw[starSp] (v)--(w1);
\draw[starSp] (v)--(w2);
\draw[starSp] (v)--(w3);

\node[text=teal!75!black,fill=white,inner sep=2pt]
  at (-1.85,.3) {$S$};
\node[text=orange!85!black,fill=white,inner sep=2pt]
  at (1.8,.3) {$S'$};
\node[align=center,anchor=north] at (0,-1.95)
  {\textcolor{teal!75!black}{$S=\{vu_1,vu_2\}$},\\[3pt]
   \textcolor{orange!85!black}{$S'=\{vw_1,vw_2,vw_3\}$}};
\end{scope}

\begin{scope}[xshift=3.75cm]
\node at (0,3.25) {$H^{\mathrm{star}}$};
\node[vertex,label=above:{$p$}] (p) at (0,2.45) {};
\node[center,label=above left:{$v$}] (v) at (0,1.25) {};
\node[auxv,label=left:{$z_S$}] (z) at (-1.7,.05) {};
\node[auxv,label=right:{$z_{S'}$}] (zp) at (1.25,.05) {};
\node[vertex,label=below:{$u_1$}] (u1) at (-2.3,-1.15) {};
\node[vertex,label=below:{$u_2$}] (u2) at (-1.1,-1.15) {};
\node[vertex,label=below:{$w_1$}] (w1) at (.15,-1.15) {};
\node[vertex,label=below:{$w_2$}] (w2) at (1.25,-1.15) {};
\node[vertex,label=below:{$w_3$}] (w3) at (2.35,-1.15) {};

\draw[origedge] (p)--node[right,fill=white,inner sep=2pt]
  {$\rep{vp}=vp$}(v);
\draw[connedge] (v)--node[above left,inner sep=2pt]
  {$c_S$}(z);
\draw[connedge] (v)--node[above right,inner sep=2pt]
  {$c_{S'}$}(zp);
\draw[starS] (z)--(u1);
\draw[starS] (z)--(u2);
\draw[starSp] (zp)--(w1);
\draw[starSp] (zp)--(w2);
\draw[starSp] (zp)--(w3);

\node[align=center,anchor=north] at (0,-1.95)
  {\textcolor{teal!75!black}{$\rep{vu_i}=z_Su_i$},\\[3pt]
   \textcolor{orange!85!black}{$\rep{vw_i}=z_{S'}w_i$}};
\end{scope}
\end{tikzpicture}
\caption{Two removable stars centered at the same vertex $v$.
The edges of $S$ are teal, and those of $S'$ are orange;
the same colors identify their images in $H^{\mathrm{star}}$.
Each star has its own auxiliary vertex and connector in
$H^{\mathrm{star}}$; the edge $vp\in N$ keeps its endpoints.
The auxiliary vertices have degrees three and four, so their degree bounds are $|S|-1=1$ and $|S'|-1=2$ in \LPone,
respectively.}
\label{fig:star-splitting}
\end{figure}

For each star $S$ centered at $v_S$, introduce a vertex $z_S$
and replace each edge $v_Su\in S$ by $z_Su$.
Add the connector $c_S:=v_Sz_S$ and keep every edge of $N$
with its endpoints unchanged. Let $H^{\mathrm{star}}=(W,A^{\mathrm{star}})$ be the resulting
connected graph, where $W:=V\cup\{z_S:S\in\mathcal S\}$.
See Figure~\ref{fig:star-splitting}.

Define the injective map $\operatorname{rep}:E(G)\to E(H^{\mathrm{star}})$ by
\[
 \rep{e}:=
 \begin{cases}
  z_Su,&e=v_Su\in S,\quad S\text{ is centered at }v_S,\\[3pt]
  e,&e\in N.
 \end{cases}
\]
Thus $\rep{e}$ is the edge of $H^{\mathrm{star}}$ obtained from $e$ in this
construction. For $B\subseteq E(G)$, write
$B_{H^{\mathrm{star}}}:=\{\rep{e}:e\in B\}$; this defines $S_{H^{\mathrm{star}}}$,
$R_{H^{\mathrm{star}}}$, and $N_{H^{\mathrm{star}}}$.
With $C_{H^{\mathrm{star}}}:=\{c_S:S\in\mathcal S\}$, we have
\[
 A^{\mathrm{star}}=R_{H^{\mathrm{star}}}\mathbin{\dot\cup}N_{H^{\mathrm{star}}}\mathbin{\dot\cup}C_{H^{\mathrm{star}}}.
\]

Define the vertex upper bounds
\begin{equation}
\label{eq:framework-1-demands}
    b^{\mathrm{star}}_w:=
    \begin{cases}
        \deg_{H^{\mathrm{star}}}(w), & w\in V,\\
        k_S-1, & w=z_S,\ S\in\mathcal{S}.
    \end{cases}
\end{equation}

Let
\[
    \Pone(G,\mathcal S):=P_{\mathrm{join}}\bigl(H^{\mathrm{star}},\odd(H^{\mathrm{star}}),b^{\mathrm{star}}\bigr).
\]
By Theorem~\ref{thm:bounded-T-join}, the feasible region of \LPone, abbreviated $\Pone$, is
integral and has the following description:
\begin{equation}
\label{eq:framework-1-polytope}
\begin{aligned}
 0\le x_a&\le1
 &&\text{for every }a\in A^{\mathrm{star}},\\[3pt]
 x(S_{H^{\mathrm{star}}})+x_{c_S}&\le k_S-1
 &&\text{for every }S\in\mathcal S,\\[3pt]
 x\bigl(\delta_{H^{\mathrm{star}}}(U)\setminus F\bigr)+|F|-x(F)&\ge1
 &&\begin{aligned}[t]
   &\text{for every }U\subseteq W,\ F\subseteq\delta_{H^{\mathrm{star}}}(U)\\
   &\text{with }|\delta_{H^{\mathrm{star}}}(U)|+|F|\text{ odd}.
 \end{aligned}
\end{aligned}
\end{equation}
Note that the degree bounds at vertices $v\in V$ follow from
$0\le x_a\le1$, 
so we omit them from the displayed system.\\

\textbf{The \LPone.}
Define
\[
    \PhiOne_{G,\mathcal S}(x)
    :=m-x(R_{H^{\mathrm{star}}})+x(N_{H^{\mathrm{star}}}),
\]
and consider
\begin{equation}
\label{eq:framework-1-lp}
    \text{\LPone}_{\mathrm{OPT}}(G,\mathcal S)
    :=\min\left\{
        \PhiOne_{G,\mathcal S}(x):x\in\Pone
    \right\}.
\end{equation}
If $\Pone=\varnothing$, we set
$\text{\LPone}_{\mathrm{OPT}}(G,\mathcal S):=+\infty$.
Thus, selecting a removable edge decreases the objective by one,
selecting a nonremovable edge increases it by one, and selecting
a connector leaves it unchanged.

\textbf{How do we obtain a tour?}
By integrality, whenever the \LPone{} is feasible,
it has an integral optimal solution $z$. Let
\[
    J:=\{e\in E:z_{\rep{e}}=1\}.
\]
Contracting all connectors in $H^{\mathrm{star}}$ gives $G$.
Since $z$ defines an $\odd(H^{\mathrm{star}})$-join, summing
the degree parities at the vertices merged by each contraction
gives
\begin{equation}
\label{eq:projected-parity}
    \odd(J)=\odd(G).
\end{equation}
The star degree bounds also give
\[
    |J\cap S|=z(S_{H^{\mathrm{star}}})\le k_S-1
    \qquad(S\in\mathcal S).
\]

Starting with $G$, delete the edges of $J\cap R$ and double
the edges of $J\cap N$. The resulting multigraph $T_z$
is connected because $\mathcal S$ is a \textsc{Star} removable system
and at least one edge of each star remains. Every degree
is even because $J$ is an $\odd(G)$-join.
Thus $T_z$ is a tour of length
\begin{equation}
\label{eq:framework-1-tour-length}
    |E(T_z)|
    =m-|J\cap R|+|J\cap N|
    =\PhiOne_{G,\mathcal S}(z).
\end{equation}
\begin{theorem}[\LPone{} tour bound]
\label{thm:framework-1-bound}
If $\Pone$ is nonempty, an integral optimal solution of
\eqref{eq:framework-1-lp} produces a tour of length
\[
    \text{\LPone}_{\mathrm{OPT}}(G,\mathcal S).
\]
Consequently,
\[
    \operatorname{tsp}(G)\le \text{\LPone}_{\mathrm{OPT}}(G,\mathcal S).
\]
More generally, every $x\in \Pone$ certifies
\[
    \operatorname{tsp}(G)\le \PhiOne_{G,\mathcal S}(x).
\]
\end{theorem}

\begin{proof}
Theorem~\ref{thm:bounded-T-join} guarantees an integral
optimal solution $z$. Apply \eqref{eq:framework-1-tour-length}
and use $\PhiOne_{G,\mathcal S}(z)=\text{\LPone}_{\mathrm{OPT}}(G,\mathcal S)\le\PhiOne_{G,\mathcal S}(x)$.
\end{proof}

\section{The \LPtwo{}: allowing fully selected stars and restoring connectivity}
\label{sec:framework-2}

As in the \LPone{}, we start with $G$, choose an
$\odd(G)$-join $J$, double its edges, and delete both copies
of every edge in $J\cap R$. The \LPtwo{} allows $J$ to contain
an entire removable star. In that case, the deletions remove
every edge of the star, so we choose an arbitrary edge of that
star and add back two copies of it.

After doing this for every fully selected star, at least one
edge of each star remains, so the \textsc{Star} removable system property
ensures connectivity. Adding two copies preserves the even
degrees and increases the tour length by two for each such
star. We include this additional cost in the \LPtwo{} objective.

Section~\ref{subsec:four-vertex-separation} gives an example where the \LPtwo{} yields a smaller tour bound
for the same \textsc{Star} removable system. We do not know whether the \LPone{} gives an approximation ratio below $4/3$ for simple
$4$-regular, $4$-vertex-connected graphs. Using the \LPtwo{},
we obtain a $13/10$ bound for this class in
Appendix~\ref{sec:quartic-framework-2}.

\subsection{The auxiliary graph and degree bounds}
\begin{figure}[!t]
\centering
\begin{tikzpicture}[scale=.9]
\node[center,label=above:{$v_S$}] (v) at (0,2) {};
\node[auxv,label=left:{$z_S$}] (z) at (0,.8) {};
\node[vertex,label=below:{$u_1$}] (a) at (-1.3,-.7) {};
\node[vertex,label=below:{$u_2$}] (b) at (0,-.7) {};
\node[vertex,label=below:{$u_{k_S}$}] (c) at (1.3,-.7) {};
\draw[connedge] (v)--node[left,font=\small] {$c_S$}(z);
\draw[origrep] (z)--(a); \draw[origrep] (z)--(b); \draw[origrep] (z)--(c);
\draw[loopedge] (z) to[out=20,in=-20,looseness=8]
 node[right,font=\small] {$\ell_S$} (z);
\node[anchor=west,align=left,font=\small] at (2.5,1.15)
 {$x(S_H)+x_{c_S}+2x_{\ell_S}\le k_S+1$\\[7pt]
  $\ell_S$ selected $\Longrightarrow |J\cap S|\le k_S-1$\\[4pt]
  $S\subseteq J\Longrightarrow \ell_S$ absent};
\end{tikzpicture}
\caption{The \LPtwo{} adds one loop $\ell_S$ at each $z_S$ in $H$.
Selecting $\ell_S$ prevents the join from selecting every edge
of $S_H$. If every edge of $S_H$ is selected, we add two copies
of an edge of $S$ in $G$.}
\label{fig:single-loop}
\end{figure}
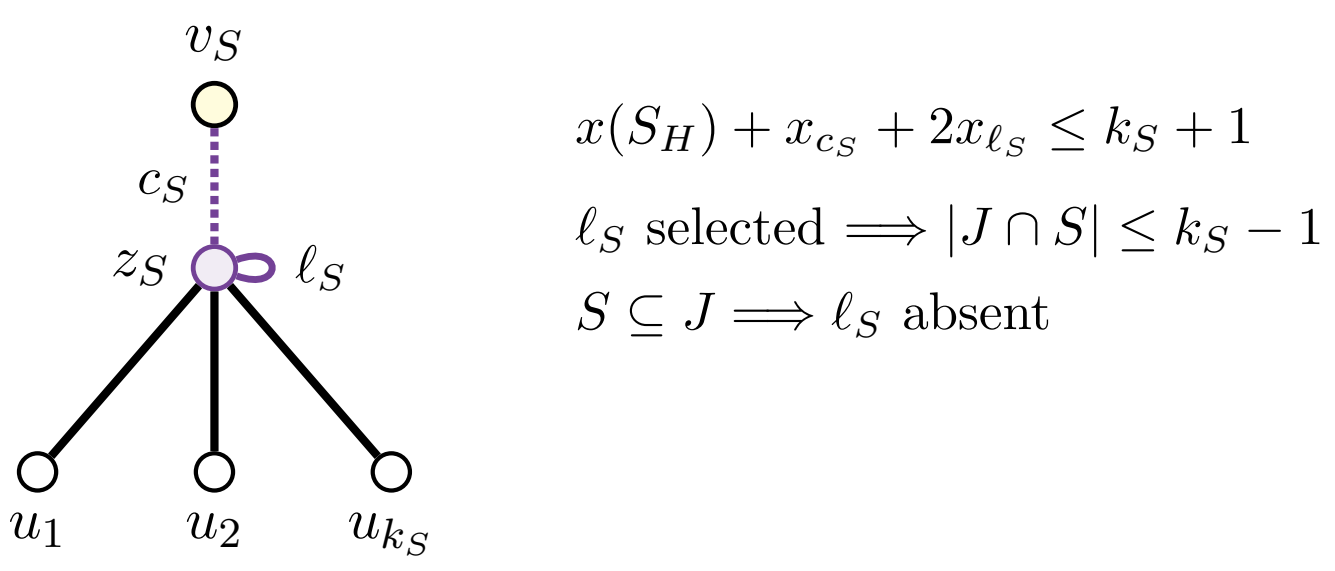

Construct $H=(W,A)$ from $H^{\mathrm{star}}$ by adding one loop $\ell_S$
at each $z_S$ (see Figure~\ref{fig:single-loop}), so
\[
 A=A^{\mathrm{star}}\mathbin{\dot\cup}\{\ell_S:S\in\mathcal S\}.
\]
These loops are part of the \LPtwo{}; the slack loops in
Theorem~\ref{thm:bounded-T-join} are used only in the reduction
to $b$-factors.

Let $H^\circ$ be the graph obtained from $H$ by deleting the
loops $\ell_S$, and write $A^\circ:=E(H^\circ)$.
Thus $H^\circ=H^{\mathrm{star}}$ and
$A^\circ=A^{\mathrm{star}}$.
Deleting loops preserves degree parity, so
$\odd(H^\circ)=\odd(H)$.
To distinguish the two formulations, we use
$H^{\mathrm{star}}$ and $A^{\mathrm{star}}$ when working
with the \LPone{}, and $H^\circ$ and $A^\circ$ for the
nonloop graph and its edge set when working with the \LPtwo{}. We use the same map $\operatorname{rep}$, now viewed as a map
from $E(G)$ to $E(H)$. Recall that $\operatorname{rep}$ maps each edge $uw\in S$, where $S$ is centered at $u$,
to $z_Sw$, and each edge $e\in N$ to itself. For $B\subseteq E(G)$,
write $B_H:=\{\rep{e}:e\in B\}=B_{H^{\mathrm{star}}}$; in particular,
this defines $S_H$, $R_H$, and $N_H$.
Write $C_H:=\{c_S:S\in\mathcal S\}$ for the set of connectors.

Define the vertex upper bounds
\begin{equation}
\label{eq:framework-2-demands}
    b^{\mathrm{SR}}_w:=
    \begin{cases}
        \deg_H(w),&w\in V,\\[3pt]
        k_S+1,&w=z_S,\quad S\in\mathcal S.
    \end{cases}
\end{equation}
Consider the degree-bounded $T$-join polytope
\[
    {P_{\mathcal R}}(G,\mathcal S):=
    P_{\mathrm{join}}\bigl(H,\odd(H),b^{\mathrm{SR}}\bigr).
\]
When $G$ and $\mathcal S$ are fixed, abbreviate
${P_{\mathcal R}}(G,\mathcal S)$ to ${P_{\mathcal R}}$.
Its projection onto the nonloop coordinates is denoted by
\[
 {\mathcal P^\circ}(G,\mathcal S)
 :=\{y\in\mathbb R^{A^\circ}:\text{there exists }
       \gamma\in\mathbb R^{\mathcal S}
       \text{ with }(y,\gamma)\in{P_{\mathcal R}}(G,\mathcal S)\},
\]
where $\gamma_S$ is the coordinate of the loop $\ell_S$ for $S\in \mathcal S$.
We abbreviate this projection to ${\mathcal P^\circ}$ when
$G$ and $\mathcal S$ are fixed.

By Theorem~\ref{thm:bounded-T-join}, ${P_{\mathcal R}}$ is integral
and consists precisely of the vectors $x\in\mathbb R^A$ satisfying
\begin{equation}
\label{eq:framework-2-polytope}
\begin{aligned}
 0\le x_a&\le1
 &&\text{for every }a\in A,\\[3pt]
 x(S_H)+x_{c_S}+2x_{\ell_S}&\le k_S+1
 &&\text{for every }S\in\mathcal S,\\[3pt]
 x\bigl(\delta_H(U)\setminus F\bigr)+|F|-x(F)&\ge1
 &&\begin{aligned}[t]
   &\text{for every }U\subseteq W,\ F\subseteq\delta_H(U)\\
   &\text{with }|\delta_H(U)|+|F|\text{ odd}.
 \end{aligned}
\end{aligned}
\end{equation}

{Note that $\mathcal P^\circ$ is the
$\odd(H^\circ)$-join polytope of $H^\circ$.
The vector in $\mathbb R^A$ with value one on $A^\circ$
and zero on every loop belongs to $P_{\mathcal R}$, so
$P_{\mathcal R}\ne\varnothing$.}

\subsection{The linear program}

Define
\[
    \Phi_H(x)
    :=m-x(R_H)+x(N_H)
      +2\sum_{S\in\mathcal S}(1-x_{\ell_S}).
\]
The following theorem bounds the tour length by $\Phi_H(x)$
for $x\in{P_{\mathcal R}}$. The linear program is
\begin{equation}
\label{eq:framework-2-lp}
    \text{\LPtwo}_{\mathrm{OPT}}(G,\mathcal S)
    :=\min\left\{
        \Phi_H(x):x\in{P_{\mathcal R}}
    \right\}.
\end{equation}

\begin{theorem}[\LPtwo{} tour bound]
\label{thm:framework-2-bound}
For every $x\in {P_{\mathcal R}}$, there is a distribution of tours of $G$
whose expected length is at most
\[
    \Phi_H(x).
\]
Consequently,
\[
    \operatorname{tsp}(G)
    \le \text{\LPtwo}_{\mathrm{OPT}}(G,\mathcal S).
\]
\end{theorem}

\begin{proof}
Let $x\in {P_{\mathcal R}}$. By Theorem~\ref{thm:bounded-T-join}, there are
edge sets $K_1,\ldots,K_t\subseteq A$ and coefficients
$\lambda_1,\ldots,\lambda_t$ such that
\begin{equation}
\label{eq:framework-2-join-decomposition}
    x=\sum_{i=1}^{t}\lambda_i\mathbf{1}_{K_i},
    \qquad \lambda_i>0,
    \qquad \sum_{i=1}^{t}\lambda_i=1,
\end{equation}
where, for every $i$,
\[
    \odd(K_i)=\odd(H),
    \qquad \deg_{K_i}(w)\le b^{\mathrm{SR}}_w\quad(w\in W).
\]
For every edge $a\in A$, the convex combination gives
\[
    \sum_{i=1}^{t}\lambda_i\mathbf{1}_{\{a\in K_i\}}=x_a.
\]
For each $i$, let
\[
    J_i:=\{e\in E:\rep{e}\in K_i\}.
\]
Deleting loops preserves degree parity, so the projection
argument used for the \LPone{} shows that $J_i$
is an $\odd(G)$-join.

If $\ell_S\in K_i$, the degree bound at $z_S$ gives
\[
    |J_i\cap S|+\mathbf{1}_{\{c_S\in K_i\}}+2\le k_S+1,
\]
and hence $|J_i\cap S|\le k_S-1$. Therefore, if $J_i$
contains every edge of $S$, then $\ell_S\notin K_i$. It follows that
\begin{equation}
\label{eq:saturation-probability}
    \sum_{i=1}^{t}\lambda_i\mathbf{1}_{\{S\subseteq J_i\}}
    \le \sum_{i=1}^{t}\lambda_i\mathbf{1}_{\{\ell_S\notin K_i\}}
    =1-x_{\ell_S}.
\end{equation}

Apply the construction above to each $J_i$, adding back two
copies of an arbitrary edge from each star $S\subseteq J_i$.
The resulting multigraph $T_i$ is connected and has even
degrees, so it is a tour of length
\[
    |E(T_i)|=m-|J_i\cap R|+|J_i\cap N|
       +2\sum_{S\in\mathcal S}\mathbf{1}_{\{S\subseteq J_i\}}.
\]
Multiplying by $\lambda_i$, summing over $i$, and using
\eqref{eq:saturation-probability}, we obtain
\[
\begin{aligned}
    \sum_{i=1}^{t}\lambda_i|E(T_i)|
    &=m-x(R_H)+x(N_H)
      +2\sum_{S\in\mathcal S}\sum_{i=1}^{t}
            \lambda_i\mathbf{1}_{\{S\subseteq J_i\}}\\
    &\le m-x(R_H)+x(N_H)
      +2\sum_{S\in\mathcal S}(1-x_{\ell_S})\\
    &=\Phi_H(x).
\end{aligned}
\]

At least one $T_i$ has length no greater than this weighted average.
Taking $x$ to minimize $\Phi_H$ over ${P_{\mathcal R}}$ proves the final bound.
\end{proof}

\subsection{The objective on nonloop coordinates}
\label{subsec:repair-nonloop}

The projection ${\mathcal P^\circ}$ consists precisely of the
vectors $y\in\mathbb R^{A^\circ}$ satisfying
\begin{equation}\label{eq:repair-nonloop-polytope}
\begin{aligned}
 0\le y_a&\le1
 &&\text{for every }a\in A^\circ,\\[3pt]
 y(C\setminus F)+|F|-y(F)&\ge1
 &&\begin{aligned}[t]
   &\text{for every }U\subseteq W,\ C=\delta_{H^\circ}(U),\ F\subseteq C\\
   &\text{with }|C\setminus F|\text{ odd}.
 \end{aligned}
\end{aligned}
\end{equation}
Every point of ${P_{\mathcal R}}$ projects to a vector satisfying
these inequalities. Conversely, any such vector extends to
a point of ${P_{\mathcal R}}$ by setting all loop coordinates to
zero. The degree bounds then follow from $y_a\le1$, and
loops do not cross cuts.

For a fixed graph $G$ and \textsc{Star} removable system $\mathcal S$, define
\[
 \Phi_{G,\mathcal S}(y)
 :=\min\{\Phi_H(x):x\in{P_{\mathcal R}}(G,\mathcal S),\ x|_{A^\circ}=y\}
 \qquad(y\in{\mathcal P^\circ}(G,\mathcal S)).
\]
We show that
\begin{equation}\label{eq:repair-nonloop-cost}
 \Phi_{G,\mathcal S}(y)
 =m-\sum_{S\in\mathcal S}(k_S-1)+y(N_H)
   +\sum_{S\in\mathcal S}
     \max\{y_{c_S},\,k_S-1-y(S_H)\}.
\end{equation}
For $y\in{\mathcal P^\circ}$, define $\widetilde y\in\mathbb R_+^A$ by
\[
 \widetilde y_a:=y_a\quad(a\in A^\circ),\qquad
 \widetilde y_{\ell_S}:=\min\left\{1,
       \frac{k_S+1-y(S_H)-y_{c_S}}2\right\}
       \quad(S\in\mathcal S).
\]
These coordinates satisfy the box and degree bounds. Since loops
cross no cut, the Blossom inequalities are unchanged, so
$\widetilde y\in{P_{\mathcal R}}$. For fixed $y$, this choice makes
each loop coordinate as large as possible and therefore minimizes
$\Phi_H$. Substituting these values and using
$y(R_H)=\sum_{S\in\mathcal S}y(S_H)$ gives
\[
\begin{aligned}
 \Phi_{G,\mathcal S}(y)
 &=\Phi_H(\widetilde y)\\
 &=m+y(N_H)-\sum_{S\in\mathcal S}y(S_H)
   +\sum_{S\in\mathcal S}
     \max\{0,\,y(S_H)+y_{c_S}-k_S+1\}\\
 &=m-\sum_{S\in\mathcal S}(k_S-1)+y(N_H)
   +\sum_{S\in\mathcal S}\max\{y_{c_S},\,k_S-1-y(S_H)\}.
\end{aligned}
\]

\begin{corollary}\label{cor:repair-nonloop-bound}
For every $y\in{\mathcal P^\circ}$, there is a distribution of tours
of $G$ whose expected length is at most $\Phi_{G,\mathcal S}(y)$.
Moreover,
\[
 \operatorname{tsp}(G)
 \le\min_{y\in{\mathcal P^\circ}(G,\mathcal S)}\Phi_{G,\mathcal S}(y)
 =\text{\LPtwo}_{\mathrm{OPT}}(G,\mathcal S).
\]
\end{corollary}
\begin{proof}
Apply Theorem~\ref{thm:framework-2-bound} to $\widetilde y$.
Every point of ${P_{\mathcal R}}$ projects to a point of
${\mathcal P^\circ}$, and the displayed choice minimizes the tour
bound for each fixed projection. This proves the equality of
the two optimal values.
\end{proof}

\begin{proposition}[Comparison of the two formulations]
\label{prop:framework-containment}
Every $x\in \Pone$ extends to a point $\widetilde x\in {P_{\mathcal R}}$ by setting
$\widetilde x_a:=x_a$ for $a\in A^{\mathrm{star}}$ and $\widetilde x_{\ell_S}:=1$ for
all $S\in\mathcal S$. Its certificate satisfies
\[
 \Phi_H(\widetilde x)=\PhiOne_{G,\mathcal S}(x).
\]
Consequently, whenever $\Pone$ is nonempty,
\[
 \text{\LPtwo}_{\mathrm{OPT}}(G,\mathcal S)
 \le \text{\LPone}_{\mathrm{OPT}}(G,\mathcal S).
\]
\end{proposition}
\begin{proof}
The auxiliary degree bound increases from $k_S-1$ to $k_S+1$,
exactly the contribution of the added loop. The Blossom
inequalities do not change, since loops belong to no cut. The
added loops each have value one, so their terms
$2(1-\widetilde x_{\ell_S})$ are zero. Hence
$\Phi_H(\widetilde x)=\PhiOne_{G,\mathcal S}(x)$.
\end{proof}

The next example shows that the \LPtwo{} can give a strictly better
tour bound than the \LPone{} for the same \textsc{Star} removable system.

\subsection{Comparing the two LPs with the circulation method}
\label{subsec:four-vertex-separation}

In this comparison, both the M\"omke--Svensson circulation method
and \hyperref[proc:algorithm]{\ALG} construct their DFS tree by
running greedy DFS on the same optimal subtour solution. We use
the same root and tie-breaking choices for both methods.
The graph in Figure~\ref{fig:four-vertex-separation} has $23$ vertices
and $30$ edges. The displayed vector $x$ is a positive optimal extreme
point of the subtour LP, with $x_e\in\{1/2,1\}$ and $x(E)=26$.
We root the DFS at $3$ and explore the children from left to right
as drawn. The resulting stars are
\[
\begin{aligned}
 S_0&=\{(0,2),(0,14),(0,16),(0,20)\},&
 S_1&=\{(1,12),(1,13)\},\\
 S_3&=\{(3,11),(2,3),(3,4)\},&
 S_6&=\{(6,7),(5,6)\},\\
 S_8&=\{(8,21),(8,9)\},
\end{aligned}
\]
where $(u,v)$ denotes the edge joining $u$ and $v$.
By Lemma~\ref{lem:DFS-removability},
$\mathcal S=\{S_0,S_1,S_3,S_6,S_8\}$ is a \textsc{Star} removable system.

For the M\"omke--Svensson method, let $f^*$ be the integral
circulation displayed in Figure~\ref{fig:separation-ms-flow},
which minimizes $C(f)$ from \eqref{eq:MS-circulation-cost}.
Write $\MSopt(f^*)$ for the minimum tour length
obtained by optimizing the parity correction on its support
with the fixed MS removable pairing specified below
\cite[proof of Theorem~12.3]{TraubVygen}; see also
\eqref{eq:MS-optimized-parity}. For these choices,
\[
 \begin{aligned}
 {\text{\SubtourLP}(G)}=26
 &<\text{\LPtwo}_{\mathrm{OPT}}(G,\mathcal S)=27\\
 &<\text{\LPone}_{\mathrm{OPT}}(G,\mathcal S)=28\\
 &<\MSopt(f^*)=29\\
 &<\frac43n+\frac23C(f^*)-\frac23=\frac{92}{3}.
 \end{aligned}
\]
Figure~\ref{fig:separation-lp-tours} shows tours attaining the two LP
bounds.

The minimum circulation cost is $1$.
Figure~\ref{fig:separation-ms-flow} shows the edges retained by
$f^*$. For this support, use the MS removable pairs
\[
 \{(0,14),(0,16)\},\qquad
 \{(1,12),(1,13)\},\qquad
 \{(6,7),(5,6)\}.
\]
The best parity correction gives the tour of length $29$ shown
in Figure~\ref{fig:separation-ms-tour}.
Table~\ref{tab:separation-comparison} also gives the standard MS
bound for this circulation.

\begin{table}[!t]
\centering
\begin{tabular}{lc}
\hline
Quantity & Value\\
\hline
{Subtour optimum $\text{\SubtourLP}(G)$} & $26$\\
\LPtwo, \ALG{} & $27$\\
\LPone, \eqref{eq:framework-1-lp} & $28$\\
MS with optimized parity correction, $\MSopt(f^*)$ & $29$\\
MS bound, \eqref{eq:MS-support-vs-cost} &
 $\dfrac43n+\dfrac23C(f^*)-\dfrac23=\dfrac{92}{3}\approx30.667$\\[3pt]
\hline
\end{tabular}
\caption{The two LP optima for the specified DFS tree and the optimized MS
tour length on the displayed support. The standard MS estimate gives
an integer bound of $30$.}
\label{tab:separation-comparison}
\end{table}

\begingroup
\tikzset{
 ex vertex/.style={circle,draw=black,fill=white,minimum size=4.8mm,
   inner sep=0pt,line width=.65pt,font=\small},
 ex unit/.style={draw=blue!65!black,line width=1.35pt},
 ex half/.style={draw=orange!90!black,line width=1.35pt},
 ex tree/.style={draw=blue!65!black,line width=1.05pt},
 ex back/.style={draw=red!70!black,line width=1.05pt},
 ex plain/.style={draw=black!60,line width=1.05pt},
 ex doubled/.style={draw=blue!70!black,line width=1.2pt},
 ex absent/.style={draw=gray!55,dashed,line width=.65pt},
 ex star0/.style={draw=orange!90!black,line width=1.55pt},
 ex star1/.style={draw=teal!85!black,line width=1.55pt},
 ex star3/.style={draw=violet!85!black,line width=1.55pt},
 ex star6/.style={draw=magenta!80!black,line width=1.55pt},
 ex star8/.style={draw=cyan!65!black,line width=1.55pt},
 ex lab/.style={font=\small,fill=white,inner sep=1pt},
 ex panel/.style={x=.57cm,y=.79cm,font=\small}}

\newcommand{\exnodes}{
 \foreach \v/\xx/\yy in {
  3/0/12,11/0/11.2,0/0/10.4,14/0/9.6,6/0/8.8,7/0/8,
  15/-1.5/7.2,1/-1.5/6.4,12/-2.5/5.6,4/-2.5/4.8,
  5/-2.5/4,13/-2.5/3.2,19/-.5/5.6,18/-.5/4.8,
  17/-.5/4,16/-.5/3.2,8/1.5/7.2,21/1.5/6.4,
  10/1.5/5.6,20/.7/4.9,22/2.5/4.8,9/2.5/4,2/2.5/3.2}{\node[ex vertex] (v\v) at (\xx,\yy) {$\v$};}}
\newcommand{\exedges}[2]{\foreach \u/\v in {#2}{\draw[#1] (v\u)--(v\v);}}
\newcommand{\exdouble}[1]{\foreach \u/\v in {#1}{\draw[ex doubled] (v\u) to[bend left=27] (v\v);
  \draw[ex doubled] (v\u) to[bend right=27] (v\v);}}
\newcommand{\excurve}[4]{\draw[preaction={draw=white,line width=2.7pt},#1] #2;
 \if\relax\detokenize{#4}\relax\else
  \path #2 node[ex lab,pos=#3] {#4};\fi}
\newcommand{\exA}[2]{\excurve{#1}{(v2)
 .. controls (6.3,3.2) and (6.3,10.4) .. (v0)}{.58}{#2}}
\newcommand{\exG}[2]{\excurve{#1}{(v16)
 .. controls (-.9,2.05) and (-4.1,2.05) .. (-4.7,3.8)
 .. controls (-5.7,5.3) and (-5.7,10.4) .. (v0)}{.58}{#2}}
\newcommand{\exH}[2]{\excurve{#1}{(v20)
 .. controls (3.5,5.2) and (3.5,10.4) .. (v0)}{.55}{#2}}
\newcommand{\exD}[2]{\excurve{#1}{(v13)
 .. controls (-4,3.2) and (-4,6.4) .. (v1)}{.5}{#2}}
\newcommand{\exRootA}[2]{\excurve{#1}{(v2)
 .. controls (7.4,3.2) and (7.4,12) .. (v3)}{.68}{#2}}
\newcommand{\exRootB}[2]{\excurve{#1}{(v4)
 .. controls (-7,4.8) and (-7,12) .. (v3)}{.58}{#2}}
\newcommand{\exSix}[2]{\excurve{#1}{(v5)
 .. controls (-5,4) and (-5,8.8) .. (v6)}{.56}{#2}}
\newcommand{\exEight}[2]{\excurve{#1}{(v9)
 .. controls (4.1,4) and (4.1,7.2) .. (v8)}{.5}{#2}}
\newcommand{\exalltree}[1]{\exedges{#1}{3/11,11/0,0/14,14/6,
 6/7,7/15,15/1,1/12,12/4,4/5,5/13,1/19,19/18,18/17,
 17/16,7/8,8/21,21/10,10/20,10/22,22/9,9/2}}

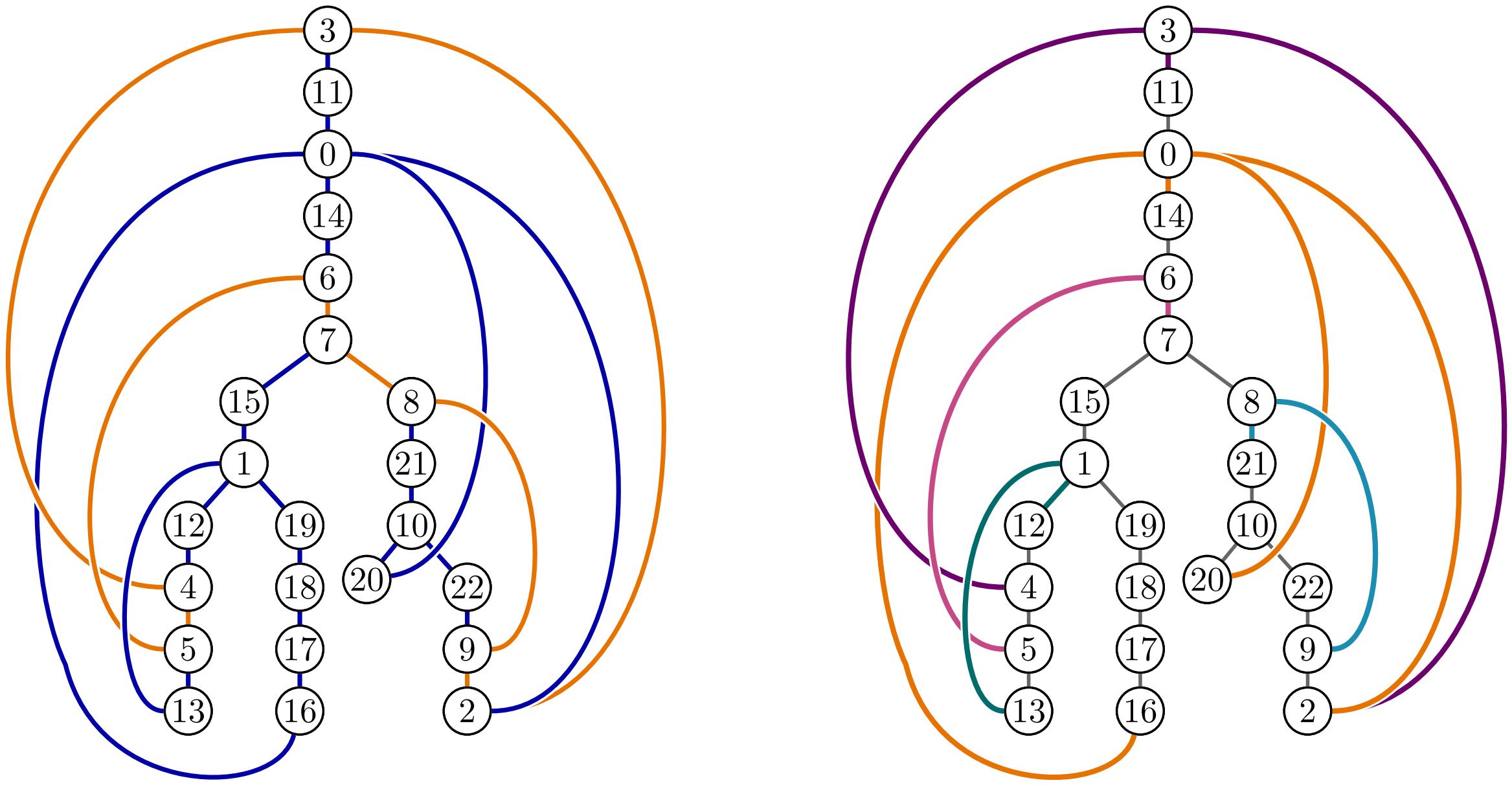
\begin{figure}[!htbp]
\centering
\begin{subfigure}[t]{.48\textwidth}\centering
\begin{tikzpicture}[ex panel]
\path[use as bounding box] (-5.9,1.95) rectangle (6.2,12.4);
\exnodes
\exedges{ex unit}{3/11,11/0,0/14,14/6,7/15,15/1,1/12,
 12/4,5/13,1/19,19/18,18/17,17/16,8/21,21/10,10/20,10/22,22/9}
\draw[ex half] (v6)--(v7);
\draw[ex half] (v4)--(v5);
\draw[ex half] (v7)--(v8);
\draw[ex half] (v9)--(v2);
\exG{ex unit}{}\exRootB{ex half}{}
\exRootA{ex half}{}\exA{ex unit}{}
\exSix{ex half}{}\exD{ex unit}{}
\exH{ex unit}{}\exEight{ex half}{}
\exnodes
\end{tikzpicture}
\caption{Optimal extreme point, ${\text{\SubtourLP}(G)}=26$.
\textcolor{blue!65!black}{Blue edges: $x_e=1$};
\textcolor{orange!90!black}{orange edges: $x_e=1/2$}.}
\label{fig:separation-input}
\end{subfigure}\hfill
\begin{subfigure}[t]{.48\textwidth}\centering
\begin{tikzpicture}[ex panel]
\path[use as bounding box] (-5.9,1.95) rectangle (6.2,12.4);
\exnodes
\exedges{ex plain}{11/0,14/6,7/15,15/1,12/4,4/5,5/13,
 1/19,19/18,18/17,17/16,7/8,21/10,10/20,10/22,22/9,9/2}
\exedges{ex star3}{3/11}\exedges{ex star0}{0/14}
\exedges{ex star1}{1/12}\exedges{ex star6}{6/7}
\exedges{ex star8}{8/21}
\exG{ex star0}{}\exRootB{ex star3}{}\exRootA{ex star3}{}
\exA{ex star0}{}\exSix{ex star6}{}\exD{ex star1}{}
\exH{ex star0}{}\exEight{ex star8}{}
\exnodes
\end{tikzpicture}
\caption{The stars \textcolor{orange!90!black}{$S_0$},
 \textcolor{teal!85!black}{$S_1$}, \textcolor{violet!85!black}{$S_3$},
 \textcolor{magenta!80!black}{$S_6$}, \textcolor{cyan!65!black}{$S_8$}.}
\label{fig:separation-stars}
\end{subfigure}
\caption{The subtour solution and the removable stars for the greedy DFS
rooted at $3$. Straight edges belong to the DFS tree; curved edges
are back edges. Each star has its own color in (b).}
\label{fig:four-vertex-separation}
\end{figure}

\begin{figure}[!htbp]
\centering
\begin{subfigure}[t]{.48\textwidth}\centering
\begin{tikzpicture}[ex panel]
\path[use as bounding box] (-5.9,1.95) rectangle (6.2,12.4);
\exnodes
\exedges{ex plain}{3/11,11/0,7/15,15/1,1/12,12/4,4/5,5/13,
 1/19,19/18,18/17,17/16,7/8,8/21,21/10,10/22,22/9,9/2}
\exdouble{14/6,10/20,6/7}
\exG{ex plain}{}\exRootA{ex plain}{}\exD{ex plain}{}
\exnodes
\end{tikzpicture}
\caption{\LPtwo: length $27$. Two copies of $(6,7)$ reconnect $S_6$.}
\label{fig:separation-framework-2}
\end{subfigure}\hfill
\begin{subfigure}[t]{.48\textwidth}\centering
\begin{tikzpicture}[ex panel]
\path[use as bounding box] (-5.9,1.95) rectangle (6.2,12.4);
\exnodes
\exedges{ex plain}{0/14,14/6,7/15,15/1,1/12,12/4,5/13,
 1/19,19/18,18/17,17/16,7/8,8/21,21/10,10/22,22/9,9/2}
\exdouble{11/0,4/5,10/20}
\exG{ex plain}{}\exRootB{ex plain}{}\exRootA{ex plain}{}
\exSix{ex plain}{}\exD{ex plain}{}
\exnodes
\end{tikzpicture}
\caption{\LPone: length $28$. At least one edge of every star is kept.}
\label{fig:separation-framework-1}
\end{subfigure}
\caption{Tours obtained from the \textsc{Star} removable system in
Figure~\ref{fig:separation-stars}. Blue parallel arcs denote two copies
of an edge; every other displayed edge is used once.}
\label{fig:separation-lp-tours}
\end{figure}
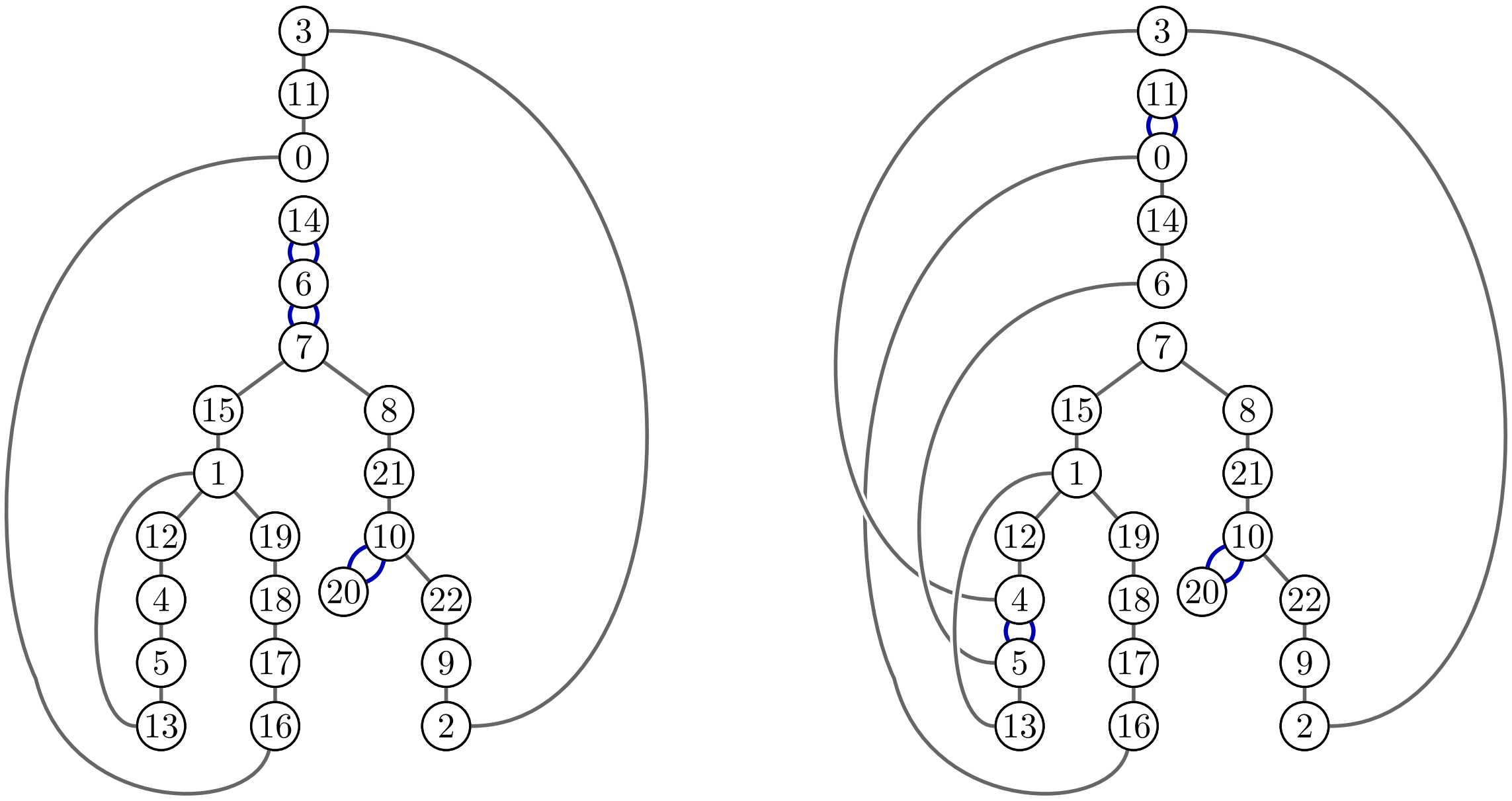

\begin{figure}[!htbp]
\centering
\begin{subfigure}[t]{.48\textwidth}\centering
\begin{tikzpicture}[ex panel]
\path[use as bounding box] (-5.9,1.95) rectangle (6.2,12.4);
\exnodes\exalltree{ex tree}
\exG{ex back}{$1$}\exRootB{ex absent}{$0$}
\exRootA{ex back}{$1$}\exA{ex absent}{$0$}
\exSix{ex back}{$1$}\exD{ex back}{$1$}
\exH{ex back}{$1$}\exEight{ex absent}{$0$}
\exnodes
\end{tikzpicture}
\caption{A minimum-cost circulation support. Back-arc flows are shown; $C(f^*)=1$.}
\label{fig:separation-ms-flow}
\end{subfigure}\hfill
\begin{subfigure}[t]{.48\textwidth}\centering
\begin{tikzpicture}[ex panel]
\path[use as bounding box] (-5.9,1.95) rectangle (6.2,12.4);
\exnodes
\exedges{ex plain}{3/11,11/0,6/7,1/12,12/4,4/5,1/19,19/18,
 18/17,17/16,7/8,8/21,21/10,10/22,22/9,9/2}
\exdouble{14/6,10/20,5/13,1/15,15/7}
\exG{ex plain}{}\exRootA{ex plain}{}\exSix{ex plain}{}
\exnodes
\end{tikzpicture}
\caption{MS: length $29$ on the displayed support.}
\label{fig:separation-ms-tour}
\end{subfigure}
\caption{A minimum-cost circulation support and its optimized tour.
Dashed edges are omitted from the support. The labels in (a) give
back-arc flow values. In (b), blue parallel arcs denote doubled edges.}
\label{fig:separation-ms-comparison}
\end{figure}
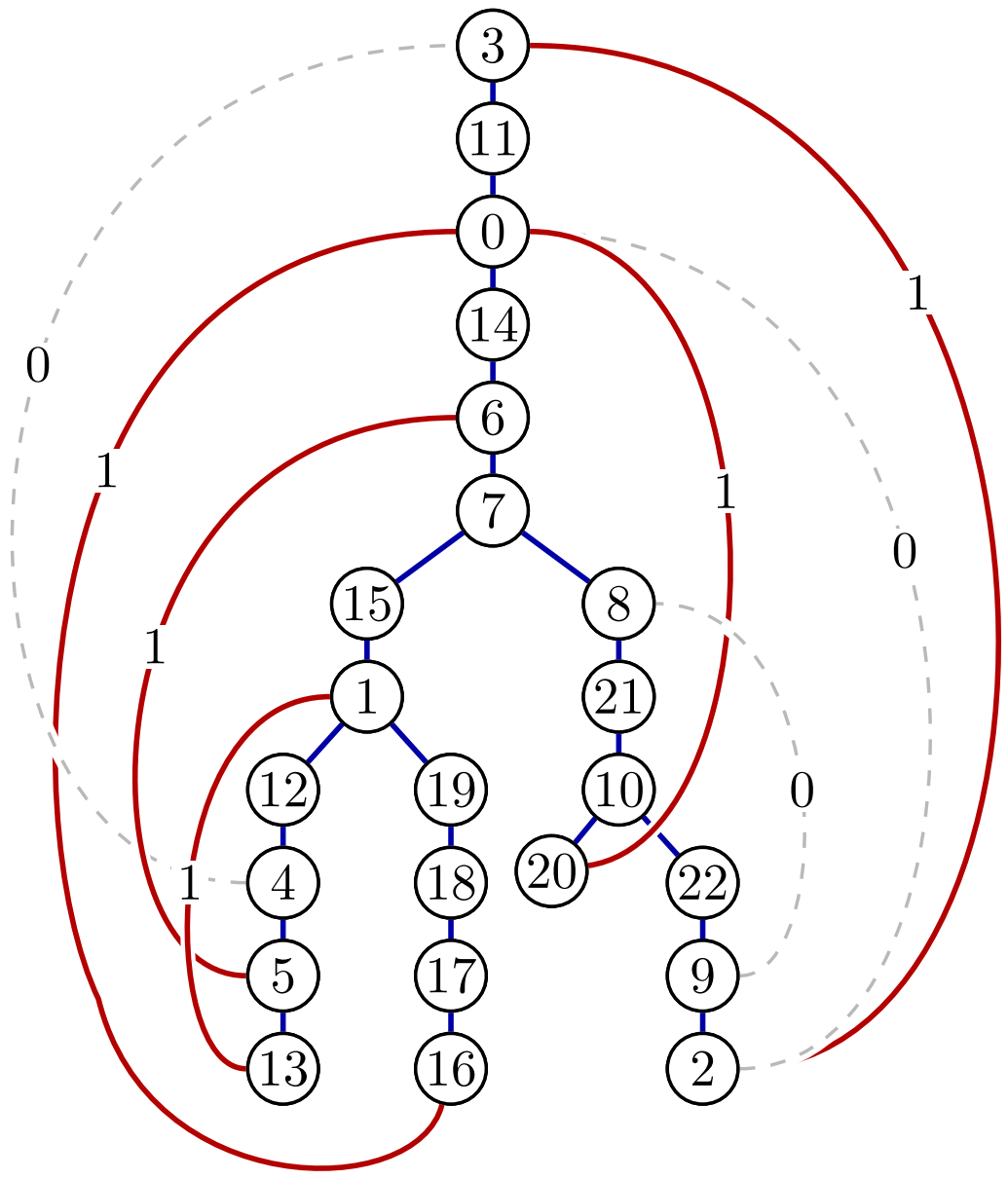
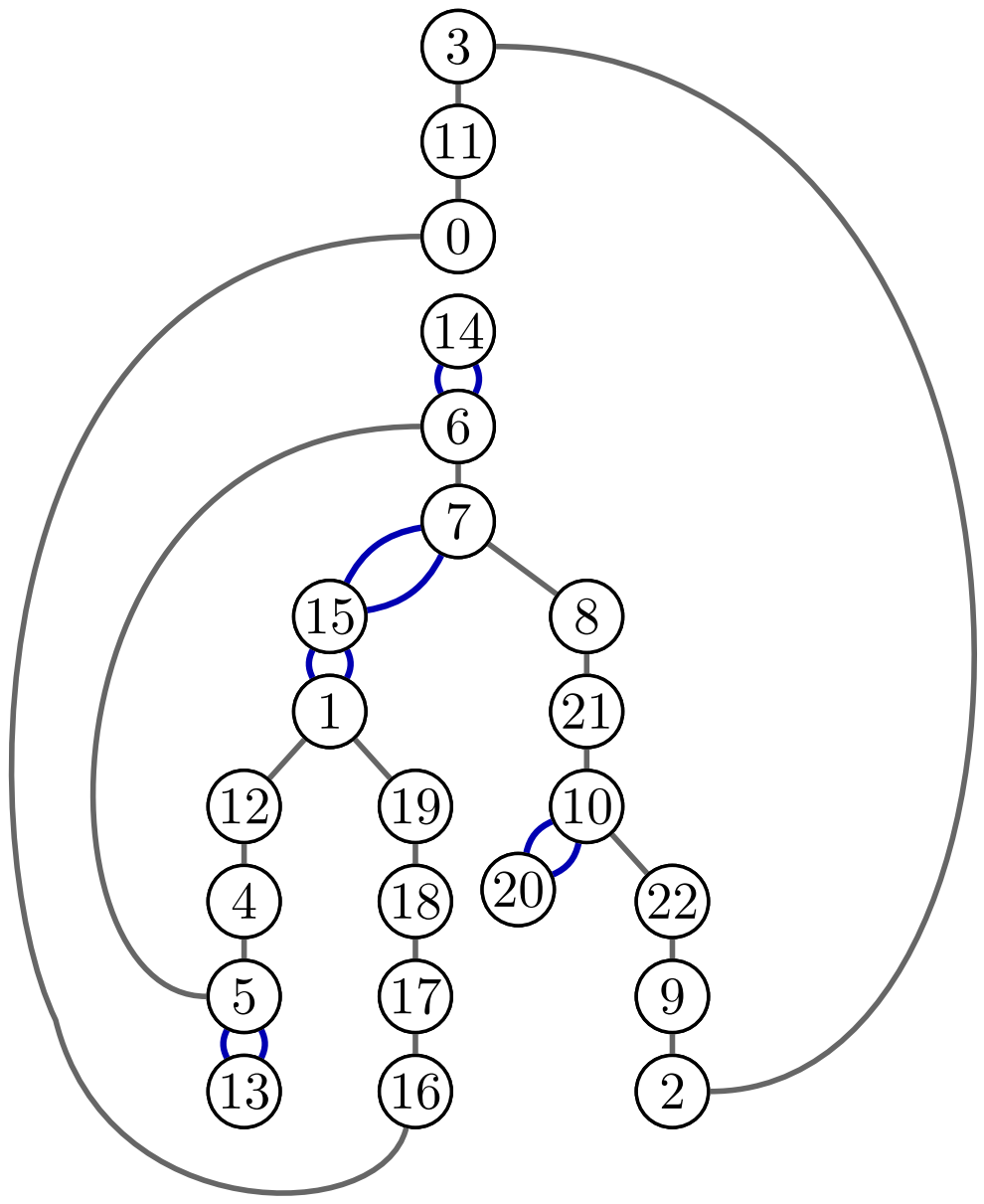
\endgroup

\FloatBarrier
\section{Removable stars from a fixed DFS tree}
\label{sec:DFS-removable-stars}

The two LPs take a valid \textsc{Star} removable system as input.
This section constructs one from any fixed DFS tree.

This construction generalizes the DFS removable-pairing method of
M\"omke and Svensson~\cite[Section~4]{MomkeSvensson}: rather than
pairing a child edge with just one back edge from that child subtree,
we group it with all such back edges ending at the same parent.

Let $G=(V,E)$ be a connected simple graph, and fix a DFS tree
$T$ rooted at $r$. Write $n:=|V|$ and let
$T_v$ denote the subtree rooted at $v$.
Every edge outside $E(T)$ joins a vertex to one of its
proper ancestors. We call these edges \emph{back edges}, with
\emph{upper endpoint} denoting the endpoint closer to $r$.

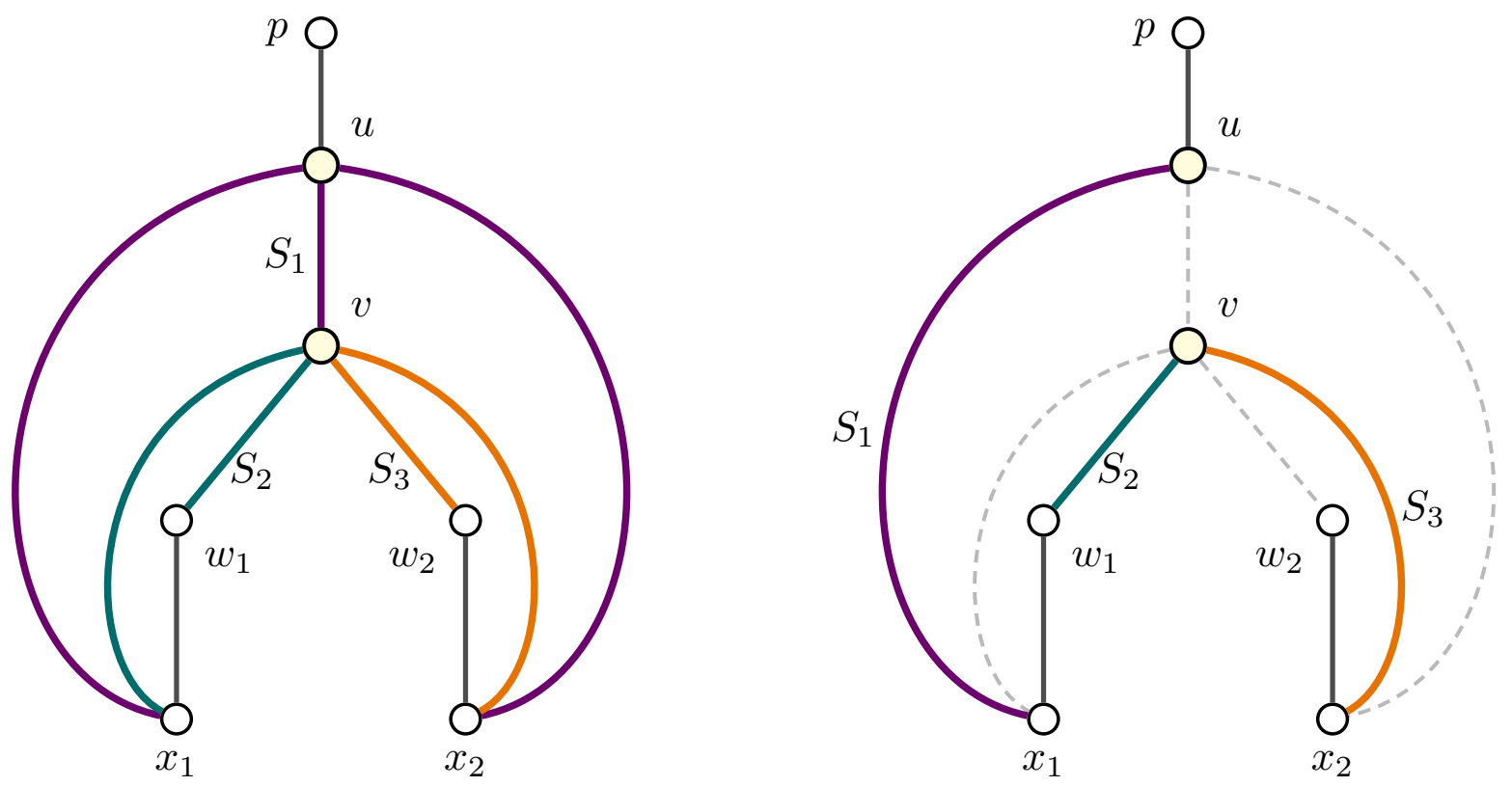
\begin{figure}[!t]
\centering
\begingroup
\tikzset{
  dfs star one/.style={draw=violet!85!black,line width=1.7pt},
  dfs star two/.style={draw=teal!85!black,line width=1.7pt},
  dfs star three/.style={draw=orange!90!black,line width=1.7pt},
  dfs outside/.style={draw=black!70,line width=1.2pt},
  dfs deleted/.style={draw=gray!55,densely dashed,line width=.9pt},
  dfs label/.style={font=\small,inner sep=2pt}
}

\begin{subfigure}[t]{.32\textwidth}
\centering
\resizebox{\linewidth}{!}{\begin{tikzpicture}[font=\small]
\path[use as bounding box] (-3.2,-2.5) rectangle (3.2,4.15);

\node[vertex,label=left:{$p$}] (p) at (0,3.7) {};
\node[center,label=above right:{$u$}] (u) at (0,2.6) {};
\node[center,label=above right:{$v$}] (v) at (0,1.1) {};
\node[vertex,label=below right:{$w_1$}] (w1) at (-1.2,-.35) {};
\node[vertex,label=below left:{$w_2$}] (w2) at (1.2,-.35) {};
\node[vertex,label=below:{$x_1$}] (x1) at (-1.2,-2) {};
\node[vertex,label=below:{$x_2$}] (x2) at (1.2,-2) {};

\draw[dfs outside] (p)--(u);
\draw[dfs outside] (w1)--(x1);
\draw[dfs outside] (w2)--(x2);

\draw[dfs star one]
  (u)--node[dfs label,left] {$S_1$}(v);
\draw[dfs star one] (u)
  .. controls (-3.1,2.15) and (-3.1,-1.6) .. (x1);
\draw[dfs star one] (u)
  .. controls (3.1,2.15) and (3.1,-1.6) .. (x2);

\draw[dfs star two]
  (v)--node[dfs label,right,pos=.75] {$S_2$}(w1);
\draw[dfs star two] (v)
  .. controls (-2.05,.65) and (-2.05,-1.55) .. (x1);

\draw[dfs star three]
  (v)--node[dfs label,left,pos=.75] {$S_3$}(w2);
\draw[dfs star three] (v)
  .. controls (2.05,.65) and (2.05,-1.55) .. (x2);
\end{tikzpicture}}
\caption{Three stars; $S_2$ and $S_3$ are centered at $v$.}
\end{subfigure}\hspace{.04\textwidth}\begin{subfigure}[t]{.32\textwidth}
\centering
\resizebox{\linewidth}{!}{\begin{tikzpicture}[font=\small]
\path[use as bounding box] (-3.2,-2.5) rectangle (3.2,4.15);

\node[vertex,label=left:{$p$}] (p) at (0,3.7) {};
\node[center,label=above right:{$u$}] (u) at (0,2.6) {};
\node[center,label=above right:{$v$}] (v) at (0,1.1) {};
\node[vertex,label=below right:{$w_1$}] (w1) at (-1.2,-.35) {};
\node[vertex,label=below left:{$w_2$}] (w2) at (1.2,-.35) {};
\node[vertex,label=below:{$x_1$}] (x1) at (-1.2,-2) {};
\node[vertex,label=below:{$x_2$}] (x2) at (1.2,-2) {};

\draw[dfs deleted] (u)--(v);
\draw[dfs deleted] (v)--(w2);
\draw[dfs deleted] (u)
  .. controls (3.1,2.15) and (3.1,-1.6) .. (x2);
\draw[dfs deleted] (v)
  .. controls (-2.05,.65) and (-2.05,-1.55) .. (x1);

\draw[dfs outside] (p)--(u);
\draw[dfs outside] (w1)--(x1);
\draw[dfs outside] (w2)--(x2);

\draw[dfs star one] (u)
  .. controls (-3.1,2.15) and (-3.1,-1.6)
  .. node[dfs label,left,pos=.48] {$S_1$}(x1);
\draw[dfs star two]
  (v)--node[dfs label,right,pos=.75] {$S_2$}(w1);
\draw[dfs star three] (v)
  .. controls (2.05,.65) and (2.05,-1.55)
  .. node[dfs label,right,pos=.45] {$S_3$}(x2);
\end{tikzpicture}}
\caption{Keep $ux_1$, $vw_1$, and $vx_2$, one from each star.}
\end{subfigure}

\caption{The DFS stars
$S_1=\{uv,ux_1,ux_2\}$,
$S_2=\{vw_1,vx_1\}$, and
$S_3=\{vw_2,vx_2\}$ have different colors.
Straight edges belong to the DFS tree; curved edges are back edges.
On the right, gray dashed edges are deleted. The remaining edges
form the spanning path $p-u-x_1-w_1-v-x_2-w_2$.}
\label{fig:DFS-star}
\endgroup
\end{figure}

For each tree edge $uv$, where $u$ is the parent of $v$, define
\begin{equation}
\label{eq:DFS-back-edge-group}
    I(u,v):=\{uw\in E\setminus E(T):
                         w\in V(T_v)\}.
\end{equation}
Whenever $I(u,v)\ne\varnothing$, form
\begin{equation}
\label{eq:DFS-full-star}
    S(u,v):=\{uv\}\cup I(u,v),
\end{equation}
centered at $u$. Let $\mathcal S$ be the family of all these stars,
and put
\[
    R:=\bigcup_{S\in\mathcal{S}}S,
    \qquad N:=E\setminus R.
\]
In particular, all back edges belong to $R$, and
\[
    N=\{uv\in E(T):I(u,v)=\varnothing\}.
\]
The construction includes the root and allows different stars
with the same center, one for each applicable child subtree.
See Figure~\ref{fig:DFS-star}.

\begin{lemma}[DFS removability]
\label{lem:DFS-removability}
\label{lem:MS-full-star-removability}
The family $\mathcal{S}$ is a \textsc{Star} removable system in $G$.
Moreover, keeping every edge of $N$ and exactly one edge of each
$S\in\mathcal{S}$ always gives a spanning tree.
\end{lemma}

\begin{proof}
Every back edge belongs to exactly one group $I(u,v)$: its upper
endpoint determines $u$, and its lower endpoint determines the
child $v$ on the tree path below $u$. Distinct groups also have
distinct tree edges. Hence the stars are pairwise edge-disjoint.

Let $D\subseteq R$ satisfy $|D\cap S|\le |S|-1$ for every
$S\in\mathcal{S}$. Enlarge $D$ to $D'\subseteq R$ so that
\[
    |D'\cap S|=|S|-1
    \qquad(S\in\mathcal{S}).
\]
It suffices to prove that $G-D'$ is connected.

We first prove the following claim:
\begin{equation}
\label{eq:DFS-cycle-claim}
    \text{Every cycle of $G$ contains two edges of the same star.}
\end{equation}

Let $C$ be a cycle. Since $T$ is a tree, $C$ contains a
back edge. Choose a back edge $uw\in E(C)$ whose upper endpoint
$u$ has minimum depth among the upper endpoints of all back edges
in $C$. Let $v$ be the child of $u$ on the tree path from $u$ to $w$.
Then $uw\in I(u,v)$ crosses the cut $\delta_G(V(T_v))$.
A cycle crosses every cut an even number of times, so there is
another edge
\[
    f\in E(C)\cap\delta_G(V(T_v)),
    \qquad f\ne uw.
\]
We show that $f$ also belongs to $S(u,v)$.

\smallskip\noindent
\textbf{Case 1: $f$ is a tree edge.}
The only tree edge crossing $\delta_G(V(T_v))$ is $uv$.
Thus $f=uv$, and both $f$ and $uw$ belong to $S(u,v)$.

\smallskip\noindent
\textbf{Case 2: $f$ is a back edge.}
By the DFS ancestor property, its endpoint outside $T_v$ is
either $u$ or a proper ancestor of $u$. The latter would
contradict the choice of $uw$, since $f$ also belongs to $C$.
Therefore, $f$ joins $u$ to a vertex of $T_v$, so
$f\in I(u,v)$. Again, $f$ and $uw$ belong to $S(u,v)$.

This proves \eqref{eq:DFS-cycle-claim}.
Figure~\ref{fig:DFS-cycle-claim} shows the two cases for $f$.

\begin{figure}[!t]
\centering
\begin{subfigure}[t]{.46\textwidth}
\centering
\begin{tikzpicture}[scale=.9]
\draw[gray,dashed,rounded corners,fill=panelbg]
 (-1.2,-1.5) rectangle (1.2,.7);
\node[center,label=right:{$u$}] (u) at (0,1.8) {};
\node[vertex,label=right:{$v$}] (v) at (0,.2) {};
\node[vertex,label=below:{$w$}] (w) at (0,-1) {};
\draw[blueedge] (u)--node[right,font=\small] {$f=uv$}(v)--(w);
\draw[rededge] (u) to[out=180,in=180,looseness=1.4]
 node[left,font=\small] {$uw$}(w);
\node[font=\small] at (0,-1.85) {$V(T_v)$};
\end{tikzpicture}
\caption{The second crossing edge is the tree edge $uv$.}
\end{subfigure}\hfill
\begin{subfigure}[t]{.46\textwidth}
\centering
\begin{tikzpicture}[scale=.9]
\draw[gray,dashed,rounded corners,fill=panelbg]
 (-1.35,-1.5) rectangle (1.35,.7);
\node[center,label=right:{$u$}] (u) at (0,1.8) {};
\node[vertex,label=right:{$v$}] (v) at (0,.2) {};
\node[vertex,label=below:{$w$}] (w) at (-.8,-1) {};
\node[vertex,label=below:{$x$}] (x) at (.8,-1) {};
\draw[blueedge,dashed,line width=.7pt] (u)--(v);
\draw[blueedge] (w)--(v)--(x);
\draw[rededge] (u) to[out=180,in=180]
 node[left,font=\small] {$uw$}(w);
\draw[rededge] (u) to[out=0,in=0]
 node[right,font=\small] {$f=ux$}(x);
\node[font=\small] at (0,-1.85) {$V(T_v)$};
\end{tikzpicture}
\caption{The second crossing edge is a back edge ending at $u$.}
\end{subfigure}
\caption{The two cases in the proof of \eqref{eq:DFS-cycle-claim}.
The solid edges form the cycle $C$, and the shaded region contains
$V(T_v)$. In both panels, $uw$ and $f$ cross its boundary and
belong to $S(u,v)$. The dashed tree edge in the right panel
does not belong to $C$.}
\label{fig:DFS-cycle-claim}
\end{figure}
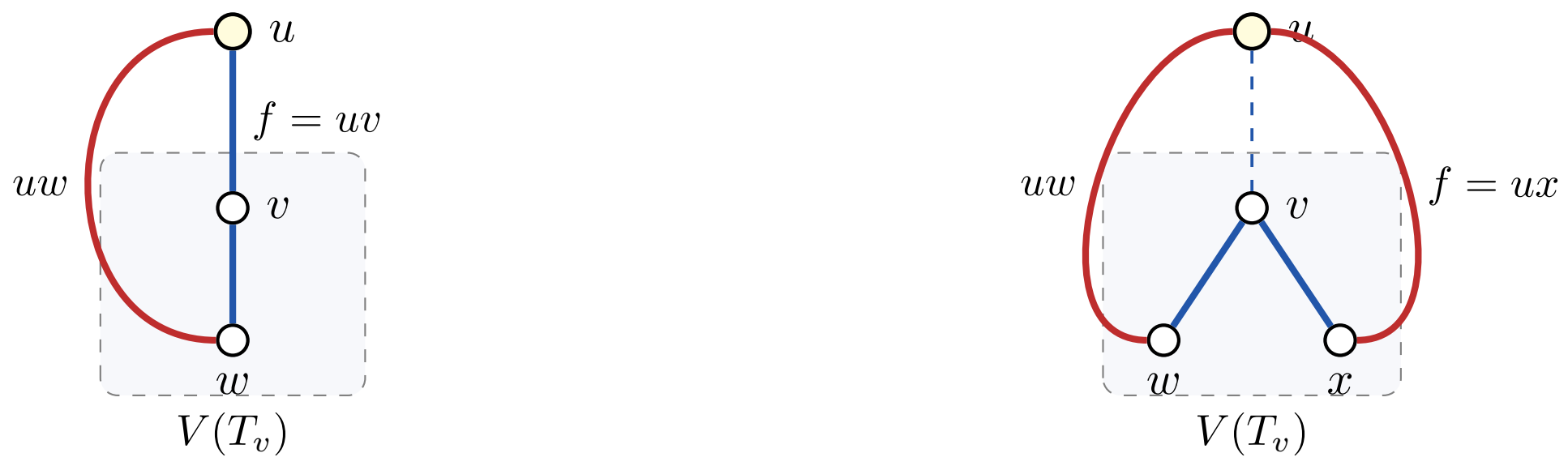

Exactly one edge of each star remains in $G-D'$. By
\eqref{eq:DFS-cycle-claim}, this graph contains no cycle.
Each star contains one tree edge and $|S|-1$ back edges,
and every back edge belongs to exactly one star. Hence
\[
    |E|=n-1+\sum_{S\in\mathcal S}(|S|-1).
\]
Since $D'$ contains exactly $|S|-1$ edges from each star,
\[
    |E(G-D')|
    =|E|-\sum_{S\in\mathcal S}(|S|-1)
    =n-1.
\]
Thus $G-D'$ is a forest on $n$ vertices with $n-1$ edges,
so it is a spanning tree. Since $D\subseteq D'$, the graph
$G-D$ is also connected. Therefore, $\mathcal S$ is a
\textsc{Star} removable system.
\end{proof}

\FloatBarrier
\section{The M\"omke--Svensson framework}
\label{sec:MS-comparison}

In this section, we recall the M\"omke--Svensson framework.
\subsection{\textsc{MS} removable pairings and the \texorpdfstring{$T$}{T}-join polytope}

We follow the auxiliary-graph construction in
\cite[Section~12.2]{TraubVygen}.
Let $G=(V,E)$ be a $2$-vertex-connected simple graph.
An \emph{\textsc{MS} removable pairing} $(R_{\mathrm{MS}},\mathcal P)$ consists
of $R_{\mathrm{MS}}\subseteq E$ and pairwise edge-disjoint pairs
$P\subseteq R_{\mathrm{MS}}$, each centered at a vertex of degree
at least three, such that $G-D$ is connected whenever
\[
 D\subseteq R_{\mathrm{MS}},\qquad |D\cap P|\le1
 \quad(P\in\mathcal P).
\]
Edges of $R_{\mathrm{MS}}$ outside the pairs may be deleted without any pair constraint.

For each pair $P=\{vw,vw'\}$ centered at $v$, introduce a
vertex $v_P$, replace $vw,vw'$ by $v_Pw,v_Pw'$, and add
the connector $vv_P$. This is the $k_S=2$ case
of Figure~\ref{fig:star-splitting}. Denote the resulting graph by
$Q$, and the representative of $e\in E$ by $e^Q$.
The proof of \cite[Theorem~12.3]{TraubVygen} shows that $Q$ is
$2$-edge-connected; each new vertex $v_P$ has degree three.

\begin{lemma}[The one-third convex combination; {\cite[proof of Theorem~12.3]{TraubVygen}}]
\label{lem:MS-distribution}
\label{lem:MS-convex-combination}
{The vector $y:=\tfrac13\mathbf{1}_{E(Q)}$
belongs to the $\odd(Q)$-join polytope.} It admits a decomposition
\begin{equation}
\label{eq:MS-join-decomposition}
 y=\sum_{i=1}^{t}\lambda_i\mathbf{1}_{K_i},
 \qquad \lambda_i>0,\qquad \sum_{i=1}^{t}\lambda_i=1,
\end{equation}
where every $K_i$ is an $\odd(Q)$-join and
$|K_i\cap\delta_Q(v_P)|=1$ for every $P\in\mathcal P$.
The projected sets $J_i:=\{e\in E:e^Q\in K_i\}$ satisfy
\begin{equation}
\label{eq:MS-projected-decomposition}
 \odd(J_i)=\odd(G),\qquad |J_i\cap P|\le1,
 \qquad \sum_{i=1}^{t}\lambda_i\mathbf{1}_{J_i}
       =\tfrac13\mathbf{1}_E.
\end{equation}
\end{lemma}

The coefficients $\lambda_i$ define the distribution: choose index
$i$ with probability $\lambda_i$.
Deleting $J_i\cap R_{\mathrm{MS}}$ and doubling
$J_i\setminus R_{\mathrm{MS}}$ gives a tour $T_i$.
Consequently, writing $m:=|E|$,
\begin{equation}
\label{eq:MS-one-third-bound}
 \sum_i\lambda_i|E(T_i)|
 =m-\tfrac13|R_{\mathrm{MS}}|
     +\tfrac13|E\setminus R_{\mathrm{MS}}|
 =\tfrac43m-\tfrac23|R_{\mathrm{MS}}|.
\end{equation}
At least one of these tours has length at most this bound.

\subsection{The circulation construction}
\label{subsec:MS-full-stars}

Suppose that $G$ is not a cycle, and fix a DFS tree $T$
rooted at a vertex $r$ with $\deg_G(r)\ge3$.

We use the circulation network described in
\cite[Section~12.4]{TraubVygen}. Orient the tree edges away
from $r$ and subdivide each arc $uv$ by a vertex $\zeta_{uv}$,
giving the arcs $(u,\zeta_{uv})$ and $(\zeta_{uv},v)$.
Replace each back edge $uw\in I(u,v)$ by the arc
$(w,\zeta_{uv})$.

Set the lower bound to one on $(u,\zeta_{uv})$ when
$u\ne r$, and to zero on every other arc.
A feasible circulation assigns nonnegative flows satisfying
these lower bounds and flow conservation.

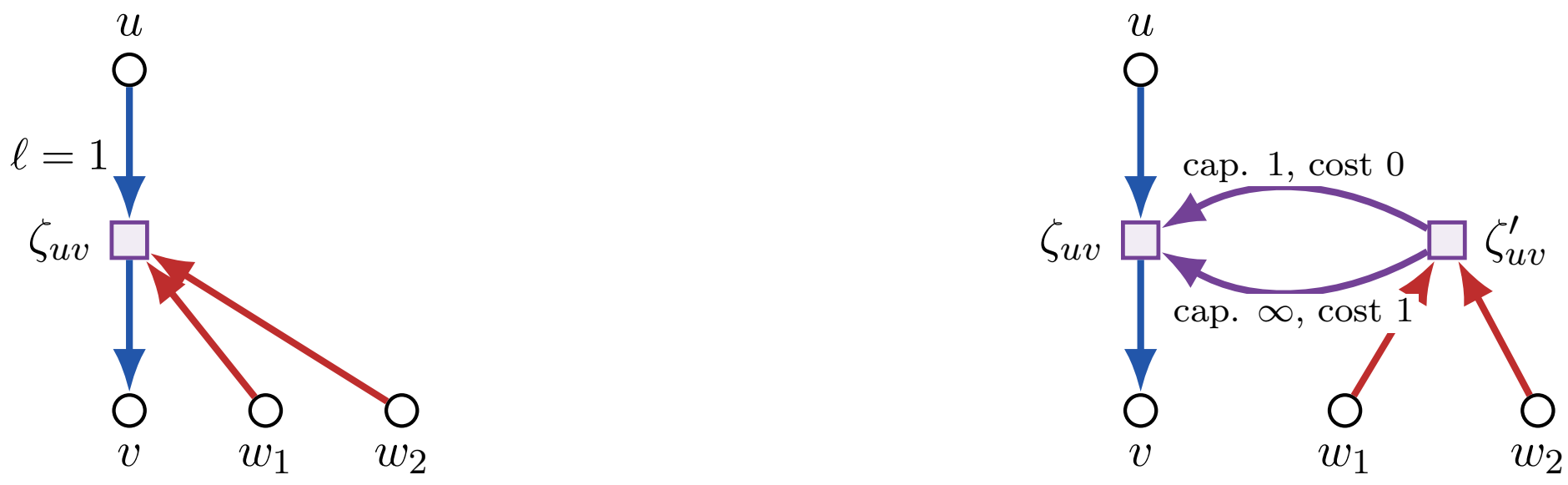
\begin{figure}[!t]
\centering
\begin{subfigure}[t]{.43\textwidth}
\centering
\begin{tikzpicture}[scale=.9,>=Latex]
\node[vertex,label=above:{$u$}] (u) at (0,2.6) {};
\node[auxv,rectangle,label=left:{$\zeta_{uv}$}] (z) at (0,1.1) {};
\node[vertex,label=below:{$v$}] (v) at (0,-.4) {};
\node[vertex,label=below:{$w_1$}] (a) at (1.2,-.4) {};
\node[vertex,label=below:{$w_2$}] (b) at (2.4,-.4) {};
\draw[blueedge,->] (u)--node[left,font=\small] {$\ell=1$}(z);
\draw[blueedge,->] (z)--(v);
\draw[rededge,->] (a)--(z); \draw[rededge,->] (b)--(z);
\end{tikzpicture}
\caption{A nonroot group in the flow network.}
\end{subfigure}\hfill
\begin{subfigure}[t]{.53\textwidth}
\centering
\begin{tikzpicture}[scale=.9,>=Latex]
\node[vertex,label=above:{$u$}] (u) at (0,2.6) {};
\node[auxv,rectangle,label=left:{$\zeta_{uv}$}] (z) at (0,1.1) {};
\node[auxv,rectangle,label=right:{$\zeta'_{uv}$}] (zp) at (2.7,1.1) {};
\node[vertex,label=below:{$v$}] (v) at (0,-.4) {};
\node[vertex,label=below:{$w_1$}] (a) at (1.8,-.4) {};
\node[vertex,label=below:{$w_2$}] (b) at (3.5,-.4) {};
\draw[blueedge,->] (u)--(z); \draw[blueedge,->] (z)--(v);
\draw[rededge,->] (a)--(zp); \draw[rededge,->] (b)--(zp);
\draw[auxedge,->] (zp) to[bend right=30]
 node[above,font=\scriptsize,fill=white,inner sep=1pt] {cap. $1$, cost $0$}(z);
\draw[auxedge,->] (zp) to[bend left=30]
 node[below,font=\scriptsize,fill=white,inner sep=1pt] {cap. $\infty$, cost $1$}(z);
\end{tikzpicture}
\caption{Linearizing the zero-cost first-unit rule.}
\end{subfigure}
\caption{The local circulation construction of Section~12.4 of
\cite{TraubVygen}. At the root, the first tree arc has lower bound zero.
}
\label{fig:MS-flow}
\end{figure}

For each $uv\in E(T)$, write
\[
 I^{\mathrm{flow}}_{uv}
 :=\{(w,\zeta_{uv}):uw\in I(u,v)\}.
\]
The first unit entering $\zeta_{uv}$ through these back arcs
is free, and each additional unit has cost one. Thus the
cost of a feasible circulation $f$ is
\begin{equation}
\label{eq:MS-circulation-cost}
 C(f):=\sum_{uv\in E(T)}
       \max\{f(I^{\mathrm{flow}}_{uv})-1,0\}.
\end{equation}
Figure~\ref{fig:MS-flow} shows how this cost is linearized.

For an integral feasible circulation $f$, let
\[
 E_f:=E(T)\cup
       \{e\in E\setminus E(T):f_e>0\},
 \qquad G_f:=(V,E_f).
\]
By \cite[Lemma~12.8]{TraubVygen}, $G_f$ is $2$-vertex-connected.
For every nonroot group with a retained back edge, pair $uv$ with
one such edge. Leave the root tree edge unpaired, as in the book.
Let $\mathcal P_f$ be these pairs, and let $R_f$ consist of all
retained back edges and all paired tree edges. Define
\begin{equation}
\label{eq:MS-support-bound}
 \Lambda_{\mathrm{MS}}(f):=\tfrac43|E_f|-\tfrac23|R_f|.
\end{equation}
For this support and fixed \textsc{MS} removable pairing, let
\begin{equation}\label{eq:MS-optimized-parity}
 \MSopt(f)
 :=\min\left\{
 |E_f|-|J\cap R_f|+|J\setminus R_f|:
 \begin{aligned}
 &J\subseteq E_f,\quad\odd(J)=\odd(G_f),\\
 &|J\cap P|\le1\quad(P\in\mathcal P_f)
 \end{aligned}\right\}.
\end{equation}
Thus $\MSopt(f)$ is the minimum tour length obtained
by optimizing the parity correction for this support and pairing.
The one-third point in Lemma~\ref{lem:MS-convex-combination}
gives the bound $\Lambda_{\mathrm{MS}}(f)$. Optimizing the parity
correction and applying \cite[Lemma~12.8]{TraubVygen} therefore gives
\begin{equation}\label{eq:MS-support-vs-cost}
 \MSopt(f)
 \le\Lambda_{\mathrm{MS}}(f)
 \le\frac43n+\frac23C(f)-\frac23,
\end{equation}
where $n:=|V|$.

\section{From the circulation bound to the \LPone{}}
\label{sec:flow-to-framework-1}

For a fixed DFS tree, we map every feasible M\"omke--Svensson
circulation to a point of $\Pone$ with no larger tour
bound. We prove this first for integral circulations; the general
case then follows from circulation integrality.

\subsection{Restoring missing pairs}
\label{subsec:MS-support-restoration}

An integral M\"omke--Svensson circulation may omit every back
edge in $I(u,v)$, even when this set is nonempty.
For a nonroot tree edge $uv$, restoring one such edge creates
an \textsc{MS} removable pair without changing the support bound.
The next lemma gives this construction formally.

\begin{lemma}[Support restoration]
\label{lem:MS-restoration}
Let $G=(V,E)$ be a simple $2$-vertex-connected graph that is
not a cycle, and let $T$ be a DFS tree rooted at $r$ with
$\deg_G(r)\ge3$.
Let $f$ be an integral feasible circulation, with support
$G_f=(V,E_f)$ and \textsc{MS} removable pairing $(R_f,\mathcal P_f)$
as in Section~\ref{subsec:MS-full-stars}.

We can restore some of the omitted back edges so that every
tree edge $uv$ with $I(u,v)\ne\varnothing$ is paired with
an edge of $I(u,v)$ in the resulting support.
Write $G_f^+=(V,E_f^+)$ for this support and
$(R_f^+,\mathcal P_f^+)$ for its \textsc{MS} removable pairing.
Then $E_f\subseteq E_f^+\subseteq E$, the DFS tree remains $T$,
and $R_f^+$ consists of all back edges in $E_f^+$ together
with the paired tree edges. Moreover, these changes do not
increase the bound, that is:
\begin{equation}
\label{eq:MS-restoration-cancellation}
 \frac43|E_f^+|-\frac23|R_f^+|
 \le \Lambda_{\mathrm{MS}}(f).
\end{equation}
\end{lemma}

\begin{figure}[!t]
\centering
\begin{subfigure}[t]{.46\textwidth}
\centering
\begin{tikzpicture}[scale=.88]
\draw[draw=gray!60,dashed,rounded corners,fill=panelbg]
 (-1.5,-1.65) rectangle (1.5,.7);
\node[font=\small,fill=panelbg] at (0,-1.65) {$V(T_v)$};
\node[vertex,label=left:{$p$}] (p) at (0,2.7) {};
\node[center,label=right:{$u$}] (u) at (0,1.6) {};
\node[vertex,label=right:{$v$}] (v) at (0,.3) {};
\node[vertex,label=below:{$w_1$}] (a) at (-.85,-.75) {};
\node[vertex,label=below:{$w_2$}] (b) at (.85,-.75) {};
\draw[blueedge] (p)--(u);
\draw[blueedge] (u)--(v);
\draw[blueedge] (v)--(a); \draw[blueedge] (v)--(b);
\draw[oldedge] (u) to[out=185,in=160] (a);
\draw[oldedge] (u) to[out=-5,in=20] (b);
\end{tikzpicture}
\caption{No back edge of $I(u,v)$ is retained.}
\end{subfigure}\hfill\begin{subfigure}[t]{.46\textwidth}
\centering
\begin{tikzpicture}[scale=.88]
\draw[draw=gray!60,dashed,rounded corners,fill=panelbg]
 (-1.5,-1.65) rectangle (1.5,.7);
\node[font=\small,fill=panelbg] at (0,-1.65) {$V(T_v)$};
\node[vertex,label=left:{$p$}] (p) at (0,2.7) {};
\node[center,label=right:{$u$}] (u) at (0,1.6) {};
\node[vertex,label=right:{$v$}] (v) at (0,.3) {};
\node[vertex,label=below:{$w_1$}] (a) at (-.85,-.75) {};
\node[vertex,label=below:{$w_2$}] (b) at (.85,-.75) {};
\draw[blueedge] (p)--(u);
\draw[blueedge] (u)--(v);
\draw[blueedge] (v)--(a); \draw[blueedge] (v)--(b);
\draw[rededge] (u) to[out=185,in=160] (a);
\draw[oldedge] (u) to[out=-5,in=20] (b);
\end{tikzpicture}
\caption{Restore $uw_1$ and pair it with $uv$.}
\end{subfigure}
\caption{A nonroot restoration. Gray dashed edges are absent from the
support; the tree edge $uv$ is retained in both panels. Restoring $uw_1$
adds one support edge and two removable edges, so the support bound
changes by $\frac43-\frac23\cdot2=0$. The DFS tree is unchanged.}
\label{fig:MS-restoration}
\end{figure}
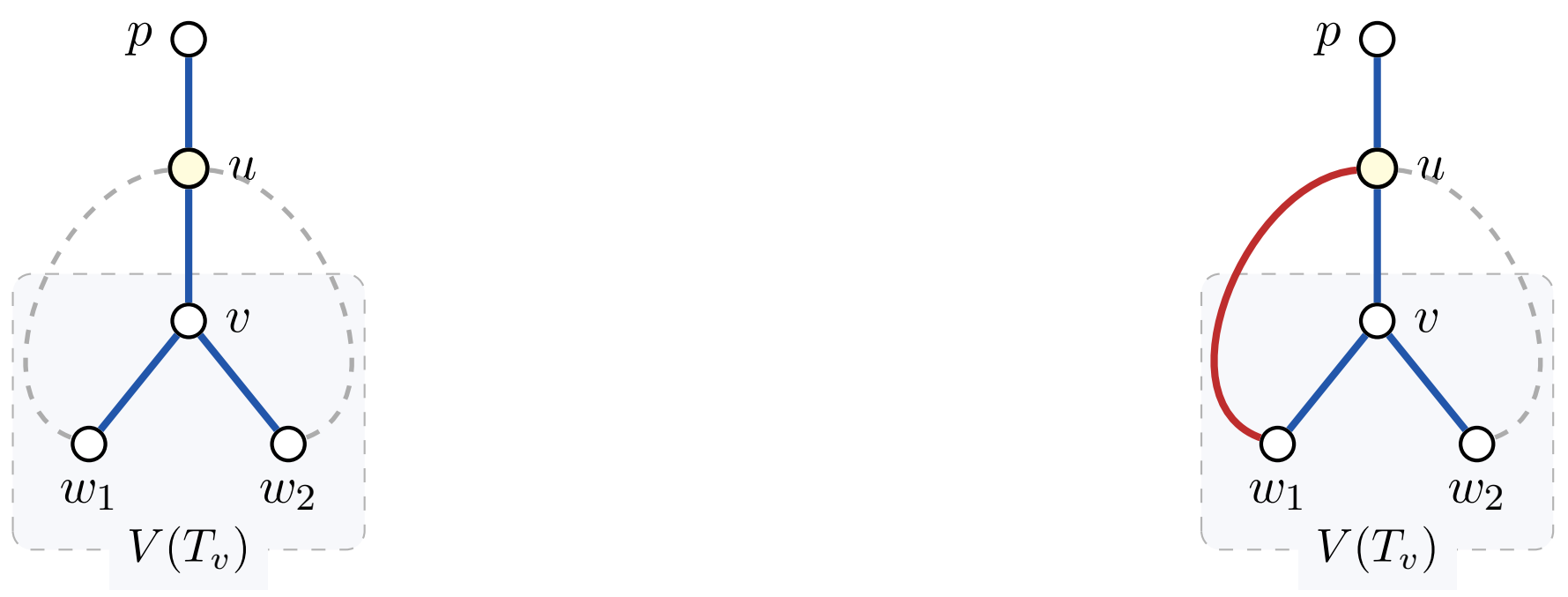

\begin{proof}
For each nonroot tree edge $uv$ with
$I(u,v)\ne\varnothing$, proceed as follows.
If an edge of $I(u,v)$ is retained, keep the
existing pair containing $uv$.
Otherwise, restore any $uw\in I(u,v)$ and add
the pair $\{uv,uw\}$.
This adds one support edge and two removable edges, so
the change in the support bound is
\[
 \frac43-\frac23\cdot2=0.
\]
Figure~\ref{fig:MS-restoration} illustrates this operation.

At the root, the tree edge is initially unpaired, while
the retained back edges are already removable
\cite[proof of Lemma~12.8]{TraubVygen}.
The support $G_f$ is $2$-vertex-connected, so at least
one root back edge is retained.
Since $\deg_G(r)\ge3$, restore a second root back edge if
necessary, then pair the root tree edge with a retained
root back edge. Restoring an edge and adding this pair
leaves the support bound unchanged; adding the pair
without restoring an edge decreases it by $2/3$.
This proves \eqref{eq:MS-restoration-cancellation}.

The pairs are edge-disjoint and their centers have degree
at least three. Removability follows from
Lemma~\ref{lem:DFS-removability}, applied to $T$
together with one back edge per pair.
\end{proof}

\subsection{Comparison with the circulation bound}
\label{subsec:MS-integral-flow-map}
\label{subsec:MS-fractional-flow-map}

To compare the \LPone{} with the circulation bound,
we first lift joins that leave an edge of every star unselected.

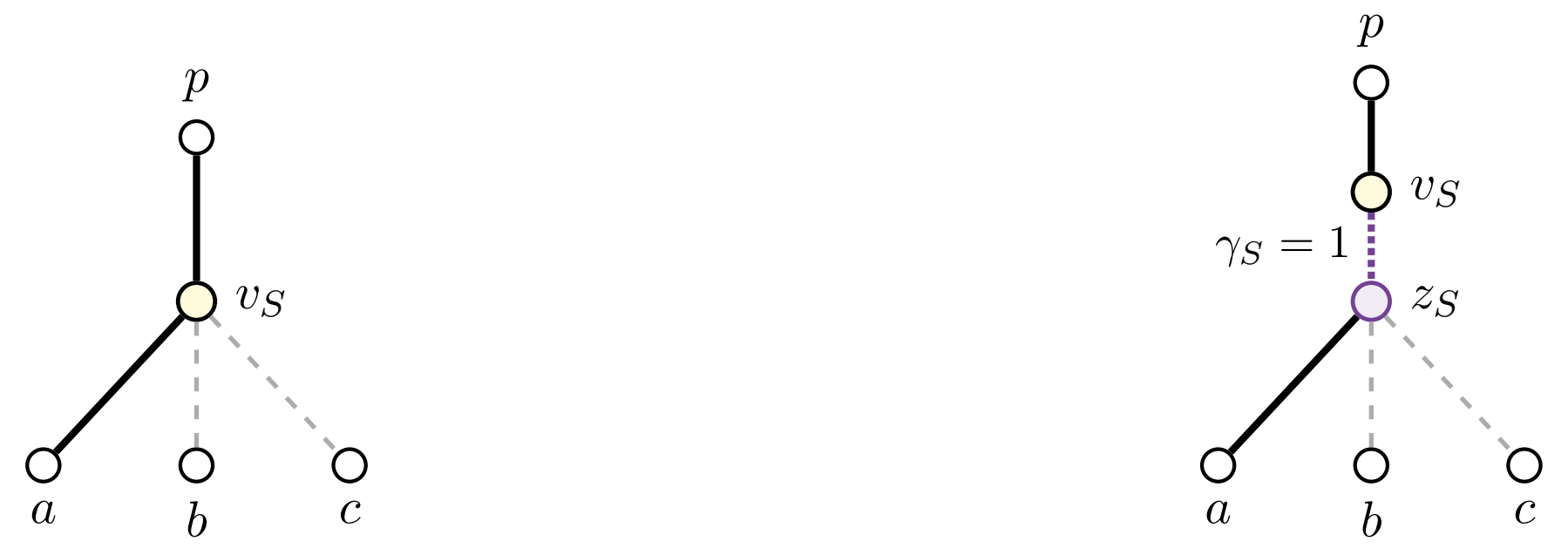
\begin{figure}[!t]
\centering
\begin{subfigure}[t]{.46\textwidth}
\centering
\begin{tikzpicture}[scale=.83]
\node[vertex,label=above:{$p$}] (p) at (0,3) {};
\node[center,label=right:{$v_S$}] (v) at (0,1.5) {};
\node[vertex,label=below:{$a$}] (a) at (-1.4,0) {};
\node[vertex,label=below:{$b$}] (b) at (0,0) {};
\node[vertex,label=below:{$c$}] (c) at (1.4,0) {};
\draw[origedge] (p)--(v)--(a);
\draw[oldedge] (v)--(b); \draw[oldedge] (v)--(c);
\end{tikzpicture}
\caption{One of the three star edges is selected.}
\end{subfigure}\hfill
\begin{subfigure}[t]{.46\textwidth}
\centering
\begin{tikzpicture}[scale=.83]
\node[vertex,label=above:{$p$}] (p) at (0,3) {};
\node[center,label=right:{$v_S$}] (v) at (0,2) {};
\node[auxv,label=right:{$z_S$}] (z) at (0,1) {};
\node[vertex,label=below:{$a$}] (a) at (-1.4,-.5) {};
\node[vertex,label=below:{$b$}] (b) at (0,-.5) {};
\node[vertex,label=below:{$c$}] (c) at (1.4,-.5) {};
\draw[origedge] (p)--(v); \draw[origrep] (z)--(a);
\draw[connedge] (v)--node[left,font=\small] {$\gamma_S=1$}(z);
\draw[oldedge] (z)--(b); \draw[oldedge] (z)--(c);
\end{tikzpicture}
\caption{The connector is selected to give even degree at $z_S$.}
\end{subfigure}
\caption{A local example of the lift with $k_S=3$ and $m_S=1$.
Black edges are selected, gray dashed edges are unselected, and the
purple connector is selected. The selected degree at $z_S$ is
$2=k_S-1$. The selected degree at $v_S$ remains even.}
\label{fig:framework-1-lift}
\end{figure}

\begin{lemma}[Lifting a join that leaves an edge of every star unselected]
\label{lem:framework-1-lift}
Every $\odd(G)$-join $J\subseteq E$ with $S\not\subseteq J$
for all $S\in\mathcal S$ has a unique lift $K\subseteq A^{\mathrm{star}}$
satisfying $\mathbf{1}_{K}\in \Pone$. Its representatives are those of $J$,
and its connector choices are determined by
\[
 \mathbf{1}_{\{c_S\in K\}}\equiv k_S+1-|J\cap S|\pmod2.
\]
\end{lemma}
\begin{proof}
For each $S\in\mathcal S$, put $m_S:=|J\cap S|$ and choose
$\gamma_S\in\{0,1\}$ with
$\gamma_S\equiv k_S+1-m_S\pmod2$.
Define $K$ by selecting the representative of every edge of $J$
and selecting $c_S$ exactly when $\gamma_S=1$.
Figure~\ref{fig:framework-1-lift} illustrates this construction.

At $z_S$, the selected degree is $m_S+\gamma_S$, whose parity
is $k_S+1=\deg_{H^{\mathrm{star}}}(z_S)$. Since $m_S\le k_S-1$, this degree is
at most $k_S$; its parity therefore makes it at most $k_S-1$,
as required by the auxiliary degree bound.

When passing from $J$ in $G$ to $K$ in $H^{\mathrm{star}}$,
each star centered at $v$ removes $m_S$ selected incidences
at $v$ and adds $\gamma_S$ selected connector incidences. Hence
\[
 \deg_K(v)=\deg_J(v)+\sum_{S:v_S=v}(\gamma_S-m_S)
 \equiv \deg_G(v)+\sum_{S:v_S=v}(1-k_S)
 =\deg_{H^{\mathrm{star}}}(v)\pmod2.
\]
Here we used that $J$ is an $\odd(G)$-join and
$\gamma_S-m_S\equiv1-k_S\pmod2$.
Thus $K$ is an $\odd(H^{\mathrm{star}})$-join. Its degree at each original
vertex is at most $\deg_{H^{\mathrm{star}}}(v)=b^{\mathrm{star}}_v$ because $K\subseteq A^{\mathrm{star}}$.
Consequently $\mathbf1_K\in\Pone$.
Finally, the parity at $z_S$ forces $\gamma_S$, so the lift is unique.
\end{proof}

\Needspace{4\baselineskip}
\begin{theorem}[Comparison with the M\"omke--Svensson bound]
\label{thm:MS-to-framework-1}
\label{thm:framework-1-dominates-MS-allflows}
Let $G=(V,E)$ be a simple $2$-vertex-connected graph, and fix
a DFS tree $T$ rooted at $r$ with $\deg_G(r)\ge3$.
Let $\mathcal S$ be the \textsc{Star} removable system constructed
from $T$ as in Appendix~\ref{sec:DFS-removable-stars}, and put
$n:=|V|$ and $m:=|E|$.

For every feasible M\"omke--Svensson circulation $f$, there
exist an integral feasible circulation $\widehat f$ with
$C(\widehat f)\le C(f)$ and a point
$x\in\Pone(G,\mathcal S)$ such that
\[
\begin{aligned}
 \operatorname{tsp}(G)
 \le \text{\LPone}_{\mathrm{OPT}}(G,\mathcal S)
 &\le \PhiOne_{G,\mathcal S}(x)\\
 &\le \Lambda_{\mathrm{MS}}(\widehat f)
 =\frac43|E_{\widehat f}|-\frac23|R_{\widehat f}|\\
 &\le \frac43n+\frac23C(f)-\frac23.
\end{aligned}
\]

\end{theorem}

\begin{proof}
By circulation integrality \cite{HoffmanKruskal1956},
we can choose an integral feasible circulation $\widehat f$
with $C(\widehat f)\le C(f)$.

Apply Lemma~\ref{lem:MS-restoration} to obtain the support
$G_{\widehat f}^+=(V,E_{\widehat f}^+)$ and its \textsc{MS} removable pairing
$(R_{\widehat f}^+,\mathcal P_{\widehat f}^+)$.
Every star of $\mathcal S$ now contains a retained pair.

Lemma~\ref{lem:MS-convex-combination} gives a convex
combination of $\odd(G_{\widehat f}^+)$-joins $J_i$ in which
each retained edge is selected with probability $1/3$ and
no $J_i$ contains both edges of a pair.
We extend each $J_i$ to an $\odd(G)$-join by adding every edge
omitted from $G_{\widehat f}^+$. Put
\[
 D:=E\setminus E_{\widehat f}^+,
 \qquad
 \widetilde J_i:=J_i\cup D.
\]
Indeed,
\[
 \odd(\widetilde J_i)
 =\odd(G_{\widehat f}^+)\mathbin{\triangle}\odd(D)
 =\odd(G).
\]
Each star contains a retained pair, and $J_i$ leaves at least
one edge of that pair unselected. Since this edge is not in $D$,
$\widetilde J_i$ also leaves at least one edge of every star
unselected. By Lemma~\ref{lem:framework-1-lift}, each $\widetilde J_i$
gives an integral point $z_i\in\Pone(G,\mathcal S)$
satisfying
\[
 (z_i)_{\rep{e}}=\mathbf{1}_{\{e\in\widetilde J_i\}}
 \qquad(e\in E).
\]
Let $\lambda_i$ be the weight of $J_i$ in the convex combination
given by Lemma~\ref{lem:MS-convex-combination}.
Using the same weights, define
\[
 x:=\sum_i\lambda_i z_i
 \in\Pone(G,\mathcal S).
\]
Since $\widetilde J_i=J_i\cup D$, we obtain
\begin{equation}
\label{eq:MS-one-third-or-one}
 x_{\rep{e}}
 =\sum_i\lambda_i\mathbf{1}_{\{e\in\widetilde J_i\}}
 =
 \begin{cases}
  1,&e\in D,\\[2pt]
  \frac13,&e\in E_{\widehat f}^+.
 \end{cases}
\end{equation}

We now compute the tour bound given by $x$.
Recall that $R$ and $N$ are the sets of removable and
nonremovable edges. Every omitted edge is a back edge and
therefore belongs to $R$. By the construction of
$R_{\widehat f}^+$,
\[
 R=D\mathbin{\dot\cup}R_{\widehat f}^+,
 \qquad
 E_{\widehat f}^+=R_{\widehat f}^+\mathbin{\dot\cup}N.
\]
Substituting the coordinates of $x$ into the objective gives
\begin{equation}
\label{eq:MS-exact-support-comparison}
\begin{aligned}
 \PhiOne_{G,\mathcal S}(x)
 &=m-|D|-\frac13|R_{\widehat f}^+|+\frac13|N|\\
 &=\frac43|E_{\widehat f}^+|-\frac23|R_{\widehat f}^+|\\
 &\le\Lambda_{\mathrm{MS}}(\widehat f)\\
 &\le\frac43n+\frac23C(\widehat f)-\frac23\\
 &\le\frac43n+\frac23C(f)-\frac23.
\end{aligned}
\end{equation}
Here the first inequality follows from
Lemma~\ref{lem:MS-restoration}, the second from
\eqref{eq:MS-support-vs-cost}, and the last from
$C(\widehat f)\le C(f)$.
The \LPone{} optimum is at most
$\PhiOne_{G,\mathcal S}(x)$, so
Theorem~\ref{thm:framework-1-bound} gives the claimed tour bound.
\end{proof}

\smallskip
\begin{lemma}[Comparison with optimized MS parity correction]
\label{lem:reconnection-MS-parity}

Let $G$ and $T$ satisfy the assumptions of
Theorem~\ref{thm:MS-to-framework-1}, and let $\mathcal S$ be
the \textsc{Star} removable system constructed from $T$
as in Appendix~\ref{sec:DFS-removable-stars}. For every integral
feasible MS circulation $f$,
\[
 \text{\LPtwo}_{\mathrm{OPT}}(G,\mathcal S)
 \le\MSopt(f).
\]
\end{lemma}

\begin{proof}
Let $J\subseteq E_f$ attain $\MSopt(f)$, and put
\[
 D:=E\setminus E_f,\qquad
 U:=(R\cap E_f)\setminus R_f,\qquad
 \widetilde J:=J\cup D.
\]
All edges of $D$ are back edges and hence belong to $R$.
Also, $R_f\subseteq R$, and $U$ consists of the tree edges
that belong to our stars but are unpaired in the MS construction.
We have
\[
 \odd(\widetilde J)
 =\odd(G_f)\mathbin{\triangle}\odd(D)=\odd(G).
\]
If $\widetilde J$ contains an entire star, its tree edge cannot
be paired in $\mathcal P_f$, since $J$ leaves at least one
edge of every pair in $\mathcal P_f$ unselected. Its tree edge therefore
belongs to $J\cap U$. Distinct stars have distinct tree edges,
so
\[
 |\{S\in\mathcal S:S\subseteq\widetilde J\}|
 \le |J\cap U|.
\]

Since $\widetilde J$ is an $\odd(G)$-join, the parity argument
in the proof of Lemma~\ref{lem:framework-1-lift} extends it
to an $\odd(H^\circ)$-join $K$ satisfying
\[
 \rep{e}\in K\quad\Longleftrightarrow\quad e\in\widetilde J
 \qquad(e\in E).
\]
This argument also applies when $\widetilde J$ contains an
entire star. Put $y:=\mathbf{1}_K\in{\mathcal P^\circ}$.

At $z_S$, the selected degree has the same parity as $k_S+1$.
It equals $k_S+1$ when $S\subseteq\widetilde J$ and is at most
$k_S-1$ otherwise. Thus \eqref{eq:repair-nonloop-cost} gives
\[
\begin{aligned}
 \Phi_{G,\mathcal S}(y)
 &=m-|\widetilde J\cap R|+|\widetilde J\cap N|
      +2|\{S\in\mathcal S:S\subseteq\widetilde J\}|\\
 &=\MSopt(f)-2|J\cap U|
      +2|\{S\in\mathcal S:S\subseteq\widetilde J\}|\\
 &\le\MSopt(f).
\end{aligned}
\]
Minimizing over ${\mathcal P^\circ}$ proves the claim.
\end{proof}

\begin{corollary}[Comparison for a fixed DFS tree]
\label{thm:framework-1-dominates-MS}
Let $G$ be a simple $2$-vertex-connected graph, and fix a
DFS tree $T$ rooted at $r$ with $\deg_G(r)\ge3$.
Let $\mathcal S$ be the \textsc{Star} removable system
constructed from $T$ as in Appendix~\ref{sec:DFS-removable-stars}.
Then
\[
 \begin{aligned}
 \operatorname{tsp}(G)
 &\le\text{\LPtwo}_{\mathrm{OPT}}(G,\mathcal S)\\
 &\le\min\left\{\text{\LPone}_{\mathrm{OPT}}(G,\mathcal S),
                 \min_f\MSopt(f)\right\}\\
 &\le\frac43n+{\frac23}\min_f C(f)-\frac23.
 \end{aligned}
\]
Here the minima are over minimum-cost integral MS circulations for
$T$, and $\MSopt(f)$ is defined in
\eqref{eq:MS-optimized-parity}. The comparison follows from
Theorem~\ref{thm:MS-to-framework-1},
Lemma~\ref{lem:reconnection-MS-parity}, and
Proposition~\ref{prop:framework-containment}.
Thus the \LPone{} gives a tour bound at least as strong as the
optimized MS circulation bound, and the \LPtwo{} also bounds
the optimized MS parity correction.
\end{corollary}

\section{A \texorpdfstring{$13/10$}{13/10} bound for \texorpdfstring{$4$}{4}-regular, \texorpdfstring{$4$}{4}-connected graphs}
\label{sec:quartic-framework-2}
Assigning edge values in $[1/3,2/3]$ makes the Blossom
inequalities easy to satisfy on cuts with at least three
edges, since each edge contributes at least $1/3$.
Smaller cuts must be checked separately.
Here we show that higher connectivity allows us to use
other edge values while still obtaining a point in
${\mathcal P^\circ}(G,\mathcal S)$.
For $4$-regular, $4$-vertex-connected graphs, this leads
to the improved approximation bound of $13/10$.

We start with an initial vector $x^{\mathrm{init}}$ for which the objective expression in
\eqref{eq:quartic-objective} equals $5n/4-1/2$.
The initial vector need not belong to ${\mathcal P^\circ}$.
We identify the Blossom cuts whose inequalities are not yet
satisfied and increase the value of an edge in each such cut
so that its inequality becomes satisfied. Some of these changes
leave the \LPtwo{} objective unchanged, while others increase
it by $1/4$. We prove that at most $(n-2)/5$ changes that increase
the objective are needed, and then verify that the resulting
vector $\bar x$ belongs to ${\mathcal P^\circ}$. This gives a tour of
length at most $13n/10-3/5$.\\

\textbf{The existing $4/3$ baseline.}
For this graph class, a $4/3$ bound already follows from Newman's
subquartic result~\cite[Theorem~1]{Newman}. Indeed, for the graph
subtour relaxation
\[
 {\text{\SubtourLP}(G)}
 :=\min\{y(E):y\ge0,\ y(\delta_G(U))\ge2
       \text{ for }\varnothing\ne U\subsetneq V\},
\]
the vector $y_e:=1/2$ is feasible because every cut has at least
four edges. Since $m=2n$, its objective is $n$, and summing the
singleton-cut constraints gives the matching lower bound
${\text{\SubtourLP}(G)}\ge n$. Hence
${\text{\SubtourLP}(G)}=n$, which is exactly the case
covered by that theorem.

Throughout this section, $G=(V,E)$ is simple, $4$-regular, and
$4$-vertex-connected, and $n:=|V|$. Fix an arbitrary DFS tree
$T$ rooted at $r$. We use the \LPtwo{} with the \textsc{Star} removable system
$\mathcal S$ constructed from $T$ as in
Appendix~\ref{sec:DFS-removable-stars}.
We abbreviate ${\mathcal P^\circ}(G,\mathcal S)$ by
${\mathcal P^\circ}$.

{In the DFS diagrams in this section, blue edges
represent edges of $T$ and red curves represent back edges;
the same colors are used for their representatives.
The DFS tree is drawn downward from $r$, so its edges are
read from parent to child and back edges from descendant
to ancestor. Dotted purple edges are connectors.}

\subsection{Vertex types and the \textsc{Star} removable system}

For a nonroot vertex $v$, let $j(v)$ be its number of DFS children and let
\[
 i(v):=\bigl|\{vw\in E\setminus E(T):
                  w\text{ is a proper descendant of }v\}\bigr|.
\]
Thus $i(v)$ counts the back edges whose upper endpoint is $v$;
we call these the back edges entering $v$. The graphs and their
back edges remain undirected. Define
\[
 B_2:=\{v\in V\setminus\{r\}:j(v)=2,\ i(v)=1\}.
\]
For any nonroot vertex $u$, its \emph{parent edge} is
$\operatorname{par}(u)u$.
A vertex $u\in B_2$ has one parent edge, two child edges, and one
back edge entering $u$, with no back edge from $u$ to a proper
ancestor. For consistency with the earlier sections, we recall the
notation. For each tree edge $uv$, let $I(u,v)$ be the set
of back edges joining $u$ to vertices in the subtree rooted
at $v$. If $I(u,v)\ne\varnothing$, we form the star
\[
 S(u,v)=\{uv\}\cup I(u,v).
\]
Write
\[
 R=\bigcup_{S\in\mathcal S}S,
 \qquad N=E(G)\setminus R.
\]
Thus $R$ consists of the removable edges, and $N$ consists
of the remaining edges, which are also called nonremovable edges.

Let $v$ be a nonroot vertex with only one DFS child $w$.
Since $G$ is $4$-regular and $v$ already has a parent edge
and the child edge $vw$, at most two back edges can end at $v$.
If none ends at $v$, no star is formed there and $vw\in N$.
If one back edge ends at $v$, it forms a star of size two
together with $vw$. If two back edges end at $v$, they form
a star of size three together with $vw$.

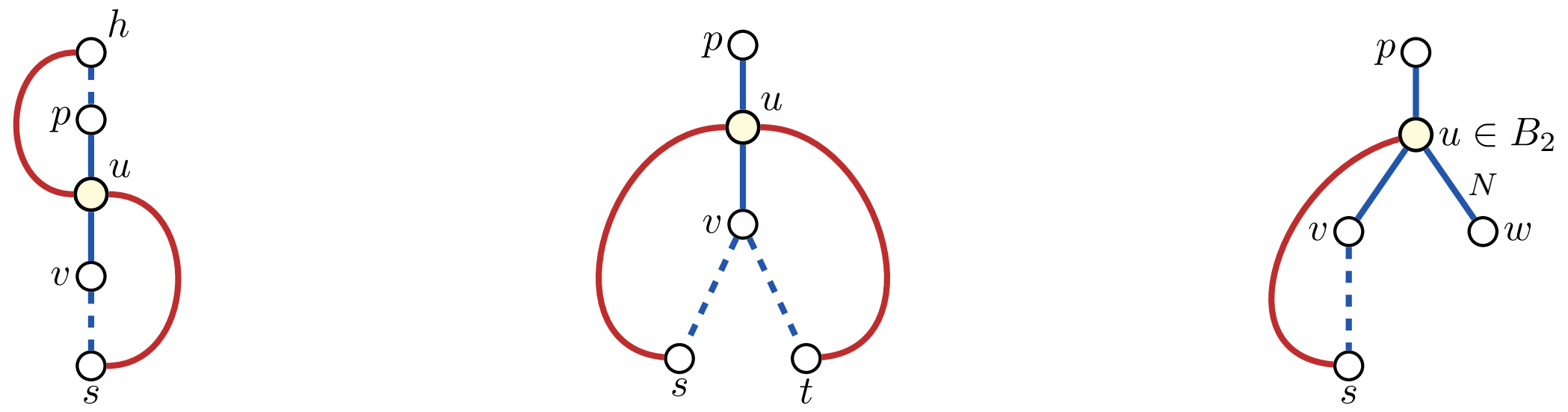
\begin{figure}[!t]
\centering
\begin{subfigure}[t]{.31\textwidth}\centering
\begin{tikzpicture}[scale=.66,font=\small,every label/.style={font=\small,fill=white,inner sep=1pt}]
\node[vertex,label=above right:{$h$}] (h) at (0,3.2) {};
\node[vertex,label=left:{$p$}] (p) at (0,2.3) {};
\node[center,label=above right:{$u$}] (u) at (0,1.3) {};
\node[vertex,label=left:{$v$}] (v) at (0,.2) {};
\node[vertex,label=below:{$s$}] (s) at (0,-1) {};
\draw[blueedge,dashed] (h)--(p);
\draw[blueedge] (p)--(u)--(v);
\draw[blueedge,dashed] (v)--(s);
\draw[rededge] (s) to[out=0,in=0,looseness=1.4] (u);
\draw[rededge] (u) to[out=180,in=180,looseness=1.4] (h);
\end{tikzpicture}
\caption{$j(u)=1$, $i(u)=1$: a star of size two.}
\end{subfigure}\hfill
\begin{subfigure}[t]{.31\textwidth}\centering
\begin{tikzpicture}[scale=.66,font=\small,every label/.style={font=\small,fill=white,inner sep=1pt}]
\node[vertex,label=left:{$p$}] (p) at (0,3.2) {};
\node[center,label=above right:{$u$}] (u) at (0,2.1) {};
\node[vertex,label=left:{$v$}] (v) at (0,.8) {};
\node[vertex,label=below:{$s$}] (s) at (-.85,-1) {};
\node[vertex,label=below:{$t$}] (t) at (.85,-1) {};
\draw[blueedge] (p)--(u)--(v);
\draw[blueedge,dashed] (v)--(s);\draw[blueedge,dashed] (v)--(t);
\draw[rededge] (s) to[out=175,in=180,looseness=1.3] (u);
\draw[rededge] (t) to[out=5,in=0,looseness=1.3] (u);
\end{tikzpicture}
\caption{$j(u)=1$, $i(u)=2$: one triple.}
\end{subfigure}\hfill
\begin{subfigure}[t]{.31\textwidth}\centering
\begin{tikzpicture}[scale=.66,font=\small,every label/.style={font=\small,fill=white,inner sep=1pt}]
\node[vertex,label=left:{$p$}] (p) at (0,3.2) {};
\node[center,label=right:{$u\in B_2$}] (u) at (0,2.1) {};
\node[vertex,label=left:{$v$}] (v) at (-.9,.8) {};
\node[vertex,label=right:{$w$}] (w) at (.9,.8) {};
\node[vertex,label=below:{$s$}] (s) at (-.9,-1) {};
\draw[blueedge] (p)--(u)--(v);
\draw[blueedge] (u)--node[right,font=\scriptsize] {$N$}(w);
\draw[blueedge,dashed] (v)--(s);
\draw[rededge] (s) to[out=175,in=195,looseness=1.3] (u);
\end{tikzpicture}
\caption{$u\in B_2$: $uv$ belongs to a star of size two and $uw\in N$.}
\end{subfigure}
\caption{Nonroot vertices with nonempty stars. Blue edges belong to
the DFS tree; red curves are back edges. Dashed blue segments denote
tree paths. In (c), the back edge ending at $u$ comes from the subtree
rooted at $v$, and $uw\in N$. }
\label{fig:quartic-types}
\end{figure}

A nonroot vertex with two DFS children either belongs to $B_2$,
in which case one child edge belongs to a star of size two
and the other belongs to $N$, or has both child edges in $N$.
A nonroot vertex with three DFS children has all three
child edges in $N$. See Figure~\ref{fig:quartic-types} for the
nonroot cases with nonempty stars. Finally, the root has exactly one DFS child and
three back edges ending at it, which together form a
star of size four.

\begin{theorem}
\label{thm:quartic-thirteen-tenths}
For every fixed DFS tree $T$ of a simple, $4$-regular,
$4$-vertex-connected graph $G$ on $n$ vertices, the \LPtwo{} with
the \textsc{Star} removable system $\mathcal S$ constructed
from $T$ as in Appendix~\ref{sec:DFS-removable-stars} certifies
\[
 \operatorname{tsp}(G)\le \text{\LPtwo}_{\mathrm{OPT}}(G,\mathcal S)
 \le \frac{13}{10}n-\frac35.
\]
A vector in ${\mathcal P^\circ}$ attaining this bound can be constructed in
polynomial time.

\end{theorem}

\subsection{The initial vector and its objective}

\begin{figure}[!t]
\centering
\begin{subfigure}[t]{.30\textwidth}\centering
\begin{tikzpicture}[x=.72cm,y=.66cm,font=\small,
 lab/.style={font=\small,fill=white,inner sep=1pt}]
\node[center,label=left:{$u$}] (u) at (0,3.3) {};
\node[auxv,label=left:{$z_S$}] (z) at (0,2.4) {};
\node[vertex,label=left:{$v$}] (v) at (0,1.3) {};
\node[vertex,label=left:{$s$}] (a) at (0,-.2) {};
\draw[connedge] (u)--node[lab,right] {$1/4$}(z);
\draw[bluerep] (z)--node[lab,left] {$1/4$}(v);
\draw[blueedge,dashed] (v)--(a);
\draw[redrep] (z) .. controls (1.7,2.4) and (1.7,-.2) .. node[lab,right,pos=.5] {$1/2$}(a);
\end{tikzpicture}
\caption{A nonroot star of size two.}
\end{subfigure}\hfill
\begin{subfigure}[t]{.32\textwidth}\centering
\begin{tikzpicture}[x=.72cm,y=.66cm,font=\small,
 lab/.style={font=\small,fill=white,inner sep=1pt}]
\node[center,label=left:{$u$}] (u) at (0,3.3) {};
\node[auxv,label=left:{$z_S$}] (z) at (0,2.4) {};
\node[vertex,label=left:{$v$}] (v) at (0,1.3) {};
\node[vertex,label=below:{$s$}] (a) at (-.8,-.2) {};
\node[vertex,label=below:{$t$}] (b) at (.8,-.2) {};
\draw[connedge] (u)--node[lab,right] {$1/4$}(z);
\draw[bluerep] (z)--node[lab,left] {$1/4$}(v);
\draw[blueedge,dashed] (v)--(a); \draw[blueedge,dashed] (v)--(b);
\draw[redrep] (z) .. controls (-1.9,2.4) and (-1.9,-.2) .. (a);
\draw[redrep] (z) .. controls (1.9,2.4) and (1.9,-.2) .. (b);
\end{tikzpicture}
\caption{A nonroot triple; red edges have value $3/4$.}
\end{subfigure}\hfill
\begin{subfigure}[t]{.34\textwidth}\centering
\begin{tikzpicture}[x=.72cm,y=.66cm,font=\small,
 lab/.style={font=\small,fill=white,inner sep=1pt}]
\node[center,label=left:{$r$}] (r) at (0,3.3) {};
\node[auxv,label=left:{$z_{S_r}$}] (z) at (0,2.4) {};
\node[vertex,label=left:{$v$}] (v) at (0,1.5) {};
\node[vertex,label=left:{$s$}] (a) at (0,.6) {};
\node[vertex,label=left:{$t$}] (b) at (0,-.3) {};
\node[vertex,label=left:{$b$}] (c) at (0,-1.2) {};
\draw[connedge] (r)--node[lab,right] {$1$}(z);
\draw[bluerep] (z)--node[lab,left] {$1/4$}(v);
\draw[blueedge,dashed] (v)--(a)--(b)--(c);
\draw[redrep] (z) .. controls (1.1,2.4) and (1.1,.6) .. (a);
\draw[redrep] (z) .. controls (2,2.4) and (2,-.3) .. (b);
\draw[redrep] (z) .. controls (2.9,2.4) and (2.9,-1.2) .. (c);
\end{tikzpicture}
\caption{The root star; red edges have value $3/4$.}
\end{subfigure}
\caption{Coordinates of $x^{\mathrm{init}}$ on $H^\circ$. Blue edges come from tree
edges, curved red edges from back edges, and dotted purple edges are connectors.
Dashed blue segments denote tree paths with omitted intermediate vertices.
Every edge in $N_H$ has value $1/4$.}
\label{fig:quartic-initial}
\end{figure}
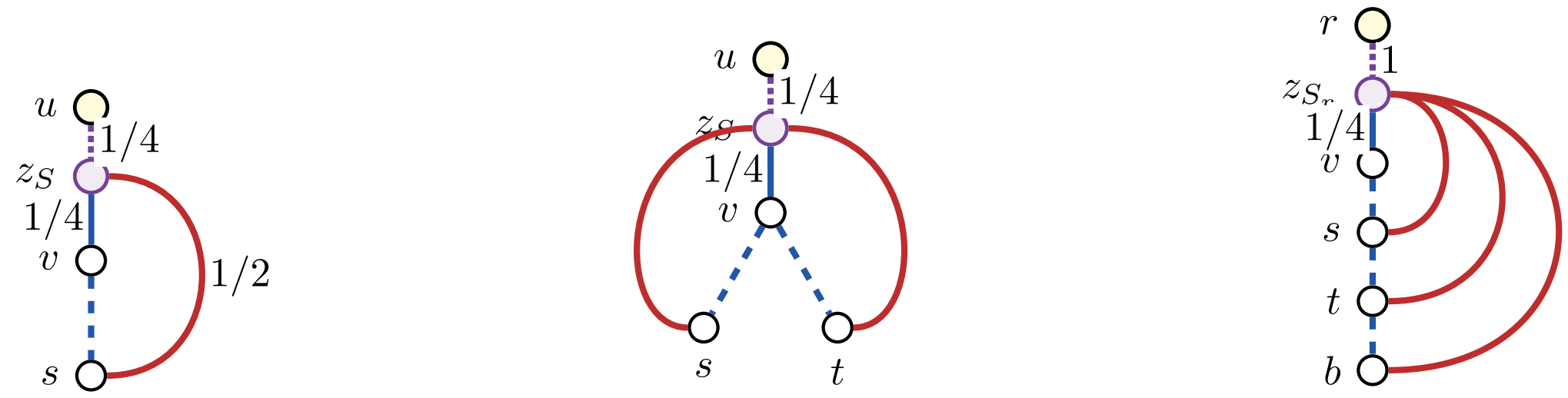

Define the initial vector $x^{\mathrm{init}}\in\mathbb R^{A^\circ}$
on $H^\circ$. Set $x^{\mathrm{init}}_e:=1/4$ for every $e\in N_H$
and use the following values on the stars (see Figure~\ref{fig:quartic-initial}):
\[
\begin{array}{c|ccc}
 S&\rep{e}\;(e\in S\cap E(T))&\rep{b}\;(b\in S\setminus E(T))
     &x^{\mathrm{init}}_{c_S}\\ \hline
 \text{nonroot star of size two}&1/4&1/2&1/4\\
 \text{nonroot triple}&1/4&3/4&1/4\\
 \text{root star}&1/4&3/4&1
\end{array}
\]
Each star $S$ contains $k_S-1$ back edges, and every back
edge belongs to exactly one star. Since there are $m-n+1$
back edges, the constant in \eqref{eq:repair-nonloop-cost} is
\[
 m-\sum_{S\in\mathcal S}(k_S-1)
 =m-(m-n+1)=n-1.
\]
For the values just assigned, every star satisfies
\[
 k_S-1-x^{\mathrm{init}}(S_H)\le x^{\mathrm{init}}_{c_S}.
\]
Thus the maximum in the initial objective equals
$x^{\mathrm{init}}_{c_S}$. Let $\bar x$ denote the vector
obtained from $x^{\mathrm{init}}$ by the correction procedure
in Section~\ref{subsec:quartic-repair-procedure}.
Since this procedure only increases coordinates, every star
satisfies $k_S-1-\bar x(S_H)\le\bar x_{c_S}$.
Its objective expression is
\begin{equation}
\label{eq:quartic-objective}
 \begin{aligned}
 &n-1+\bar x(N_H)+\sum_{S\in\mathcal S}
   \max\{\bar x_{c_S},\,k_S-1-\bar x(S_H)\}\\
 &\qquad=n-1+\bar x(N_H)+\sum_{S\in\mathcal S}\bar x_{c_S}.
 \end{aligned}
\end{equation}

Increasing
the coordinate of $\rep{e}$ for $e\in R$ leaves the objective
unchanged. Increasing a connector coordinate or the coordinate
of $\rep{e}$ for $e\in N$ by $1/4$ increases it by $1/4$.

In $x^{\mathrm{init}}$, every edge of $N_H$ and every nonroot
connector has value $1/4$, while the root connector has value one.
Using $|N|+|\mathcal S|=n-1$, the initial objective expression is
\begin{equation}
\label{eq:quartic-baseline}
 \begin{aligned}
 n-1+x^{\mathrm{init}}(N_H)
       +\sum_{S\in\mathcal S}x^{\mathrm{init}}_{c_S}
 &=n-1+\frac14|N|+\frac14(|\mathcal S|-1)+1\\
 &=n-1+\frac14(n-2)+1\\
 &=\frac54n-\frac12.
 \end{aligned}
\end{equation}
It remains to satisfy the Blossom inequalities while bounding
the number of increases that raise this value.

\subsection{The feasibility correction procedure}
\label{subsec:quartic-repair-procedure}

We show that only two types of cuts in $H^\circ$ need
to be corrected:
cuts around certain vertices and cuts around certain triangles.
In each case, the three boundary edges have value $1/4$
in $x^{\mathrm{init}}$. Increasing at least one of them to $1/2$ satisfies
the Blossom inequalities for that cut.

\textbf{Objective-preserving parent-edge changes.}
Each nonroot vertex centers at most one star: two edge-disjoint
stars, together with its parent edge, would require at least
five incident edges, contradicting $4$-regularity.
For each vertex $v$ that centers a star of size two, let
$S_v$ denote this star. Its auxiliary vertex and connector
are $z_{S_v}$ and $c_{S_v}$.

In $G$, for $u\in B_2$, let $w$ be the child whose subtree contains no
back edge ending at $u$. The edge $uw$ belongs to $N$, and the
three edges incident with the original copy of $u$ in $H^\circ$ are
\[
    \delta_{H^\circ}(u)
    =
    \left\{
        \rep{\operatorname{par}(u)u},
        \rep{uw},
        c_{S_u}
    \right\}.
\]
Initially, all three have value $1/4$, so
\[
    x^{\mathrm{init}}(\delta_{H^\circ}(u))=\frac34<1.
\]
Thus, the cut inequality $\bar x(\delta_{H^\circ}(u))\ge1$ requires an
increase of at least $1/4$ in this boundary sum.

Whenever $\operatorname{par}(u)u\in R$, increase its
representative from $1/4$ to $1/2$.
This raises the boundary sum to one without changing
\eqref{eq:quartic-objective}.

The vertices of $B_2$ whose cut inequalities remain unsatisfied after
these changes are precisely those whose parent edges belong
to $N$. Define
\begin{equation}
\label{eq:quartic-U}
    \mathcal U
    :=
    \{u\in B_2:\operatorname{par}(u)u\in N\}.
\end{equation}
For every $u\in\mathcal U$, the three incident coordinates
are still $1/4$. We call this a \textbf{vertex demand}: at least
one of these coordinates must be increased. Increasing an edge of
$N_H$ by $1/4$ increases \eqref{eq:quartic-objective} by $1/4$.

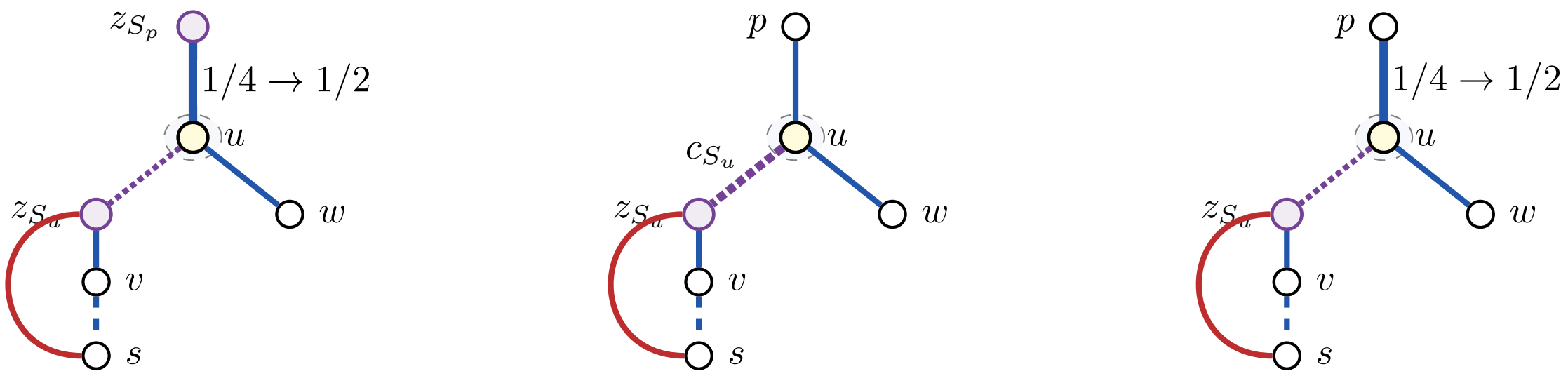
\begin{figure}[!t]
\centering
\begin{subfigure}[t]{.32\textwidth}\centering
\begin{tikzpicture}[x=.73cm,y=.58cm,font=\small]
\draw[gray,dashed,fill=panelbg] (0,1.7) circle (.37);
\node[auxv,label=left:{$z_{S_p}$}] (p) at (0,3.5) {};
\node[center,label=right:{$u$}] (u) at (0,1.7) {};
\node[auxv,label=left:{$z_{S_u}$}] (z) at (-1.25,.45) {};
\node[vertex,label=right:{$w$}] (w) at (1.25,.45) {};
\node[vertex,label=right:{$v$}] (v) at (-1.25,-.65) {};
\node[vertex,label=right:{$s$}] (a) at (-1.25,-1.85) {};
\draw[bluerep,line width=2.3pt] (p)--node[right,fill=white,inner sep=1pt] {$1/4\to1/2$}(u);
\draw[connedge] (u)--(z);
\draw[blueedge] (u)--(w);
\draw[bluerep] (z)--(v);
\draw[redrep] (z) .. controls (-2.7,.45) and (-2.7,-1.85) .. (a);
\draw[blueedge,dashed] (v)--(a);
\end{tikzpicture}
\caption{$pu\in R$: cost $0$.}
\end{subfigure}\hfill
\begin{subfigure}[t]{.32\textwidth}\centering
\begin{tikzpicture}[x=.73cm,y=.58cm,font=\small]
\draw[gray,dashed,fill=panelbg] (0,1.7) circle (.37);
\node[vertex,label=left:{$p$}] (p) at (0,3.5) {};
\node[center,label=right:{$u$}] (u) at (0,1.7) {};
\node[auxv,label=left:{$z_{S_u}$}] (z) at (-1.25,.45) {};
\node[vertex,label=right:{$w$}] (w) at (1.25,.45) {};
\node[vertex,label=right:{$v$}] (v) at (-1.25,-.65) {};
\node[vertex,label=right:{$s$}] (a) at (-1.25,-1.85) {};
\draw[blueedge] (p)--(u);
\draw[connedge,line width=2.3pt] (u)--node[above left,fill=white,inner sep=1pt] {$c_{S_u}$}(z);
\draw[blueedge] (u)--(w);
\draw[bluerep] (z)--(v);
\draw[redrep] (z) .. controls (-2.7,.45) and (-2.7,-1.85) .. (a);
\draw[blueedge,dashed] (v)--(a);
\end{tikzpicture}
\caption{Increase $c_{S_u}$: cost $1/4$.}
\end{subfigure}\hfill
\begin{subfigure}[t]{.32\textwidth}\centering
\begin{tikzpicture}[x=.73cm,y=.58cm,font=\small]
\draw[gray,dashed,fill=panelbg] (0,1.7) circle (.37);
\node[vertex,label=left:{$p$}] (p) at (0,3.5) {};
\node[center,label=right:{$u$}] (u) at (0,1.7) {};
\node[auxv,label=left:{$z_{S_u}$}] (z) at (-1.25,.45) {};
\node[vertex,label=right:{$w$}] (w) at (1.25,.45) {};
\node[vertex,label=right:{$v$}] (v) at (-1.25,-.65) {};
\node[vertex,label=right:{$s$}] (a) at (-1.25,-1.85) {};
\draw[blueedge,line width=2.3pt] (p)--node[right,fill=white,inner sep=1pt] {$1/4\to1/2$}(u);
\draw[connedge] (u)--(z);
\draw[blueedge] (u)--(w);
\draw[bluerep] (z)--(v);
\draw[redrep] (z) .. controls (-2.7,.45) and (-2.7,-1.85) .. (a);
\draw[blueedge,dashed] (v)--(a);
\end{tikzpicture}
\caption{$pu\in N$: cost $1/4$.}
\end{subfigure}
\caption{Three ways to satisfy the cut around $u\in B_2$.
In each panel, increase the thick edge from $1/4$ to $1/2$.
The shaded set is $\{u\}$. In (a), the parent edge belongs to a
star centered at $p$; the same change applies if that star has
size three or is the root star.}
\label{fig:quartic-changes}
\end{figure}

Figure~\ref{fig:quartic-changes} illustrates the
objective-preserving change and the two types of changes
that increase \eqref{eq:quartic-objective}.

\textbf{Triangle demands.}
The initial vector $x^{\mathrm{init}}$ can also violate Blossom
inequalities for cuts separating a triangle from the rest of the
graph. These cuts also require a coordinate increase.

In $G$, consider a vertex $u\in B_2$ and put $p:=\operatorname{par}(u)$.
Let $v$ be the child whose subtree contains the lower endpoint
$s$ of the unique back edge $su$ ending at $u$, and let $w$ be
the other child. Suppose that $\{pu,pw\}$ is the star centered
at $p$, so $pw$ is a back edge with upper endpoint $p$ and
lower endpoint $w$.
Suppose also that $w$ has exactly one DFS child $a$ and exactly one
back edge $tw$ entering it, with $t\in V(T_a)$.
The back edges $pw$ and $tw$ are distinct: the first joins $w$
to an ancestor, and the second joins it to a descendant.
The three stars are
\[
    \{pu,pw\},\qquad
    \{uv,su\},\qquad
    \{wa,tw\},
\]
centered at $p,u,w$, respectively.

\begin{figure}[!t]
\centering
\begin{subfigure}[t]{.46\textwidth}\centering
\begin{tikzpicture}[x=.85cm,y=.70cm,font=\small]
\node[center,label=above:{$p$}] (p) at (0,3.5) {};
\node[center,label=above right:{$u$}] (u) at (0,2) {};
\node[vertex,label=left:{$v$}] (v) at (-1.5,.5) {};
\node[center,label=left:{$w$}] (w) at (1,.5) {};
\node[vertex,label=below:{$s$}] (s) at (-1.5,-1) {};
\node[vertex,label=left:{$a$}] (a) at (1,-.8) {};
\node[vertex,label=below:{$t$}] (t) at (1,-2) {};
\draw[blueedge] (p)--(u)--(v);
\draw[blueedge] (u)--(w)--(a);
\draw[blueedge,dashed] (v)--(s);
\draw[blueedge,dashed] (a)--(t);
\draw[rededge] (s) .. controls (-3,-.8) and (-2,2) .. (u);
\draw[rededge] (w) .. controls (3,1) and (2.5,3.5) .. (p);
\draw[rededge] (t) .. controls (2.7,-1.8) and (2.7,.4) .. (w);
\end{tikzpicture}
\caption{In $G$: the stars are $\{pu,pw\}$, $\{uv,us\}$,
and $\{wa,wt\}$.}
\end{subfigure}\hfill
\begin{subfigure}[t]{.51\textwidth}\centering
\begin{tikzpicture}[x=.76cm,y=.55cm,font=\small,
 lab/.style={font=\small,fill=white,inner sep=1pt}]
\draw[gray,dashed,fill=panelbg,rounded corners=4pt] (-.5,1.0) rectangle (3.1,4.9);
\node[center,label=above:{$p$}] (p) at (0,5.6) {};
\node[auxv,label=left:{$z_{S_p}$}] (zp) at (0,4.4) {};
\node[center,label=above left:{$u$}] (u) at (0,3) {};
\node[center,label=left:{$w$}] (w) at (1.5,1.5) {};
\node[auxv,label=left:{$z_{S_u}$}] (zu) at (-1.5,1.5) {};
\node[auxv,label=right:{$z_{S_w}$}] (zw) at (1.5,.3) {};
\node[vertex,label=left:{$v$}] (v) at (-1.5,.4) {};
\node[vertex,label=below:{$s$}] (s) at (-1.5,-1) {};
\node[vertex,label=left:{$a$}] (a) at (1.5,-.8) {};
\node[vertex,label=below:{$t$}] (t) at (1.5,-2.2) {};
\draw[connedge] (p)--node[lab,right] {$c_{S_p}$}(zp);
\draw[bluerep] (zp)--(u);
\draw[blueedge] (u)--(w);
\draw[redrep] (zp) .. controls (3.5,4.4) and (3.5,1.5) .. (w);
\draw[connedge] (u)--node[lab,above left] {$c_{S_u}$}(zu);
\draw[connedge] (w)--node[lab,above right] {$c_{S_w}$}(zw);
\draw[bluerep] (zu)--(v);
\draw[redrep] (zu) .. controls (-3.2,1.5) and (-3.2,-1) .. (s);
\draw[blueedge,dashed] (v)--(s);
\draw[bluerep] (zw)--(a);
\draw[redrep] (zw) .. controls (3.2,.3) and (3.2,-2.2) .. (t);
\draw[blueedge,dashed] (a)--(t);
\end{tikzpicture}
\caption{In $H^\circ$: the three connectors form the boundary.}
\end{subfigure}
\caption{A triangle demand at $u$. The shaded vertices in $H^\circ$
are $X_u=\{z_{S_p},u,w\}$; its boundary consists of the three
connectors $c_{S_p},c_{S_u},c_{S_w}$. Dashed blue segments denote
tree paths.}
\label{fig:triangle-demand-construction}
\end{figure}
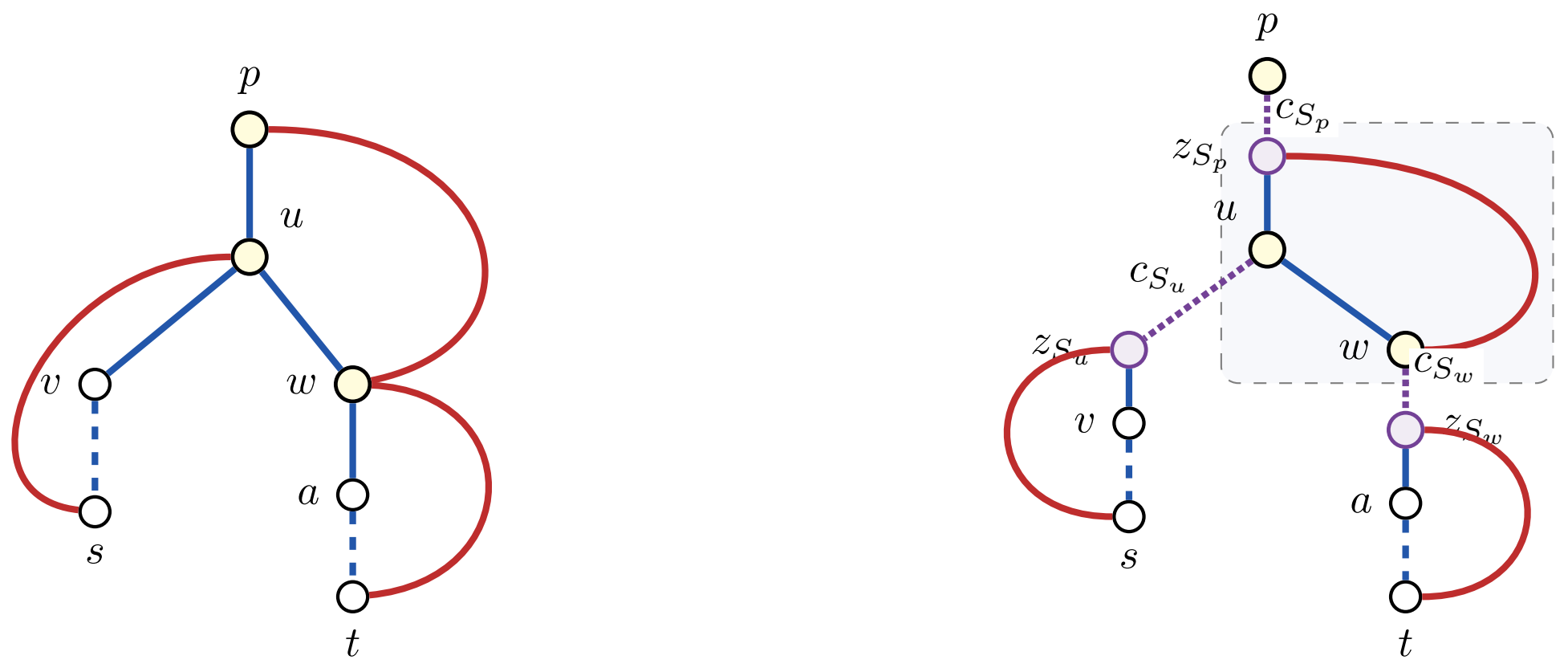

In $H^\circ$, the vertices
\[
    X_u:=\{z_{S_p},u,w\}
\]
form a triangle, with
\[
\begin{aligned}
    E(H^\circ[X_u])&=\{\rep{pu},\rep{uw},\rep{wp}\}
              =\{z_{S_p}u,uw,wz_{S_p}\},\\
    \delta_{H^\circ}(X_u)&=\{c_{S_p},c_{S_u},c_{S_w}\}.
\end{aligned}
\]
The representatives of $wa$ and $tw$ are incident with $z_{S_w}$.
The connector $c_{S_w}$ joins the original copy of $w$ to $z_{S_w}$.
Figure~\ref{fig:triangle-demand-construction} shows the construction.
Let $\mathcal D$ be the set of vertices $u\in B_2$ for which this
configuration occurs. Since $pu\in R$, every such $u$ lies
outside $\mathcal U$.

\begin{figure}[!t]
\centering
\begin{subfigure}[t]{.48\textwidth}\centering
\begin{tikzpicture}[scale=.72,
 lab/.style={font=\scriptsize,fill=white,inner sep=1.5pt}]
\draw[draw=gray!65,dashed,fill=panelbg,rounded corners=5pt] (-1.95,-1.85) rectangle (1.6,1.45);
\node[auxv,label=left:{$z_{S_p}$}] (zp) at (-.7,1.05) {};
\node[center,label=above right:{$u$}] (u) at (1.05,.05) {};
\node[center,label=below left:{$w$}] (w) at (-.25,-1.45) {};
\node[center,label=above:{$p$}] (p) at (-1.3,2.5) {};
\node[auxv,label=above:{$z_{S_u}$}] (zu) at (2.6,-.65) {};
\node[auxv,label=below:{$z_{S_w}$}] (zw) at (.85,-2.65) {};
\draw[bluerep] (zp)--node[lab,above right] {$1/2$}(u);
\draw[blueedge] (u)--node[lab,below right] {$1/4$}(w);
\draw[redrep] (w) .. controls (-2.1,-1.45) and (-2.1,1.05) .. node[lab,left,pos=.5] {$1/2$}(zp);
\draw[connedge] (p)--node[lab,left] {$1/4$}(zp);
\draw[connedge] (u)--node[lab,above] {$1/4$}(zu);
\draw[connedge] (w)--node[lab,above right] {$1/4$}(zw);
\node[font=\scriptsize] at (.0,-.15) {$X_u$};
\end{tikzpicture}
\caption{Increasing $\rep{pu}$ leaves the boundary sum at $3/4$.}
\end{subfigure}\hfill
\begin{subfigure}[t]{.48\textwidth}\centering
\begin{tikzpicture}[scale=.72,
 lab/.style={font=\scriptsize,fill=white,inner sep=1.5pt}]
\draw[draw=gray!65,dashed,fill=panelbg,rounded corners=5pt] (-1.95,-1.85) rectangle (1.6,1.45);
\node[auxv,label=left:{$z_{S_p}$}] (zp) at (-.7,1.05) {};
\node[center,label=above right:{$u$}] (u) at (1.05,.05) {};
\node[center,label=below left:{$w$}] (w) at (-.25,-1.45) {};
\node[center,label=above:{$p$}] (p) at (-1.3,2.5) {};
\node[auxv,label=above:{$z_{S_u}$}] (zu) at (2.6,-.65) {};
\node[auxv,label=below:{$z_{S_w}$}] (zw) at (.85,-2.65) {};
\draw[bluerep] (zp)--node[lab,above right] {$1/2$}(u);
\draw[blueedge] (u)--node[lab,below right] {$1/4$}(w);
\draw[redrep] (w) .. controls (-2.1,-1.45) and (-2.1,1.05) .. node[lab,left,pos=.5] {$1/2$}(zp);
\draw[connedge] (p)--node[lab,left] {$1/4$}(zp);
\draw[connedge,line width=2.2pt] (u)--node[lab,above] {$1/2$}(zu);
\draw[connedge] (w)--node[lab,above right] {$1/4$}(zw);
\node[font=\scriptsize] at (.0,-.15) {$X_u$};
\end{tikzpicture}
\caption{Connector increased: boundary sum $1$.}
\end{subfigure}
\caption{An internal change does not satisfy the triangle-cut inequality.
On the right, increasing the value on $c_{S_u}$ by $1/4$ meets the triangle
demand and increases \eqref{eq:quartic-objective} by $1/4$. The same boundary correction can
be made using $c_{S_p}$ or $c_{S_w}$.}
\label{fig:triangle-demand-repair}
\end{figure}
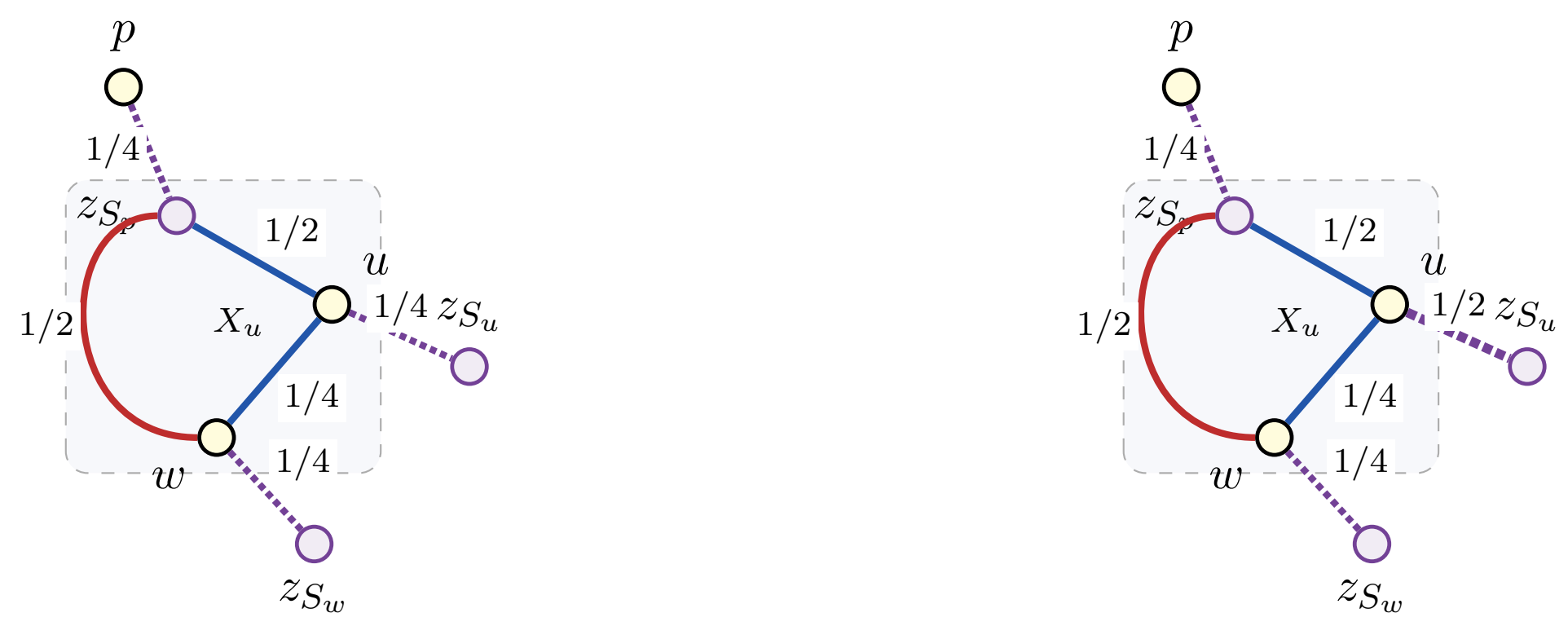

The objective-preserving change has already increased the
coordinate of $\rep{pu}$ to $1/2$. This fixes the cut around $u$,
but $\rep{pu}$ is internal to $X_u$ and does not affect its
boundary sum. The three boundary connectors retain their
initial values, so this sum remains
\[
    x^{\mathrm{init}}(\delta_{H^\circ}(X_u))
    =x^{\mathrm{init}}_{c_{S_p}}
     +x^{\mathrm{init}}_{c_{S_u}}
     +x^{\mathrm{init}}_{c_{S_w}}
    =\frac34<1.
\]
We call this the \emph{triangle demand at $u$}.
To meet it, increase any one of these three connector coordinates
from $1/4$ to $1/2$. The boundary sum then becomes one.
Figure~\ref{fig:triangle-demand-repair} contrasts the internal
parent-edge change with a boundary correction.

\textbf{Satisfying two demands with one increase.}
Each remaining correction increases a connector or an edge of
$N_H$ from $1/4$ to $1/2$. By \eqref{eq:quartic-objective},
this increases the objective by $1/4$; we call it a
\textit{paid increase}.

\begin{figure}[!t]
\centering
\begin{subfigure}[t]{.46\textwidth}\centering
\begin{tikzpicture}[scale=.66,
 lab/.style={font=\scriptsize,fill=white,inner sep=1.5pt}]
\draw[draw=gray!65,dashed,fill=panelbg] (-.25,1.15) circle (.40);
\draw[draw=gray!65,dashed,fill=panelbg] (-.25,-.95) circle (.40);
\node[center,label=left:{$u\in\mathcal U$}] (u) at (-.25,1.15) {};
\node[center,label=left:{$w\in\mathcal U$}] (w) at (-.25,-.95) {};
\node[vertex,label=above:{$p$}] (p) at (-1.2,2.55) {};
\node[auxv,label=right:{$z_{S_u}$}] (zu) at (1.3,.35) {};
\node[vertex] (t) at (-1.15,-2.4) {};
\node[auxv,label=right:{$z_{S_w}$}] (zw) at (1.3,-1.55) {};
\draw[blueedge] (p)--(u);
\draw[connedge] (u)--(zu);
\draw[blueedge,line width=2.4pt] (u)--node[lab,left=3pt] {$1/4\to1/2$}(w);
\draw[blueedge] (w)--(t);
\draw[connedge] (w)--(zw);
\end{tikzpicture}
\caption{Increasing $\rep{uw}$, where $uw\in N$, satisfies both
vertex demands.}
\end{subfigure}\hfill
\begin{subfigure}[t]{.49\textwidth}\centering
\begin{tikzpicture}[scale=.66,
 lab/.style={font=\scriptsize,fill=white,inner sep=1.5pt}]
\draw[draw=gray!65,dashed,fill=panelbg] (-.7,2.7) circle (.38);
\draw[draw=gray!65,dashed,fill=panelbg,rounded corners=5pt] (-1.95,-1.85) rectangle (1.6,1.4);
\node[center,label=above right:{$u\in\mathcal U$}] (u) at (-.7,2.7) {};
\node[auxv,label=left:{$z_{S_u}$}] (z) at (-.7,1.05) {};
\node[center,label=above right:{$v\in\mathcal D$}] (v) at (1.05,.05) {};
\node[center,label=below left:{$w$}] (w) at (-.25,-1.4) {};
\node[auxv,label=right:{$z_{S_v}$}] (zv) at (2.5,-.6) {};
\node[auxv,label=below:{$z_{S_w}$}] (zw) at (.9,-2.6) {};
\draw[blueedge] (u)--(-1.6,3.7);
\draw[blueedge] (u)--(-2,1.7);
\draw[connedge,line width=2.4pt] (u)--node[lab,right=3pt] {$1/4\to1/2$}(z);
\draw[bluerep] (z)--(v);
\draw[blueedge] (v)--(w);
\draw[redrep] (w) .. controls (-2.1,-1.4) and (-2.1,1.05) .. (z);
\draw[connedge] (v)--(zv);
\draw[connedge] (w)--(zw);
\end{tikzpicture}
\caption{The connector $c_{S_u}$ meets the vertex demand at $u$ and the triangle demand at $v\in\mathcal D$.}
\end{subfigure}
\caption{A single increase from $1/4$ to $1/2$ satisfies both shaded
cuts. The thick edge is shared by two vertex demands on the left,
and by a vertex demand and a triangle demand on the right.}
\label{fig:demand-shared-repairs}
\end{figure}
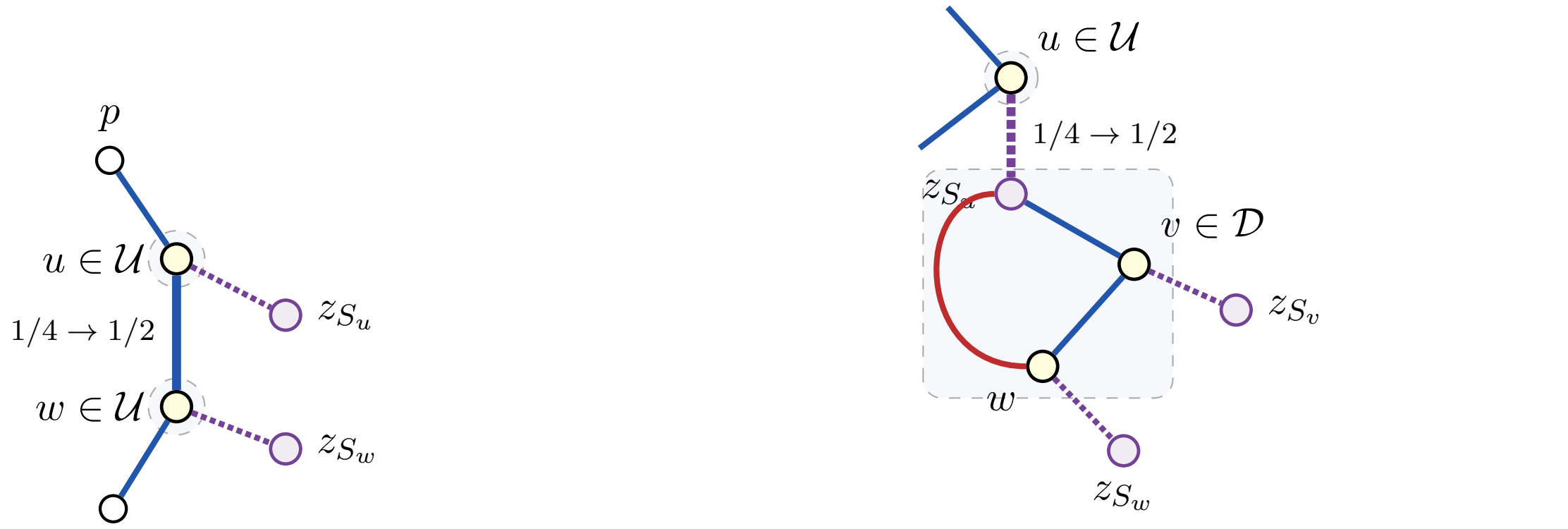

We can reduce the number of paid increases by choosing
edges that belong to the boundaries of two demands.
For example, suppose that $u,w\in\mathcal U$, where $w$
is a child of $u$ and $uw\in N$. Increasing $\rep{uw}$
satisfies both vertex demands, with a total objective
increase of $1/4$.
A connector can also belong to two demand boundaries,
as shown in Figure~\ref{fig:demand-shared-repairs}.

We next construct a graph recording these shared edges
and use a matching to choose the corrections.

\textbf{Selecting the paid increases.}
For each demand $u$, let $\mathcal E_u\subseteq E(H^\circ)$
be its three boundary edges.
More explicitly, put $p:=\operatorname{par}(u)$ and let $w$ be the
child whose subtree contains no back edge ending at $u$. Then
\[
\begin{aligned}
 \mathcal E_u&=\{\rep{pu},c_{S_u},\rep{uw}\},
     &&u\in\mathcal U,\\
 \mathcal E_u&=\{c_{S_p},c_{S_u},c_{S_w}\},
     &&u\in\mathcal D.
\end{aligned}
\]
Construct a graph $K$ whose vertices are the demands in
$\mathcal U\mathbin{\dot\cup}\mathcal D$.
Join two distinct demands when their boundary sets share an edge.
An edge of $K$ therefore records an increase that satisfies both demands.

Let $M$ be a maximum matching in $K$. For each $uu'\in M$,
we satisfy both demands by choosing an edge in
$\mathcal E_u\cap\mathcal E_{u'}$ and increasing its value
from $1/4$ to $1/2$. For every demand $u$ not covered by
$M$, we increase $c_{S_u}$ to $1/2$. If an edge is chosen more
than once, we increase its value only once. \\
Let $\bar x$ denote
the vector obtained from $x^{\mathrm{init}}$ after all
objective-preserving and paid increases.

An edge of $N_H$ is increased only when it belongs to the
boundaries of two matched vertex demands. Both endpoints of such
an edge must therefore lie in $\mathcal U$.

Now consider a nonroot vertex $v$ of $G$ with one DFS child. Since every
vertex in $\mathcal U$ has two DFS children, $v\notin\mathcal U$. Therefore,
its parent edge cannot join two vertex demands
and is not chosen for such an increase. All other paid
increases are on connectors, so the value on
$\rep{\operatorname{par}(v)v}$ remains unchanged. We will use this observation
later when verifying the Blossom inequalities.

We use one increase for each edge of $M$ and at most one
for each of the $|V(K)|-2|M|$ unmatched demands.
Thus the number of paid increases is at most
\begin{equation}
\label{eq:quartic-repair-number}
 \rho:=|M|+\bigl(|V(K)|-2|M|\bigr)
      =|V(K)|-\nu(K),
\end{equation}
where $\nu(K):=|M|$ is the maximum matching size.
We first prove that $\bar x\in{\mathcal P^\circ}$.
We then show that $K$ is a forest and use its structure
to prove $\rho\le(n-2)/5$.

\subsection{Feasibility of the final vector}
\label{subsec:quartic-feasibility}

Let $\bar x$ be the vector obtained from $x^{\mathrm{init}}$
by the corrections in
Section~\ref{subsec:quartic-repair-procedure}.
Every coordinate of $\bar x$ belongs to $\{1/4,1/2,3/4\}$,
except $\bar x_{c_{S_r}}=1$. To prove
$\bar x\in{\mathcal P^\circ}$, it remains to verify the
Blossom inequalities in
\eqref{eq:repair-nonloop-polytope}:
\begin{equation}
\label{eq:quartic-cut-goal}
 \bar x\bigl(\delta_{H^\circ}(X)\setminus F\bigr)+|F|-\bar x(F)\ge1
 \quad\begin{aligned}[t]
 &\text{for every }X\subseteq W,\\
 &F\subseteq\delta_{H^\circ}(X)
   \text{ with }|\delta_{H^\circ}(X)|+|F|\text{ odd}.
 \end{aligned}
\end{equation}

\begin{figure}[!t]
\centering
\begin{subfigure}[t]{.46\textwidth}\centering
\begin{tikzpicture}[x=.76cm,y=.58cm,font=\small]
\node[vertex,label=left:{$\bar p$}] (p) at (0,3.2) {};
\node[center,label=left:{$v$}] (v) at (0,2.2) {};
\node[auxv,label=left:{$z_S$}] (z) at (0,1.2) {};
\node[vertex] (a) at (0,.1) {};
\node[vertex] (b) at (-.85,-1.4) {};
\node[vertex] (c) at (.85,-1.4) {};
\draw[bluerep] (p)--node[right,font=\small] {$1/4$}(v);
\draw[connedge] (v)--node[right,font=\small] {$1/4$}(z);
\draw[bluerep] (z)--(a);
\draw[blueedge,dashed] (a)--(b); \draw[blueedge,dashed] (a)--(c);
\draw[redrep] (z) .. controls (-2,1.2) and (-2,-1.4) .. (b);
\draw[redrep] (z) .. controls (2,1.2) and (2,-1.4) .. (c);
\draw[draw=gray!70,dashed] (v) circle (.3);
\end{tikzpicture}
\caption{In $H^\circ$, the original vertex $v$ has degree two.}
\end{subfigure}\hfill
\begin{subfigure}[t]{.46\textwidth}\centering
\begin{tikzpicture}[x=.76cm,y=.58cm,font=\small]
\node[vertex,label=left:{$\bar p$}] (p) at (0,3.2) {};
\node[auxv,label=left:{$z_S$}] (z) at (0,1.2) {};
\node[vertex] (a) at (0,.1) {};
\node[vertex] (b) at (-.85,-1.4) {};
\node[vertex] (c) at (.85,-1.4) {};
\draw[bluerep] (p)--node[right,font=\small] {$1/4$}(z);
\draw[bluerep] (z)--(a);
\draw[blueedge,dashed] (a)--(b); \draw[blueedge,dashed] (a)--(c);
\draw[redrep] (z) .. controls (-2,1.2) and (-2,-1.4) .. (b);
\draw[redrep] (z) .. controls (2,1.2) and (2,-1.4) .. (c);
\end{tikzpicture}
\caption{In $Q$, $v$ is suppressed and $z_S$ has degree four.}
\end{subfigure}
\caption{Suppressing a degree-two vertex preserves the common value
$1/4$ on the replacement edge. Here $\bar p$ is the other endpoint
of the parent edge in $H^\circ$. The root connector, of value one,
is deleted together with its degree-one endpoint. Dashed blue segments
denote tree paths.}
\label{fig:quartic-suppression}
\end{figure}
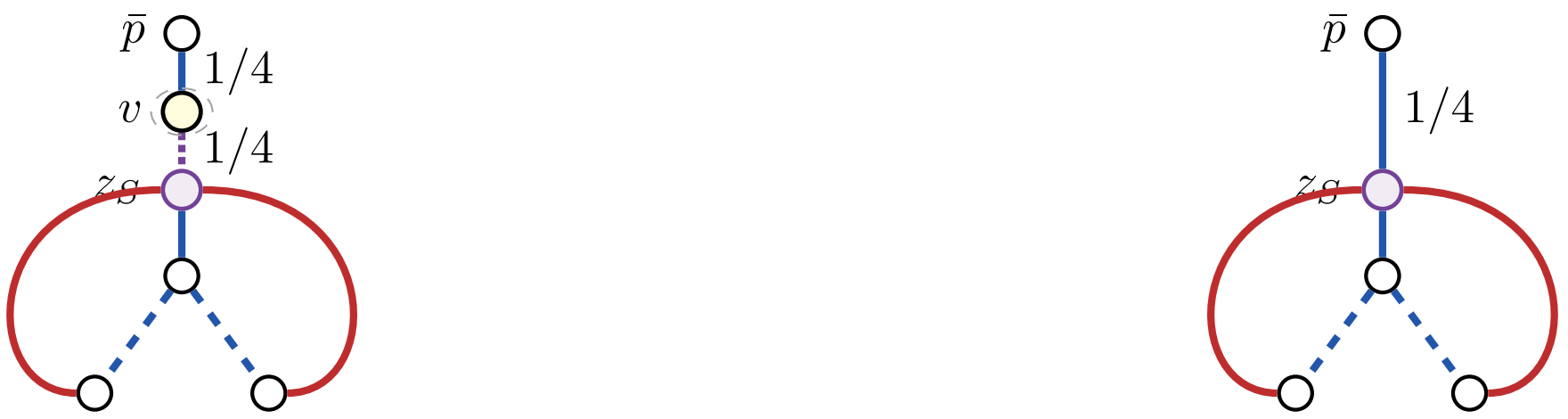

At every degree-two vertex of $H^\circ$, both incident
edges have value $1/4$. Suppress each such vertex by
replacing its two incident edges with a single edge.
Also delete the root connector and its degree-one endpoint
$r$. Denote the resulting graph by $Q$, and define
$\bar x^Q\in\mathbb R^{E(Q)}$ by assigning value $1/4$ to each
replacement edge and retaining $\bar x_a$ on every unchanged
edge $a$. See Figure~\ref{fig:quartic-suppression}.
We also write $\rep{e}$ for the corresponding edge in $Q$, with
each replacement edge inheriting the label of the parent edge
in the path it replaces.
We use $Q$ and $\bar x^Q$ only to verify the Blossom inequalities.

\begin{proposition}[An equivalent cut verification]
\label{lem:quartic-reduction}
{The vector $\bar x$ belongs to $\mathcal P^\circ$
if and only if $\bar x^Q$ belongs to the
$\odd(Q)$-join polytope.}
Moreover, $Q$ is simple and every vertex has degree three
or four. Its degree-three vertices are the auxiliary
vertices of stars of size two, the vertices in $B_2$,
and the vertices of $G$ with one DFS child and one back
edge ending at them.
\end{proposition}

\begin{proof} 
For a suppressed vertex, crossing one incident edge gives
the same contribution as the replacement edge.
If both edges cross and exactly one belongs to $F$, their
contribution is $1/4+3/4=1$. Otherwise, removing both
preserves the parity condition and only decreases the
left-hand side.

A crossing root connector contributes one if it is outside
$F$, satisfying the inequality, and zero otherwise.
In the latter case, deleting it preserves the parity
condition. Conversely, every reduced cut inequality lifts
unchanged.

After relabeling the surviving auxiliary vertices of the
root star and stars of size three by their centers,
$Q$ is obtained from $G$ by applying the construction of
$H^\circ$ only to the stars of size two.
This preserves simplicity and gives the stated degrees
and vertex types.
\end{proof}

\subsection{Verifying the small cuts}

Every coordinate of $\bar x^Q$ lies in $[1/4,3/4]$.
Hence, for $C:=\delta_Q(X)$ and $F\subseteq C$,
\begin{equation}
\label{eq:quartic-large-cuts}
 \bar x^Q(C\setminus F)+|F|-\bar x^Q(F)
 \ge\frac14|C|
 \ge1
 \qquad\text{if }|C|\ge4.
\end{equation}
It remains to consider cuts with at most three edges.
The next lemma shows that these cuts isolate a vertex
or a triangle.

\begin{lemma}[Cuts with at most three edges]
\label{lem:quartic-small-cuts}
The graph $Q$ has no bridge or two-edge cut.
Every three-edge cut has one of the following forms:
\[
 \begin{cases}
 \delta_Q(\{u\}),
   &\deg_Q(u)=3,\\[3pt]
 \delta_Q(\{u,v,w\}),
   &uv,vw,uw\in E(Q),\quad
    \deg_Q(u)=\deg_Q(v)=\deg_Q(w)=3.
 \end{cases}
\]
\end{lemma}

\begin{proof}
Let $C:=\delta_Q(X)$ have at most three edges, with both
shores nonempty, and let $s$ be the number of connectors
in $C$. For each connector $c_S=uz_S\in C$, delete
the vertex $u$ from $G$. These are $s$ distinct vertices.
Also delete the edges of $G$ corresponding to the
nonconnector edges of $C$, namely
\[
 \{e\in E(G):\rep{e}\in C\}.
\]
Denote the resulting graph by $G_C$.

After the vertex deletions, the graph is
$(4-s)$-vertex-connected and hence
$(4-s)$-edge-connected. We then delete at most $3-s$
additional edges, so $G_C$ is connected.

Now, in $Q$, delete both $u$ and $z_S$ for each connector
$c_S=uz_S\in C$, together with the nonconnector edges
of $C$. Contracting the remaining connectors gives
exactly $G_C$, which we have just shown is connected.
Since all edges crossing the cut were deleted and each
contracted connector lies within one shore, all remaining
vertices must lie on the same side of the cut.
Replacing $X$ by its complement if necessary, we may
therefore assume that every vertex of $X$ was deleted.
Each crossing connector has one endpoint in $X$, so
\[
 1\le |X|\le s\le3.
\]

If $|X|=1$, its vertex must have degree three.
If $|X|=2$, simplicity gives
\[
 |\delta_Q(X)|\ge3+3-2=4,
\]
which is impossible. Finally, if $|X|=3$, then
\[
 |\delta_Q(X)|
 =\sum_{v\in X}\deg_Q(v)-2|E(Q[X])|
 \ge9-6=3.
\]
Equality holds only when the three vertices form a
triangle and each has degree three.
This proves both assertions.
\end{proof}

\begin{lemma}[Vertex cuts]
\label{lem:quartic-vertex-cuts}
The vector $\bar x^Q$ satisfies all Blossom inequalities for
cuts of the form $\delta_Q(\{u\})$.
\end{lemma}

\begin{proof}
Cuts around degree-four vertices satisfy
\eqref{eq:quartic-large-cuts}.
For a three-edge cut $C=\{e_1,e_2,e_3\}$, the Blossom
parity condition allows $|F|=0$ or $2$, so the inequalities are
\begin{equation}
\label{eq:quartic-three-cut-test}
 \begin{aligned}
 \bar x^Q_{e_1}+\bar x^Q_{e_2}+\bar x^Q_{e_3}&\ge1,\\
 \bar x^Q_{e_i}+\bar x^Q_{e_j}-\bar x^Q_{e_k}&\le1
 \qquad(\{i,j,k\}=\{1,2,3\}).
 \end{aligned}
\end{equation}

Initially, the only unsatisfied vertex cuts are those
around $u\in B_2$, where all three incident values are
$1/4$. The correction procedure raises at least one of
them to $1/2$. Every other degree-three vertex already
has an incident value of at least $1/2$.
Thus the first inequality holds at every vertex.

At any degree-three vertex, at most one incident value
is $3/4$, and the others are at most $1/2$.
Since every value is at least $1/4$, the second inequality
follows from
\[
 \frac34+\frac12-\frac14=1.
\]
\end{proof}

It remains to check the triangle cuts.
The next lemma shows that they are exactly the cuts
used to define the triangle demands.

\begin{lemma}[Triangle cuts]
\label{lem:quartic-triangles}
Every triangle of degree-three vertices in $Q$ has the form
\[
 Q[\{z_{S_p},u,w\}],
 \qquad p:=\operatorname{par}(u),\quad u\in\mathcal D,
\]
where $w$ is the child of $u$ whose subtree contains no
back edge ending at $u$.
Its boundary is $\{c_{S_p},c_{S_u},c_{S_w}\}$, and $\bar x^Q$ satisfies
all Blossom inequalities for this cut.
\end{lemma}

\begin{proof}
A triangle cannot contain a connector: its other two
edges would correspond to parallel edges in $G$.
Thus each vertex of the triangle has its connector
on the boundary and its other two incident edges inside.

A vertex in $B_2$ has two incident tree edges.
Every other degree-three vertex has one tree edge and
one back edge, in addition to its connector.
Let $b$ denote the number of vertices of the triangle that
belong to $B_2$. The number of tree edges in the triangle is
\[
 \frac{2b+(3-b)}2=\frac{3+b}{2}.
\]
Since this is an integer, $b=1$ or $b=3$.
The latter would make all three triangle edges tree edges,
which is impossible. Hence the triangle contains exactly
one vertex $u\in B_2$.

The two triangle edges at $u$ correspond to its parent
edge and the edge $uw\in N$, where $w$ is its child.
Since $w\notin B_2$ and has degree three in $Q$, it has
exactly one DFS child and one back edge ending at it.

Put $p:=\operatorname{par}(u)$. We now show that $pu$ belongs
to a star of size two centered at $p$, so the third vertex of the
triangle is $z_{S_p}$.

Suppose first that $pu\in N$. Then the third vertex is $p$,
which has degree three in $Q$. By
Proposition~\ref{lem:quartic-reduction}, either $p\in B_2$
or $p$ centers a star of size two and has exactly one DFS child.
If $p\in B_2$, the triangle would contain two vertices
of $B_2$, namely $p$ and $u$, contradicting $b=1$.
In the latter case, $u$ is the only DFS child of $p$, so $pu$
belongs to this star, contradicting $pu\in N$.

Thus $pu$ belongs to a star centered at $p$. This star must
have size two, since a star of size three or the root star would give a
vertex of degree four in $Q$. The third triangle vertex
is therefore $z_{S_p}$. The edge joining $z_{S_p}$ to $w$
corresponds to the back edge $pw$, so
\[
 S_p=\{pu,pw\}.
\]

This is exactly the configuration defining $u\in\mathcal D$,
shown in Figure~\ref{fig:triangle-demand-construction}, with
\[
 \delta_Q(\{z_{S_p},u,w\})
 =\{c_{S_p},c_{S_u},c_{S_w}\}.
\]
The correction procedure increases at least one of these
connector values from $1/4$ to $1/2$.
All three values belong to $\{1/4,1/2\}$, so their sum is
at least one and the sum of any two is at most one.
Thus \eqref{eq:quartic-three-cut-test} holds.
\end{proof}
\begin{proposition}[The final vector]
\label{prop:quartic-completed-vector}
The vector $\bar x$ obtained by the feasibility correction
procedure belongs to ${\mathcal P^\circ}$.
\end{proposition}

\begin{proof}
By Lemma~\ref{lem:quartic-small-cuts}, every cut of $Q$
with at most three edges is a vertex cut or a triangle
cut. Their Blossom inequalities hold by
Lemmas~\ref{lem:quartic-vertex-cuts}
and~\ref{lem:quartic-triangles}.
Larger cuts satisfy \eqref{eq:quartic-large-cuts}.
{Hence $\bar x^Q$ belongs to the
$\odd(Q)$-join polytope}, and
Proposition~\ref{lem:quartic-reduction} gives
$\bar x\in{\mathcal P^\circ}$.
\end{proof}

\subsection{The structure of shared corrections}

It remains to bound how many paid increases the procedure uses.
Recall that $K$ has one vertex for each remaining demand:
\[
 V(K)=\mathcal U\mathbin{\dot\cup}\mathcal D.
\]
Two distinct demands are adjacent when their boundaries share
an edge. Formally, for $u,v\in V(K)$ with $u\ne v$,
\[
 uv\in E(K)
 \quad\Longleftrightarrow\quad
 \mathcal E_u\cap\mathcal E_v\ne\varnothing,
\]
where $\mathcal E_u$ is the set of boundary edges of the demand
at $u$. Locate a vertex demand at its vertex and a triangle demand
at its middle vertex. For adjacent demands, call the one higher
in the DFS tree the \emph{predecessor} and the lower one its
\emph{successor}. The proof below shows that the two locations
are always comparable in the DFS tree.
We use this structure to bound $\rho=|V(K)|-\nu(K)$.

\begin{lemma}[The correction forest]
\label{lem:quartic-repair-forest}
The graph $K$ is a forest, and every vertex has at most one
predecessor. A vertex in $\mathcal U$ has at most one successor
in $\mathcal U$ and one in $\mathcal D$. A vertex in
$\mathcal D$ has at most two successors, both in $\mathcal D$.
In particular, $K[\mathcal U]$ is a disjoint union of paths.
\end{lemma}

\begin{proof}
To understand the edges of $K$, we determine which two
demands can share a correction. We show that every edge
of $K$ comes from one of the following three cases.
We then use these cases to bound the number of predecessors
and successors of each demand.

Recall that a triangle demand at $u$ has boundary
\[
 \{c_{S_p},c_{S_u},c_{S_w}\},
 \qquad p=\operatorname{par}(u),
\]
where $w$ is the child of $u$ whose subtree contains no
back edge ending at $u$. This vertex $w$ has exactly
one DFS child.

\smallskip\noindent
\textbf{Case 1: Two demands share a nonremovable tree edge.}
Let $uw\in N$, where $w$ is a child of $u$.
The edge $\rep{uw}$ can belong only to vertex-demand
boundaries, since triangle boundaries contain only connectors.
The two demands must therefore be at its endpoints,
so $u,w\in\mathcal U$ (see Figure~\ref{fig:quartic-case-one}).
Conversely, if $u,w\in\mathcal U$, their boundaries both
contain $\rep{uw}$, giving the edge $uw$ in $K$.
Here $w$ is the child of $u$ whose subtree contains no
back edge ending at $u$.

\begin{figure}[!t]
\centering
\begin{tikzpicture}[x=.82cm,y=.72cm,font=\small,
 lab/.style={font=\small,fill=white,inner sep=2pt}]
\draw[gray,dashed,fill=panelbg] (0,1.2) circle (.38);
\draw[gray,dashed,fill=panelbg] (0,-.8) circle (.38);
\node[center,label=left:{$u\in\mathcal U$}] (u) at (0,1.2) {};
\node[center,label=left:{$w\in\mathcal U$}] (w) at (0,-.8) {};
\node[vertex,label=above:{$p$}] (p) at (-.9,2.6) {};
\node[auxv,label=right:{$z_{S_u}$}] (zu) at (1.7,.3) {};
\node[vertex] (a) at (-.9,-2.2) {};
\node[auxv,label=right:{$z_{S_w}$}] (zw) at (1.7,-1.5) {};
\draw[blueedge] (p)--(u);
\draw[connedge] (u)--(zu);
\draw[blueedge,line width=2.4pt] (u)--node[lab,left=2pt] {$\rep{uw}$}(w);
\draw[blueedge] (w)--(a);
\draw[connedge] (w)--(zw);
\node[align=center] at (6,2.05) {The corresponding edge of $K$};
\node[circle,draw=edgeblue,fill=edgeblue!8,minimum size=7mm] (ku) at (5,.1) {$u$};
\node[circle,draw=edgeblue,fill=edgeblue!8,minimum size=7mm] (kw) at (7,.1) {$w$};
\draw[line width=1.3pt] (ku)--(kw);
\node[align=center] at (6,-1.35) {$\mathcal E_u\cap\mathcal E_w=\{\rep{uw}\}$};
\end{tikzpicture}
\caption{Case~1. The nonremovable edge $uw$ joins two vertices in
$\mathcal U$. Its copy in $H^\circ$ crosses both singleton cuts,
so one increase satisfies both vertex demands.}
\label{fig:quartic-case-one}
\end{figure}
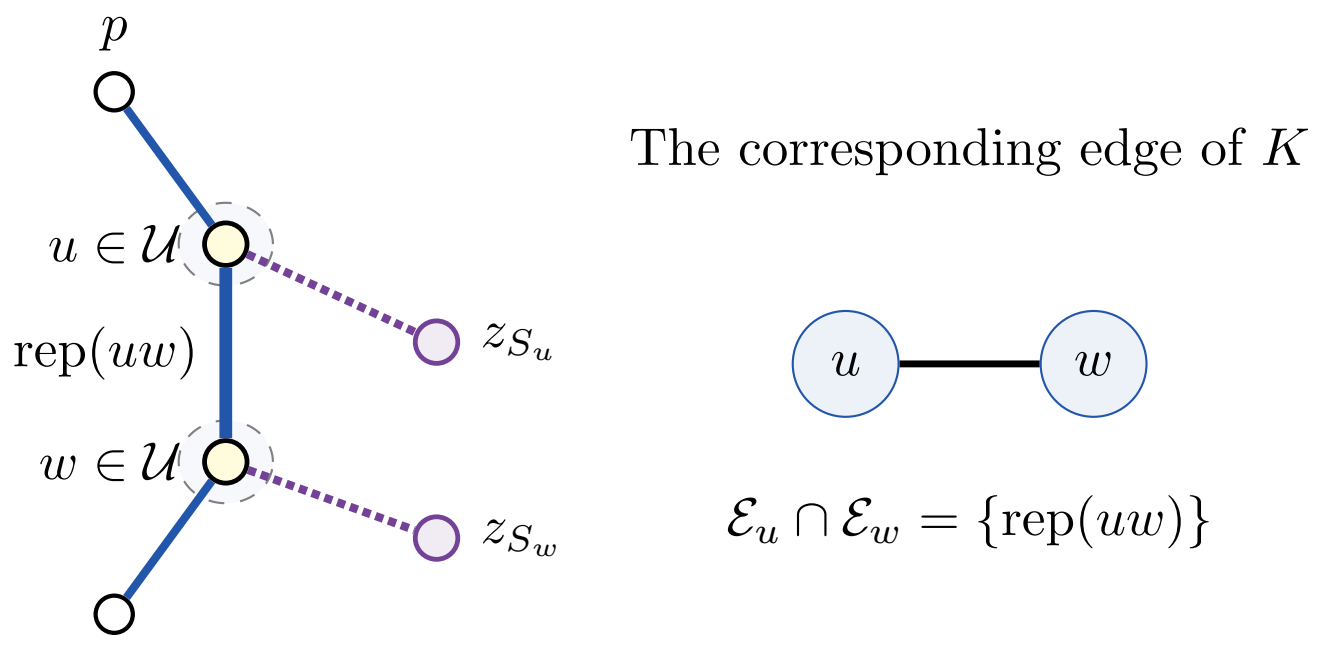

\smallskip\noindent
\textbf{Case 2: Two demands share a connector centered at $u\in B_2$.}
Let $v$ be the child of $u$ for which $uv\in S_u$.
The connector $c_{S_u}$ belongs to the boundary of the
vertex demand at $u$ if $u\in\mathcal U$, and to the boundary
of the triangle demand at $u$ if $u\in\mathcal D$.
These demands cannot both occur at $u$: a triangle demand
requires $\operatorname{par}(u)u\in R$, whereas
$u\in\mathcal U$ requires $\operatorname{par}(u)u\in N$.

The connector can also belong to the boundary of the
triangle demand at $v$, when $v\in\mathcal D$.
Thus it can be shared by a vertex demand and a triangle
demand, or by two triangle demands (see Figure~\ref{fig:quartic-case-two}).

There are no other demands whose boundaries contain
$c_{S_u}$. Indeed, a triangle demand whose parent is $u$
must be at $v$, because its triangle contains the tree
edge of $S_u$, namely $uv$.
For any triangle demand at $t$, its boundary is
\[
 \{c_{S_p},c_{S_t},c_{S_w}\},
 \qquad p:=\operatorname{par}(t),
\]
where $w$ is a child of $t$ with exactly one DFS child.
Since $u\in B_2$ has two DFS children, $u$ cannot be
this vertex $w$.

Thus the only demands whose boundaries can contain
$c_{S_u}$ are the demand at $u$ and the triangle demand
at $v$. They share this connector when both demands exist.
Therefore, $c_{S_u}$ gives the edge $uv$ in $K$ exactly when
\[
 u\in\mathcal U\cup\mathcal D
 \qquad\text{and}\qquad v\in\mathcal D.
\]

\begin{figure}[!t]
\centering
\begin{subfigure}[t]{.48\textwidth}
\centering
\begin{tikzpicture}[x=.75cm,y=.52cm,font=\small,
 lab/.style={font=\small,fill=white,inner sep=1pt}]

\draw[gray,dashed,fill=panelbg]
  (0,5.5) circle (.36);
\draw[gray,dashed,fill=panelbg,rounded corners=4pt]
  (.65,.55) rectangle (4,4.5);

\node[vertex,label=above:{$\operatorname{par}(u)$}]
  (p) at (0,6.8) {};
\node[center,label=right:{$u\in\mathcal U$}]
  (u) at (0,5.5) {};
\node[vertex] (a) at (-1.6,4) {};
\node[auxv,label=left:{$z_{S_u}$}]
  (z) at (1.2,4) {};
\node[center,label=left:{$v$}]
  (v) at (1.2,2.5) {};
\node[center,label=left:{$w$}]
  (w) at (2.5,1) {};
\node[auxv,label=left:{$z_{S_v}$}]
  (zv) at (-.3,1) {};
\node[auxv,label=right:{$z_{S_w}$}]
  (zw) at (2.5,-.4) {};

\draw[blueedge] (p)--(u);
\draw[blueedge] (u)--(a);
\draw[connedge,line width=2.4pt]
  (u)--node[lab,left] {$c_{S_u}$}(z);
\draw[bluerep] (z)--(v);
\draw[blueedge] (v)--(w);
\draw[redrep] (z)
  .. controls (4.4,4) and (4.4,1) .. (w);
\draw[connedge] (v)--(zv);
\draw[connedge] (w)--(zw);

\end{tikzpicture}
\caption{$u\in\mathcal U$, $v\in\mathcal D$.}
\end{subfigure}\hfill
\begin{subfigure}[t]{.48\textwidth}
\centering
\begin{tikzpicture}[x=.75cm,y=.52cm,font=\small,
 lab/.style={font=\small,fill=white,inner sep=1pt}]

\draw[gray,dashed,fill=panelbg,rounded corners=4pt]
  (-.55,2.45) rectangle (2.8,6.1);
\draw[gray,dashed,fill=panelbg,rounded corners=4pt]
  (-4.1,-.35) rectangle (-1.65,3.35);

\node[auxv,label=left:{$z_{S_p}$}]
  (zp) at (0,5.6) {};
\node[center,label=left:{$u$}]
  (u) at (0,4.3) {};
\node[center,label=left:{$a$}]
  (a) at (1.3,2.9) {};
\node[auxv,label=above left:{$z_{S_u}$}]
  (zu) at (-2.2,2.9) {};
\node[center,label=right:{$v$}]
  (v) at (-2.2,1.5) {};
\node[center,label=right:{$b$}]
  (b) at (-2.2,.1) {};
\node[center,label=left:{$p$}]
  (p) at (0,6.8) {};
\node[auxv,label=right:{$z_{S_a}$}]
  (za) at (1.3,1.6) {};
\node[auxv,label=right:{$z_{S_v}$}]
  (zv) at (-.7,.2) {};
\node[auxv,label=right:{$z_{S_b}$}]
  (zb) at (-2.2,-1.1) {};

\draw[connedge] (p)--(zp);
\draw[bluerep] (zp)--(u);
\draw[blueedge] (u)--(a);
\draw[redrep] (zp)
  .. controls (3.2,5.6) and (3.2,2.9) .. (a);
\draw[connedge] (a)--(za);
\draw[connedge,line width=2.4pt]
  (u)--node[lab,above left] {$c_{S_u}$}(zu);
\draw[bluerep] (zu)--(v);
\draw[blueedge] (v)--(b);
\draw[redrep] (zu)
  .. controls (-4.4,2.9) and (-4.4,.1) .. (b);
\draw[connedge] (v)--(zv);
\draw[connedge] (b)--(zb);

\end{tikzpicture}
\caption{$u,v\in\mathcal D$.}
\end{subfigure}

\caption{Case~2. Both configurations give the edge $uv$ in $K$.
The shared connector $c_{S_u}$ crosses the two shaded cuts.
On the left they are $\{u\}$ and $\{z_{S_u},v,w\}$;
on the right they are $\{z_{S_p},u,a\}$ and $\{z_{S_u},v,b\}$.
The vertex $u$ has two DFS children in both configurations.}
\label{fig:quartic-case-two}
\label{fig:quartic-shared}
\end{figure}
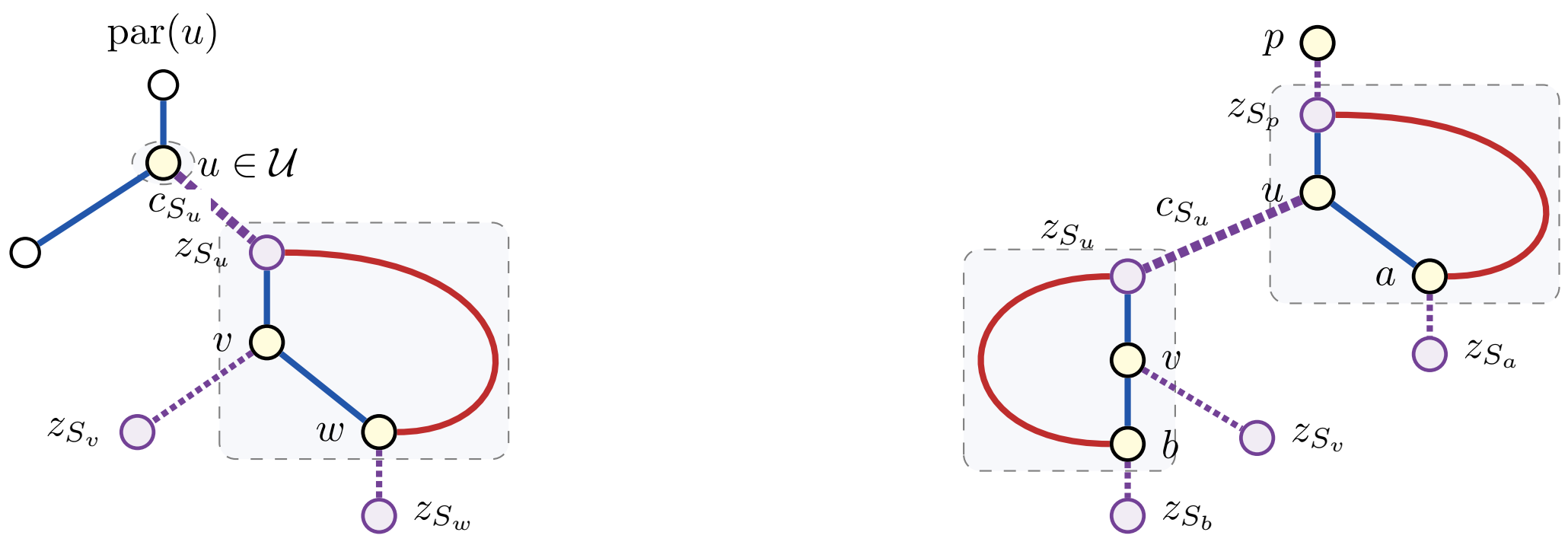
\smallskip\noindent
\textbf{Case 3: Two triangle demands share a connector at a vertex
with exactly one DFS child.}
Let $p:=\operatorname{par}(w)$, and let $v$ be the only DFS
child of $w$. Suppose that $w$ is the center of a star
$S_w$ of size two.
There is no demand at $w$, since every vertex in
$\mathcal U\cup\mathcal D$ has two DFS children.

The connector $c_{S_w}$ can belong to the boundary of a
triangle demand at $p$, when $w$ is the child of $p$ in
that triangle. It can also belong to the boundary of a
triangle demand at $v$, whose parent is $w$.
These are the only possibilities. When both occur,
$c_{S_w}$ is shared by the two triangle demands and gives
the edge $pv$ in $K$; see Figure~\ref{fig:quartic-case-three}.

\begin{figure}[!t]
\centering
\begin{tikzpicture}[x=.85cm,y=.49cm,font=\small,
 lab/.style={font=\small,fill=white,inner sep=1pt}]
\draw[gray,dashed,fill=panelbg,rounded corners=4pt] (-.55,3.05) rectangle (2.9,6.7);
\draw[gray,dashed,fill=panelbg,rounded corners=4pt] (.75,-1.15) rectangle (4.2,2.6);
\node[auxv,label=left:{$z_{S_a}$}] (za) at (0,6.2) {};
\node[center,label=left:{$p$}] (p) at (0,4.9) {};
\node[center,label=left:{$w$}] (w) at (1.3,3.5) {};
\node[auxv,label=left:{$z_{S_w}$}] (zw) at (1.3,2.1) {};
\node[center,label=left:{$v$}] (v) at (1.3,.7) {};
\node[center,label=left:{$b$}] (b) at (2.6,-.7) {};
\node[center,label=left:{$a$}] (a) at (0,7.5) {};
\node[auxv,label=left:{$z_{S_p}$}] (zp) at (-1.5,3.5) {};
\node[auxv,label=left:{$z_{S_v}$}] (zv) at (-.2,-.7) {};
\node[auxv,label=right:{$z_{S_b}$}] (zb) at (2.6,-2) {};
\draw[connedge] (a)--(za);
\draw[bluerep] (za)--(p);
\draw[blueedge] (p)--(w);
\draw[redrep] (za) .. controls (3.3,6.2) and (3.3,3.5) .. (w);
\draw[connedge] (p)--(zp);
\draw[connedge,line width=2.4pt] (w)--node[lab,left] {$c_{S_w}$}(zw);
\draw[bluerep] (zw)--(v);
\draw[blueedge] (v)--(b);
\draw[redrep] (zw) .. controls (4.6,2.1) and (4.6,-.7) .. (b);
\draw[connedge] (v)--(zv);
\draw[connedge] (b)--(zb);
\node[align=center] at (7.5,6.7) {In the DFS tree};
\node[center,label=right:{$p\in\mathcal D$}] (tp) at (6.8,5.3) {};
\node[vertex,label=left:{$w$}] (tw) at (6.8,3.1) {};
\node[center,label=right:{$v\in\mathcal D$}] (tv) at (6.8,.9) {};
\draw[blueedge] (tp)--(tw)--(tv);
\node[align=left,anchor=west] at (5.4,-1.2) {$w$ has exactly one DFS child;\\the edge in $K$ is $pv$.};
\end{tikzpicture}
\caption{Case~3. Here $a=\operatorname{par}(p)$ and
$p=\operatorname{par}(w)$. The shared connector $c_{S_w}$ crosses
the cuts around $\{z_{S_a},p,w\}$ and $\{z_{S_w},v,b\}$.
The demands are at $p$ and $v$; there is no demand at $w$.}
\label{fig:quartic-case-three}
\end{figure}
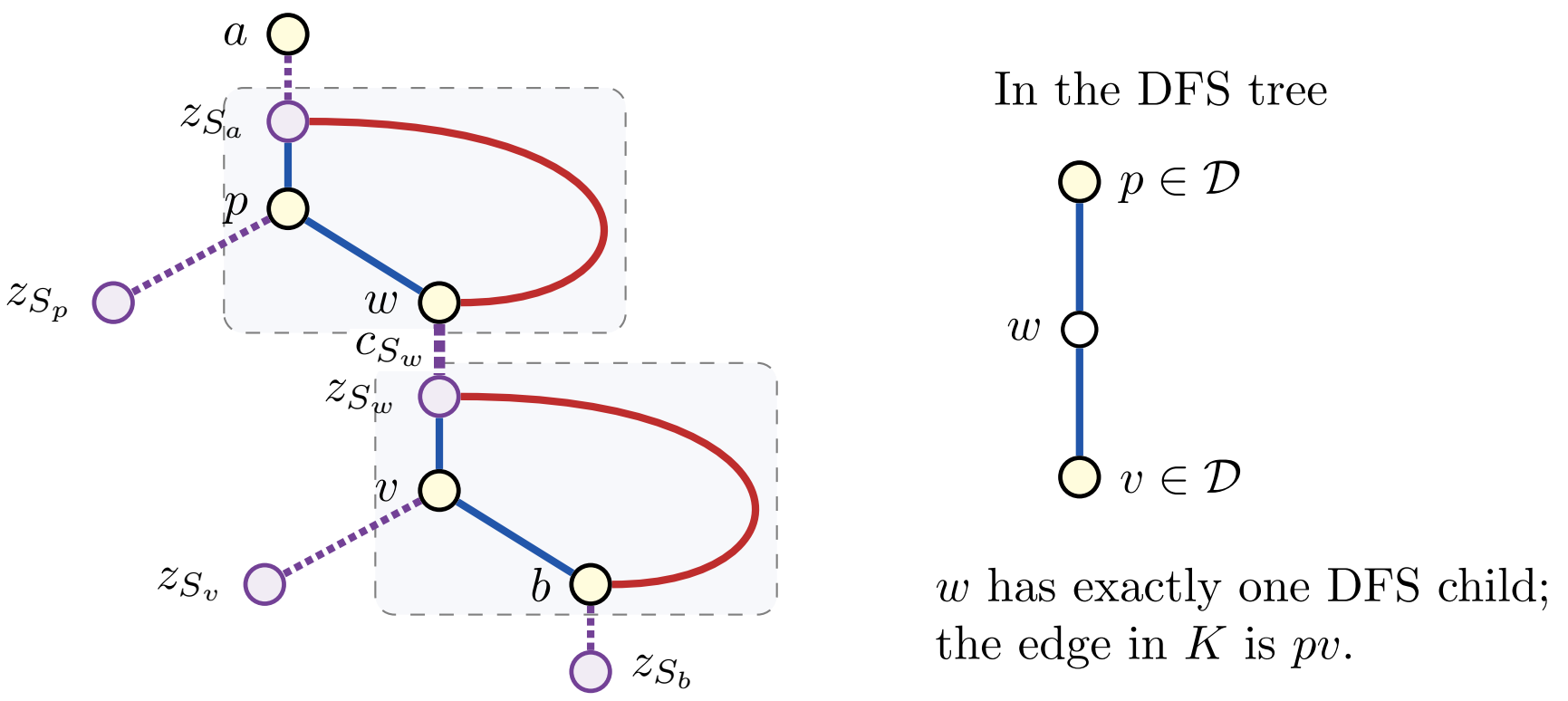

These cases cover every possible shared correction.
They also show that adjacent demands are located at
vertices where one is a descendant of the other.

For a vertex demand at $u$, the only possible predecessor
comes from its nonremovable parent edge, by Case~1.
Its possible successors are a vertex demand at the child
joined to $u$ by a nonremovable edge, and a triangle demand
at its other child, by Case~2.
Thus it has at most one predecessor and at most one
successor of each type.

Now consider a triangle demand at $u$, with boundary
$\{c_{S_p},c_{S_u},c_{S_w}\}$.
Only $c_{S_p}$ can be shared with a predecessor.
If $p\in B_2$, this predecessor must be the demand at $p$.
If $p$ has one DFS child, it must be a triangle demand
at $\operatorname{par}(p)$ whose lower vertex is $p$.
These alternatives are exclusive, so there is at most
one predecessor.
Each of $c_{S_u}$ and $c_{S_w}$ can be shared with at most
one successor, and Cases~2 and~3 show that both successors
must be triangle demands.

Finally, suppose that $K$ contains a cycle.
A vertex of greatest DFS depth on that cycle would have
two predecessors, contradicting the bound just proved.
Hence $K$ is a forest.
Every vertex of $K[\mathcal U]$ has at most one predecessor
and one successor, so its components are paths.
\end{proof}

\begin{figure}[!t]
\centering
\begin{tikzpicture}[x=.75cm,y=.56cm,
 un/.style={circle,draw=edgeblue,fill=edgeblue!8,minimum size=7mm,font=\small},
 dn/.style={rectangle,rounded corners=2pt,draw=connector,fill=connector!8,minimum size=7mm,font=\small},
 ke/.style={draw=gray!65,line width=.8pt},
 km/.style={draw=black,line width=2.4pt}]
\node[un] (u1) at (0,3.0) {$U$};
\node[un] (u2) at (-2,1.3) {$U$};
\node[un] (u3) at (-3.1,-.4) {$U$};
\node[dn] (d1) at (2,1.3) {$D$};
\node[dn] (d2) at (-.9,-.4) {$D$};
\node[dn] (d3) at (1.05,-.4) {$D$};
\node[dn] (d4) at (3.0,-.4) {$D$};
\node[dn] (d5) at (-1.8,-2.1) {$D$};
\node[dn] (d6) at (0,-2.1) {$D$};
\draw[ke] (u1)--(u2);\draw[km] (u1)--(d1);
\draw[km] (u2)--(u3);\draw[ke] (u2)--(d2);
\draw[ke] (d1)--(d3);\draw[ke] (d1)--(d4);
\draw[km] (d2)--(d5);\draw[ke] (d2)--(d6);
\node[dn] (d7) at (6.5,3.0) {$D$};
\node[dn] (d8) at (5.5,1.3) {$D$};\node[dn] (d9) at (7.5,1.3) {$D$};
\draw[km] (d7)--(d8);\draw[ke] (d7)--(d9);
\node[align=left,font=\small,anchor=west] at (4.4,-.8)
 {Circle: vertex demand\\Rectangle: triangle demand\\Thick edge: shared correction};
\end{tikzpicture}
\caption{A correction forest with a maximum matching shown by thick
edges. Each matched pair uses one correction; each unmatched demand
uses its own connector. Circles denote vertex demands and rectangles
triangle demands.}
\label{fig:quartic-forest}
\end{figure}
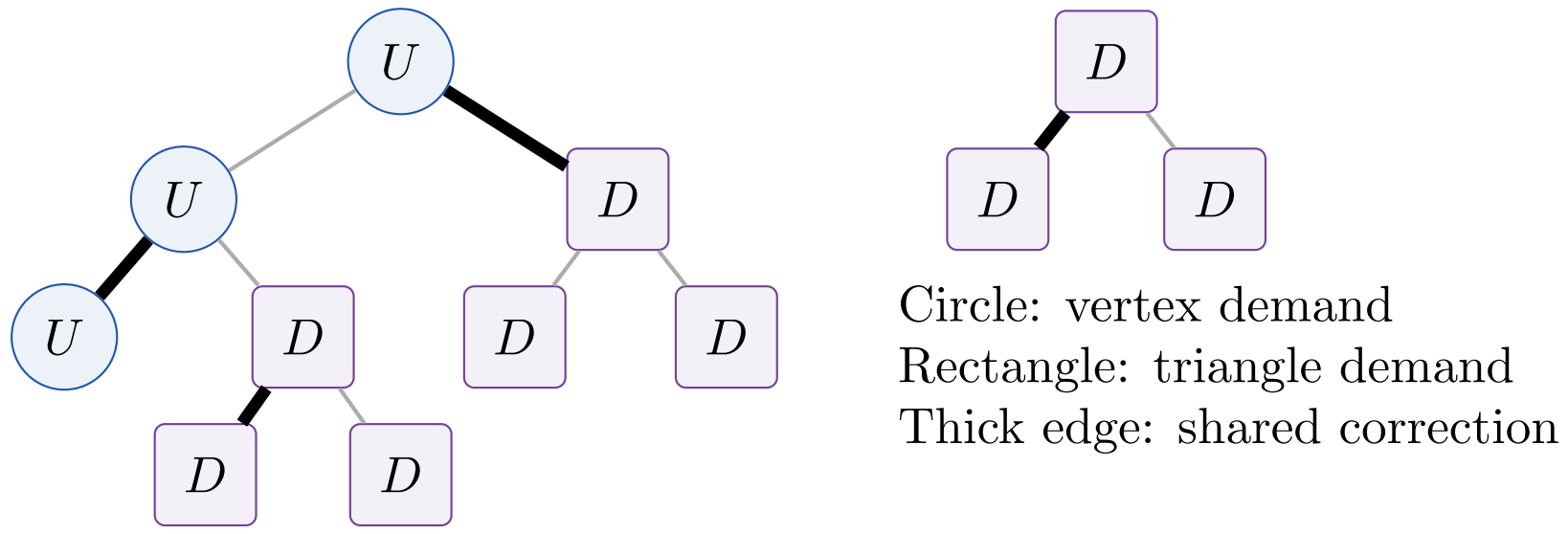

The procedure in Section~\ref{subsec:quartic-repair-procedure} uses
at most $\rho=|V(K)|-\nu(K)$ paid increases. Since $K$ is a forest,
K\"onig's theorem gives
\[
 \rho=\alpha(K),
\]
where $\alpha(K)$ is its independence number.
Figure~\ref{fig:quartic-forest} illustrates the matching used by
the procedure.

\subsection{Counting the corrections}

The correction procedure uses at most
\[
 \rho=|V(K)|-\nu(K)
\]
paid increases. We want to prove that $\rho\le(n-2)/5$.

Since $K$ is a forest, K\"onig's theorem implies that $\rho$
is the maximum size of an independent set of $K$.
It therefore suffices to prove $5|I|\le n-2$ for every
independent set $I$.

Let $c_{\mathcal U}$ be the number of connected components
of $K[\mathcal U]$, each of which is a path, possibly
consisting of a single vertex.
Let $c_{\mathcal D}$ be the number of connected components
of $K$ whose root belongs to $\mathcal D$.
We prove
\[
 5|I|
 \le 3|\mathcal U|+4|\mathcal D|
       +2c_{\mathcal U}+c_{\mathcal D}
 \le n-2.
\]
The first inequality uses the structure of $K$.
The second follows by counting the edges and vertices
of the DFS tree required by these demands.

\smallskip\noindent
\textbf{Bounding the size of an independent set.}
Fix an independent set $I$.
On each path of $K[\mathcal U]$, at most one more vertex
belongs to $I$ than to its complement. Thus
\[
 |I\cap\mathcal U|
 \le|\mathcal U\setminus I|+c_{\mathcal U},
\]
or equivalently,
\[
 2|I\cap\mathcal U|\le|\mathcal U|+c_{\mathcal U}.
\]

Every vertex in $I\cap\mathcal D$ either is the root of
its component or has a predecessor outside $I$.
A predecessor in $\mathcal U$ has at most one successor
in $\mathcal D$, and a predecessor in $\mathcal D$
has at most two. Therefore,
\[
 |I\cap\mathcal D|
 \le|\mathcal U\setminus I|
   +2|\mathcal D\setminus I|+c_{\mathcal D}.
\]
Using $|\mathcal U\setminus I|=|\mathcal U|-|I\cap\mathcal U|$
and $|\mathcal D\setminus I|=|\mathcal D|-|I\cap\mathcal D|$,
we obtain
\[
 |I\cap\mathcal D|
 \le|\mathcal U|-|I\cap\mathcal U|
   +2|\mathcal D|-2|I\cap\mathcal D|+c_{\mathcal D}.
\]
Rearranging gives
\[
 |I\cap\mathcal U|+3|I\cap\mathcal D|
 \le|\mathcal U|+2|\mathcal D|+c_{\mathcal D}.
\]
Combining these two inequalities with
$|I\cap\mathcal D|\le|\mathcal D|$, we obtain
\[
 \begin{aligned}
 5|I|
 &=2\bigl(2|I\cap\mathcal U|\bigr)
   +\bigl(|I\cap\mathcal U|+3|I\cap\mathcal D|\bigr)
   +2|I\cap\mathcal D|\\
 &\le
 2\bigl(|\mathcal U|+c_{\mathcal U}\bigr)
 +|\mathcal U|+2|\mathcal D|+c_{\mathcal D}
 +2|\mathcal D|\\
 &=3|\mathcal U|+4|\mathcal D|
   +2c_{\mathcal U}+c_{\mathcal D}.
 \end{aligned}
\]

\smallskip\noindent
\textbf{Relating the demands to the size of $G$.}
We now prove the second inequality,
\[
 3|\mathcal U|+4|\mathcal D|
 +2c_{\mathcal U}+c_{\mathcal D}\le n-2.
\]

For $i\in\{1,2\}$, let $n_i$ count the nonroot vertices
with exactly one DFS child and exactly $i$ back edges
ending at them. Of the $n+1$ back edges, three end at
the root. Counting the remaining upper endpoints gives
\begin{equation}
\label{eq:quartic-back-count}
 n-2=n_1+2n_2+|B_2|.
\end{equation}
There are $1+n_1+n_2+|B_2|$ stars, including the root
star. Each contains one tree edge, and every other tree
edge belongs to $N$. Hence
\[
 n-1=|N|+1+n_1+n_2+|B_2|.
\]
Comparing this with \eqref{eq:quartic-back-count} gives
\begin{equation}
\label{eq:quartic-N-count}
 n_2=|N|,
 \qquad
 n-2=2|N|+n_1+|B_2|.
\end{equation}
We bound $|N|$ and $n_1+|B_2|$ separately.

\smallskip\noindent
\textbf{The nonremovable edges.}
We claim that
\begin{equation}
\label{eq:quartic-U-roots}
 |N|\ge|\mathcal U|+|\mathcal D|+c_{\mathcal U}.
\end{equation}
Every demand center belongs to $B_2$ and has two DFS
children. The edge to one child belongs to its star;
the edge to the other is nonremovable. These give
$|\mathcal U|+|\mathcal D|$ distinct edges of $N$.

For each path of $K[\mathcal U]$, let $u$ be its highest
vertex and put $p:=\operatorname{par}(u)$.
Since $u\in\mathcal U$, the edge $pu$ is nonremovable.
We show that $p$ is not a demand center, so this edge
has not already been counted.

If $p\in\mathcal U$, then $pu$ gives an edge of
$K[\mathcal U]$, contradicting the choice of $u$.
If $p\in\mathcal D$, then $pu\in N$ makes $u$ the lower
vertex of the triangle demand at $p$. That vertex has
exactly one DFS child, whereas $u\in\mathcal U$ has two.
Both possibilities are therefore excluded.

Different path roots have different parent edges.
These give the additional $c_{\mathcal U}$ edges in
\eqref{eq:quartic-U-roots}.

\smallskip\noindent
\textbf{Counting vertices.}
We show that
\begin{equation}
\label{eq:quartic-D-roots}
 n_1+|B_2|
 \ge|\mathcal U|+2|\mathcal D|+c_{\mathcal D}.
\end{equation}
For each $u\in\mathcal D$, let $w_u$ be the lower vertex
of its triangle demand. Define
\[
 \begin{aligned}
 A_1&:=\mathcal U\cup\mathcal D,\\
 A_2&:=\{w_u:u\in\mathcal D\},\\
 A_3&:=\{\operatorname{par}(u):u\in\mathcal D,\ 
          u\text{ is a component root in }K\}.
 \end{aligned}
\]

All vertices of $A_1$ belong to $B_2$, and
$|A_1|=|\mathcal U|+|\mathcal D|$.
Every vertex of $A_2$ has exactly one DFS child and one
back edge ending there, so $A_1\cap A_2=\varnothing$.
Since $\operatorname{par}(w_u)=u$, different triangle
demands have different lower vertices. Thus $|A_2|=|\mathcal D|$.

Consider $p=\operatorname{par}(u)\in A_3$, where the
demand at $u$ is a component root in $K$.
The star $S_p=\{pu,pw_u\}$ has size two, so either
$p\in B_2$ or $p$ has exactly one DFS child and one
back edge ending there.
If $p\in A_1$, the demands at $p$ and $u$ share
$c_{S_p}$, making $p$ a predecessor of $u$ in $K$.
If $p\in A_2$, write $p=w_a$ for some $a\in\mathcal D$.
Then $a=\operatorname{par}(p)$, and the demands at $a$
and $u$ share $c_{S_p}$, making $a$ a predecessor of $u$.
Both contradict the root condition. Hence
$A_3\cap(A_1\cup A_2)=\varnothing$.

Different root triangle demands also have different
parents: the unique star at a parent contains exactly
one tree edge, which determines the center of the
triangle below it. Therefore $|A_3|=c_{\mathcal D}$.

The three sets are pairwise disjoint. Every vertex of
$(A_1\cup A_2\cup A_3)\setminus B_2$ has exactly one DFS
child and one back edge ending there, so
\[
 |(A_1\cup A_2\cup A_3)\setminus B_2|\le n_1.
\]
Consequently,
\[
 n_1+|B_2|
 \ge |A_1|+|A_2|+|A_3|
 =|\mathcal U|+2|\mathcal D|+c_{\mathcal D},
\]
as required.

Using the two counts in \eqref{eq:quartic-N-count},
we obtain
\begin{equation}
\label{eq:quartic-forest-budget}
 \begin{aligned}
 n-2
 &=2|N|+n_1+|B_2|\\
 &\ge
 2\bigl(|\mathcal U|+|\mathcal D|+c_{\mathcal U}\bigr)
 +|\mathcal U|+2|\mathcal D|+c_{\mathcal D}\\
 &=3|\mathcal U|+4|\mathcal D|
   +2c_{\mathcal U}+c_{\mathcal D}.
 \end{aligned}
\end{equation}
Together with the independent-set bound, this gives
$5|I|\le n-2$. Taking an independent set of maximum
size proves
\begin{equation}
\label{eq:quartic-repair-budget}
 \rho\le\frac{n-2}{5}.
\end{equation}

\subsection{The resulting tour bound}
We now complete the proof of
Theorem~\ref{thm:quartic-thirteen-tenths}.

\begin{proof}
The final vector $\bar x$ belongs to ${\mathcal P^\circ}$ by
Proposition~\ref{prop:quartic-completed-vector}.
The objective expression at $x^{\mathrm{init}}$ is $5n/4-1/2$.
The changes to edges in $R_H$ leave it unchanged,
and each paid increase raises it by $1/4$.
Since at most $(n-2)/5$ paid increases are needed,
\[
 \begin{aligned}
 \operatorname{tsp}(G)
 &\le\text{\LPtwo}_{\mathrm{OPT}}(G,\mathcal S)
 \le\Phi_{G,\mathcal S}(\bar x)\\
 &\le\frac54n-\frac12+\frac14\rho\\
 &\le\frac54n-\frac12+\frac{n-2}{20}\\
 &=\frac{13}{10}n-\frac35.
 \end{aligned}
\]
The demands, the graph $K$, and a maximum matching in $K$
can all be found in polynomial time. The \LPtwo{} tour
recovery then produces a tour satisfying this bound.
\end{proof}

\section{A \texorpdfstring{$4/3$}{4/3} approximation for \texorpdfstring{$k$}{k}-connected graphs with \texorpdfstring{$\Delta\le k+1$}{Delta at most k+1}}
\label{sec:k-connected-degree-bounded}

We now construct a point in the feasible region of \LPone{} for simple
$k$-vertex-connected graphs of maximum degree at most $k+1$,
where $k\ge3$. This includes $k$-connected, $k$-regular graphs.
For $k=3$ the class contains the $3$-connected cubic graphs, for which
stronger bounds are known~\cite{DKM17,WYY23}, and for large $k$ a randomized
$(1+O(1/\sqrt{k}))$-approximation follows from~\cite[Theorem~5]{FRS14}; see
also~\cite{Vis12} for regular graphs. The bound~\eqref{eq:k-connected-four-thirds}
given below is deterministic and
holds for every $k\ge 3$ and every DFS tree.

\begin{theorem}
\label{thm:k-connected-four-thirds}
Let $k\ge3$ and let $G=(V,E)$ be a simple $k$-vertex-connected graph
with $\Delta(G)\le k+1$. Fix any DFS tree $T$, and let
$\mathcal S$ be the \textsc{Star} removable system constructed from $T$
as in Appendix~\ref{sec:DFS-removable-stars}. Then
\begin{equation}
\label{eq:k-connected-four-thirds}
 \begin{aligned}
 \operatorname{tsp}(G)
 &\le \text{\LPtwo}_{\mathrm{OPT}}(G,\mathcal S)\\
 &\le \text{\LPone}_{\mathrm{OPT}}(G,\mathcal S)
 \le \frac43n-\frac23.
 \end{aligned}
\end{equation}
A tour satisfying this bound can be computed in polynomial time.
\end{theorem}
\begin{proof}
We construct $y\in\Pone(G,\mathcal S)$ and prove that
$\PhiOne_{G,\mathcal S}(y)=4n/3-2/3$.

\smallskip\noindent
\textbf{The edge values.}
Let $r$ be the DFS root. For a star $S=S(u,v)$, write
$t_S:=|I(u,v)|$, so $k_S=t_S+1$. If $u\ne r$, the parent edge
of $u$ and its child edge $uv$ are distinct from these back
edges. Hence
\[
 1\le t_S\le\deg_G(u)-2\le k-1.
\]
Since $G$ is $2$-vertex-connected, the root has exactly one DFS
child. Its star $S_r$ therefore contains all edges incident to
$r$, and
\[
 t_{S_r}=\deg_G(r)-1\in\{k-1,k\}.
\]
In particular, $t_{S_r}\ge2$.

For $e\in E(G)$ and $S\in\mathcal S$, define
\begin{equation}
\label{eq:k-reciprocal-values}
 \begin{aligned}
 y_{\rep{e}}
 &:=\begin{cases}
    \dfrac13,&e\in E(T),\\[4pt]
    1-\dfrac{2}{3t_S},
       &e\in I(u,v),\quad S=S(u,v)\ne S_r,\\[6pt]
    1-\dfrac{4}{3t_{S_r}},&e\in S_r\setminus E(T),
   \end{cases}\\[6pt]
 y_{c_S}
 &:=\begin{cases}
    \dfrac13,&S\ne S_r,\\[3pt]
    1,&S=S_r.
   \end{cases}
 \end{aligned}
\end{equation}
Every coordinate lies in $[1/3,1]$.

\smallskip\noindent
\textbf{The auxiliary degree bounds.}
The different values at the root ensure that every star satisfies
its degree bound with equality:
\[
 \begin{aligned}
 y(S_{H^{\mathrm{star}}})+y_{c_S}
 &=\frac13+t_S\left(1-\frac{2}{3t_S}\right)+\frac13
   =t_S=k_S-1
   &&(S\ne S_r),\\[3pt]
 y((S_r)_{H^{\mathrm{star}}})+y_{c_{S_r}}
 &=\frac13+t_{S_r}\left(1-\frac{4}{3t_{S_r}}\right)+1
   =t_{S_r}=k_{S_r}-1.
 \end{aligned}
\]

\smallskip\noindent
\textbf{The Blossom inequalities.}
We first record a bound used in the cut calculations. For a back
edge $b$ in a nonroot star $S$,
\[
 1-y_{\rep{b}}=\frac{2}{3t_S}\ge\frac{2}{3(k-1)}.
\]
For a back edge in the root star,
\[
 1-y_{\rep{b}}=\frac{4}{3t_{S_r}}
 \ge\frac{4}{3k}\ge\frac{2}{3(k-1)}.
\]
Thus every back edge satisfies
\begin{equation}
\label{eq:k-cut-coverage}
 1-y_{\rep{b}}\ge\frac{2}{3(k-1)}.
\end{equation}

Consider a cut $C:=\delta_{H^{\mathrm{star}}}(U)$ and $F\subseteq C$ with
$|C\setminus F|$ odd. If $|C\setminus F|\ge3$, then
$y(C\setminus F)\ge1$, which proves its Blossom inequality.
Otherwise, $C\setminus F=\{f\}$, and the required inequality
is equivalent to
\[
 \sum_{a\in C}(1-y_a)\ge2(1-y_f).
\]
Since $y_f\ge1/3$, it suffices to show that the sum is at least
$4/3$.

The root connector has value one, so the inequality is immediate
if $f=c_{S_r}$. Assume otherwise. The original root $r$ is
incident only to $c_{S_r}$ in $H^{\mathrm{star}}$. If this connector crosses the
cut, move $r$ to the shore containing $z_{S_r}$. This changes no
other crossing edge and leaves the displayed sum unchanged.
We may therefore contract $c_{S_r}$. The resulting cut is still
nonempty, since $f$ still crosses it.

The edges $\rep{e}$ with $e\in E(T)$ and the remaining connectors
form a spanning tree, and each has value $1/3$. We call this the
expanded tree and consider the number of its edges crossing the cut.

\smallskip\noindent
\textbf{Case 1: At least two expanded-tree edges cross.}
These two edges alone give
\[
 \sum_{a\in C}(1-y_a)\ge\frac23+\frac23=\frac43.
\]

\smallskip\noindent
\textbf{Case 2: Exactly one expanded-tree edge crosses.}
The two shores are the vertex sets of the two components obtained
by deleting this edge from the expanded tree. We show that at
least $k-1$ back edges also cross.

If the crossing edge represents a tree edge $uv$, consider the
original subtree $T_v$. The other crossing edges correspond
exactly to the back edges in $\delta_G(V(T_v))$. Since $G$ is
$k$-edge-connected, this original cut has at least $k$ edges.
Only $uv$ is a tree edge, so at least $k-1$ back edges cross.

If the crossing edge is a connector $c_S=uz_S$, where
$S=S(u,v)$ and $u\ne r$, the component below it contains
$z_S$ and the subtree corresponding to $T_v$ in
$H^{\mathrm{star}}$. For every $b\in I(u,v)$, both endpoints
of $\rep{b}$ lie in this component. Thus, apart from $c_S$,
the crossing edges correspond exactly to the back edges
of $G$ from $V(T_v)$ to proper ancestors of $u$.

To count these edges, delete $u$ from $G$. They form the cut
between $V(T_v)$ and $V\setminus(V(T_v)\cup\{u\})$ in $G-u$.
Both sides are nonempty: one contains $v$, and the other contains
$r$. The graph $G-u$ is $(k-1)$-vertex-connected and hence
$(k-1)$-edge-connected. Thus at least $k-1$ back edges cross
in this case as well.

In either case, the single crossing expanded-tree edge and
\eqref{eq:k-cut-coverage} give
\[
 \begin{aligned}
 \sum_{a\in C}(1-y_a)
 &\ge\frac23+(k-1)\frac{2}{3(k-1)}\\
 &=\frac43.
 \end{aligned}
\]
These cases prove every Blossom inequality. Together with the
coordinate and degree bounds, they show that
$y\in\Pone(G,\mathcal S)$.

\smallskip\noindent
\textbf{The objective.}
Every back edge belongs to exactly one star, and every star
contains exactly one tree edge. Consequently,
\[
 \sum_{S\in\mathcal S}t_S=m-n+1,
 \qquad |N|+|\mathcal S|=n-1.
\]
Using the degree equalities $y(S_{H^{\mathrm{star}}})=t_S-y_{c_S}$ and
$y(N_{H^{\mathrm{star}}})=|N|/3$, we obtain
\[
 \begin{aligned}
 \PhiOne_{G,\mathcal S}(y)
 &=m+\frac{|N|}{3}
      -\sum_{S\in\mathcal S}(t_S-y_{c_S})\\
 &=n-1+\frac{|N|}{3}+\sum_{S\in\mathcal S}y_{c_S}\\
 &=n-1+\frac{|N|}{3}+1+\frac{|\mathcal S|-1}{3}\\
 &=\frac43n-\frac23.
 \end{aligned}
\]
Since $\text{\LPone}_{\mathrm{OPT}}(G,\mathcal S)\le\PhiOne_{G,\mathcal S}(y)$,
Theorem~\ref{thm:framework-1-bound} and
Proposition~\ref{prop:framework-containment} give
\eqref{eq:k-connected-four-thirds}. The point is computable
in polynomial time, as are \LPone{} optimization and
the corresponding tour recovery.
\end{proof}

\section{\texorpdfstring{{Analysis for arbitrary subtour value}}{Analysis for arbitrary subtour value}}
\label{s12pair:general}
\label{sec:general-L}
\label{s10g:main-subsection}

Let $(G_0,x^*)$ be the input to \ALG{}, where $x^*$ is an
optimal extreme subtour solution whose positive support is
$2$-vertex-connected. Write $G=(V,E)$ for this support,
{$n:=|V|$, and $m:=|E|$.}
If $G$ is a cycle, \ALG{} returns it directly. Otherwise,
let $x$ be the box-constrained \SubtourLP{} solution obtained
from $x^*$ by \ALG{} on the same positive support, with
\[
 0<x_e\le1\quad(e\in E),
 \qquad x(E)=x^*(E)={\text{\SubtourLP}(G)}.
\]
We use this vector $x$ throughout the section. The extremality
of $x^*$ gives $m\le2n-3$, although $x$ need not be extreme.

\begin{theorem}\label{s12pair:theorem}\label{s10g:theorem}
Let $T$ be the greedy DFS tree constructed by
\hyperref[proc:algorithm]{\ALG} on $G$, and let
$\mathcal S$ be the \textsc{Star} removable system constructed
from $T$ as in Appendix~\ref{sec:DFS-removable-stars}. Then
\begin{equation}\label{s10g:strong-bound}
 \begin{aligned}
 \operatorname{tsp}(G)
 &\le \text{\LPtwo}_{\mathrm{OPT}}(G,\mathcal S)\\
 &\le \frac{89}{64}n+\frac98({\text{\SubtourLP}(G)}-n)-\frac{13}{16}\\
 &=\frac{89}{64}{\text{\SubtourLP}(G)}-\frac{17}{64}({\text{\SubtourLP}(G)}-n)-\frac{13}{16}.
 \end{aligned}
\end{equation}
Together with the standard block reduction, this gives a
polynomial-time $89/64$-approximation relative to the subtour
optimum for every connected simple graph.
\end{theorem}

\begin{figure}[!t]
\centering
{\begin{tikzpicture}[x=1cm,y=1cm]
\draw[s10shore] (-1.25,-2.2) rectangle (3.15,.2);
\node[s10note,anchor=south east] at (3.05,-2.1) {$T_v$};
\node[s10v] (a) at (0,3.45) {$a$};
\node[s10v] (h) at (0,2.35) {$h$};
\node[s10c] (u) at (0,1.15) {$u$};
\node[s10v] (v) at (0,-.3) {$v$};
\node[s10v] (w) at (-.55,-1.6) {$w$};
\node[s10v] (x) at (2,-1.6) {$x$};
\draw[s10t] (a)--(h);
\draw[s10t] (h)--node[s10lab,left] {$x_{hu}$}(u);
\draw[s10t] (u)--node[s10lab,left] {$q_S$}(v);
\draw[s10t] (v)--(w); \draw[s10t] (v)--(x);
\draw[s10b] (u) .. controls (2.3,1.05) and (3,-.65) .. (x);
\draw[s10b] (u) .. controls (2,.7) and (1.2,-1.85) .. (w);
\draw[s10b] (v) .. controls (-2.65,.05) and (-2.6,3.45) .. node[s10lab,left,pos=.57] {$o(v)$} (a);
\node[s10note,anchor=west] at (.45,2.35) {$h=\operatorname{par}(u)$};
\node[s10note,anchor=west] at (.45,1.4) {center of $S$};
\node[s10note,anchor=west,text width=3.6cm] (amass) at (3.55,.25) {$a_S=x(I(u,v))$:\\back-edge weight\\from $T_v$ to $u$};
\draw[gray!65] (amass.west)--(2.5,-.1);
\node[s10note,anchor=west,text width=3.6cm] at (3.55,-1.25) {$o(v)$: total weight\\of back edges from $v$\\to its proper ancestors};
\node[s10note] at (.95,-2.6) {$t_S=|I(u,v)|,\qquad k_S=t_S+1,\qquad e_S=(a_S-1)_+$};
\end{tikzpicture}}
\caption{{The star $S=S(u,v)$ in $G$, with the subtree $T_v$ shaded.
Blue edges are tree edges; red edges are back edges.
The tree edge $uv$ has value $q_S=x_{uv}$, and the back edges
in $I(u,v)$ have total $x$-value $a_S$.}}
\label{fig:s10-master}
\end{figure}
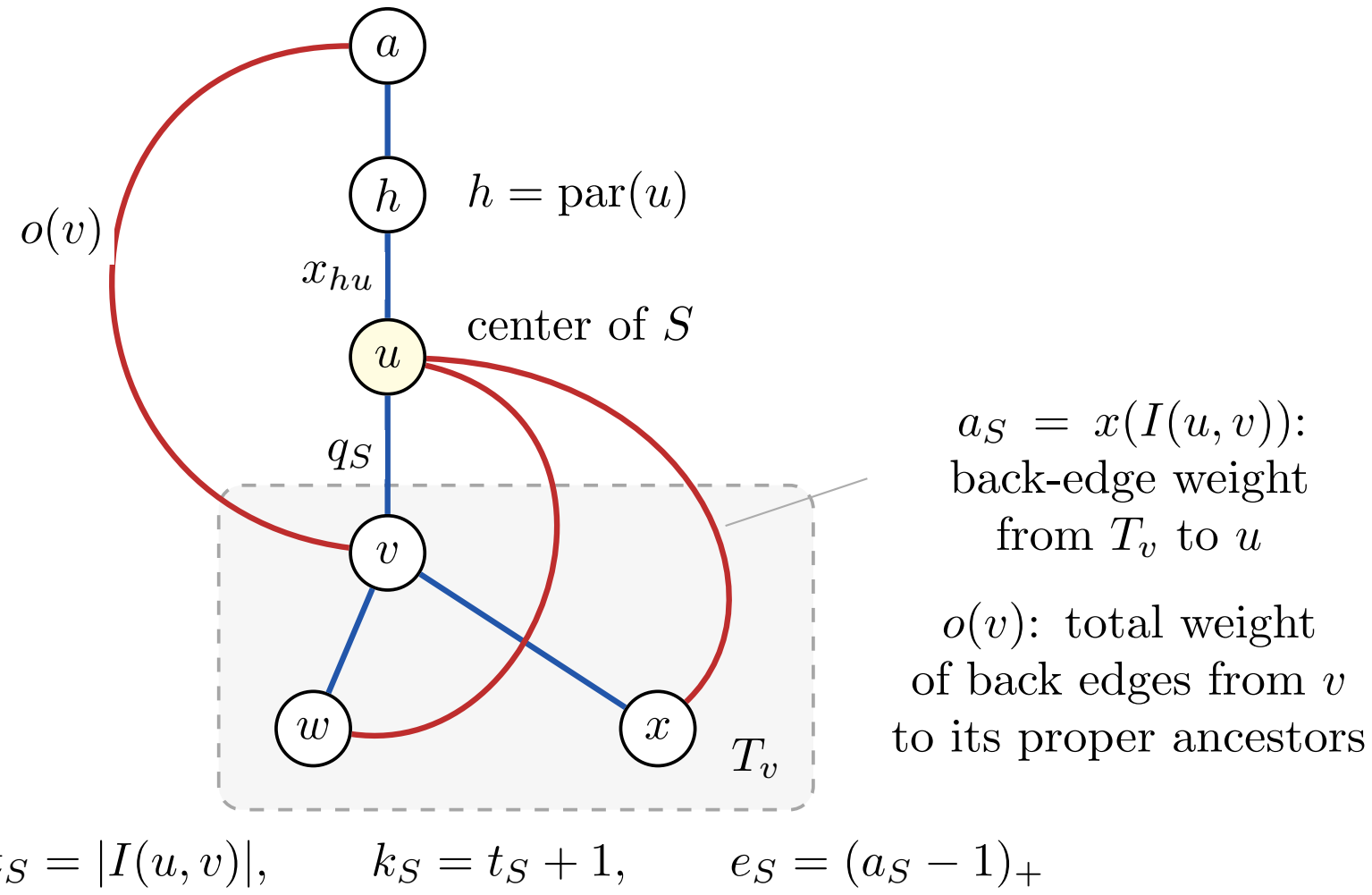

The algorithm roots the greedy DFS at a vertex $r$ with
$\deg_G(r)\ge3$.
{For the calculations, write $L:=\text{\SubtourLP}(G)$}
and put
\[
 \begin{aligned}
 \Delta&:=L-n,\\
 \eta_v&:=x(\delta_G(v))-2\ge0,
 &\sum_{v\in V}\eta_v&=2\Delta.
 \end{aligned}
\]
{For a star $S=S(u,v)$, put
\[
 q_S:=x_{uv},\qquad a_S:=x(I(u,v)),\qquad
 t_S:=|I(u,v)|,\qquad k_S:=t_S+1,
 \qquad e_S:=(a_S-1)_+,
\]
where $(z)_+:=\max\{z,0\}$. Thus $q_S$ is the value of the
unique tree edge of $S$, while $a_S$ and $t_S$ are the total
$x$-value and number of its back edges. Call $S$ positive when
$e_S>0$. Greedy DFS gives $a_S\le t_Sq_S$.
Let $o(v)$ be the total $x$-value of back edges starting at $v$
and ending at its proper ancestors. For consecutive stars
$S=S(u,v)$ and $S'=S(v,w)$, the value of the parent edge
of the center $v$ is $q_S=x_{uv}$.
{Figure~\ref{fig:s10-master}
illustrates the quantities associated with $S$.}}

{Every back edge belongs to exactly one star,
and each star contains one tree edge. Hence
\[
 |N|+|\mathcal S|=n-1,
 \qquad m-\sum_{S\in\mathcal S}(k_S-1)=n-1.
\]
By \eqref{eq:repair-nonloop-cost}, the objective is therefore
\begin{equation}\label{eq:general-join-cost}
 \Phi_{G,\mathcal S}(y)
 =n-1+y(N_H)+\sum_{S\in\mathcal S}
   \max\{y_{c_S},\,k_S-1-y(S_H)\}.
\end{equation}
Corollary~\ref{cor:repair-nonloop-bound} bounds the tour returned
by \ALG{} by this value for every $y\in\mathcal P^\circ$.}

The root has one DFS child and at least two back edges.
Every back edge belongs to exactly one star, so
\begin{equation}\label{s12pair:sparsity}
 \sum_{S\ne S_r}t_S
 =m-n+1-t_{S_r}
 \le n-4.
\end{equation}

\subsection{The point \texorpdfstring{${Q(d)}$}{Q(d)} and its connector values}

For every nonroot star $S$, define
\[
 \bar e_S:=\min\{e_S,1-q_S\}.
\]
{
Define $Q:\R_+^{\mathcal S}\to\R_+^{E(H^\circ)}$ by specifying
its values on the nonconnector edges and then on the connectors.
We call $d\in\R_+^{\mathcal S}$ admissible when
\[
 d_{S_r}=0,\qquad 0\le d_S\le\bar e_S\quad(S\ne S_r).
\]
For any $d\in\R_+^{\mathcal S}$, set
}
\[
 {Q(d)}_{\rep{e}}:=
 \begin{cases}
 \dfrac13+\dfrac23d_S,
   &e=uv,\quad S=S(u,v)\ne S_r,\\[5pt]
 1-\dfrac23x_e,&e\notin E(T),\\[5pt]
 \dfrac13,&\text{otherwise}.
 \end{cases}
\]
{The connector assignment in
\eqref{proc:following-connector} is specific to $\SubtourLP(G)=n$.
For arbitrary $L$, we use minimum cuts to choose the connector
values and prove the Blossom inequalities.}

Set ${Q(d)}_{c_{S_r}}:=1$ and let $K$ be the graph obtained
by contracting the root connector in $H^\circ$.
{
Using the values $Q(d)_{\rep{e}}$ already defined above,
assign capacity $1-Q(d)_{\rep{e}}$ to each edge $\rep{e}$
of $K$ and capacity $2/3$ to each remaining connector.
These capacities are nonnegative when $d$ is admissible.}

{For each nonroot star $S$ centered at $v$, define $\kappa_S(d)$
by the following finite minimum over cuts separating $v$ from $z_S$
in $K-c_S$, where $c_S=vz_S$:
}
\begin{equation}\label{s12pair:mincut-completion}
 \kappa_S(d):=
 \min_{\substack{U\subseteq V(K):\\v\in U,\ z_S\notin U}}
 \left\{
 \begin{aligned}
 &\sum_{\substack{e\in E(G):\\
                  \rep{e}\in\delta_{K-c_S}(U)}}
       \bigl(1-{Q(d)}_{\rep{e}}\bigr)\\[-2pt]
 &\quad+\frac23
 \bigl|\{S'\in\mathcal S\setminus\{S_r\}:
             c_{S'}\in\delta_{K-c_S}(U)\}\bigr|
 \end{aligned}
 \right\}.
\end{equation}
{
This minimum is well-defined for every $d\in\R_+^{\mathcal S}$;
for admissible $d$, it is a minimum-cut value with nonnegative capacities.
Complete the definition of $Q(d)$ by setting
}
\[
 {Q(d)}_{c_S}:=
 \begin{cases}
  1,&S=S_r,\\[3pt]
  \max\left\{\dfrac13,\,1-\kappa_S(d)\right\},&S\ne S_r.
 \end{cases}
\]
A nonroot connector therefore remains at $1/3$ when
$\kappa_S(d)\ge2/3$. Otherwise, we increase its value so that
\[
 1-{Q(d)}_{c_S}=\kappa_S(d).
\]
{
For arbitrary nonnegative $d$, these formulas define a nonnegative
vector. The next lemma proves feasibility for admissible $d$.
}

\begin{lemma}\label{s12pair:connector-completion}
{For every admissible $d\in\R_+^{\mathcal S}$,
the vector $Q(d)$ belongs to $\mathcal P^\circ$.}
\end{lemma}

\begin{proof}
{
For admissible $d$ and $S=S(u,v)\ne S_r$, the restriction $d_S\le1-q_S$ gives
}
\[
 1-{Q(d)}_{\rep{uv}}
 =\frac23(1-d_S)
 \ge\frac23q_S
 =\frac23x_{uv}.
\]
The same inequality holds with equality for back edges,
and follows from $x_e\le1$ for the remaining edges. Thus
\[
 1-{Q(d)}_{\rep{e}}\ge\frac23x_e
 \qquad(e\in E(G)).
\]
We use this inequality to bound the terms in the Blossom
inequalities for ${Q(d)}$ using the subtour cut constraints for $x$.

Each $\kappa_S(d)$ is the value of a minimum cut with nonnegative
capacities.
All coordinates of ${Q(d)}$ lie in $[1/3,1]$.
By \eqref{eq:repair-nonloop-polytope}, it remains to
prove the Blossom inequalities.

Consider $C:=\delta_{H^\circ}(U)$ and $F\subseteq C$ with
$|C\setminus F|$ odd. If $|C\setminus F|\ge3$, then
\[
 {Q(d)}(C\setminus F)+|F|-{Q(d)}(F)
 \ge\frac{|C\setminus F|}{3}\ge1.
\]
We therefore need only consider $C\setminus F=\{f\}$,
when the Blossom inequality is equivalent to
\begin{equation}\label{s12pair:balanced}
 1-{Q(d)}_f\le
 \sum_{a\in C\setminus\{f\}}\bigl(1-{Q(d)}_a\bigr).
\end{equation}
If $f=c_{S_r}$, its left-hand side is zero.
Otherwise, place the degree-one vertex $r$ on the same side
of the cut as $z_{S_r}$. This changes no crossing edge except
possibly $c_{S_r}$, whose complement is zero. It therefore
suffices to prove \eqref{s12pair:balanced} for cuts of $K$.

Suppose the inequality fails for some edge $f$.
Among the cuts violating it for this same $f$, choose
$C=\delta_K(U)$ with the fewest crossing connectors.
We consider three cases.

\smallskip\noindent
\textbf{Case 1: No connector crosses $C$.}
Contracting all connectors identifies $C$ with the edges of
$H^\circ$ corresponding to a nontrivial cut $\delta_G(U_G)$ of $G$.
The subtour constraint therefore gives
\[
 \begin{aligned}
 \sum_{a\in C}\bigl(1-{Q(d)}_a\bigr)
 &=\sum_{e\in\delta_G(U_G)}\bigl(1-{Q(d)}_{\rep{e}}\bigr)\\
 &\ge\frac23x(\delta_G(U_G))\ge\frac43.
 \end{aligned}
\]
Since $1-{Q(d)}_f\le2/3$, we obtain
\[
 \begin{aligned}
 \sum_{a\in C\setminus\{f\}}\bigl(1-{Q(d)}_a\bigr)
 &\ge\frac43-\bigl(1-{Q(d)}_f\bigr)\\
 &\ge1-{Q(d)}_f.
 \end{aligned}
\]
This contradicts the choice of $C$ as a violating cut.

\smallskip\noindent
\textbf{Case 2: A connector crosses $C$, and every connector in
$C\setminus\{f\}$ has value $1/3$.}
Every connector $c_{S'}\in C\setminus\{f\}$ satisfies
\[
 1-{Q(d)}_{c_{S'}}=\frac23,
\]
which equals its capacity in the minimum-cut calculation.
For edges of the form $\rep{e}$, the capacity is
$1-{Q(d)}_{\rep{e}}$ by definition. Thus every edge in
$C\setminus\{f\}$ has capacity equal to its complement.

Suppose first that $f=\rep{e}$ for some $e\in E(G)$.
Choose a connector $c_{S'}$ crossing $C$. Since
$c_{S'}\ne f$, the sum over $C\setminus\{f\}$ contains
the term $1-{Q(d)}_{c_{S'}}=2/3$. Also, ${Q(d)}_f\ge1/3$,
so
\[
 \sum_{a\in C\setminus\{f\}}\bigl(1-{Q(d)}_a\bigr)
 \ge1-{Q(d)}_{c_{S'}}
 =\frac23
 \ge1-{Q(d)}_f.
\]

Now suppose that $f=c_S=vz_S$. Since $C=\delta_K(U)$
contains $c_S$, the cut
\[
 C\setminus\{f\}=\delta_{K-c_S}(U)
\]
separates $v$ from $z_S$ in $K-c_S$.
Every connector $c_{S'}\in C\setminus\{f\}$ has value
$1/3$, so its complement $1-{Q(d)}_{c_{S'}}=2/3$ equals
its assigned capacity. The capacity of each representative
edge also equals its complement by definition.

Recall that $\kappa_S(d)$ is the minimum capacity of a cut
separating $v$ from $z_S$ in $K-c_S$, as defined in
\eqref{s12pair:mincut-completion}. Therefore,
\[
 \begin{aligned}
 \sum_{a\in C\setminus\{f\}}\bigl(1-{Q(d)}_a\bigr)
 &\ge\kappa_S(d)\\
 &\ge\min\left\{\frac23,\kappa_S(d)\right\}\\
 &=1-{Q(d)}_f,
 \end{aligned}
\]
where the equality follows from the definition of
${Q(d)}_{c_S}$. Thus \eqref{s12pair:balanced} holds in both
situations (see Figure~\ref{fig:s12pair:case-two}), contradicting the choice of $C$.

\begin{figure}[!t]
\centering
\begin{subfigure}[t]{.48\textwidth}
\centering
\begin{tikzpicture}[x=.95cm,y=.85cm]
\draw[s10shore,fill=gray!12]
  (-2.4,3.85)--(.5,3.85)--(.5,1.85)
  --(-.6,.45)--(-2.4,.45)--cycle;
\node[s10note] at (-1.9,3.4) {$U$};
\node[s10z,fill=gray!20] (p) at (0,3.4) {$z_{S''}$};
\node[s10v,fill=gray!20] (u) at (0,2.2) {$u$};
\node[s10z,fill=gray!20] (s) at (-1.2,1) {$z_S$};
\node[s10z] (sp) at (1.2,1) {$z_{S'}$};
\node[s10v] (v) at (-1.2,-.2) {$v$};
\node[s10v] (w) at (1.2,-.2) {$w$};
\node[s10v] (x) at (-1.2,-1.3) {$x_1$};
\node[s10v] (y) at (1.2,-1.3) {$x_2$};
\draw[s10t] (p)--(u);
\draw[s10con] (u)--node[s10lab,above left] {$c_S$}(s);
\draw[s10con,line width=1.8pt]
  (u)--node[s10lab,above right] {$c_{S'}$}(sp);
\draw[s10t] (s)--node[s10lab,left] {$f$}(v)--(x);
\draw[s10t] (sp)--(w)--(y);
\draw[s10b] (s) .. controls (-2.3,.6) and (-2.3,-1) .. (x);
\draw[s10b] (sp) .. controls (2.3,.6) and (2.3,-1) .. (y);
\node[s10note,text width=6cm] at (0,-2.5)
 {$\displaystyle
   \sum_{a\in C\setminus\{f\}}(1-{Q(d)}_a)
   \ge\frac23\ge1-{Q(d)}_f$};
\end{tikzpicture}
\caption{$f=\rep{uv}$ and ${Q(d)}_{c_{S'}}=1/3$.}
\label{fig:s12pair:representatives-preserved}
\end{subfigure}\hfill
\begin{subfigure}[t]{.48\textwidth}
\centering
\begin{tikzpicture}[x=.95cm,y=.85cm]
\draw[s10shore,fill=gray!12]
  (-.55,2.7)--(2.4,2.7)--(2.4,.45)
  --(.6,.45)--(-.55,1.6)--cycle;
\node[s10note] at (1.9,2.3) {$U$};
\node[s10z] (p) at (0,3.4) {$z_{S''}$};
\node[s10v,fill=gray!20] (v) at (0,2.2) {$v$};
\node[s10z] (s) at (-1.2,1) {$z_S$};
\node[s10z,fill=gray!20] (sp) at (1.2,1) {$z_{S'}$};
\node[s10v] (w1) at (-1.2,-.2) {$w_1$};
\node[s10v] (w2) at (1.2,-.2) {$w_2$};
\node[s10v] (x) at (-1.2,-1.3) {$x_1$};
\node[s10v] (y) at (1.2,-1.3) {$x_2$};
\draw[s10t] (p)--(v);
\draw[s10con,draw=gray!65]
  (v)--node[s10lab,above left] {$f=c_S$}(s);
\draw[black,line width=.9pt]
  (-.73,1.48)--(-.47,1.72)
  (-.73,1.72)--(-.47,1.48);
\draw[s10con] (v)--node[s10lab,above right] {$c_{S'}$}(sp);
\draw[s10t] (s)--(w1)--(x);
\draw[s10t] (sp)--(w2)--(y);
\draw[s10b] (s) .. controls (-2.3,.6) and (-2.3,-1) .. (x);
\draw[s10b] (sp) .. controls (2.3,.6) and (2.3,-1) .. (y);
\node[s10note,text width=6cm] at (0,-2.5)
 {$\displaystyle
   \sum_{a\in C\setminus\{f\}}(1-{Q(d)}_a)
   \ge\kappa_S(d)\ge1-{Q(d)}_f$};
\end{tikzpicture}
\caption{$C\setminus\{f\}=\delta_{K-c_S}(U)$.}
\label{fig:s12pair:connector-mincut}
\end{subfigure}
\caption{The two situations in Case~2. Shading marks $U$.
In (b), $c_S$ is deleted for the minimum-cut calculation.
Blue edges and dotted connectors follow the DFS tree;
red curves correspond to back edges.
}
\label{fig:s12pair:case-two}
\end{figure}

\smallskip\noindent
\textbf{Case 3: A connector in $C\setminus\{f\}$ has value
greater than $1/3$.}
Choose such a connector $e=c_S$. Let $W$ attain the minimum
in \eqref{s12pair:mincut-completion}, and put $D:=\delta_K(W)$.
Then $e\in D$, and $D\setminus\{e\}$ has capacity $\kappa_S(d)$.
Since ${Q(d)}_e>1/3$, the definition of its coordinate gives
\[
 \kappa_S(d)=1-{Q(d)}_e<\frac23.
\]
Every connector is assigned capacity $2/3$ in the minimum-cut
construction. Since the total capacity of $D\setminus\{e\}$
is less than $2/3$, this set contains only representative
edges. Their capacities equal their complements, so
\[
 \sum_{a\in D\setminus\{e\}}\bigl(1-{Q(d)}_a\bigr)
 =\kappa_S(d)=1-{Q(d)}_e.
\]

We use this equality to show that $f\notin D$.
Since $e\in C\setminus\{f\}$ and $C$ violates
\eqref{s12pair:balanced},
\[
 1-{Q(d)}_f
 >\sum_{a\in C\setminus\{f\}}\bigl(1-{Q(d)}_a\bigr)
 \ge1-{Q(d)}_e.
\]
On the other hand, the equality above gives
\[
 1-{Q(d)}_a\le1-{Q(d)}_e
 \qquad(a\in D\setminus\{e\}).
\]
Thus $f\notin D\setminus\{e\}$, and since $f\ne e$,
we obtain $f\notin D$.

Now replace $C$ by the cut
\[
 C':=C\mathbin{\triangle}D
    =\delta_K(U\mathbin{\triangle}W).
\]
The edge $f$ remains in the cut, while $e$ leaves it
(see Figure~\ref{fig:s12pair:cut-exchange}). Moreover,
\[
 \begin{aligned}
 \sum_{a\in C'\setminus\{f\}}\bigl(1-{Q(d)}_a\bigr)
 &\le\sum_{a\in C\setminus\{f\}}\bigl(1-{Q(d)}_a\bigr)
       -\bigl(1-{Q(d)}_e\bigr)\\
 &\qquad+\sum_{a\in D\setminus\{e\}}\bigl(1-{Q(d)}_a\bigr)\\
 &=\sum_{a\in C\setminus\{f\}}\bigl(1-{Q(d)}_a\bigr)\\
 &<1-{Q(d)}_f.
 \end{aligned}
\]
Thus $C'$ still violates \eqref{s12pair:balanced} for $f$.
But $D$ contains no connector other than $e$, so $C'$ has
exactly one fewer crossing connector than $C$. This contradicts
the choice of $C$.

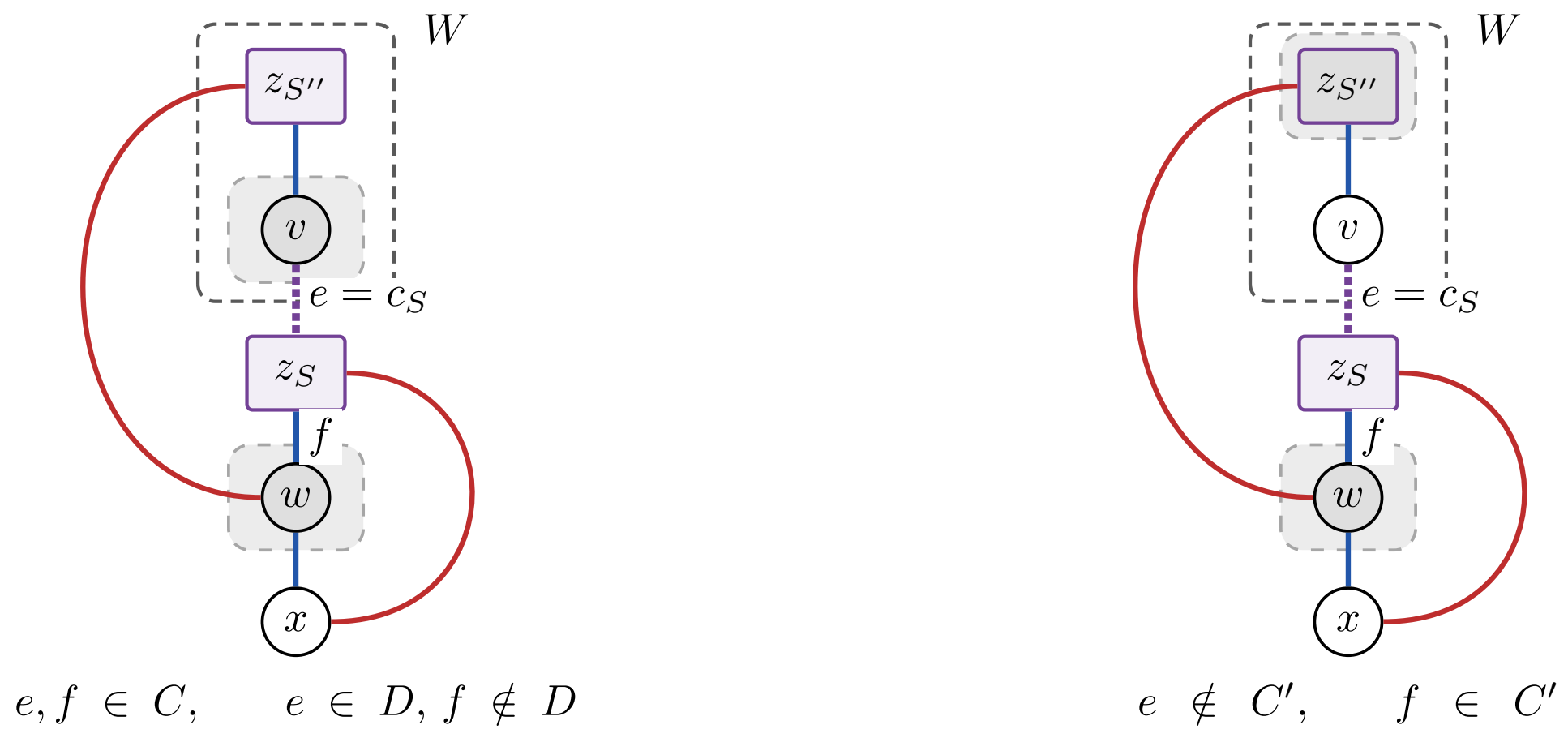
\begin{figure}[!t]
\centering
\begin{subfigure}[t]{.48\textwidth}
\centering
\begin{tikzpicture}[x=1cm,y=.78cm,
  witness boundary/.style={draw=black!65,densely dashed,
    rounded corners=4pt,line width=.8pt}]
\draw[s10shore,fill=gray!15]
  (-.55,1.25) rectangle (.55,2.35);
\draw[s10shore,fill=gray!15]
  (-.55,-1.55) rectangle (.55,-.45);
\draw[witness boundary] (-.8,1.05) rectangle (.8,3.95);
\node[s10note,anchor=west] at (.9,3.9) {$W$};
\node[s10z] (p) at (0,3.3) {$z_{S''}$};
\node[s10v,fill=gray!25] (v) at (0,1.8) {$v$};
\node[s10z] (s) at (0,.3) {$z_S$};
\node[s10v,fill=gray!25] (w) at (0,-1) {$w$};
\node[s10v] (x) at (0,-2.3) {$x$};
\draw[s10t] (p)--(v);
\draw[s10con,line width=1.8pt]
  (v)--node[s10lab,right] {$e=c_S$}(s);
\draw[s10t,line width=1.5pt]
  (s)--node[s10lab,right] {$f$}(w);
\draw[s10t] (w)--(x);
\draw[s10b] (p)
  .. controls (-2.2,3.3) and (-2.2,-1) .. (w);
\draw[s10b] (s)
  .. controls (1.8,.3) and (1.8,-2.3) .. (x);
\node[s10note,text width=5.8cm] at (0,-3.15)
 {$e,f\in C$,\qquad $e\in D$, $f\notin D$};
\end{tikzpicture}
\caption{$C=\delta_K(U)$.}
\end{subfigure}\hfill
\begin{subfigure}[t]{.48\textwidth}
\centering
\begin{tikzpicture}[x=1cm,y=.78cm,
  witness boundary/.style={draw=black!65,densely dashed,
    rounded corners=4pt,line width=.8pt}]
\draw[s10shore,fill=gray!15]
  (-.55,2.75) rectangle (.55,3.85);
\draw[s10shore,fill=gray!15]
  (-.55,-1.55) rectangle (.55,-.45);
\draw[witness boundary] (-.8,1.05) rectangle (.8,3.95);
\node[s10note,anchor=west] at (.9,3.9) {$W$};
\node[s10z,fill=gray!25] (p) at (0,3.3) {$z_{S''}$};
\node[s10v] (v) at (0,1.8) {$v$};
\node[s10z] (s) at (0,.3) {$z_S$};
\node[s10v,fill=gray!25] (w) at (0,-1) {$w$};
\node[s10v] (x) at (0,-2.3) {$x$};
\draw[s10t] (p)--(v);
\draw[s10con,line width=1.8pt]
  (v)--node[s10lab,right] {$e=c_S$}(s);
\draw[s10t,line width=1.5pt]
  (s)--node[s10lab,right] {$f$}(w);
\draw[s10t] (w)--(x);
\draw[s10b] (p)
  .. controls (-2.2,3.3) and (-2.2,-1) .. (w);
\draw[s10b] (s)
  .. controls (1.8,.3) and (1.8,-2.3) .. (x);
\node[s10note,text width=5.8cm] at (0,-3.15)
 {$e\notin C'$,\qquad $f\in C'$};
\end{tikzpicture}
\caption{$C'=\delta_K(U\mathbin{\triangle}W)$.}
\end{subfigure}
\caption{Case~3. The shaded sets are $U$ and $U\mathbin{\triangle}W$,
respectively; the dashed outline marks $W$. The connector $e$ stops crossing, while $f$ remains
in the cut.}
\label{fig:s12pair:cut-exchange}
\end{figure}

The three cases cover every possibility for $C$, and each
contradicts its choice as a violating cut. Thus
\eqref{s12pair:balanced} holds for all cuts of $K$ and, by
the root-connector reduction above, for all cuts of $H^\circ$.
All Blossom inequalities hold, so ${Q(d)}\in{\mathcal P^\circ}$.
\end{proof}

\subsection{How much can a connector increase?}

For a nonroot star $S$ centered at $v$, define
\begin{equation}\label{s12pair:tau}
 \tau_S:=
 \bigl(1-\max\{0,2-q_S-a_S,q_S+a_S-\eta_v\}\bigr)_+.
\end{equation}
We use the subtour constraints to bound the capacities of
the cuts in \eqref{s12pair:mincut-completion}. Together with
the cuts that isolate $v$ or $z_S$ after contracting the
other connectors, these give the following bound.

\begin{lemma}\label{s12pair:connector-estimate}
{Let $d\in\R_+^{\mathcal S}$ be admissible.}
Let $S$ be a nonroot star centered at $v$, and let
$u:=\operatorname{par}(v)$. Then
\begin{equation}\label{s12pair:connector-bound}
 {Q(d)}_{c_S}\le
 \begin{cases}
  \dfrac13+\dfrac23\max\{\tau_S,(d_{S^{\mathrm{par}}}-o(v))_+\},
  &\begin{aligned}
    &\text{if $v$ has exactly one DFS child and}\\[4pt]
    &S^{\mathrm{par}}=S(u,v)\in\mathcal S\setminus\{S_r\},
   \end{aligned}\\[16pt]
  \dfrac13+\dfrac23\tau_S,
  &\begin{aligned}
    &\text{if $v$ has at least two DFS children}\\[4pt]
    &\text{or }uv\in N\cup S_r.
   \end{aligned}
 \end{cases}
\end{equation}
\end{lemma}

\begin{proof}
By the definition of the connector coordinates,
$1-{Q(d)}_{c_S}=\min\{2/3,\kappa_S(d)\}$. We therefore prove
the lemma by bounding the sums in the minimum defining
$\kappa_S(d)$ in \eqref{s12pair:mincut-completion}.

Let $C:=\delta_{K-c_S}(U)$ be any cut with $v\in U$ and
$z_S\notin U$. If $C$ contains a connector, its term in
that sum is $2/3$, which already gives the required bound.
We may therefore suppose that $C$ contains no connector.
The sum to be bounded is then
$\sum_{a\in C}(1-{Q(d)}_a)$.

If $v$ has at least two DFS children or $uv\in N\cup S_r$,
we prove
\[
 \sum_{a\in C}\bigl(1-{Q(d)}_a\bigr)
 \ge\frac23(1-\tau_S).
\]
If $v$ has exactly one DFS child and its parent edge belongs
to a nonroot star $S^{\mathrm{par}}=S(u,v)$, it suffices to prove
\[
 \sum_{a\in C}\bigl(1-{Q(d)}_a\bigr)
 \ge\frac23\left(1-\max\{\tau_S,(d_{S^{\mathrm{par}}}-o(v))_+\}\right).
\]
Taking the minimum over the cuts gives the required bound
on $\kappa_S(d)$ and hence on ${Q(d)}_{c_S}$.

Contract all remaining connectors in $K-c_S$. None crosses
$C$, so the sum above does not change. The resulting graph is obtained from $G$ by adding $z_S$
and replacing each edge $vw\in S$ by $z_Sw$. Write the two sides as $A\cup\{v\}$ and
$B\cup\{z_S\}$, where $A,B$ partition
$V(G)\setminus\{v\}$. We consider three cases.

\smallskip\noindent
\textbf{Case 1: $A\ne\varnothing$ and $B\ne\varnothing$.}
The subtour constraints for $B$ and $A$, together with
$x(\delta_G(v))=2+\eta_v$, give
\[
 \begin{aligned}
 x(E_G(A,B))+x(E_G(\{v\},B))&\ge2,\\
 x(E_G(A,B))+2+\eta_v-x(E_G(\{v\},B))&\ge2.
 \end{aligned}
\]
After the contractions, the edges of $S_H$ are incident to
$z_S$, while the edges $\rep{e}$ with
$e\in\delta_G(v)\setminus S$ are incident to $v$.
Thus an edge of $S_H$ crosses $C$ precisely when its other
endpoint lies in $A$, and an edge incident to $v$ crosses
precisely when its other endpoint lies in $B$.
The edges corresponding to $E_G(A,B)$ also cross $C$;
see Figure~\ref{fig:s12pair:split-cut}.
Using $1-{Q(d)}_{\rep{e}}\ge2x_e/3$, we obtain
\[
 \begin{aligned}
 \sum_{a\in C}\bigl(1-{Q(d)}_a\bigr)
 \ge\frac23\Bigl(
   &x(E_G(A,B))
    +x(S\cap E_G(\{v\},A))
   +x(E_G(\{v\},B)\setminus S)
 \Bigr).
 \end{aligned}
\]

To prove the lemma in this case, it suffices to show that
\[
 \begin{aligned}
 1-\tau_S
 &\le\max\{0,2-q_S-a_S,q_S+a_S-\eta_v\}\\
 &\le x(E_G(A,B))+x(S\cap E_G(\{v\},A))+x(E_G(\{v\},B)\setminus S).
 \end{aligned}
\]
We bound each term in the maximum separately.

\smallskip\noindent
\textbf{The term $0$.}
Every term on the right is nonnegative.

\smallskip\noindent
\textbf{The term $2-q_S-a_S$.}
The cut $\delta_G(B)$ consists of the edges between $A$
and $B$ and the edges between $v$ and $B$. Therefore,
\[
 \begin{aligned}
 &x(E_G(A,B))
  +x(S\cap E_G(\{v\},A))
  +x(E_G(\{v\},B)\setminus S)\\
 &\quad\ge x(E_G(A,B))
           +x(E_G(\{v\},B)\setminus S)\\
 &\quad=x(\delta_G(B))-x(S\cap E_G(\{v\},B))\\
 &\quad\ge2-x(S)\\
 &\quad=2-q_S-a_S.
 \end{aligned}
\]
Here we used the subtour constraint for $B$ and
$S\cap E_G(\{v\},B)\subseteq S$.

\smallskip\noindent
\textbf{The term $q_S+a_S-\eta_v$.}
We use the cut $\delta_G(A)$ in the same way.
The edges incident to $v$ outside $S$ have total $x$-value
\[
 x(\delta_G(v)\setminus S)
 =x(\delta_G(v))-x(S)
 =2+\eta_v-q_S-a_S.
\]
Consequently,
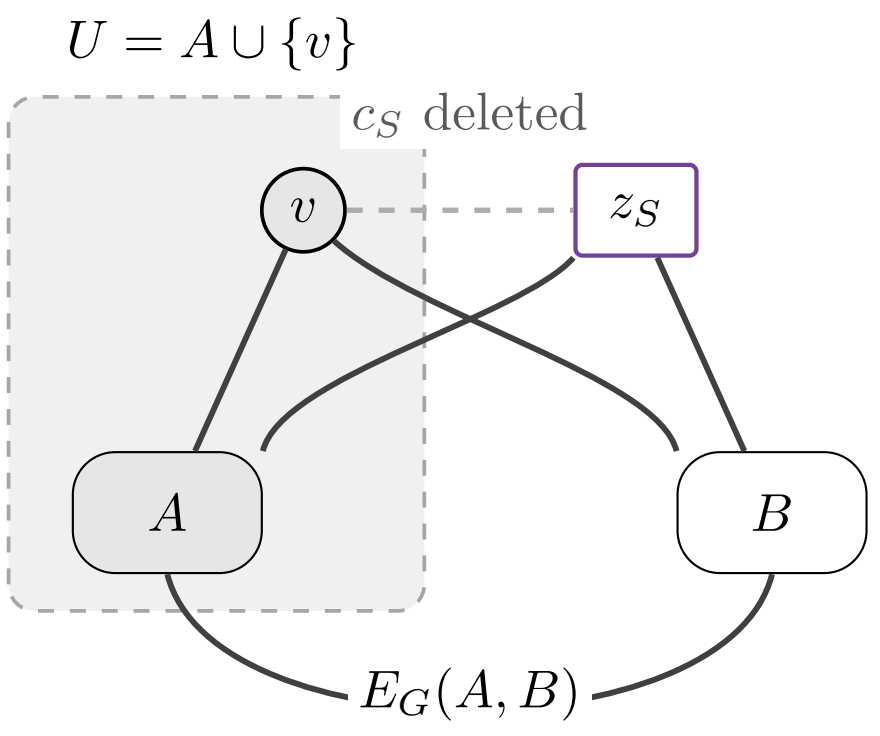
\[
 \begin{aligned}
 &x(E_G(A,B))
  +x(S\cap E_G(\{v\},A))
  +x(E_G(\{v\},B)\setminus S)\\
 &\quad\ge x(E_G(A,B))
           +x(S\cap E_G(\{v\},A))\\
 &\quad=x(\delta_G(A))-x(E_G(\{v\},A)\setminus S)\\
 &\quad\ge2-x(\delta_G(v)\setminus S)\\
 &\quad=2-(2+\eta_v-q_S-a_S)\\
 &\quad=q_S+a_S-\eta_v.
 \end{aligned}
\]
Here we used the subtour constraint for $A$ and
$E_G(\{v\},A)\setminus S\subseteq\delta_G(v)\setminus S$.

Combining these bounds with the preceding estimate on the
capacity of $C$ gives
\[
 \begin{aligned}
 \sum_{a\in C}\bigl(1-{Q(d)}_a\bigr)
 &\ge\frac23\max\{0,2-q_S-a_S,q_S+a_S-\eta_v\}\\
 &\ge\frac23(1-\tau_S).
 \end{aligned}
\]

\begin{figure}[!t]
\centering
\begin{tikzpicture}[x=1cm,y=1cm,
 group/.style={draw,rounded corners=8pt,minimum width=1.25cm,minimum height=.8cm,
               inner sep=3pt,font=\small},
 original/.style={draw=black!75,line width=1.1pt}]
\draw[s10shore,fill=gray!12] (-3.05,-1.3) rectangle (-.3,2.1);
\node[s10note] at (-1.7,2.45) {$U=A\cup\{v\}$};
\node[s10v,fill=gray!20] (v) at (-1.1,1.35) {$v$};
\node[s10z,fill=white] (z) at (1.1,1.35) {$z_S$};
\node[group,fill=gray!20] (a) at (-2,-.65) {$A$};
\node[group,fill=white] (b) at (2,-.65) {$B$};
\draw[gray!65,dashed,line width=1pt] (v)--(z);
\node[s10lab,text=black!65] at (0,1.98) {$c_S$ deleted};
\draw[original] (v)--(a);
\draw[original] (z)--(b);
\draw[original] (v.south east)
 .. controls (-.35,.65) and (1.25,.25) .. (b.north west);
\draw[original] (z.south west)
 .. controls (.35,.65) and (-1.25,.25) .. (a.north east);
\draw[original] (a.south)
 .. controls (-1.7,-2.25) and (1.7,-2.25) .. (b.south);
\node[s10lab] at (0,-1.85) {$E_G(A,B)$};
\node[s10note,anchor=west,text width=5.8cm,align=left] at (3.3,1.25)
 {The edges from $z_S$ to $A$ belong to $S_H$.\\[5pt]
  The edges from $v$ to $B$ correspond to edges outside $S$.};
\node[s10note,anchor=west,text width=5.8cm,align=left] at (3.3,-.65)
 {$\displaystyle
   \sum_{a\in C}\bigl(1-{Q(d)}_a\bigr)
   \ge\frac23(1-\tau_S)$.};
\end{tikzpicture}
\caption{Case~1 after contracting the other connectors and
deleting $c_S$. Shading marks $A\cup\{v\}$; boxes denote sets
of vertices and solid lines denote sets of edges.
The subtour constraints for $A$ and $B$ give the capacity bound.}
\label{fig:s12pair:split-cut}
\end{figure}

\smallskip\noindent
\textbf{Case 2: The cut isolates $z_S$.}
Here $B=\varnothing$, and the crossing edges are exactly $S_H$.
See Figure~\ref{fig:s12pair:singleton-cuts} for this case and Case~3 below.
Their total capacity is
\[
 \begin{aligned}
 \sum_{a\in S_H}\bigl(1-{Q(d)}_a\bigr)
 &=\frac23(1-d_S+a_S)\\
 &\ge\frac23,
 \end{aligned}
\]
because $d_S\le\bar e_S\le e_S\le a_S$.

\smallskip\noindent
\textbf{Case 3: The cut isolates $v$ after contraction.}
Here $A=\varnothing$, and the crossing edges are
\[
 \{\rep{e}:e\in\delta_G(v)\setminus S\}.
\]
Suppose $v$ has exactly one DFS child and
$S^{\mathrm{par}}=S(u,v)\in\mathcal S\setminus\{S_r\}$.
The edges of $\delta_G(v)\setminus S$ are then $uv$ and
the back edges starting at $v$. By the definition of ${Q(d)}$,
\[
 \begin{aligned}
 \sum_{e\in\delta_G(v)\setminus S}
       \bigl(1-{Q(d)}_{\rep{e}}\bigr)
 &=\bigl(1-{Q(d)}_{\rep{uv}}\bigr)
   +\sum_{\substack{b\notin E(T):\\b\text{ starts at }v}}
       \bigl(1-{Q(d)}_{\rep{b}}\bigr)\\
 &=\frac23(1-d_{S^{\mathrm{par}}})+\frac23o(v)\\
 &=\frac23(1-d_{S^{\mathrm{par}}}+o(v)).
 \end{aligned}
\]
If $uv\in N\cup S_r$, then $\rep{uv}$ crosses the cut
and $1-{Q(d)}_{\rep{uv}}=2/3$. Thus the cut has capacity
at least $2/3$.

Finally, suppose $v$ has at least two DFS children, and
choose a child $w$ other than the child defining $S$.
If $vw\in N$, then $\rep{vw}$ crosses the cut and has
capacity $2/3$. If $S''=S(v,w)\in\mathcal S$, its connector
has been contracted, so every edge of $S''_H$ crosses the cut.
Consequently,
\[
 \begin{aligned}
 \sum_{e\in\delta_G(v)\setminus S}
       \bigl(1-{Q(d)}_{\rep{e}}\bigr)
 &\ge\sum_{a\in S''_H}\bigl(1-{Q(d)}_a\bigr)\\
 &=\frac23(1-d_{S''}+a_{S''})\\
 &\ge\frac23.
 \end{aligned}
\]
\begin{figure}[!t]
\centering
\begin{tikzpicture}[x=.9cm,y=.9cm]
\draw[s10shore,fill=gray!12] (-.65,-.55) rectangle (.65,.55);
\node[s10v,fill=gray!20] (v) at (0,0) {$v$};
\node[s10v] (u) at (0,1.9) {$u$};
\node[s10v] (a) at (-1.45,3.1) {};
\node[s10z] (s) at (0,-1.7) {$z_S$};
\draw[s10t] (u)--node[s10lab,right] {$\rep{uv}$}(v);
\draw[s10b] (v) .. controls (-2.2,.2) and (-2.2,2.7) .. (a);
\draw[gray,dashed] (v)--(s);
\node[s10note,text width=4cm] at (0,3.85) {(a) Exactly one DFS child};
\node[s10note,text width=3.9cm] at (0,-2.7)
 {$\displaystyle\frac23(1-d_{S^{\mathrm{par}}}+o(v))$};
\begin{scope}[shift={(5,0)}]
\draw[s10shore,fill=gray!12] (-.65,-.55) rectangle (.65,.55);
\node[s10v,fill=gray!20] (v) at (0,0) {$v$};
\node[s10z] (s) at (-1.2,-1.65) {$z_S$};
\node[s10v] (w) at (1.2,-1.4) {$w$};
\node[s10v] (x) at (1.2,-2.3) {$x$};
\draw[gray,dashed] (v)--(s);
\draw[s10t] (v)--(w)--(x);
\draw[s10b] (v) .. controls (2.65,-.2) and (2.65,-2.1) .. (x);
\node[s10note,text width=4cm] at (0,3.85) {(b) At least two DFS children};
\node[s10note,text width=4.5cm] at (.3,-3.15)
 {$\displaystyle\sum_{a\in S''_H}
      (1-{Q(d)}_a)\ge\frac23$};
\end{scope}
\begin{scope}[shift={(10.2,0)}]
\draw[s10shore,fill=gray!12] (-.7,-.55) rectangle (.7,.55);
\node[s10z,fill=gray!20] (s) at (0,0) {$z_S$};
\node[s10v] (v) at (0,1.9) {$v$};
\node[s10v] (w) at (0,-1.35) {$w$};
\node[s10v] (x) at (1.1,-2.2) {$x$};
\draw[gray,dashed] (v)--(s);
\draw[s10t] (s)--(w)--(x);
\draw[s10b] (s) .. controls (2,-.25) and (2,-1.45) .. (x);
\node[s10note] at (0,3.85) {(c) The cut at $z_S$};
\node[s10note,text width=4cm] at (0,-3.15)
 {$\displaystyle\frac23(1-d_S+a_S)\ge\frac23$};
\end{scope}
\end{tikzpicture}
\caption{The singleton cuts after contracting the other
connectors. Gray dashed edges mark the deleted connector $c_S$.
The expressions below the panels give the cut capacity in
(a) and (c), and a lower bound in (b).
Curved edges may represent several back edges.}
\label{fig:s12pair:singleton-cuts}
\end{figure}
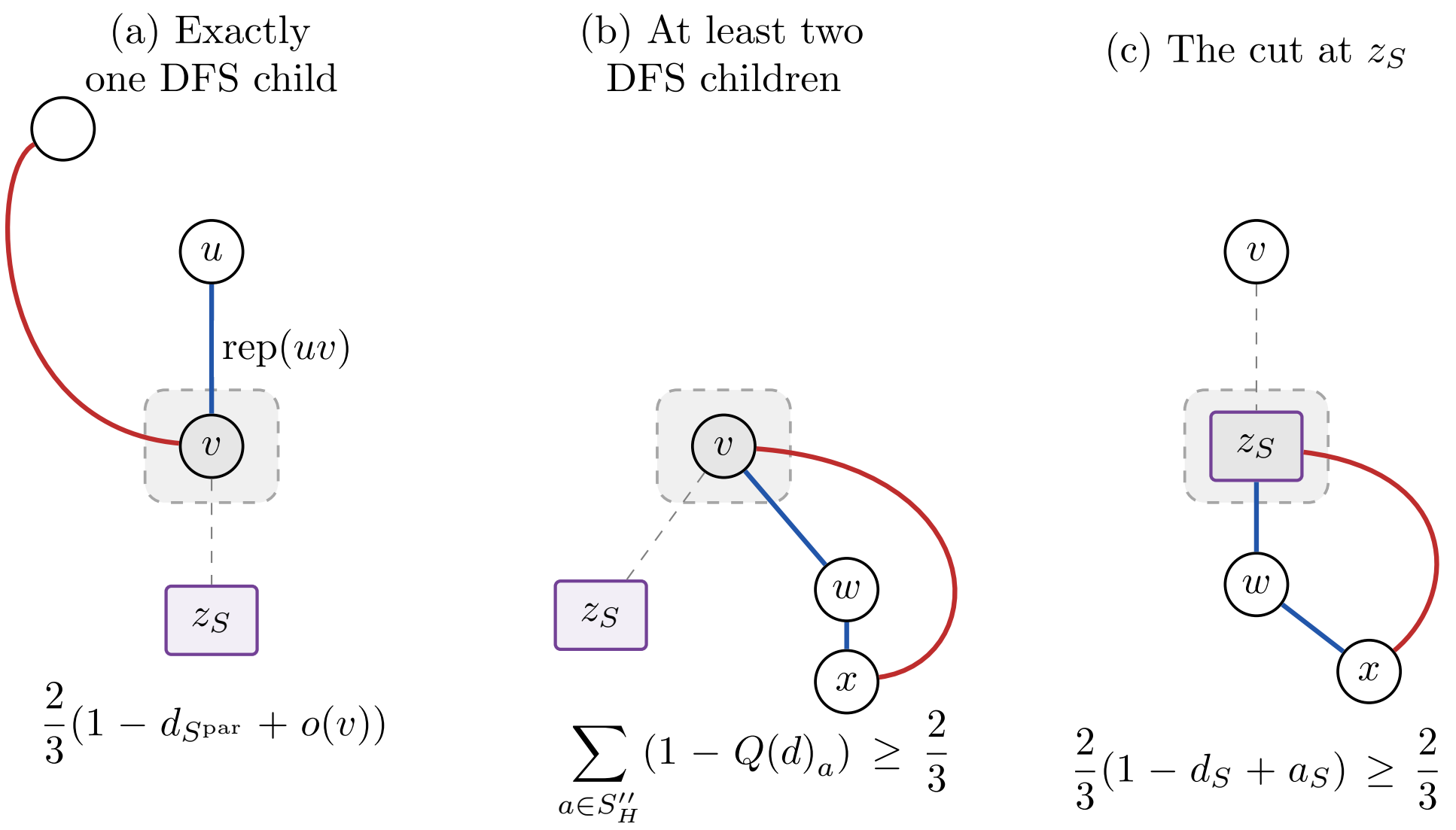

These cases cover every cut in the minimum defining
$\kappa_S(d)$. Taking the minimum of their capacity bounds
and using
\[
 {Q(d)}_{c_S}
 =1-\min\left\{\frac23,\kappa_S(d)\right\}
\]
gives \eqref{s12pair:connector-bound}.
\end{proof}

When $\eta_v=0$, we have
$\max\{2-q_S-a_S,q_S+a_S\}\ge1$, so $\tau_S=0$.
Whenever the bound in \eqref{s12pair:connector-bound}
exceeds $1/3$, the singleton cut at $v$ attains the minimum
defining $\kappa_S(d)$, and hence the connector attains
that bound. When the bound equals $1/3$, equality follows
from ${Q(d)}_{c_S}\ge1/3$.
{Thus \eqref{s12pair:connector-bound} is an equality
when $\eta_v=0$.}

\subsection{The point \texorpdfstring{${Q^{(2)}}$}{Q(2)}}
\label{s12pair:half-subtour}

\begin{lemma}[Lifting a join from $G$ to $H^\circ$]
\label{s10n:join-lifting}
{{Let $y$ belong to the
$\odd(G)$-join polytope.}
Then there exists $\widetilde y\in\mathcal P^\circ$ such that
\[
 \widetilde y_{\rep{e}}=y_e\qquad(e\in E(G)).
\]}
\end{lemma}
\begin{proof}
{By Theorem~\ref{thm:bounded-T-join}, write
\[
 y=\sum_i\lambda_i\mathbf 1_{J_i},\qquad
 \lambda_i\ge0,\qquad\sum_i\lambda_i=1,
\]
where each $J_i$ is an $\odd(G)$-join.
For each $J_i$, form an edge set $K_i\subseteq E(H^\circ)$
by selecting $\rep{e}$ exactly when $e\in J_i$
and choosing the connectors by
\begin{equation}\label{s10n:connector-parity}
 \mathbf 1_{\{c_S\in K_i\}}
 \equiv k_S+1-|J_i\cap S|\pmod2.
\end{equation}
The parity argument in the proof of
Lemma~\ref{lem:framework-1-lift} shows that $K_i$ is an
$\odd(H^\circ)$-join. Set
\[
 \widetilde y:=\sum_i\lambda_i\mathbf 1_{K_i}.
\]
Convexity gives $\widetilde y\in\mathcal P^\circ$, and
\[
 \widetilde y_{\rep{e}}
 =\sum_i\lambda_i\mathbf 1_{\{e\in J_i\}}=y_e.
\]
Each integral join has a unique lift, whereas the fractional
lift may depend on the chosen convex decomposition of $y$.}
\end{proof}

{Define $Q^{(2),G}_e:=x_e/2$ for $e\in E(T)$
and $Q^{(2),G}_e:=1-x_e/2$ otherwise. Since $0\le x_e\le1$, the definition of $Q^{(2),G}$ gives
\[
 \min\{Q^{(2),G}_e,\,1-Q^{(2),G}_e\}
 =\frac{x_e}{2}\ge0
 \qquad(e\in E(G)).
\]Consequently, for every
nontrivial cut $C$ of $G$ and every $F\subseteq C$,
\[
 \sum_{e\in C\setminus F}Q^{(2),G}_e
 +\sum_{e\in F}(1-Q^{(2),G}_e)
 \ge\frac{x(C)}2\ge1.
\]
{Thus $Q^{(2),G}$ belongs to the
$\odd(G)$-join polytope.}
Lemma~\ref{s10n:join-lifting} allows us to lift $Q^{(2),G}$
to a point of $\mathcal P^\circ$ whose value on each edge
$\rep{e}$ is $Q^{(2),G}_e$ for $e\in E(G)$. This lift need not be unique;
fix any such lift and denote it by $Q^{(2)}$.
We next use local Blossom cut inequalities to bound its
connector values. These bounds hold for every such lift.
The cut at the DFS root $r$ in $H^\circ$ consists only of
$c_{S_r}$. Its Blossom inequality gives
$Q^{(2)}_{c_{S_r}}\ge1$, which implies $Q^{(2)}_{c_{S_r}}=1$.}

{For a nonroot star $S=S(v,w)$, take
$C:=\delta_{H^\circ}(\{z_S\})$ and
$F:=\{c_S\}\cup\{\rep{b}:b\in I(v,w)\}$.
Since $C\setminus F=\{\rep{vw}\}$, the Blossom inequality gives
\[
 1\le\frac{q_S}{2}+(1-Q^{(2)}_{c_S})+\frac{a_S}{2}.
\]
Hence
\begin{equation}\label{s10:Q-connector-basic}
 Q^{(2)}_{c_S}\le\frac{q_S+a_S}{2}.
\end{equation}
When $v$ has even degree in $T$, take}
\[
 U:=\{v\}\cup
   \{z_{S'}:S'\ne S\text{ is centered at }v\},
 \qquad
 C:=\delta_{H^\circ}(U).
\]
The cut contains $c_S$ and the edges $\rep{e}$ with
$e\in\delta_G(v)\setminus S$. Choose
\[
 F:=\{c_S\}\cup
   \{\rep{b}\in C:b\in E(G)\setminus E(T)\}.
\]
The edges in $C\setminus F$ correspond to the edges of $T$
incident to $v$, except for the one belonging to $S$.
Consequently, $|C\setminus F|=\deg_T(v)-1$ is odd, so
we can apply the Blossom inequality in
\eqref{eq:repair-nonloop-polytope} for this choice of $C$ and $F$.

Each edge $\rep{e}\in C\setminus\{c_S\}$ contributes
$x_e/2$ to this inequality: through ${Q^{(2)}}_{\rep{e}}$ when
$e\in E(T)$, and through $1-{Q^{(2)}}_{\rep{e}}$ otherwise.
Using $x(\delta_G(v))=2+\eta_v$, we obtain
\[
 \begin{aligned}
 1
 &\le {Q^{(2)}}(C\setminus F)+|F|-{Q^{(2)}}(F)\\
 &=1-{Q^{(2)}}_{c_S}
   +\frac12\sum_{e\in\delta_G(v)\setminus S}x_e\\
 &=1-{Q^{(2)}}_{c_S}
   +\frac{2+\eta_v-q_S-a_S}{2}.
 \end{aligned}
\]
Therefore,
\[
 {Q^{(2)}}_{c_S}\le\frac{2+\eta_v-q_S-a_S}{2}.
\]

The two estimates imply that every nonroot star $S$
centered at $v$ satisfies
\begin{equation}\label{s12pair:Q-connector-bounds}
 {Q^{(2)}}_{c_S}\le
 \begin{cases}
  \min\left\{\dfrac{q_S+a_S}{2},
             \dfrac{2+\eta_v-q_S-a_S}{2}\right\},
      &\text{if $v$ has even degree in $T$},\\[10pt]
  \dfrac{q_S+a_S}{2},
      &\text{if $v$ has odd degree in $T$}.
 \end{cases}
\end{equation}
These bounds follow from the Blossom inequalities for ${Q^{(2)}}$,
so they hold simultaneously for all stars.

\subsection{A convex combination and the contribution of each star}

\begin{lemma}\label{s11p:scalar}
{For every nonroot star $S$,
\[
 \bar e_S\le\frac{t_S-1}{t_S+1},
 \qquad
 \frac{\bar e_S}{2}-\frac14\le\frac{t_S+1}{64}.
\]}
\end{lemma}
\begin{proof}
{If $\bar e_S=0$, both inequalities follow from
$t_S\ge1$. Otherwise $e_S>0$, and greedy DFS together with
$\bar e_S\le e_S$ and $\bar e_S\le1-q_S$ gives
\[
 1+\bar e_S\le1+e_S=a_S
 \le t_Sq_S\le t_S(1-\bar e_S).
\]
Rearranging proves the first inequality. For the second inequality, use the first to obtain
\[
 \begin{aligned}
 \frac{\bar e_S}{2}-\frac14
 &\le\frac{t_S-1}{2(t_S+1)}-\frac14\\
 &=\frac{t_S-3}{4(t_S+1)}.
 \end{aligned}
\]
Moreover, since $t_S\ge1$,
\[
 \begin{aligned}
 \frac{t_S+1}{64}-\frac{t_S-3}{4(t_S+1)}
 &=\frac{(t_S+1)^2-16(t_S-3)}{64(t_S+1)}\\
 &=\frac{(t_S-7)^2}{64(t_S+1)}
 \ge0.
 \end{aligned}
\]
Combining these inequalities proves
$\frac {\bar e_S}{2}-\frac14\le\frac{(t_S+1)}{64}$.
}
\end{proof}

Let $\bar d\in\R_+^{\mathcal S}$ be defined by
\[
 \bar d_S:=
 \begin{cases}
 0,&S=S_r,\\[4pt]
 \displaystyle\left(\bar e_S-\frac{q_S}{2}
                              -\frac{t_S}{32}\right)_+,
                                       &S\ne S_r.
 \end{cases}
\]
{
Set $Q^{(1)}:=Q(\bar d)$. Its coordinates are
}
\[
 \begin{aligned}
 {Q^{(1)}}_{\rep{e}}
 &=
 \begin{cases}
 \displaystyle\frac13+\frac23
       \left(\bar e_S-\frac{q_S}{2}-\frac{t_S}{32}\right)_+,
          &e=uv,\quad S=S(u,v)\ne S_r,\\[6pt]
 \displaystyle1-\frac23x_e,&e\notin E(T),\\[4pt]
 \displaystyle\frac13,&\text{otherwise},
 \end{cases}\\[7pt]
 {Q^{(1)}}_{c_S}
 &=
 \begin{cases}
 1,&S=S_r,\\[3pt]
 \max\{1/3,\,1-\kappa_S(\bar d)\},&S\ne S_r.
 \end{cases}
 \end{aligned}
\]
Since $0\le \bar d_S\le\bar e_S$ for every nonroot star,
Lemma~\ref{s12pair:connector-completion}
gives ${Q^{(1)}}\in{\mathcal P^\circ}$. Set
\[
 Y:=\frac34{Q^{(1)}}+\frac14{Q^{(2)}}.
\]
By convexity, $Y\in{\mathcal P^\circ}$.

Recall that
\[
 \Phi_{G,\mathcal S}(Y)
 =n-1+Y(N_H)
   +\sum_{S\in\mathcal S}
       \max\{Y_{c_S},\,k_S-1-Y(S_H)\}.
\]
For each nonroot star $S$, define
\[
 \chi_S:=\max\{(q_S+a_S-2)_+,\tau_S\}.
\]
The next lemma bounds each nonroot star's contribution
$\max\{Y_{c_S},\,k_S-1-Y(S_H)\}$ to the objective.

\begin{lemma}\label{s12pair:local-charge}
Every nonroot star $S$ satisfies
\[
 \max\{Y_{c_S},\,k_S-1-Y(S_H)\}
 \le\frac14+\frac{q_S+a_S}{8}+\frac{t_S}{64}
       +\frac{\chi_S}{2}.
\]
Moreover, at every nonroot vertex $v$,
\begin{equation}\label{s12pair:vertex-charge}
 \sum_{S\text{ centered at }v}\chi_S\le\eta_v.
\end{equation}
\end{lemma}

\begin{proof}
We bound both terms in the maximum.

\smallskip\noindent
\textbf{The term $k_S-1-Y(S_H)$.}
Recall that $S$ contains one edge of $T$ and $t_S$ back edges.
The edge of $T$ has $x$-value $q_S$, and the back edges
have total $x$-value $a_S$.
By the definitions of ${Q^{(1)}}$ and ${Q^{(2)}}$, 
\[
 \begin{aligned}
 {Q^{(1)}}(S_H)
 &=\frac13+\frac23\bar d_S+t_S-\frac23a_S,\\[4pt]
 {Q^{(2)}}(S_H)
 &=\frac{q_S}{2}+t_S-\frac{a_S}{2}.
 \end{aligned}
\]
Since $Y=\frac34{Q^{(1)}}+\frac14{Q^{(2)}}$, we obtain
\[
 \begin{aligned}
 Y(S_H)
 &=\frac34\left(
      \frac13+\frac23\bar d_S+t_S-\frac23a_S
    \right)
   +\frac14\left(
      \frac{q_S}{2}+t_S-\frac{a_S}{2}
    \right)\\[4pt]
 &=t_S+\frac14+\frac{\bar d_S}{2}
       +\frac{q_S}{8}-\frac{5a_S}{8}.
 \end{aligned}
\]
{Using $k_S-1=t_S$ gives}
\[
 \begin{aligned}
 k_S-1-Y(S_H)
 &=-\frac14-\frac{\bar d_S}{2}
       -\frac{q_S}{8}+\frac{5a_S}{8}\\[4pt]
 &=\frac14+\frac{q_S+a_S}{8}
   +\frac{a_S-1-\bar d_S}{2}-\frac{q_S}{4}.
 \end{aligned}
\]
We must therefore bound the last two terms by
$t_S/64+\chi_S/2$. The choice of $\bar d_S$ gives
\[
 \bar e_S-\bar d_S
 =\min\left\{\bar e_S,\frac{q_S}{2}+\frac{t_S}{32}\right\}
 \le\frac{q_S}{2}+\frac{t_S}{32}.
\]
Since $e_S=(a_S-1)_+$, $q_S\le1$, and
$\bar e_S=\min\{e_S,1-q_S\}$, we have
\[
 e_S-\bar e_S=(q_S+a_S-2)_+.
\]
Using $a_S-1\le e_S$, we obtain
\[
 \begin{aligned}
 \frac{a_S-1-\bar d_S}{2}-\frac{q_S}{4}
 &\le\frac{\bar e_S-\bar d_S}{2}-\frac{q_S}{4}+\frac{(q_S+a_S-2)_+}{2}\\
 &\le\frac{t_S}{64}+\frac{(q_S+a_S-2)_+}{2}\\
 &\le\frac{t_S}{64}+\frac{\chi_S}{2}.
 \end{aligned}
\]
This proves the required bound for $k_S-1-Y(S_H)$, including
when $e_S=0$.

We now bound $Y_{c_S}$. By
Lemma~\ref{s12pair:connector-estimate}, there are two cases
for the connector bound in ${Q^{(1)}}$. We consider each case
separately and combine this bound with the corresponding
estimate for ${Q^{(2)}}_{c_S}$.

\smallskip\noindent
\smallskip\noindent
\textbf{Case 1: ${Q^{(1)}}_{c_S}\le\frac13+\frac23\tau_S$.}
This case includes every star whose center $v$ has at least
two DFS children. If $v$ has exactly one DFS child, it also
applies when its parent edge belongs to $N$ or to $S_r$,
or belongs to a nonroot star $S^{\mathrm{par}}=S(u,v)$ satisfying
\[
 \bar d_{S^{\mathrm{par}}}\le o(v)+\tau_S.
\]
The connector estimate and the basic bound for ${Q^{(2)}}$ give
\[
 \begin{aligned}
 Y_{c_S}
 &\le\frac34\left(\frac13+\frac23\tau_S\right)
       +\frac14\frac{q_S+a_S}{2}\\
 &=\frac14+\frac{q_S+a_S}{8}+\frac{\tau_S}{2}\\
 &\le\frac14+\frac{q_S+a_S}{8}+\frac{\chi_S}{2}.
 \end{aligned}
\]

\smallskip\noindent
\textbf{Case 2: ${Q^{(1)}}_{c_S}>\frac13+\frac23\tau_S$.}
Write $S=S(v,w)$, where $w$ is the only DFS child of its
center $v$. Its parent edge $uv$ belongs to a nonroot star
\[
 S^{\mathrm{par}}:=S(u,v),\qquad u:=\operatorname{par}(v),
\]
and $\bar d_{S^{\mathrm{par}}}-o(v)>\tau_S\ge0$.
We bound $Y_{c_S}$ using the increase chosen at $S^{\mathrm{par}}$.
Lemma~\ref{s12pair:connector-estimate} gives
\[
 {Q^{(1)}}_{c_S}\le\frac13+\frac23(\bar d_{S^{\mathrm{par}}}-o(v)).
\]
The center $v$ has one parent and one child, so it has degree two in $T$.
Using the stronger bound for ${Q^{(2)}}$, we obtain
\begin{equation}\label{s12pair:local-max}
 \begin{aligned}
 Y_{c_S}
 &\le\frac34\left(\frac13+\frac23(\bar d_{S^{\mathrm{par}}}-o(v))\right)
       +\frac14\left(\frac{2+\eta_v-q_S-a_S}{2}\right)\\
 &=\frac14+\frac{q_S+a_S}{8}
       +\frac{\bar d_{S^{\mathrm{par}}}-o(v)}{2}
       -\frac{q_S+a_S-1-\eta_v/2}{4}.
 \end{aligned}
\end{equation}
It remains to bound the last two terms by $t_S/64$.

{Since $q_{S^{\mathrm{par}}}=x_{uv}$, the degree equation at $v$ is}
\[
 q_{S^{\mathrm{par}}}+q_S+a_S+o(v)=2+\eta_v.
\]
Thus
\[
 q_S+a_S-1-\frac{\eta_v}{2}
 =1+\frac{\eta_v}{2}-q_{S^{\mathrm{par}}}-o(v).
\]
Also, $\bar d_{S^{\mathrm{par}}}>o(v)\ge0$, so $S^{\mathrm{par}}$ is positive and
\[
 \bar d_{S^{\mathrm{par}}}=\bar e_{S^{\mathrm{par}}}-\frac{q_{S^{\mathrm{par}}}}{2}-\frac{t_{S^{\mathrm{par}}}}{32}.
\]
Substituting these identities gives
\[
 \begin{aligned}
 &\frac{\bar d_{S^{\mathrm{par}}}-o(v)}{2}
       -\frac{q_S+a_S-1-\eta_v/2}{4}\\
 &\qquad=\frac{\bar e_{S^{\mathrm{par}}}}{2}-\frac14
       -\frac{t_{S^{\mathrm{par}}}}{64}-\frac{o(v)}4-\frac{\eta_v}{8}\\
 &\qquad\le\frac{\bar e_{S^{\mathrm{par}}}}{2}-\frac14-\frac{t_{S^{\mathrm{par}}}}{64}.
 \end{aligned}
\]
{Lemma~\ref{s11p:scalar} now gives
\[
 \frac{\bar e_{S^{\mathrm{par}}}}2-\frac14
   -\frac{t_{S^{\mathrm{par}}}}{64}
 \le\frac1{64}\le\frac{t_S}{64}.
\]
}
The last inequality uses $t_S\ge1$; the successor $S$ need
not be positive. This proves the connector bound in Case~2.
Together, the two cases and the bound for the representative
edges prove the first part of the lemma.

\smallskip\noindent
\textbf{The sum of the additional terms at a center $v$.}
We now prove \eqref{s12pair:vertex-charge}. Fix a nonroot vertex $v$. We want to prove
\[
 \sum_{S\text{ centered at }v}\chi_S\le\eta_v.
\]
Recall that
\[
 \begin{aligned}
 x(\delta_G(v))&=2+\eta_v,\\
 \tau_S
 &=\bigl(1-\max\{0,2-q_S-a_S,q_S+a_S-\eta_v\}\bigr)_+,\\
 \chi_S&=\max\{(q_S+a_S-2)_+,\,\tau_S\}.
 \end{aligned}
\]
For every star $S$ centered at $v$, we have
\[
 q_S+a_S=x(S)\le x(\delta_G(v))=2+\eta_v,
\]
and hence $(q_S+a_S-2)_+\le\eta_v$.

To bound $\tau_S$, suppose first that $\tau_S>0$.
We obtain the following inequalities from the definition of $\tau_S$:
\[
 \begin{aligned}
 \tau_S&\le1-(2-q_S-a_S)=q_S+a_S-1,\\
 \tau_S&\le1-(q_S+a_S-\eta_v)=1+\eta_v-q_S-a_S.
 \end{aligned}
\]
Adding gives $2\tau_S\le\eta_v$. This inequality also holds
when $\tau_S=0$. Thus both terms defining $\chi_S$ are at
most $\eta_v$, and
\[
 \chi_S\le\eta_v.
\]

\smallskip\noindent
\textbf{Case 1: At most one star centered at $v$ has $\chi_S>0$.}
All other terms in the sum are zero, so the individual bound
above gives
\[
 \sum_{S\text{ centered at }v}\chi_S\le\eta_v.
\]

\smallskip\noindent
\smallskip\noindent
\textbf{Case 2: At least two stars centered at $v$ have $\chi_S>0$.}
We show that each positive term satisfies
\[
 \chi_S\le q_S+a_S-1.
\]

Consider a star $S$ with $\chi_S>0$. We must have
$q_S+a_S>1$, since otherwise both terms in
\[
 \chi_S=\max\{(q_S+a_S-2)_+,\,\tau_S\}
\]
would be zero. Therefore,
\[
 (q_S+a_S-2)_+\le q_S+a_S-1.
\]
The other term satisfies the same bound: when $\tau_S>0$,
we proved above that $\tau_S\le q_S+a_S-1$; when $\tau_S=0$,
the inequality follows from $q_S+a_S>1$. Thus
$\chi_S\le q_S+a_S-1$.

We now sum over the stars with $\chi_S>0$. Their edge sets
are disjoint and contained in $\delta_G(v)$, so
\[
 \sum_{\substack{S\text{ centered at }v\\\chi_S>0}}
       (q_S+a_S)
 \le x(\delta_G(v))=2+\eta_v.
\]
There are at least two such stars. Consequently,
\[
 \begin{aligned}
 \sum_{S\text{ centered at }v}\chi_S
 &\le
 \sum_{\substack{S\text{ centered at }v\\\chi_S>0}}
       (q_S+a_S-1)\\
 &=
 \sum_{\substack{S\text{ centered at }v\\\chi_S>0}}
       (q_S+a_S)
 -
 \sum_{\substack{S\text{ centered at }v\\\chi_S>0}}1\\
 &\le(2+\eta_v)-2\\
 &=\eta_v.
 \end{aligned}
\]
This proves \eqref{s12pair:vertex-charge}.
\end{proof}

\subsection{The resulting tour bound}
\label{subsec:general-aggregation}

{Put $L:=\text{\SubtourLP}(G)$.}
We now sum the bounds from Lemma~\ref{s12pair:local-charge}.
The additional terms from the nonroot stars satisfy
\[
 \begin{aligned}
 \sum_{S\ne S_r}\left(\frac{t_S}{64}+\frac{\chi_S}{2}\right)
 &\le\frac{n-4}{64}+\frac12\sum_{v\ne r}\eta_v\\
 &=\frac{n-4}{64}+\Delta-\frac{\eta_r}{2},
 \end{aligned}
\]
where we used \eqref{s12pair:sparsity} and
$\sum_v\eta_v=2\Delta$.

For every $e\in N$, we have $Y_{\rep{e}}=1/4+x_e/8$.
The edges in $N$ and the nonroot stars partition
$E\setminus S_r$. Since the root has one DFS child,
$q_{S_r}+a_{S_r}=2+\eta_r$, and thus
\[
 |N|+|\mathcal S|-1=n-2,
 \qquad
 x(E\setminus S_r)=L-2-\eta_r.
\]
Consequently,
\[
 \begin{aligned}
 Y(N_H)+\sum_{S\ne S_r}\left(\frac14+\frac{q_S+a_S}{8}\right)
 &=\frac{n-2}{4}+\frac{L-2-\eta_r}{8}\\
 &=\frac38(n-2)+\frac{\Delta}{8}-\frac{\eta_r}{8}.
 \end{aligned}
\]

It remains to include the root star. Its connector has value one
in both ${Q^{(1)}}$ and ${Q^{(2)}}$, and the coordinate of its tree edge is unchanged
in ${Q^{(1)}}$. Substituting $a_{S_r}=2+\eta_r-q_{S_r}$ gives
\[
 \begin{aligned}
 k_{S_r}-1-Y((S_r)_H)
 &=-\frac14-\frac{q_{S_r}}8+\frac{5a_{S_r}}8\\
 &=1+\frac58\eta_r-\frac34q_{S_r}.
 \end{aligned}
\]
Its contribution is therefore
\[
 \begin{aligned}
 \max\{Y_{c_{S_r}},\,k_{S_r}-1-Y((S_r)_H)\}
 &=1+\left(\frac58\eta_r-\frac34q_{S_r}\right)_+\\
 &\le1+\frac58\eta_r.
 \end{aligned}
\]

Using the formula for $\Phi_{G,\mathcal S}(Y)$,
we substitute the bound for the root star and the bounds for
the nonroot stars, and then apply the estimates for their sums:
\[
\begin{aligned}
\Phi_{G,\mathcal S}(Y)
&=n-1+Y(N_H)
  +\sum_{S\in\mathcal S}
     \max\{Y_{c_S},\,k_S-1-Y(S_H)\}\\[5pt]
&\le n+\frac58\eta_r+Y(N_H)
  +\sum_{S\ne S_r}
     \max\{Y_{c_S},\,k_S-1-Y(S_H)\}\\[5pt]
&\le n+\frac58\eta_r+Y(N_H)
  +\sum_{S\ne S_r}
     \left(\frac14+\frac{q_S+a_S}{8}
           +\frac{t_S}{64}+\frac{\chi_S}{2}\right)\\[5pt]
&=\frac{11}{8}n-\frac34+\frac{\Delta}{8}
  +\frac{\eta_r}{2}
  +\frac1{64}\sum_{S\ne S_r}t_S
  +\frac12\sum_{S\ne S_r}\chi_S\\[5pt]
&\le\frac{11}{8}n-\frac34+\frac{\Delta}{8}
  +\frac{\eta_r}{2}+\frac{n-4}{64}
  +\frac12\sum_{S\ne S_r}\chi_S\\[5pt]
&\le\frac{11}{8}n-\frac34+\frac{\Delta}{8}
  +\frac{\eta_r}{2}+\frac{n-4}{64}
  +\frac12\sum_{v\ne r}\eta_v\\[5pt]
&=\frac{11}{8}n-\frac34+\frac{\Delta}{8}
  +\frac{\eta_r}{2}+\frac{n-4}{64}
  +\frac{2\Delta-\eta_r}{2}\\[5pt]
&=\frac{11}{8}n-\frac34+\frac98\Delta+\frac{n-4}{64}\\[5pt]
&=\frac{89}{64}n+\frac98\Delta-\frac{13}{16}.
\end{aligned}
\]
Since $Y\in{\mathcal P^\circ}$
and $L=n+\Delta$, we obtain
\[
\begin{aligned}
\operatorname{tsp}(G)
&\le\text{\LPtwo}_{\mathrm{OPT}}(G,\mathcal S)\\
&\le\Phi_{G,\mathcal S}(Y)\\
&\le\frac{89}{64}n+\frac98\Delta-\frac{13}{16}\\
&=\frac{89}{64}L-\frac{17}{64}\Delta-\frac{13}{16}\\
&\le\frac{89}{64}L.
\end{aligned}
\]

\subsection{\texorpdfstring{{Specialization to $L=n$}}{Specialization to L=n}}
\label{subsec:value-n-specialization}

\begin{proposition}[{Specialization to $L=n$}]
\label{prop:value-n-specialization}
{Assume $L=n$. Then $\eta_v=0$ for every vertex
and $\bar e_S=e_S$ for every nonroot star. For every vector
$d$ satisfying $d_{S_r}=0$ and $0\le d_S\le e_S$,
the point $Q(d)$ constructed in this appendix agrees exactly
with \eqref{proc:increase-choice} and
\eqref{proc:following-connector}, and belongs to
$\mathcal P^\circ$. Its objective value is
\begin{equation}\label{eq:value-n-objective}
 \Phi_{G,\mathcal S}(Q(d))
 =\frac43n-\frac23+
 \sum_{S\ne S_r}\max\left\{
 Q(d)_{c_S}-\frac13,\,\frac23(e_S-d_S)\right\}.
\end{equation}
The point $Q^{(2)}$ satisfies the connector bounds in
\eqref{proc:connector-bounds}. The general choice of $\bar d$
reduces to
\[
 \bar d_{S_r}=0,\qquad
 \bar d_S=\left(e_S-\frac{q_S}{2}-\frac{t_S}{32}\right)_+
 \quad(S\ne S_r).
\]
With $Q^{(1)}=Q(\bar d)$ and
$Y=\frac34Q^{(1)}+\frac14Q^{(2)}$, we have
$Y\in\mathcal P^\circ$ and, for every nonroot star $S$,
\begin{equation}\label{eq:value-n-local}
 \max\{Y_{c_S},\,k_S-1-Y(S_H)\}
 \le\frac14+\frac{q_S+a_S}{8}+\frac{t_S}{64}.
\end{equation}
Consequently,
\begin{equation}\label{eq:value-n-bound}
 \Phi_{G,\mathcal S}(Y)\le\frac{89n-52}{64},
\end{equation}
and \ALG{} returns a tour satisfying this bound.}
\end{proposition}

\begin{proof}
{Since $L=n$, we have $\eta_v=0$ for every vertex.
Thus every singleton cut is tight.
For a nonroot star $S=S(u,v)$, we obtain
$q_S+a_S\le x(\delta_G(u))=2$.
Together with $q_S\le1$, this gives $e_S\le1-q_S$,
and hence $\bar e_S=e_S$.
We can therefore replace $\bar e_S$ by $e_S$ in
Lemma~\ref{s11p:scalar} and in the definition of $\bar d$.
The admissibility conditions are exactly those in
Section~\ref{proc:increases}, and
Lemma~\ref{s12pair:connector-completion} shows that
$Q(d)\in\mathcal P^\circ$.}

{We next check that the two constructions
give the same point $Q(d)$.
For every nonroot star $S'$ centered at $v$, we have
$\eta_v=0$ and hence $\tau_{S'}=0$.
The equality case proved after
Lemma~\ref{s12pair:connector-estimate} then gives precisely
the nonroot connector values in
\eqref{proc:following-connector}.
The root connector has value one in both constructions,
and the values on the nonconnector edges agree with
\eqref{proc:increase-choice}.}

{For every nonroot star $S$, the definition
of $Q(d)$ gives
\[
 Q(d)(S_H)=\frac13+\frac23d_S+t_S-\frac23a_S.
\]
When $a_S>1$, we have $e_S=a_S-1$.
When $a_S\le1$, admissibility gives $d_S=0$, and
$Q(d)_{c_S}\ge1/3$. Thus replacing $a_S-1$ by $e_S$
in the maximum defining the contribution of $S$ does
not change its value.
The root star contributes one, every edge in $N_H$ has
value $1/3$, and $|N|+|\mathcal S|=n-1$.
Substituting into \eqref{eq:general-join-cost} gives
\eqref{eq:value-n-objective}.
Setting $d=0$ also gives \eqref{proc:initial-cost}.}

{Setting $\eta_v=0$ in
\eqref{s12pair:Q-connector-bounds} gives the connector
bounds in \eqref{proc:connector-bounds}.
Also, $q_S+a_S\le2$ and $\tau_S=0$ imply $\chi_S=0$
for every nonroot star.
Lemma~\ref{s12pair:local-charge} therefore gives
\eqref{eq:value-n-local}.
Finally, applying the calculation in
Appendix~\ref{subsec:general-aggregation} with
$\Delta=\eta_r=0$ gives \eqref{eq:value-n-bound}.
Corollary~\ref{cor:repair-nonloop-bound} then gives a tour
satisfying this bound.}
\end{proof}

\section{Further improvements}
\label{sec:further-improvements}

We describe two refinements of the \LPtwo{} analysis that can
improve the upper bound on the integrality ratio of the subtour LP
for graphic TSP: a third point $Z$ obtained from a
circulation and a sharper bound on the number of back edges.
We give the constructions and the resulting estimates, but do not
present their proofs or their combined use in the approximation
analysis in full. {Throughout this section, we use
the setup of Appendix~\ref{sec:general-L} with $L=n$, as
specialized in Proposition~\ref{prop:value-n-specialization}.}
In this case, the input
extreme solution $x^*$ already satisfies $x^*_e\le1$, so we
choose $x=x^*$ in \ALG{}.

\subsection{The point \texorpdfstring{$Z$}{Z} and its circulation}
\label{sec:further-Z}

In the M\"omke--Svensson circulation of
Section~\ref{subsec:MS-full-stars}, the first chosen back edge
entering a nonroot star is free, and each additional one has
unit cost. The \LPtwo{} allows the second chosen back edge
to be free as well, provided that a chosen back edge starts
at the child of the star's center.

More precisely, let $S=S(u,v)$ be a nonroot star. If at least
two chosen back edges belong to $S$ and a chosen back edge
starts at $v$, we can increase $Z_{\rep{uv}}$ from $1/3$ to $1$
while preserving feasibility. This decreases
$k_S-1-Z(S_H)$ by $2/3$, which pays for the second chosen
back edge in $S$.

We therefore use the following costs. At each nonroot star,
the first incoming unit is free. A second unit is free if it
continues along a back edge starting at $v$; all further
incoming units have cost one. At the root star, the first
two incoming units are free. As in the M\"omke--Svensson
construction, we require at least one unit of flow through the arcs
corresponding to every nonroot connector and every edge of $N_H$.

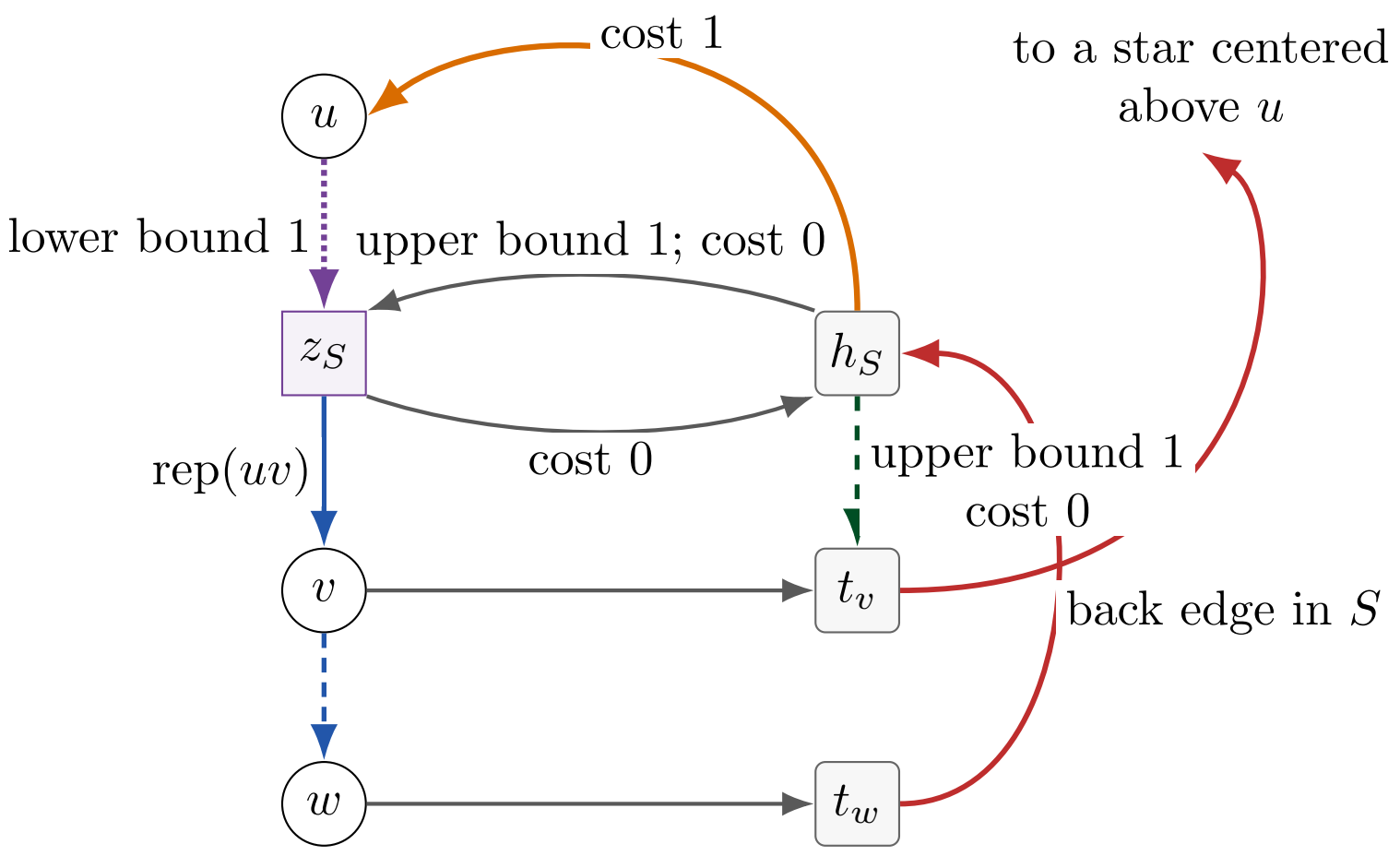
\begin{figure}[!t]
\centering
\begin{tikzpicture}[
 x=.85cm,y=.85cm,>=Latex,
 fz vertex/.style={
   circle,draw,fill=white,minimum size=6mm,
   inner sep=1pt,font=\small},
 fz auxiliary/.style={
   rectangle,draw=connector,fill=connector!7,
   minimum size=6mm,inner sep=2pt,font=\small},
 fz added/.style={
   rectangle,rounded corners=2pt,draw=black!60,
   fill=black!3,minimum size=6mm,
   inner sep=2pt,font=\small},
 fz tree/.style={
   ->,draw=edgeblue,line width=1pt},
 fz connector/.style={
   ->,draw=connector,densely dotted,line width=1.2pt},
 fz back/.style={
   ->,draw=edgered,line width=1.1pt},
 fz free/.style={
   ->,draw=black!65,line width=.8pt},
 fz second/.style={
   ->,draw=green!65!black,dashed,line width=1.1pt},
 fz paid/.style={
   ->,draw=orange!85!black,line width=1.2pt},
 fz label/.style={
   fill=white,inner sep=2pt,font=\small,align=center},
 fz note/.style={
   font=\small,align=center}]

\node[fz vertex] (u) at (0,4) {$u$};
\node[fz auxiliary] (z) at (0,2) {$z_S$};
\node[fz vertex] (v) at (0,0) {$v$};
\node[fz vertex] (w) at (0,-1.8) {$w$};

\node[fz added] (h) at (4.5,2) {$h_S$};
\node[fz added] (tv) at (4.5,0) {$t_v$};
\node[fz added] (tw) at (4.5,-1.8) {$t_w$};

\draw[fz connector]
  (u)--node[fz label,left] {lower bound $1$}(z);
\draw[fz tree]
  (z)--node[fz label,left] {$\rep{uv}$}(v);
\draw[fz tree,densely dashed] (v)--(w);

\draw[fz free] (v)--(tv);
\draw[fz free] (w)--(tw);

\draw[fz back] (tw.east)
  .. controls (6.6,-1.8) and (6.6,2)
  .. node[fz label,right,pos=.45]
     {back edge in $S$} (h.east);

\draw[fz back] (tv.east)
  .. controls (8.2,0) and (8.2,3.2)
  .. (7.4,3.7);
\node[fz note,anchor=south] at (7.4,3.8)
  {to a star centered\\above $u$};

\draw[fz free] (h.north west)
  .. controls (3,2.75) and (1.5,2.75)
  .. node[fz label,above]
     {upper bound $1$; cost $0$} (z.north east);

\draw[fz free] (z.south east)
  .. controls (1.5,1.25) and (3,1.25)
  .. node[fz label,below]
     {cost $0$} (h.south west);

\draw[fz paid] (h.north)
  .. controls (4.5,4.9) and (1.5,4.9)
  .. node[fz label,above,pos=.5]
     {cost $1$} (u.east);

\draw[fz second] (h)--node[fz label,right,pos=.55]
  {upper bound $1$\\cost $0$}(tv);

\end{tikzpicture}
\caption{The gadget at $S=S(u,v)$. The green arc permits the
second free unit. Unlabelled arcs have cost zero and no upper
bound; the dashed blue segment denotes a downward path.}
\label{fig:Z-circulation}
\end{figure}

Figure~\ref{fig:Z-circulation} gives a circulation formulation
of these costs. Back arcs belonging to $S$ enter $h_S$,
while back arcs starting at $v$ leave $t_v$.
The first incoming unit can return through $h_S\to z_S$
at cost zero. A second unit can use $h_S\to t_v$ at cost
zero if it continues along a back arc starting at $v$.
Each of these arcs has upper bound one, and the remaining
flow returns through $h_S\to u$ at unit cost.
If no chosen back edge covers $c_S$, the cycle
$u\to z_S\to h_S\to u$ satisfies its lower bound at cost one.
We denote the total circulation cost by $C^Z(f)$.

Choose an optimal integral circulation $f$, and let $B$ be
the set of back edges whose arcs carry positive flow.

The edges $\rep{e}$ with $e\in E(T)$ and the connectors
form a tree in $H^\circ$. For each $b\in B$, consider
the cycle formed by $\rep{b}$ and the path joining its
endpoints in this tree. Assign value $1/3$ to every edge
belonging to at least one of these cycles, and value one
to all remaining edges. Thus the root connector has value
one, and the flow lower bounds ensure that every edge
of $N_H$ has value $1/3$.

For each nonroot star $S=S(u,v)$ with
$|B\cap I(u,v)|\ge2$ and one unit of flow on $h_S\to t_v$,
increase $Z_{\rep{uv}}$ to one. Flow conservation at $t_v$
implies that a back edge in $B$ starts at $v$, so this
increase satisfies the condition described above.

The resulting point belongs to ${\mathcal P^\circ}$ and satisfies
\begin{equation}\label{eq:Z-total-cost}
 \Phi_{G,\mathcal S}(Z)
 \le \frac{4n-2}{3}+\frac23 C^Z(f).
\end{equation}

\FloatBarrier
\subsection{Laminarity and unpartitionable vertices}
\label{subsec:further-laminarity}

We call a vertex $v$ \emph{partitionable} if there is a set
$F\subseteq\delta_G(v)$ with $x(F)=1$, and \emph{unpartitionable}
otherwise. Since $x(\delta_G(v))=2$, a partitionable vertex has
its incident edges partitioned into two sets of weight one.
Let $b$ be the number of nonroot unpartitionable vertices.

The uncrossing argument of
\cite{CornuejolsFonluptNaddef} gives a laminar family
$\mathcal L$ of tight cut sets whose cut vectors span all
tight-cut vectors on the support of $x$. We choose these
sets to exclude $r$ and include every nonroot singleton
and $V\setminus\{r\}$.

In the inclusion tree of $\mathcal L$, the parent of a set
is the smallest set in the family that properly contains it.
The root is $V\setminus\{r\}$, the leaves are the $n-1$
nonroot singletons, and the children of each nonsingleton
set partition that set.

An unpartitionable singleton $\{v\}$ cannot have a parent
with exactly two children. Indeed, if the other child is $A$,
tightness of $\{v\}$, $A$, and $A\cup\{v\}$ gives
\[
 2x(E_G(\{v\},A))
 =x(\delta_G(v))+x(\delta_G(A))
  -x(\delta_G(A\cup\{v\}))=2.
\]
This would give a set of edges incident to $v$ with total
weight one, contradicting unpartitionability.

Thus each of the $b$ unpartitionable singletons has a parent
with at least three children. Counting the nodes of the
inclusion tree gives at most $2n-3-\lceil b/3\rceil$ sets.
Since $x$ is positive and extreme, its tight-cut vectors
have rank $m$, and hence $\mathcal L$ contains at least
$m$ sets. Therefore,
\begin{equation}\label{eq:further-laminar-count}
 m\le2n-3-\left\lceil\frac b3\right\rceil.
\end{equation}
Every back edge belongs to one star, and the root star contains
at least two back edges. Consequently,
\begin{equation}\label{eq:further-back-count}
 \sum_{S\ne S_r}t_S
 \le n-4-\left\lceil\frac b3\right\rceil.
\end{equation}
This improves the count used when summing the local star bounds.
Partitionable centers also give a stronger local estimate: for a
positive star $S$ centered at such a vertex, greedy DFS gives
\[
 e_S\le\left\lfloor\frac{t_S-1}{2}\right\rfloor q_S.
\]
Thus partitionable centers improve the local estimates, while
unpartitionable vertices give a saving in the total number of
back edges.

\end{document}